\documentclass[usegraphicx,useAMS,usenatbib]{mn2e}

\usepackage{amsmath}
\usepackage{aas_macros}

\usepackage{maobspack}
\usepackage{maobsabbrevs}

\usepackage[T1]{fontenc}
\usepackage{ae,aecompl}

\usepackage{newtxtext,newtxmath}

\usepackage{lineno}
\modulolinenumbers[5]

\newcommand{\modl}[1]{model \texttt{\mbox{#1}}}
\newcommand{\modls}[1]{models \texttt{#1}}
\newcommand{\Modl}[1]{Model \texttt{#1}}

\newcommand{\modelname}[1]{\texttt{#1}}

\newcommand{\Rpthree}{\modelname{35OC-Rp3}\xspace}
\newcommand{\Rpfour}{\modelname{35OC-Rp4}\xspace}

\newcommand{\mRptwo}{\modl{35OC-Rp2}\xspace}
\newcommand{\mRpthree}{\modl{35OC-Rp3}\xspace}
\newcommand{\mRpfour}{\modl{35OC-Rp4}\xspace}
\newcommand{\mRs}{\modl{\mbox{35OC-Rs}}\xspace}

\newcommand{\nusp}{neutrinosphere\xspace}
\newcommand{\nusps}{neutrinospheres\xspace}
\newcommand{\panel}[1]{{\mbox{\textit{(#1)}}}}
\newcommand{\banel}[1]{\textit{#1}}
\newcommand{\Erot}{\mathcal{T}}
\newcommand{\Egrav}{\mathcal{W}}
\newcommand{\Erotshear}{\mathcal{F}}
\newcommand{\Emag}{\mathcal{B}}
\newcommand{\Emagunit}{\mathcal{B}_{47}}
\newcommand{\ZAMS}{\textsc{zams}}
\newcommand{\pnss}{\textsc{pns}}
\newcommand{\bhs}{\textsc{bh}}
\newcommand{\Erotpns}{{\cal T}^{\mathrm{\pnss}}}
\newcommand{\Erotunit}{{\cal T}_{49}}
\newcommand{\Erotshearpns}{{\cal F}^{\mathrm{\pnss}}}
\newcommand{\Tpns}{{\cal T}^{\mathrm{\pnss}}_{52}}
\newcommand{\Ebindingunit}{|E_{52}^{\rm bd}|}
\newcommand{\Ebinding}{E^{\rm bd}}
\newcommand{\Mdisk}{M_\textsc{df}}
\newcommand{\tdf}{t_\textsc{df}}
\newcommand{\tpb}{t_\mathrm{pb}}
\newcommand{\MpreSN}{M_{\mathrm{pre}\textsc{sn}}}
\newcommand{\Mmax}{M_{\rm bry}^{\rm max}}
\newcommand{\epsilonB}{\epsilon_{\textsc{b}}}
\newcommand{\epsilonBF}{\epsilon_{\textsc{bf}}}
\newcommand{\betaG}{\beta_{\textsc{g}}}

\let\oldhref\href
\renewcommand{\href}[2]{\oldhref{#1}{\hbox{#2}}}

\newcommand*{\DEBUG}{}%

\ifdefined\DEBUG

\else
\renewcommand{\rplc}[2]{{#2}}

\renewcommand{\delt}[1]{}
\fi

\newcommand{\miRyC}{RYC2018-024938-I}

\defcitealias{Obergaulinger_Aloy__2019}{Paper I}
\defcitealias{Obergaulinger_Aloy_2020__mnras_PaperIII}{Paper III}
\defcitealias{Metzger_et_al__2011__mnras__Theprotomagnetarmodelforgamma-raybursts}{M11}

\begin{document}

\title[Formation of protomagnetars and collapsars]{Magnetorotational core collapse of possible GRB
  progenitors. II. Formation of protomagnetars and collapsars.}

\author[Aloy \&  Obergaulinger]{
  M.\'A.~Aloy$^1$\thanks{E-mail: miguel.a.aloy@uv.es},  M.~Obergaulinger$^{1,2}$\thanks{E-mail: martin.obergaulinger@uv.es}
  \\
  $^1$ Departament d{\'{}}Astronomia i Astrof{\'i}sica, Universitat de
  Val{\`e}ncia, \\ Edifici d{\'{}}Investigaci{\'o} Jeroni Munyoz, C/
  Dr.~Moliner, 50, E-46100 Burjassot (Val{\`e}ncia), Spain 
  \\
  $^2$ Institut f{\"u}r Kernphysik, Theoriezentrum, S2|11
  Schlo{\ss}gartenstr.~2, 64289 Darmstadt, Germany
}
\label{firstpage}
\pagerange{\pageref{firstpage}--\pageref{lastpage}}
\maketitle

\begin{abstract}
  We assess the variance of the post-collapse evolution remnants of
  compact, massive, low-metallicity stars, under small changes in the
  degrees of rotation and magnetic field of selected pre-supernova
  cores. These stellar models are commonly considered progenitors of
  long gamma-ray bursts. The fate of the proto-neutron star (PNS)
  formed after collapse, whose mass may continuously grow due to
  accretion, critically depends on the poloidal magnetic field
  strength at bounce. Should the poloidal magnetic field be
  sufficiently weak, the PNS collapses to a black hole (BH) within a
  few seconds. Models on this evolutionary track contain promising
  collapsar engines. Poloidal magnetic fields smooth over large radial
  scales (\eg dipolar fields) or slightly augmented with respect to
  the original pre-supernova core yield long-lasting PNSs.  In these
  models, BH formation is avoided or staved off for a long time,
  hence, they may produce proto-magnetars (PMs). Some of our PM
  candidates have been run for $\lesssim 10\,$s after core bounce, but
  they have not entered the Kelvin-Helmholtz phase yet. Among these
  models, some display episodic events of spin-down during which we
  find properties broadly compatible with the theoretical expectations
  for PMs ($M_\pnss \approx 1.85\,\Msol - 2.5\,\Msol$,
  $\bar{P}_\pnss \approx 1.5 - 4\,$ms, and
  $b^{\rm surf}_\pnss \lesssim 10^{15}\,$G) and their very collimated
  supernova ejecta has nearly reached the stellar surface with (still
  growing) explosion energies $\gtrsim \zehnh{2}{51}\,\erg$.
\end{abstract}



\section{Introduction}
\label{Sek:Intro}

The collapse of the core of a massive and sufficiently fast rotating
star is needed to form the central engine of a long gamma-ray burst
\citep[GRB; for a review, see,
\eg,][]{Kumar_Zhang_2015PhR...561....1}.  Likely, the first stage,
namely the formation of a \pns (PNS) and the generation of a shock
wave at its surface is common to all except for possibly the most
massive stars that directly produce black holes (BHs).  During the
subsequent period of up to several seconds, neutrinos streaming out of
the PNS transfer energy to the gas behind the stalled shock wave and,
together with hydrodynamic instabilities, rotation and magnetic
fields, favour shock revival.  If no explosion sets in, or if it
proceeds asymmetrically, the PNS may accrete matter until its mass
exceeds the upper limit for stability against self-gravity and it
collapses to a BH that will continue to accrete the inner layers of
the star.  The delay before this secondary collapse is negatively
correlated with the mass of the PNS at its birth and with the mass
accretion rate, both of which depend on the progenitor structure.
Otherwise, the PNS formed in the collapse will gradually cool and
transform into a young neutron star.

The long-term evolution can branch into additional directions besides
those of failed or successful core-collapse supernovae (CCSNe) with a
PNS or a BH at the centre.  Both a rapidly rotating and strongly
magnetized PNS, commonly termed proto-magnetar (PM; see, \eg
\citealt{Metzger_et_al__2011__mnras__Theprotomagnetarmodelforgamma-raybursts,Metzger_et_al__2015__mnras__Thediversityoftransientsfrommagnetarbirthincorecollapsesupernovae}),
and a system consisting of a BH and an accretion torus generated by a
core with sufficiently high angular momentum (a collapsar;
\citealt{Woosley_et_al__1993__apj__Theevolutionofmassivestarsincludingmassloss-Presupernovamodelsandexplosion})
can launch collimated, relativistic outflows that will produce a GRB
after they breakout from the stellar surface.

Stellar evolution modelling suggests that the conditions tend to be
most favourable for long GRBs in stars of fairly high masses
\citep[\eg][]{Woosley_Heger__2006__apj__TheProgenitorStarsofGamma-RayBursts,Yoon_2006A&A...460..199}. It
should be noted, though, that the results of current stellar evolution
calculations do not lend themselves to simple rules connecting, \eg
the zero-age main-sequence mass, $M_\ZAMS$, of the star and its
pre-collapse profiles of density, composition, and temperature
\citep[but see, \eg][]{Woosley_2020ApJ...896...56}.  Those profiles
and, in particular, the size of the iron core and the surrounding
shells at collapse play an important role in determining the fate of
the stellar remnant, as demonstrate systematic studies of core
collapse across a large range of stellar masses in spherically
symmetric and multi-dimensional simulations
\citep{OConnor_Ott__2011__apj__BlackHoleFormationinFailingCore-CollapseSupernovae,Janka__2012__ARNPS__ExplosionMechanismsofCore-CollapseSupernovae,Ugliano_et_al__2012__apj__Progenitor-explosionConnectionandRemnantBirthMassesforNeutrino-drivenSupernovaeofIron-coreProgenitors,Sukhbold_et_al__2016__apj__Core-collapseSupernovaefrom9to120SolarMassesBasedonNeutrino-poweredExplosions,Nakamura_et_al__2015__pasj__Systematicfeaturesofaxisymmetricneutrino-drivencore-collapsesupernovamodelsinmultipleprogenitors,Bruenn_et_al__2016__apj__TheDevelopmentofExplosionsinAxisymmetricAbInitioCore-collapseSupernovaSimulationsof12-25MStars,Ertl_2020ApJ...890...51,Woosley_2020ApJ...896...56}.
\cite{OConnor_Ott__2011__apj__BlackHoleFormationinFailingCore-CollapseSupernovae}
found a useful, yet simple (one dimensional) criteria for
distinguishing cores that are likely to explode from those where a
failed explosion leads to BH formation in terms of the compactness of
the core. However, the compactness criterion neither accounts for
rotation nor for the dynamical effects of magnetic fields in the
pre-collapsed stellar cores.

One of the most important constraints on the formation of the central
engine of long GRBs comes from the requirement of high angular
momentum in the progenitor star.  Including rotation in stellar
evolution modelling, which is mostly based on spherically symmetric
calculations, is non-trivial and depends on many approximations.
Nevertheless, several sets of models for rotating pre-collapse cores
exists, among which the ones by
\cite{Woosley_Heger__2006__apj__TheProgenitorStarsofGamma-RayBursts}
are most valuable for our purpose as they account for the effect of
magnetic fields redistributing angular momentum.  Other groups have
also computed the evolution of fast-rotating main sequence, low
metallicity, single stars systematically predicting rather
fast-rotating pre-collapse cores
\citep[\eg][]{Yoon_Langer_2005A&A...443..643,Woosley_Heger__2006__apj__TheProgenitorStarsofGamma-RayBursts,Meynet_Maeder_2007A&A...464L..11,Ekstrom_2012A&A...537A.146,Aguilera-Dena_et_al__2018__apj__RelatedProgenitorModelsforLong-durationGamma-RayBurstsandTypeIcSuperluminousSupernovae}.
Therefore, we focus here on low-metallicity stars.  Rapid rotation can
add centrifugal support to the PNS and stabilise it beyond the maximum
mass for non-rotating PNSs, besides leading to global asymmetries of
the PNS.  These effects are most pronounced when rotation is combined
with a strong magnetic field that can tap into the rotational energy
\citep[see,
\eg][]{Bisnovatyi-Kogan_Popov_Samokhin__1976__APSS__MHD_SN,Symbalisty__1984__ApJ_MHD_SN,Akiyama_etal__2003__ApJ__MRI_SN,Kotake_etal__2004__Apj__SN-magrot-neutrino-emission,Thompson_Quataert_Burrows__2004__ApJ__Vis_Rot_SN,
  Obergaulinger_Aloy_Mueller__2006__AA__MR_collapse,
  Obergaulinger_et_al__2006__AA__MR_collapse_TOV,
  Moiseenko_et_al__2006__mnras__A_MR_CC_model_with_jets,
  Dessart_et_al__2007__apj__MagneticallyDrivenExplosionsofRapidlyRotatingWhiteDwarfsFollowingAccretion-InducedCollapse,Burrows_etal__2007__ApJ__MHD-SN,Winteler_et_al__2012__apjl__MagnetorotationallyDrivenSupernovaeastheOriginofEarlyGalaxyr-processElements,Sawai_et_al__2013__apjl__GlobalSimulationsofMagnetorotationalInstabilityintheCollapsedCoreofaMassiveStar,Mosta_et_al__2014__apjl__MagnetorotationalCore-collapseSupernovaeinThreeDimensions,Moesta_et_al__2015__nat__Alarge-scaledynamoandmagnetoturbulenceinrapidlyrotatingcore-collapsesupernovae,Obergaulinger_Aloy__2017__mnras__Protomagnetarandblackholeformationinhigh-massstars,Obergaulinger_Aloy__2019,Bugli_2020MNRAS.492...58,Kuroda_et_al__2020__arXiveprints__MagnetorotationalExplosionofaMassiveStarSupportedbyNeutrinoHeatinginGeneralRelativisticThreeDimensionalSimulations}.
Large-scale fields are most effective at generating jet-like
explosions
\citep[\eg][]{Wheeler_et_al__2000__apj__Asymmetric_SNe_Pulsars_Magnetars_and_GRBs,Uzdensky_MacFadyen__2007__apj__Magnetar-Driven_Magnetic_Tower_as_a_Model_for_Gamma-Ray_Bursts_and_Asymmetric_Supernovae,
  Dessart_et_al__2008__apjl__TheProto-NeutronStarPhaseoftheCollapsarModelandtheRoutetoLong-SoftGamma-RayBurstsandHypernovae,Bugli_2020MNRAS.492...58,Obergaulinger_Aloy_2020__mnras_PaperIII},
whereas small-scale fields such as those amplified by turbulence
driven by, \eg the magnetorotational instability \citep[MRI,
\eg][]{Balbus_Hawley__1998__RMP__MRI}, can act as effective viscosity
and enhance the heating of the post-shock gas
\citep{Thompson_Quataert_Burrows__2004__ApJ__Vis_Rot_SN}.  It should,
however, be noted that the dynamic relevance of the magnetic field
depends crucially on the ratio of the magnetic energy to the kinetic
energy, which in most, though not all, typical pre-collapse cores is
expected to be rather small
\citep[\eg][]{Meier_etal__1976__ApJ__MHD_SN,Obergaulinger_et_al__2014__mnras__Magneticfieldamplificationandmagneticallysupportedexplosionsofcollapsingnon-rotatingstellarcores}.
Hence, processes that amplify the seed field such as flux-freezing
compression, winding by the differential rotation, dynamos driven by
the MRI or hydrodynamic instabilities are important ingredients to the
overall picture
\citep{Akiyama_etal__2003__ApJ__MRI_SN,Obergaulinger_etal__2009__AA__Semi-global_MRI_CCSN,Masada_et_al__2012__apj__LocalSimulationsoftheMagnetorotationalInstabilityinCore-collapseSupernovae,Moesta_et_al__2015__nat__Alarge-scaledynamoandmagnetoturbulenceinrapidlyrotatingcore-collapsesupernovae,Rembiasz_et_al__2016__JournalofPhysicsConferenceSeries__TerminationoftheMRIviaparasiticinstabilitiesincore-collapsesupernovae:influenceofnumericalmethods,Rembiasz_et_al__2016__mnras__Onthemaximummagneticfieldamplificationbythemagnetorotationalinstabilityincore-collapsesupernovae,Raynaud_et_al__2020__ScienceAdvances__MagnetarFormationthroughaConvectiveDynamoinProtoneutronStars,ReboulSalze_et_al__2020__arXiveprints__AGlobalModeloftheMagnetorotationalInstabilityinProtoneutronStars}.

Most studies involving the central engines of long GRBs start from
cores where a PM or a system consisting of a BH and an accretion torus
are set up by hand in a stellar core rather than from the results of
self-consistent simulations of the processes leading up to the
formation of those central engines \citep[though with some remarkable
  exceptions;][]{MacFadyen_Woosley__1999__ApJ__Collapsar,MacFadyen__2001__apj__Supernovae_Jets_and_Collapsars}.
Starting from these initial conditions, the GRB engines are then investigated
using theoretical analysis and numerical simulations with typically
somewhat simplified microphysics in the case of collapsars 
\citep[][]{Aloy_etal__2000__ApJL__Collapsar,Zhang_et_al__2003__apj__Relativistic_Jets_in_Collapsars,Zhang_2004ApJ...608..365Z, Proga_et_al__2003__apjl__Axisymmetric_MHD_Collapsar,Proga__2005__apj__On_MHD_Jet_Production_in_Collapsing_and_Rotating_Envelope, Mizuta_etal_2006ApJ...651..960M,Lee_Ramirez-Ruiz__2006__apj__Accretion_Modes_in_CollapsarsProspects_for_Gamma-Ray_Burst_Production,Morsony_2007,Morsony_et_al__2010__apj__TheOriginandPropagationofVariabilityintheOutflowsofLong-durationGamma-rayBursts, Nagataki_et_al__2007__apj__GRB_Jet_Formation_in_Collapsars,Barkov_Komissarov__2008__mnras__StellarexplosionspoweredbytheBlandford-Znajekmechanism,Mizuta_Aloy__2009__apj__Angular_Energy_Distribution_of_Collapsar-Jets, Harikae_et_al__2009__apj__Long-Term_Evolution_of_Slowly_Rotating_Collapsar_in_SRMHD,Nagakura_et_al__2011__apj__Jet_Propagations_Breakouts_and_Photospheric_Emissions_in_Collapsing_Massive_Progenitors_of_Long-duration_GRBs, Nagakura__2013__apj__ThePropagationofNeutrino-drivenJetsinWolf-RayetStars, Lazzati_2013, Lopez_Camara_2013,Lopez-Camara_2014,  Lopez-Camara_2016ApJ...826..180, Cuesta_et_al__2015__mnras__Numericalmodelsofblackbody-dominatedgamma-raybursts-IHydrodynamicsandtheoriginofthethermalemission,Ito_2015, Cuesta_et_al_2015_MNRAS_446_1737,Batta_Lee__2016__mnras__Innerengineshutdownfromtransitionsintheangularmomentumdistributionincollapsars, Bromberg_Tchekhovskoy_2016MNRAS.456.1739, Aloy_etal_2018MNRAS.478.3576}
and of PMs 
\citep{Wheeler_et_al__2000__apj__Asymmetric_SNe_Pulsars_Magnetars_and_GRBs,Zhang_Meszaros__2001__apjl__Gamma-RayBurstAfterglowwithContinuousEnergyInjection:SignatureofaHighlyMagnetizedMillisecondPulsar,Thompson_et_al__2004__apj__Magnetar_Spin-Down_Hyperenergetic_SNe_and_GRBs,Burrows_etal__2007__ApJ__MHD-SN,Metzger_et_al__2007__apj__Proto-NeutronStarWindswithMagneticFieldsandRotation,Bucciantini_et_al__2007__mnras__Magnetar-driven_bubbles_and_the_origin_of_collimated_outflows_in_GRBs,Bucciantini_et_al__2008__mnras__Relativistic_jets_and_lGRBs_from_the_birth_of_magnetars,Bucciantini_et_al__2009__mnras__Magnetizedrelativisticjetsandlong-durationGRBsfrommagnetarspin-downduringcore-collapsesupernovae,Metzger_et_al__2011__mnras__Theprotomagnetarmodelforgamma-raybursts,Bucciantini_et_al__2012__mnras__Shortgamma-rayburstswithextendedemissionfrommagnetarbirth:jetformationandcollimation,Metzger_et_al__2015__mnras__Thediversityoftransientsfrommagnetarbirthincorecollapsesupernovae}.

In a first paper of this series \citep[][Paper
I]{Obergaulinger_Aloy__2019}, we have explored the outcomes of the
stellar collapse of cores that may produce the central engine of long
GRBs. There, we have focused on the criteria that decides whether
small variations of the initial magnetorotational conditions in the
stellar progenitor shape the ensuing supernova explosion (or the lack
thereof).  In this paper the goal is to understand whether relatively
small variations of the magneto-rotational profile in the progenitor
star determine the final fate of the compact remnant left after core
collapse. We restrict our analysis here to rapidly rotating,
low-metallicity stars. More specifically, we focus on the following
issues:
\begin{enumerate}
\item How sensitive is the type of central engine to \emph{small}
  variations in the magnetic field topology and strength? Do
  \emph{small} variations of the pre-collapse rotational profile
  change the type of the post-collapse compact remnant?
\item Can collapsars or PMs be produced given the rotation rates
  and, in particular for PMs, also magnetic fields predicted by
  stellar evolution modelling?
\end{enumerate}

The open issues at the centre of our interest depend sensitively on a
multitude of closely coupled physical effects, suggesting that they
may be addressed by numerical simulations incorporating an accurate
treatment of the evolution of MHD and neutrino transport.  Therefore,
our simulations are based on a state-of-the-art code combining
high-order methods for solving the hyperbolic terms of the MHD and
transport equations with a post-Newtonian treatment of gravity, as
well as a spectral two-moment neutrino transport including corrections
due to the velocity (Doppler shifts, aberration) and the gravitation
field (gravitational blue/redshift) and the relevant reactions between
neutrinos and matter.  The neutrino transport, though less costly
than, \eg, a full Boltzmann solver, is the computationally most
expensive part of the simulations.  Its high cost together with the
long simulation times we want to reach and the variety of models limit
us in most cases to axisymmetric models.  Aware that the final answers
can only come from unrestricted three-dimensional (3D) models, in
order to qualitatively assess the validity of our axisymmetric
simulations, we will show the results of low-resolution 3D models for
two cases, which are prototypes of collapsar- and PM-forming central
engines.

The work by
\cite{Dessart_et_al__2012__apj__TheArduousJourneytoBlackHoleFormationinPotentialGamma-RayBurstProgenitors}
represents a milestone in determining the viability of the formation
of a GRB engine from the class of stars we are considering here.  It
presented simulations of the collapse of a star of
$M_\ZAMS = 35 \, \msol$ (in fact, one of the stars we are studying)
including magnetohydrodynamics (MHD) and neutrino transport.  The
results suggested a high likelihood of fairly early magnetorotational
explosions that inhibit the further growth of the PNS, thus preventing
the formation of a collapsar for rapidly rotating cores.  Here, we
extend the
\citeauthor{Dessart_et_al__2012__apj__TheArduousJourneytoBlackHoleFormationinPotentialGamma-RayBurstProgenitors}
study by including additional models, by using improved neutrino
physics, and by considering very long simulation times.

This article is organised as follows. We begin describing our initial
models and parameters in Sect.\,\ref{Sek:Init}, carefully justifying
the modifications on the original stellar progenitor properties
performed (see also App.\,\ref{sek:Variance}). Section\,\ref{Sek:Res}
describes our results separating models into two classes, namely, BH
forming and NS forming models. The long term evolution after core
bounce is specifically considered in Sect.\,\ref{Sek:PNS-GRB}, again
distinguishing between BH and NS forming models. For the latter ones,
we carefully assess whether the PM model of
\cite{Metzger_et_al__2011__mnras__Theprotomagnetarmodelforgamma-raybursts}
\citepalias[][hereafter]{Metzger_et_al__2011__mnras__Theprotomagnetarmodelforgamma-raybursts}
is fulfilled. We provide a small overview of two 3D models that have
been run at relatively low resolution and until less than a second
post-bounce (Sect.\,\ref{Sek:3d}). Finally, Sect.\,\ref{Sek:SumCon}
contains our conclusions and discussion of our results.

\section{Initial models and parameters}
\label{Sek:Init}

The simulations that follow were performed using the neutrino-MHD code
presented in
\cite{Just_et_al__2015__mnras__Anewmultidimensionalenergy-dependenttwo-momenttransportcodeforneutrino-hydrodynamics}.
With respect to the description of the algorithms, implementation, and
tests given there \citep[and
in][]{Rembiasz_et_al__2017__apjs__OntheMeasurementsofNumericalViscosityandResistivityinEulerianMHDCodes}
and to the previous application in simulations of magnetized core
collapse
\citep{Obergaulinger_et_al__2014__mnras__Magneticfieldamplificationandmagneticallysupportedexplosionsofcollapsingnon-rotatingstellarcores},
we made several modifications for the purpose of running the present
set of models. These modifications have already been used in the
closely related publication
\citep{Obergaulinger_Aloy__2017__mnras__Protomagnetarandblackholeformationinhigh-massstars},
as well as in the exploration of magneto-rotational collapse of lower
mass and solar metallicity progenitors
\cite{Obergaulinger_Aloy_Just__2018} and the effects of the multipolar
structure of large scale magnetic fields in the outcome of core
collapse of low-metallicity, high mass stars
\citep{Bugli_2020MNRAS.492...58}. We refer to
\citetalias{Obergaulinger_Aloy__2019} for a detailed list of all these
modifications and an extended description of the microphysics and numerical methods. Our code includes a general relativistic approximate gravitational potential, $\Phi$ as in \cite{Marek_etal__2006__AA__TOV-potential}. This allows us to define a lapse function as $\alpha:=\exp{(\Phi/c^2)}$, where $c$ is the light speed in vacuum. Operatively, we say that a BH has formed when either $\alpha < \alpha_{\rm th}:=0.4$ at $r=0$ or the maximum density, $\rho_{\rm max}$ is larger than $\rho_{\rm th}:=\zehnh{1.1}{15}\,\gccm$. We note that when  $\alpha$ and $\rho_{\rm max}$ approach the operative threshold values $\alpha_{\rm th}$ and $\rho_{\rm th}$, respectively, the dynamics at the centre speeds up very quickly. In a matter of a few time steps $\alpha$ may sink to values below $\zehn{-8}$ and the central density rise to values $\sim \zehn{16}\,\gccm$, where the equation of state employed \citepalias[LS220 with a suitable extension to low densities below $\rho_{\rm low}=\zehnh{6}{7}\,\gccm$;][Sec.\,2.1]{Obergaulinger_Aloy__2019} is no longer causal.

We focus on model 35OC, which was computed by
\cite{Woosley_Heger__2006__apj__TheProgenitorStarsofGamma-RayBursts}
as a model for a rapidly rotating star of zero-age main-sequence mass
$M_\ZAMS = 35 \, \Msol$ including the redistribution of angular
momentum by magnetic fields according to the model of
\cite{Spruit__2002__AA__Dynamo}.  At collapse, the star retains a mass
of $\MpreSN \simeq 28.07 \, \Msol$ and possesses an iron core of mass
$M_{\mathrm{Fe}} = 2.02 \, \Msol$, radius
$R_{\mathrm{Fe}} \approx 3000 \,\km$, and central density of
$\rho_{\mathrm{c}} \approx \zehnh{2.4}{9} \, \gccm$.  The core rotates
differentially with angular velocities of
$\Omega_{\mathrm{PC,c}} \approx 1.98 \, s^{-1}$ at the centre and
$\Omega_{\mathrm{PC,Fe}} \approx 0.1 \, s^{-1}$ at the surface of the
iron core.  Another initial model with the same ZAMS mass as model
35OC has been employed, namely model 35OB
\citep[Secs.\,\ref{sSek:35OB-RO} and \ref{sSek:35OC-RRw}][
$\MpreSN \simeq
21.2\msol$]{Woosley_Heger__2006__apj__TheProgenitorStarsofGamma-RayBursts}
The latter model is not supposed to produce a collapsar central engine
(it does not rotate fast enough). However we place it among the grid
of computed models because it provides an initial profile, which
differs \emph{sufficiently} from \modelname{35OC} so as to assess the
influence of the stellar progenitor structure in the final outcome,
especially, on the formation of BHs accompanied with successful SN
explosions.

We map the pre-collapse structure of the core (computed in one spatial
dimension), viz.~the hydrodynamic variables such as density, electron
fraction, temperature, and rotational velocity, onto our simulation
grid.  The magnetic field of the model is the result of MHD
instabilities.  Hence, it is not a global, \eg dipole field
encompassing the entire star, but rather confined to several shells.
We simulate a series of models based on the field as given by the
pre-collapse model and an additional series of models in which we
replace the field by a global dipole field and a toroidal component.%
\footnote{Large scale dipolar fields have been inferred from spectropolarimetric observations of massive stars, \eg in the O stars $\theta^{1}$ Ori C \citep{Donati_etal_2002MNRAS.333...55} or HD 191612 \citep{Donati_etal_2006MNRAS.365L...6}.}
The details on the exact initialisation of the magnetic field can be
found in \citetalias{Obergaulinger_Aloy__2019} as well as in
App.\,\ref{sek:Variance}.  Within each of the two series, we vary the
normalisation of the poloidal and toroidal components, and, in the
first series, also their angular distribution.  In order to assess the
importance of rotation, we add a version of the same model with an
artificially reduced angular velocity. Table~\ref{Tab:models} provides
a summary of the most important properties of our models.

Modifying the stellar profiles of rotation or of magnetic field is an
artificial manipulation intended to probe the impact of these two
factors on the dynamics.  In reality, a change of the rotational
velocity would most likely entail changes of the density,
thermodynamical, and composition profiles of the core. However, given
the uncertainties in stellar evolution modelling regarding the
treatment of mass loss, angular momentum transport and other magnetic
processes \citep[see,
\eg][]{Maeder_Meynet_2012RvMP...84...25,Keszthelyi_2019MNRAS.tmp..744},
we assume that there is some room for (small) variations of the
aforementioned rotational and magnetic profiles. A more consistent
treatment would require computing a relatively dense grid of stellar
evolution models \emph{ab initio}, changing slightly the parameters
that control the aforementioned effects of rotation, magnetic fields
and mass loss, but this is beyond the scope of this paper. We stress,
however, that even without modifying the magnetic field in the stellar
evolution models, the mapping from these one-dimensional progenitors
to our multidimensional computational grid is not unique and may
introduce a significant variations in the post-bounce evolution (see
App.\,\ref{sek:Variance}).

The numerical grids and initialisation employed here are the same as
in \citetalias{Obergaulinger_Aloy__2019}. All models have been
simulated on spherical grids. In axially symmetric models, the mesh
consisted of $n_{\theta} = 128$ zones in $\theta$-direction and
$n_r = 400$ radial zones with a width given in terms of a parameter
$(\delta_r)_0 = 600 \, \mathrm{m}$,
$\delta r = \max \{ (\delta r)_0, r \pi / n_{\theta} \}$.  In energy
space, we used $n_{\epsilon} = 10$ energy bins distributed
logarithmically between $\epsilon_{\mathrm{min}} = 3 \, \MeV$ and
$\epsilon_{\mathrm{max}} = 240 \, \MeV$. In 3D, we use a grid with $(n_r,n_\theta,n_\phi)=(300,64,128)$ zones, i.e. with a similar effective radial resolution in the PNS and its surroundings but with a coarser angular grid than in the axisymmetric models.

\begin{table*}
  \centering
  \begin{tabular}{|l|l|l@{\hskip 0.2cm}l|ll|ccc|ccc|cc|cc|}
    \hline
    name & star & $\Omega(r)$ & B-field & fate & BH
    &$\tdf$& $\Mdisk$ & $M_{\rm e}^{\rm max}$&  $\Emagunit$& $\Erotunit$ & $\Ebindingunit$ &$M_{\rm
      exp}$& $E^{\rm exp}_{51}$
    \\
    &&(Hz)&($\zehn{11}$\,G)&&&(s)&$(\Msol)$& $(\Msol)$&& &&$(\Msol)$&
\\
    \hline
    \modelname{35OC-RO} & \modelname{35OC}
    & $1.98$ (Or) & 1.1, Or & MR & $\surd$ & $9.3$ & $7.5$ &$20.6$& $70.0$&$17.4$&$2.84$ & $0.18^{\dagger}$&$1.03^{\dagger}$
    \\
    \modelname{35OC-RO2} & \modelname{35OC}
    & $1.98$ (Or) & $2.2,\,\mathrm{p},2\mathrm{t}$ & MR & $\surd$& $9.3$ & $7.5$&$20.6$& $279$&$17.4$&$2.84$&$0.32^{\dagger}$&$1.39^{\dagger}$
    \\
    \modelname{35OC-Rp2} & \modelname{35OC}
    & $1.98$ (Or) & $1.3,\,2\mathrm{p},1\mathrm{t}$ & MR & $\times$& $9.3$ & $7.5$&
    $20.6$&$82.5$&$17.4$&$2.84$& $0.44$& $1.96$
    \\
    \modelname{35OC-Rp3} & \modelname{35OC}
    & $1.98$ (Or) & $1.5,\,3\mathrm{p},1\mathrm{t}$ & MR & $\times$& $9.3$ & $7.5$&
    $20.6$&$103$&$17.4$&$2.84$& $0.57$ & $2.66$
    \\
    \modelname{35OC-Rp4} & \modelname{35OC}
    & $1.98$ (Or) & $1.8,\,4\mathrm{p},1\mathrm{t}$ & MR & $\times$& $9.3$ & $7.5$&
    $20.6$&$132$&$17.4$&$2.84$ & $0.57$ & $3.22$
    \\
    \modelname{35OC-Rw} & \modelname{35OC}
    & $1.98$ (Or) & 0.10, \,$a(10,10)$ & $\nu$-$\Omega$ & ? & $9.3$ & $7.5$&
    $20.6$&$0.0134$&$17.4$&$2.84$ & $0.21^{\dagger}$ & $0.67^{\dagger}$
    \\
    \modelname{35OC-Rs} & \modelname{35OC}
    & $1.98$ (Or) & $10, \, a(12,12)$ & MR & $\times$&$9.3$ & $7.5$ &$20.6$&$135$&
    $17.4$&$2.84$ & $1.42$ & $5.60$
    \\
    \modelname{35OC-Sw} & \modelname{35OC}
    & $0.49$ ($\times \frac{1}{4}$) & $10^{-3}, \, a(8,10)$ & $\nu$ & $\surd$& $42.9$ &
    $22.8$&$5.27$&$0.0134$&$1.1$&$2.89$ &$0.26^{\dagger}$ &$1.67^{\dagger}$
    \\
    \modelname{35OC-RRw} & \modelname{35OC}
    & $2.97$ ($\times 1.5$) & $1.1\times10^{-6},\, $Or$/10^6$ & $\nu$-$\Omega$ & $\times$& $2.9$ & $3.14$&
    $24.9$ & $3.5\times\zehn{-12}$&$69.5$&$2.84$ & $0.034$ & $0.21$
   \\
    \hline
    \modelname{35OB-RO} & \modelname{35OB}
    & $1.54$ (Or)& $0.64$,\, Or & $\nu$-$\Omega$ &$\surd$& $37.4$ & $17.3$&
    $3.9$& $34.7$ &$0.47$&$1.11$ & $0.13^{\dagger}$ & $0.82^{\dagger}$
    \\
    \modelname{35OB-RRw} & \modelname{35OB}
    & $3.09$ ($\times 2$) & $6.4\times 10^{-7},\, $Or$/10^{6}$ & $\times$ &  $+$& $15.3$ &
    $10.0$& $11.2$&$1.2\times\zehn{-12}$&$3.88$&$1.12$ & $\times$ & $\times$
    \\
    \hline
  \end{tabular}
  \caption{
    List of our axisymmetric models. 
    Each simulation is listed with its name and the progenitor star.
    The third column indicates the rotational frequency at $r=0$ and the type of the
    rotation profile (in parenthesis): ``Or'' stands for the original profile taken from the
    stellar evolution calculation, and $\times n$ means that we multiplied
    the original angular velocity by a uniform factor $n$.
    The fourth column similarly shows the magnetic field strength at $r=0$ in units of $\zehn{8}\,$G, as well as the type of magnetic field:
    ``Or'' indicates the magnetic field profile of the original
    stellar evolution model, $x\mathrm{p},y\mathrm{t}$ means that the
    original poloidal and toroidal fields have been multiplied by
    factors $x$ and $y$, respectively, and $\mathrm{a}(x,y)$ stands
    for an artificial dipolar field with maximum poloidal and toroidal field
    components of $10^x$ and $10^y$ G, respectively.  
    The fifth column, ``fate'', gives a brief indication of the
    evolution of the model: $\nu$ means a standard neutrino-driven
    shock revival, $\nu$-$\Omega$ one strongly affected by rotation, MR a
    magnetorotational explosion, and $\times$ a failed explosion.
    The sixth column shows the sign $\surd$ if a BH formed during the
    simulation, $+$ if it did not, but we consider its formation
    likely on time scales of seconds after the end of the simulation,
    and $\times$ if no BH was formed and we estimate the final remnant
    to be a NS. The fate of \modl{35OC-Rw} is unclear, hence we annotate
    it with a  question mark.
    The seventh to ninth columns provide the accretion disc formation time
    (Eq.\,\ref{eq:tdf}), the corresponding Lagrangian mass
    coordinate for disc formation ($\Mdisk$), and the maximum mass
    available for the ejecta computed as $M_{\rm e}^{\rm max}:=\MpreSN - \Mdisk$,
    respectively. 
    The tenth and eleventh columns list the pre-SN magnetic and rotational
    energy in units of $\zehn{47}\,$erg and $\zehn{49}\,$erg,
    respectively. The twelfth column correspond to the binding energy of the pre-SN envelope (defined by the layers of the star at a distance larger than 2000\,km from the centre).
    The last two columns provide proxy values for the explosion mass
    ($M_{\rm exp}$) and of the explosion energy $E^{\rm exp}$ (in
    units of $\zehn{51}\,$erg) two seconds after bounce or,
    alternatively, at the final time of the simulation if BH formation
    occurs before $t = 2 \sek$. Values in these two columns annotated
    with $^{\dagger}$ are shown by the time of BH formation.  }
  \label{Tab:models}
\end{table*}

\begin{figure*}
  \centering
    \begin{tikzpicture}
  \pgftext{%
  \includegraphics[width=0.99\linewidth]{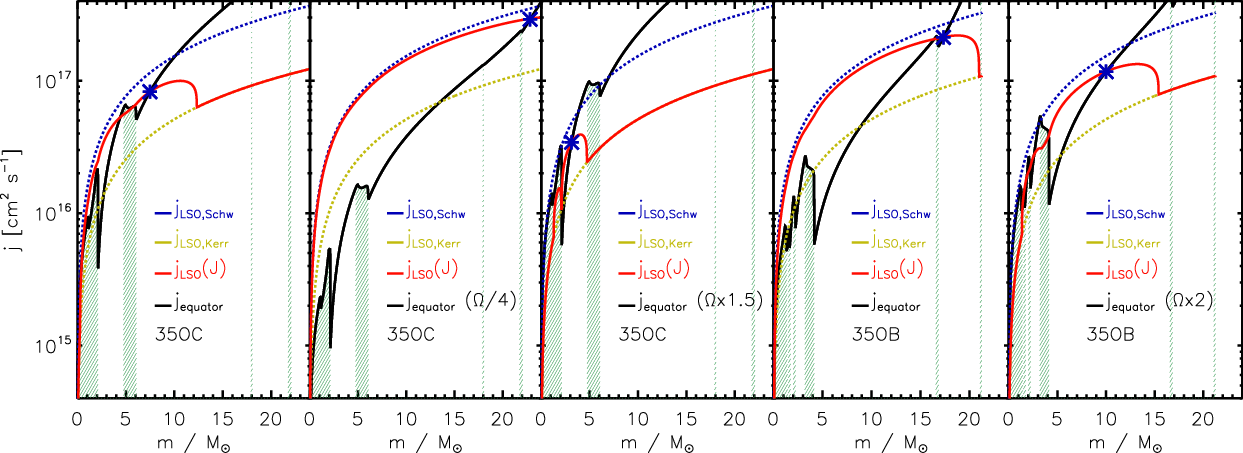}
         }%
   \node at (-7.4,+2.9) {\large (a)};
   \node at (-4.1,+2.9) {\large (b)};
   \node at (-0.85,2.9) {\large (c)};
    \node at (2.4,2.9) {\large (d)};
    \node at (5.72,2.9) {\large (e)};
  \end{tikzpicture}
  \caption{
    Equatorial profile of the initial specific angular momentum
      (black lines) of the models with the same rotational profile as
      the progenitor star 35OC \panel{a}, with one fourth of the
      rotational frequency of the stellar progenitor (\panel{b};
      \eg corresponding to \modl{35OC-Sw}), and with $1.5$ times the
      rotational frequency of the stellar progenitor (\panel{c};
      \eg corresponding to \modl{35OC-RRw}). Panels \panel{d} and \panel{e} correspond to the specific angular momentum of models
      \modelname{35OB-RO} and \modelname{35OB-RRw}, respectively. In
      each panel, the blue dashed lines denote the angular momentum
      needed to support matter at the LSO for a Schwarzschild BH,
      while the yellow dashed lines are for a Kerr BH with
      dimensionless spin $a=1$.  The red lines indicate the specific
      angular momentum at the LSO for a BH with the mass and angular
      momentum inside the displayed mass coordinate in the pre-SN
      star.  The green-hatched parts of the plot denote
      the mass shells of the pre-SN star with non-zero magnetic
      field.
    }
  \label{Fig:progenitor-profile}
\end{figure*}

We display in \figref{Fig:progenitor-profile} the distribution of
specific angular momentum, $j$, for the different variants of the
stellar progenitor models \modelname{35OC} and \modelname{35OB}. The
original profile of angular momentum
(\figref{Fig:progenitor-profile}\panel{a}) is employed in, \eg models
\modelname{35OC-RO}, \modelname{35OC-RO2}, \modelname{35OC-Rp2},
\modelname{35OC-Rp3}, etc. (see Tab.\,\ref{Tab:models}), while slower
and faster rotation are assumed in \modls{35OC-Sw}
(\figref{Fig:progenitor-profile}\panel{b}) and \modelname{35OC-RRw}
(\figref{Fig:progenitor-profile}\panel{c}), respectively. In each of
the panels we also display the angular momentum needed to support
matter in the last stable (circular) orbit (LSO) for either a
non-rotating (blue dashed lines) or maximally rotating Kerr BH (dashed
yellow lines) with a mass equal to that enclosed by the mass
coordinate annotated in the abscissa axis. Also, a red line marks the
specific angular momentum at the LSO for a rotating Kerr BH with a
mass and angular momentum equal to those enclosed by the mass
coordinate in the progenitor star.\footnote{We employ the formulae of
  \cite{Bardeen_1972ApJ...178..347} to compute the corresponding
  angular momentum.}  It is (roughly) expected that where the red line
lies below the black line, matter in the star may have enough angular
momentum to form an accretion disc
\citep{Woosley_Heger__2006__apj__TheProgenitorStarsofGamma-RayBursts}. We
denote the Lagrangian mass coordinate of this point $\Mdisk$. This
estimation of $\Mdisk$ does not account for the post-bounce dynamics,
which may transport angular momentum altering the pre-SN specific
angular momentum profile.

If disc formation takes place after the central compact remnant
collapses to a BH, a collapsar engine may result. In order to estimate
the disk formation time, $\tdf$, we assume it equals twice the
free-fall time of the innermost mass element that reaches a Keplerian
velocity
\citep[\eg][]{Dessart_et_al__2012__apj__TheArduousJourneytoBlackHoleFormationinPotentialGamma-RayBurstProgenitors},
thus
\begin{equation}
    \tdf =2\pi\sqrt{\frac{r^{3}_{\mathrm{pre}\textsc{sn}}(\Mdisk)}{8G\Mdisk}},
    \label{eq:tdf}
\end{equation}
where $r_{\mathrm{pre}\textsc{sn}} (\Mdisk)$ is the radius of the
disk-forming Lagrangian mass element in the pre-SN model. The value of
$\tdf$ for the outermost Lagrangian mass coordinate that satisfies the
criteria to form an accretion disc are listed in
Tab.\,\ref{Tab:models}.

In \modl{35OC-RO}, with the original specific angular momentum
distribution, there are two mass shells where a disc may form, namely
$4.4\lesssim \Mdisk/\msol \lesssim 5.7$ or for
$\Mdisk\gtrsim 7.5\msol$. We mark the outermost point where the
standard disc formation criterion holds with a blue asterisk in
\figref{Fig:progenitor-profile}. In the model with reduced angular
frequency (\figref{Fig:progenitor-profile}\panel{b}),
$\Mdisk\sim 22.8\msol$, while the model with faster rotational speed
than the pre-SN star the whole progenitor beyond $\Mdisk\sim 2.9\msol$
may produce an accretion disk.
The mass of the star above $r_{\mathrm{pre}\textsc{sn}} (\Mdisk)$ is
available for the SN ejecta. We list the maximum available SN ejecta
mass (corresponding to $\MpreSN - \Mdisk$) in
Tab.\,\ref{Tab:models}. Note that this is only a rough estimate since,
(i) a fair fraction of the aforementioned mass may fall back onto the
central compact object on sufficiently long time scales (especially if
the SN explosion is asymmetric and allows for simultaneous mass
accretion and ejection), and (ii) not necessarily all the mass up to
$\Mdisk$ may necessarily end up in the central compact object; a
fraction of it may be incorporated into the outflow ejecta by the
action of Maxwell stresses or neutrino heating.  This means that
between $\sim 20.6\msol$ and $\sim 23.6\msol$ may be ejected for the
original rotational profile of \modl{35OC-RO}. The \emph{exact} amount
depending on whether the accretion disc forms above $4.4\msol$ or
$7.5\msol$. For the fast rotating version of the pre-SN star, a larger
fraction of the star ($\sim 24.9\msol$) could be ejected. In the case
of the slowly rotating \modl{35OC-Sw}, only $\sim 5.3\msol$ could be
unbound as SN ejecta. Considering that the ejecta mass in typical Type
Ic SN may feature a broad range \citep[say,
$M_{\rm ejecta}\sim 1\msol - 10\msol$; \eg][]{Taubenberger_2006MNRAS},
and the fact that the mass remaining above the potential disc
formation mass coordinate is only an upper bound for the SN ejecta
mass (see above), even the models with modified rotational profiles
may yield an ejecta mass \emph{broadly} compatible with observational
estimates. In Tab.\,\ref{Tab:models}, we show the proxy values
computed for the mass ejected by the successful SN explosions. These
values are still small compared to the rough estimates provided
above. In part, this is because, with the exception of \mRptwo and
\Rpthree, the rest of the models have been evolved only for
$\tpb:=t-t_{\rm bounce}\lesssim 3\,\sek$ after bounce. The strong
magnetorotational explosion of \mRs stands out of the rest with an
ejecta mass $\sim 1.4\msol$ after only $\tpb\approx 2.3\,\sek$.

We finally notice that the pre-SN models display an alternated pattern
of magnetised (green hatched regions in
\figref{Fig:progenitor-profile}) and unmagnetised mass shells. This is
due to the presence of convective layers in the evolved stars, where
the standard angular momentum transport by magnetic torques have not
been applied. Indeed, in our stellar models $\lesssim 15\%$ of the
total pre-SN mass contains non-zero magnetic field. Most of the
magnetised mass shells are located within
$M_{\textsc{b}}\lesssim 6\msol$ ($\lesssim 4\msol$) in \modl{35OC}
(\modelname{35OB}), which means that they could free-fall in times
$\lesssim 7\,$s ($\lesssim 3.9\,$s), as obtained from application of
Eq.\,\ref{eq:tdf} to the corresponding mass shell (i.e. replacing
$\Mdisk$ by $M_{\textsc{b}}$).

\section{Results}
\label{Sek:Res}

In this section we describe the dynamics of the two-dimensional
(axisymmetric) models grouping them into models that form or may form
a BH and models which do not form BHs.
\begin{figure*}
  \centering
  \begin{tikzpicture}
  \pgftext{\vbox{
  \hbox{
  \includegraphics[width=0.32\linewidth]{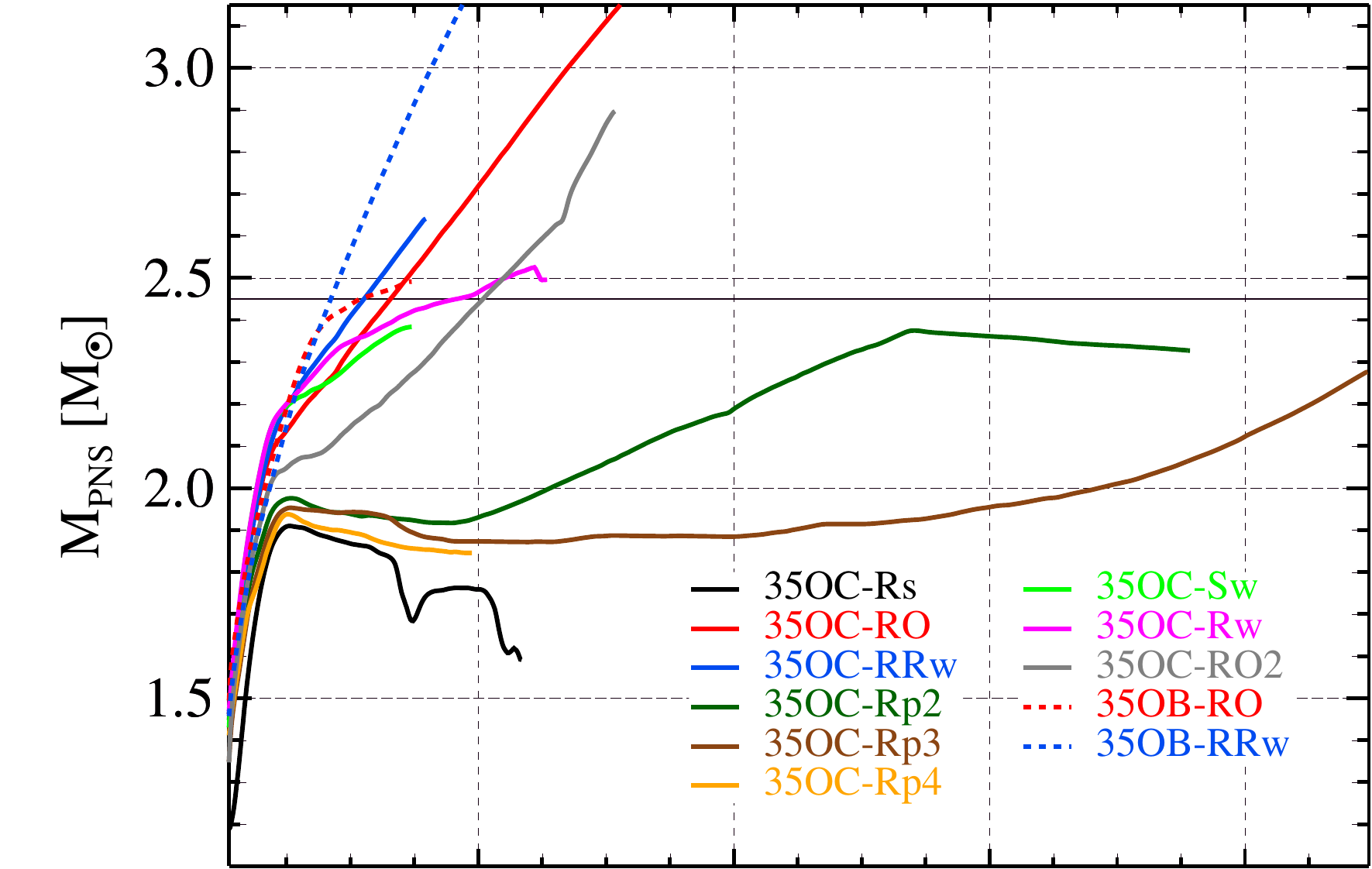}
  \includegraphics[width=0.32\linewidth]{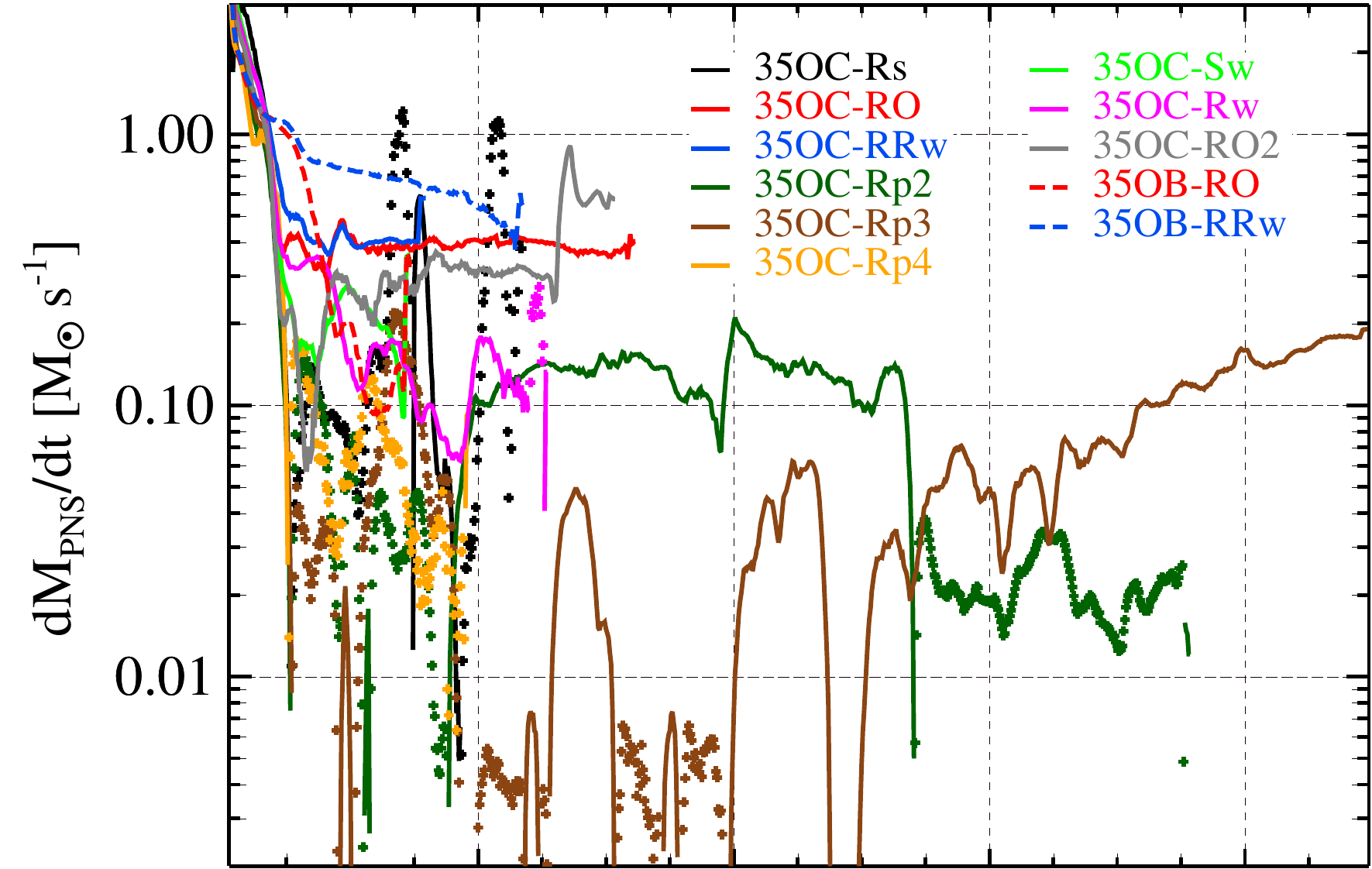}
  \includegraphics[width=0.32\linewidth]{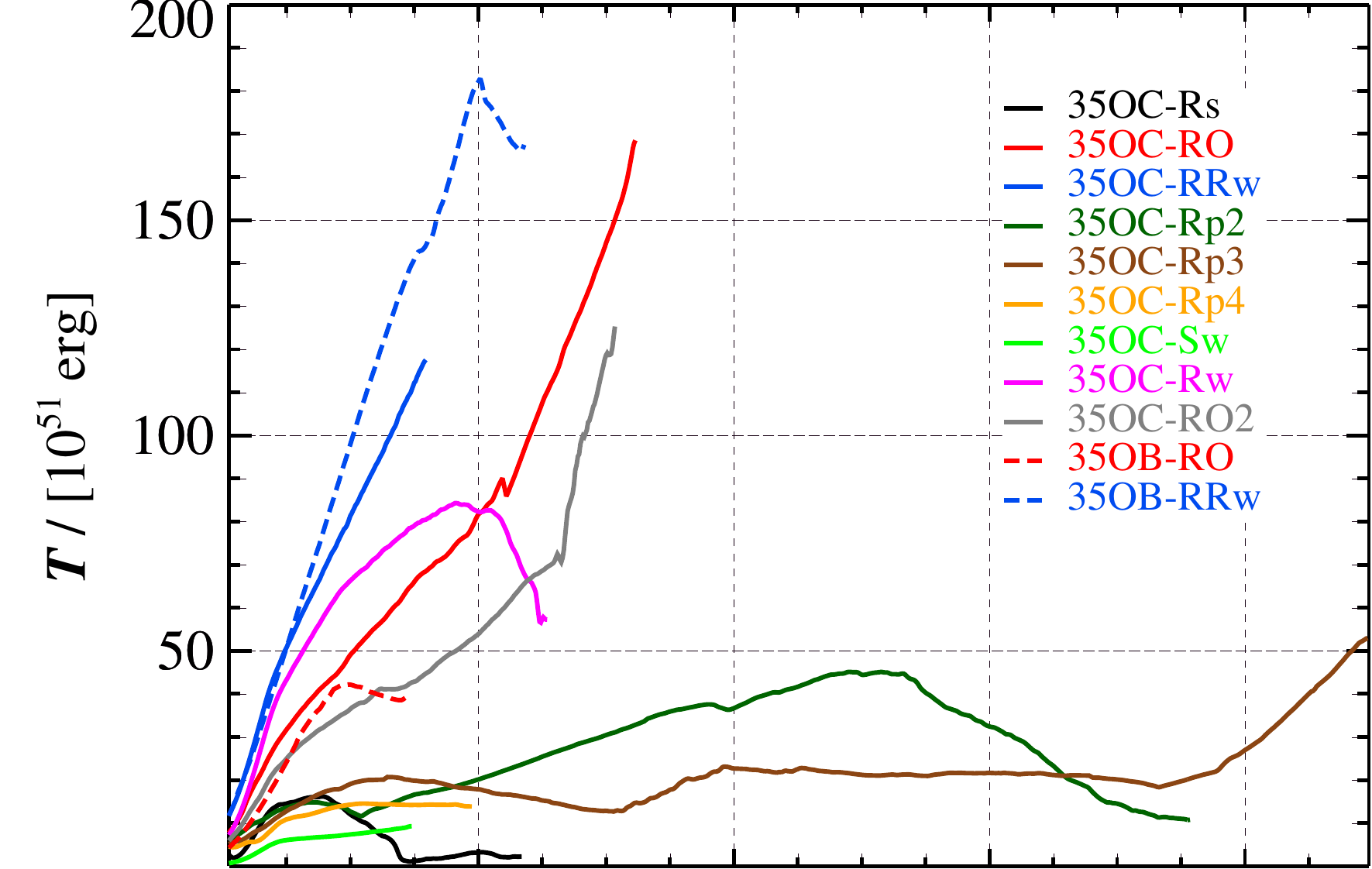}
  }
  \vspace{-0.08cm}\hbox{
  \includegraphics[width=0.32\linewidth]{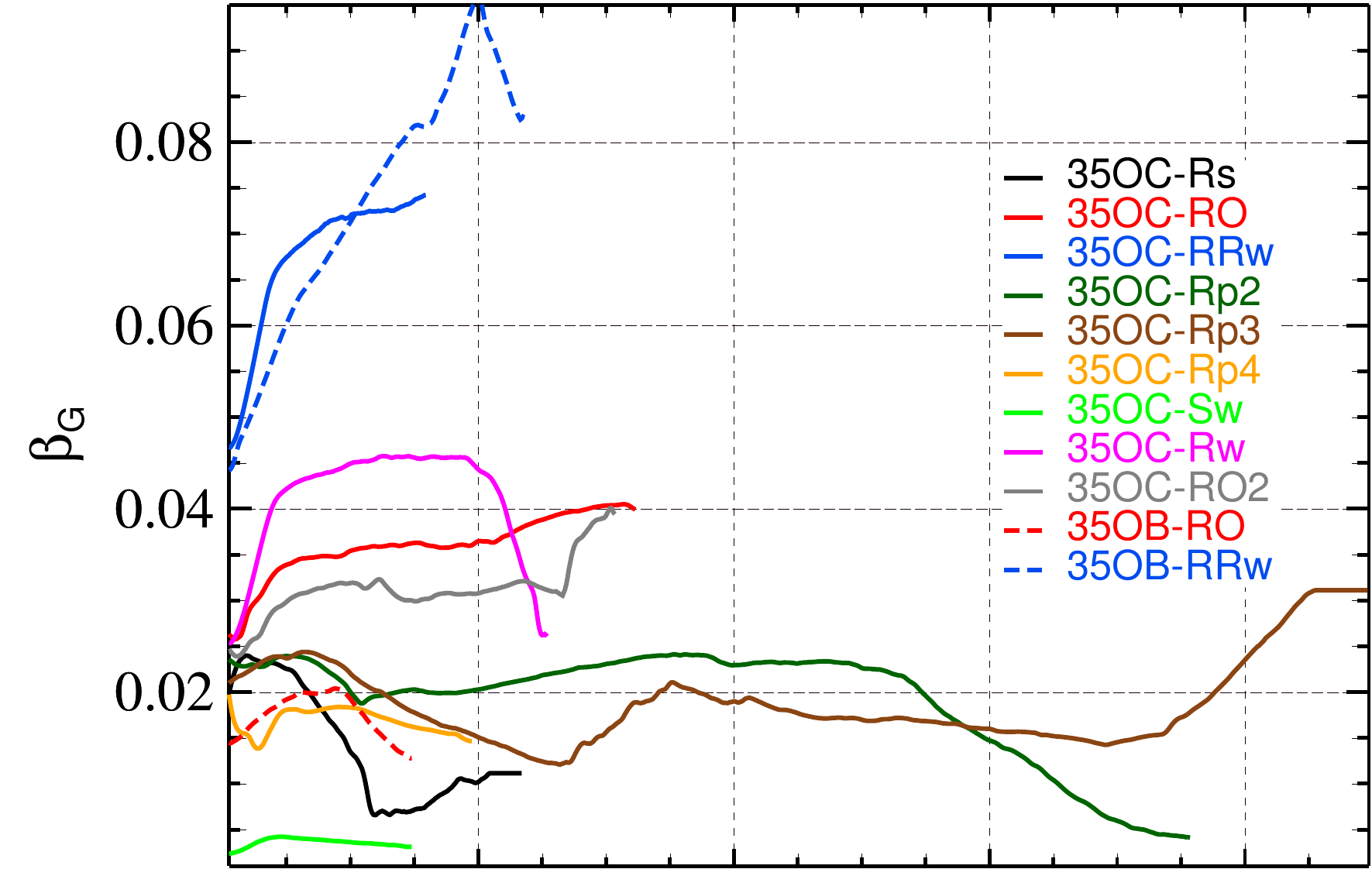}
  \includegraphics[width=0.32\linewidth]{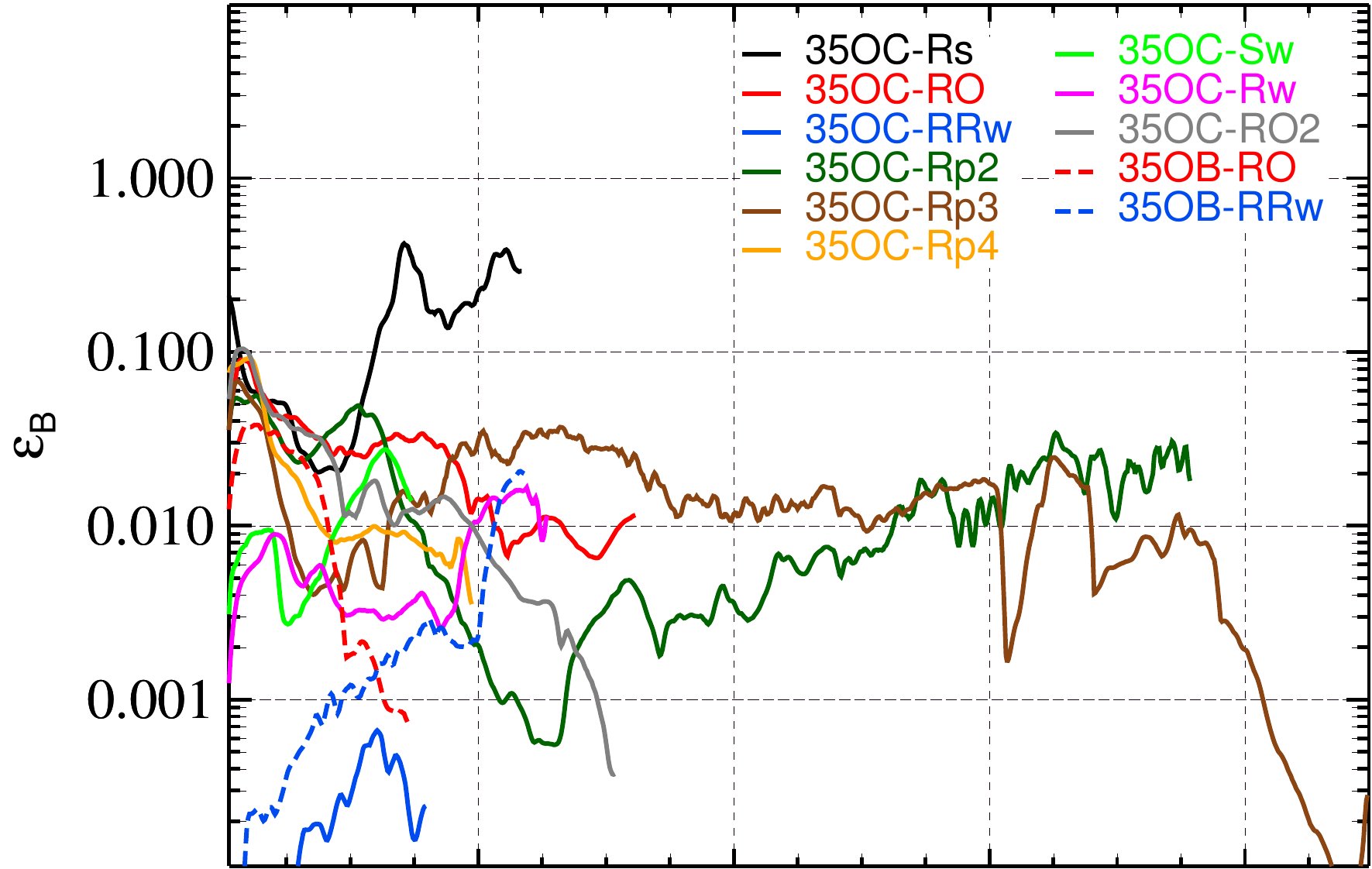}
  \includegraphics[width=0.32\linewidth]{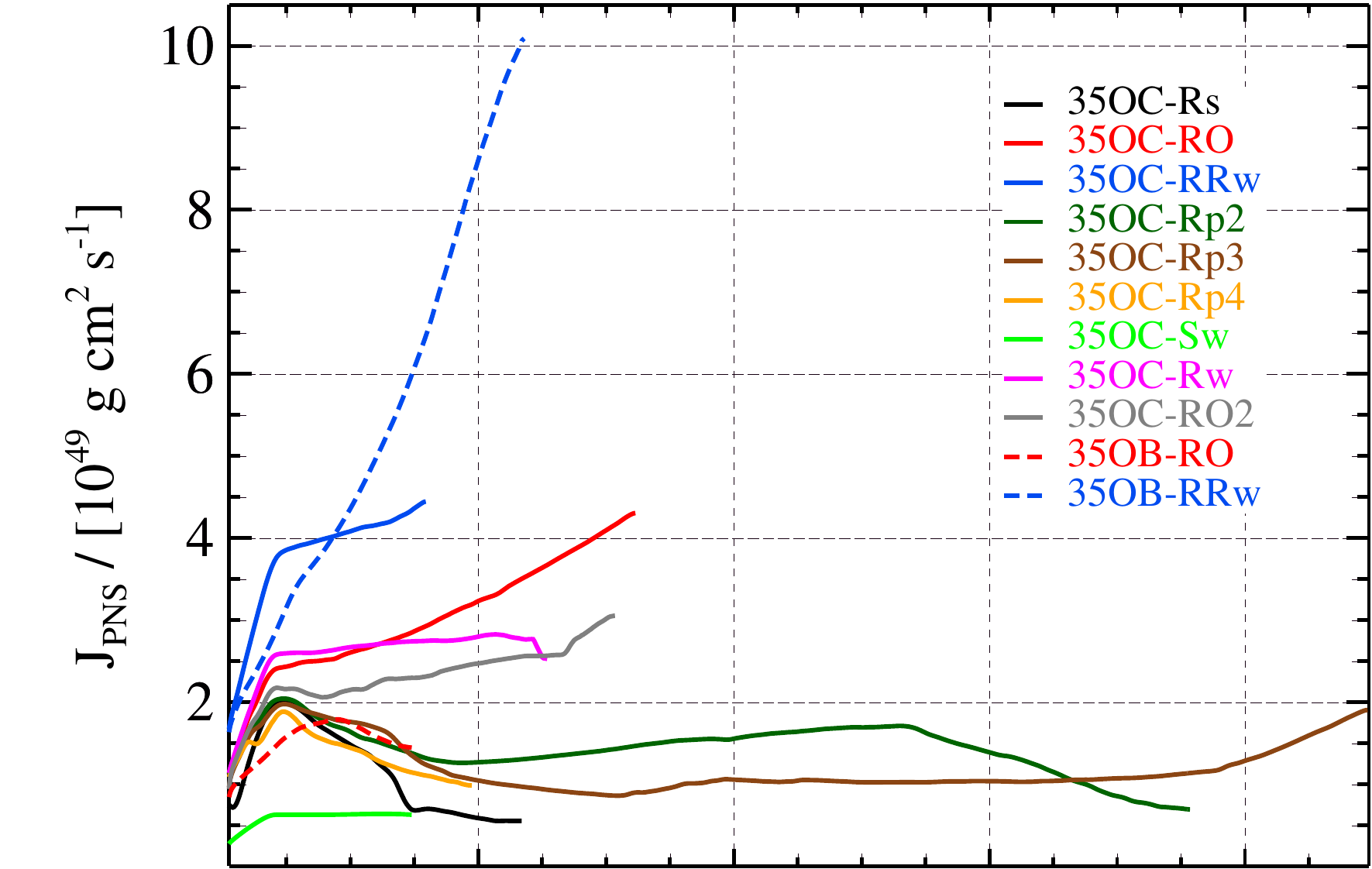}
  }
 \vspace{-0.08cm}\hbox{
  \includegraphics[width=0.32\linewidth]{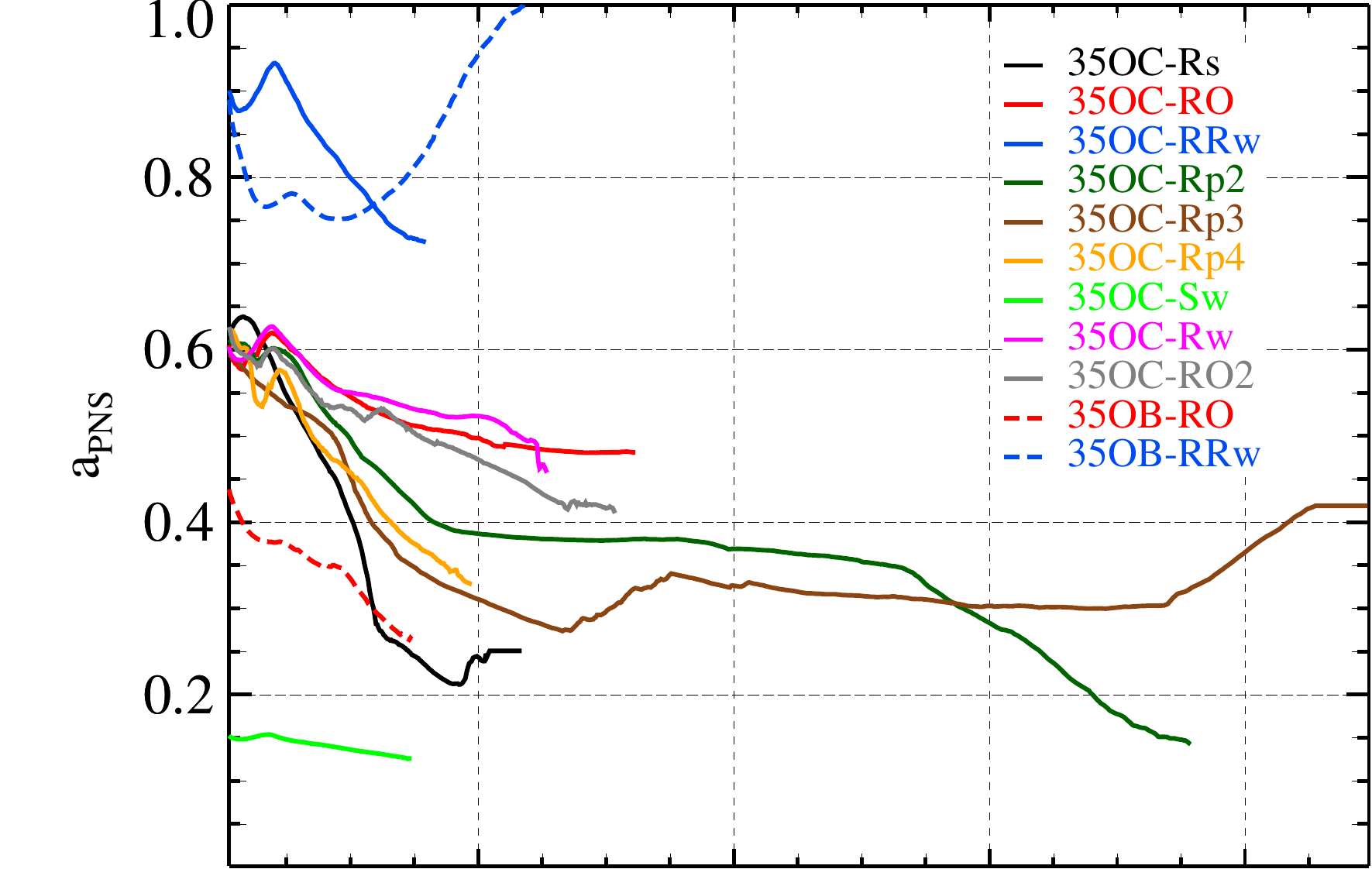}
  \includegraphics[width=0.32\linewidth]{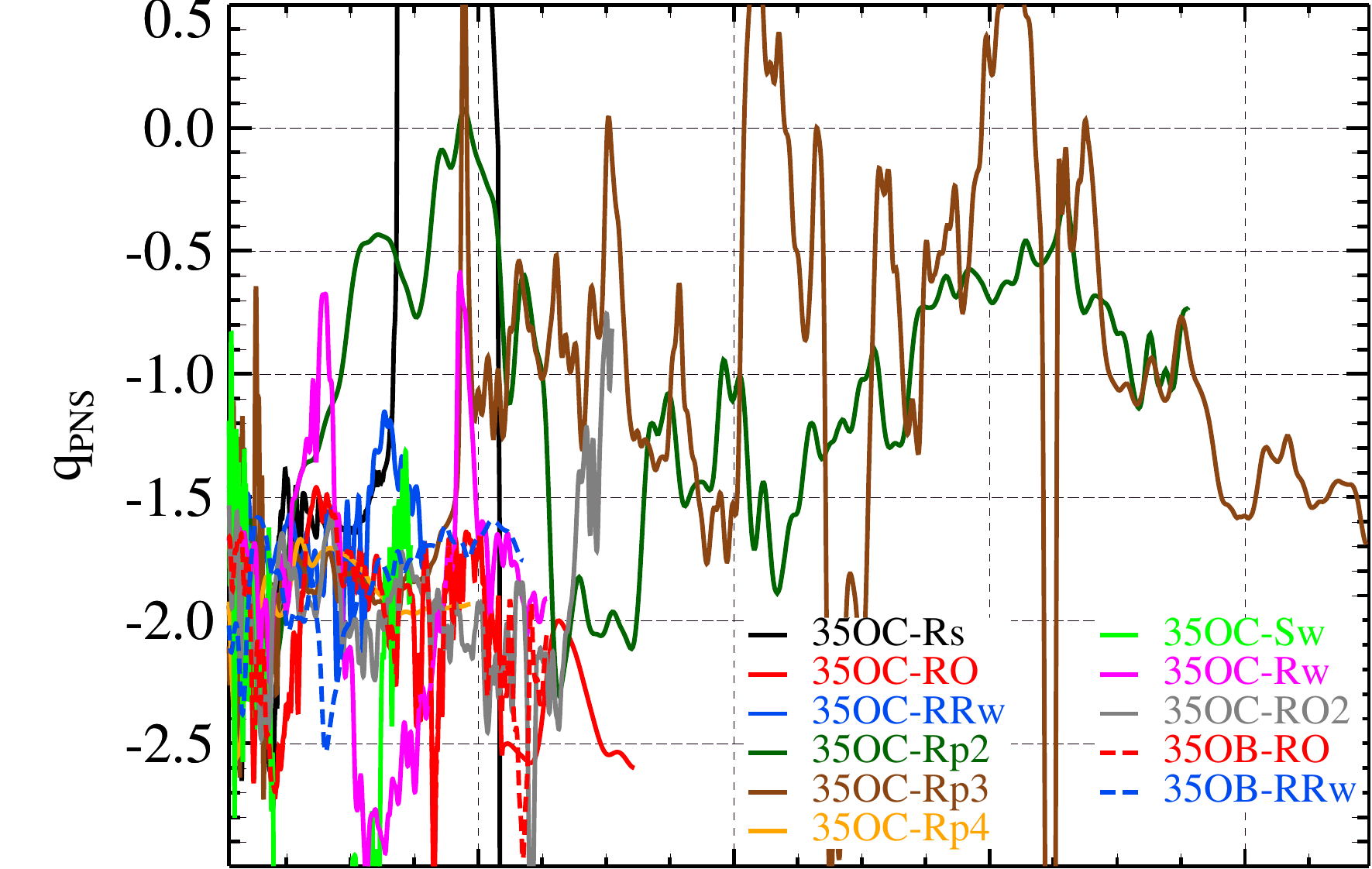}
  \includegraphics[width=0.32\linewidth]{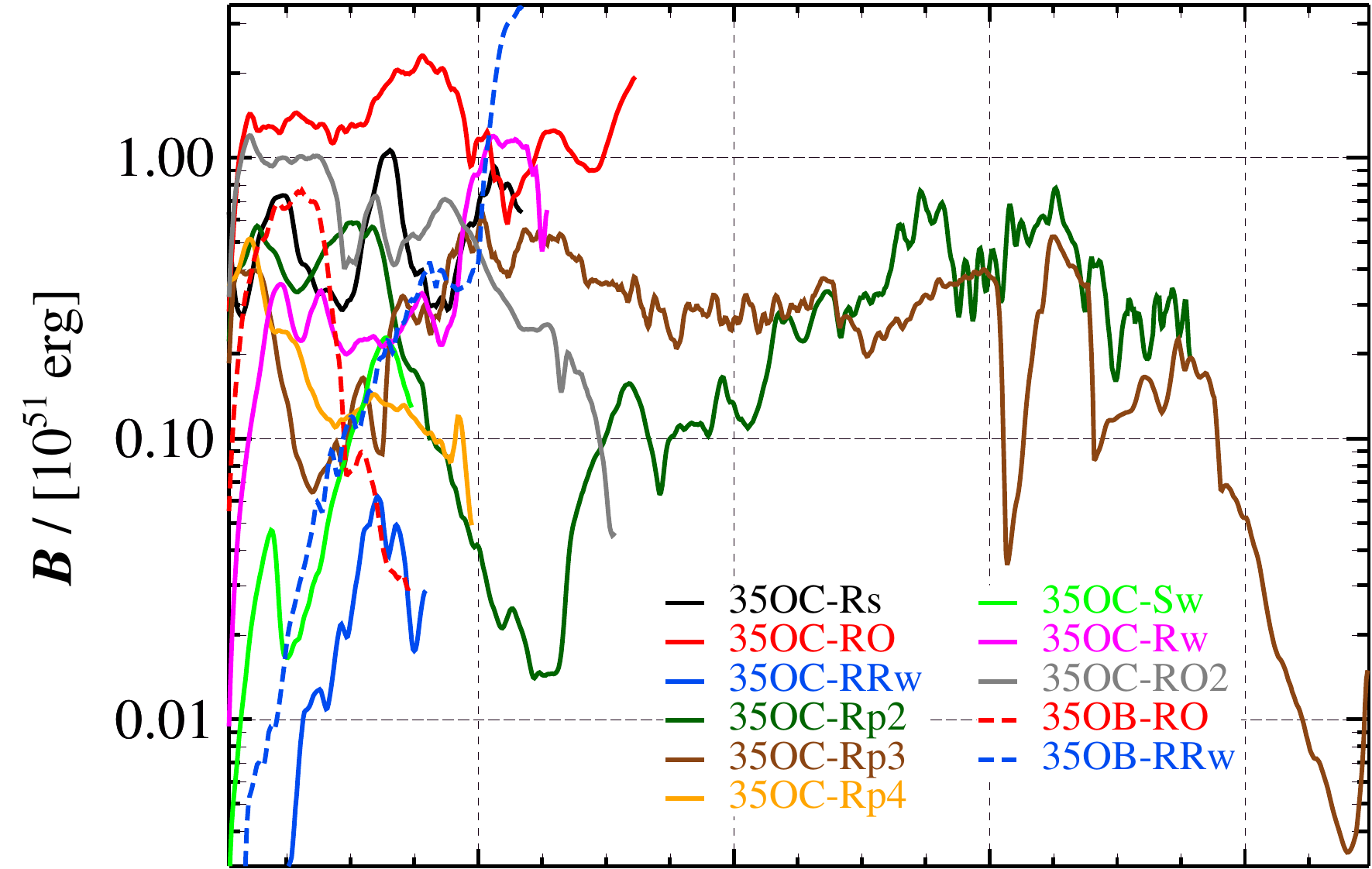}
  }
  \vspace{-0.08cm}\hbox{
  \includegraphics[width=0.32\linewidth]{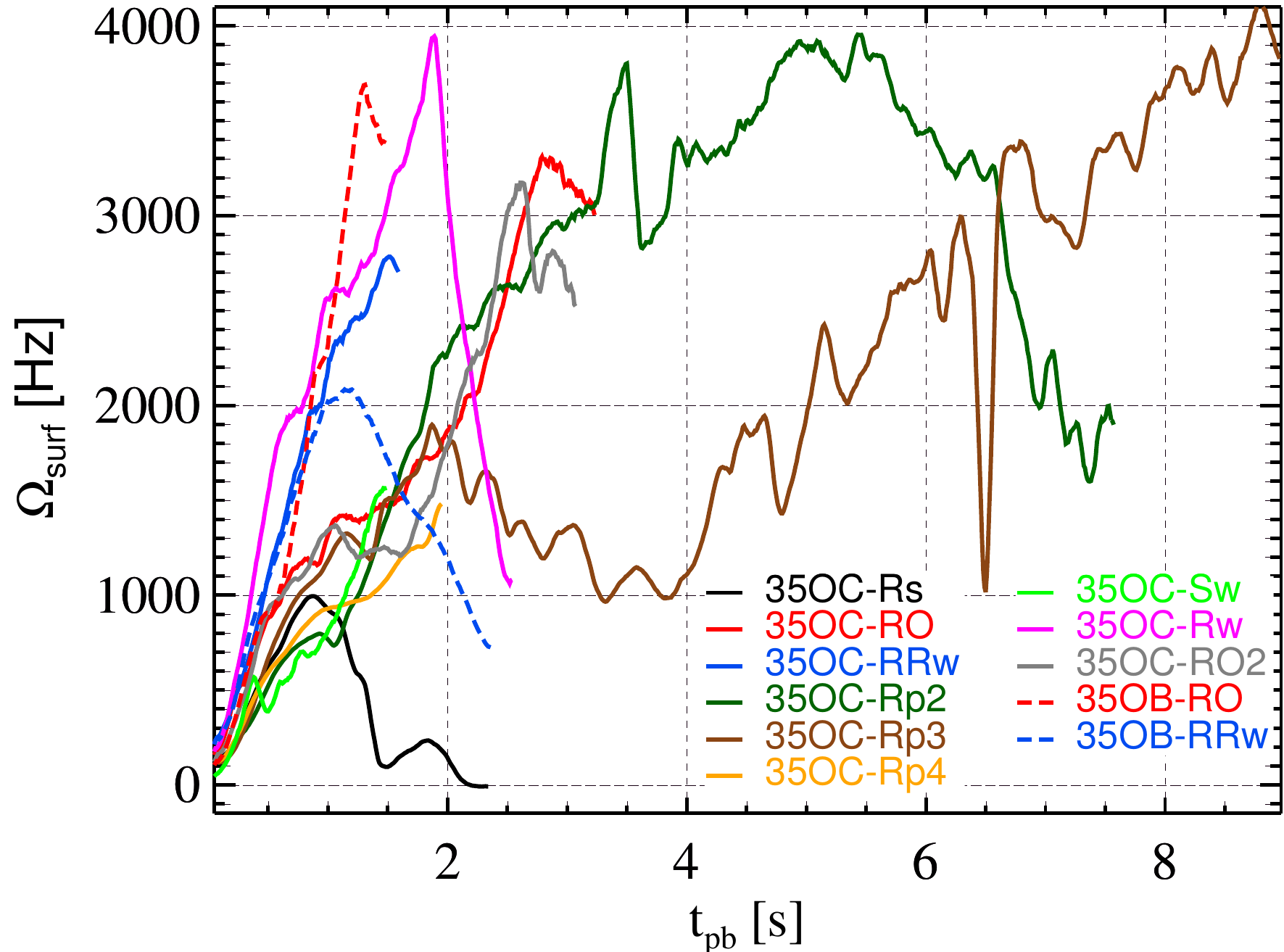}
  \includegraphics[width=0.32\linewidth]{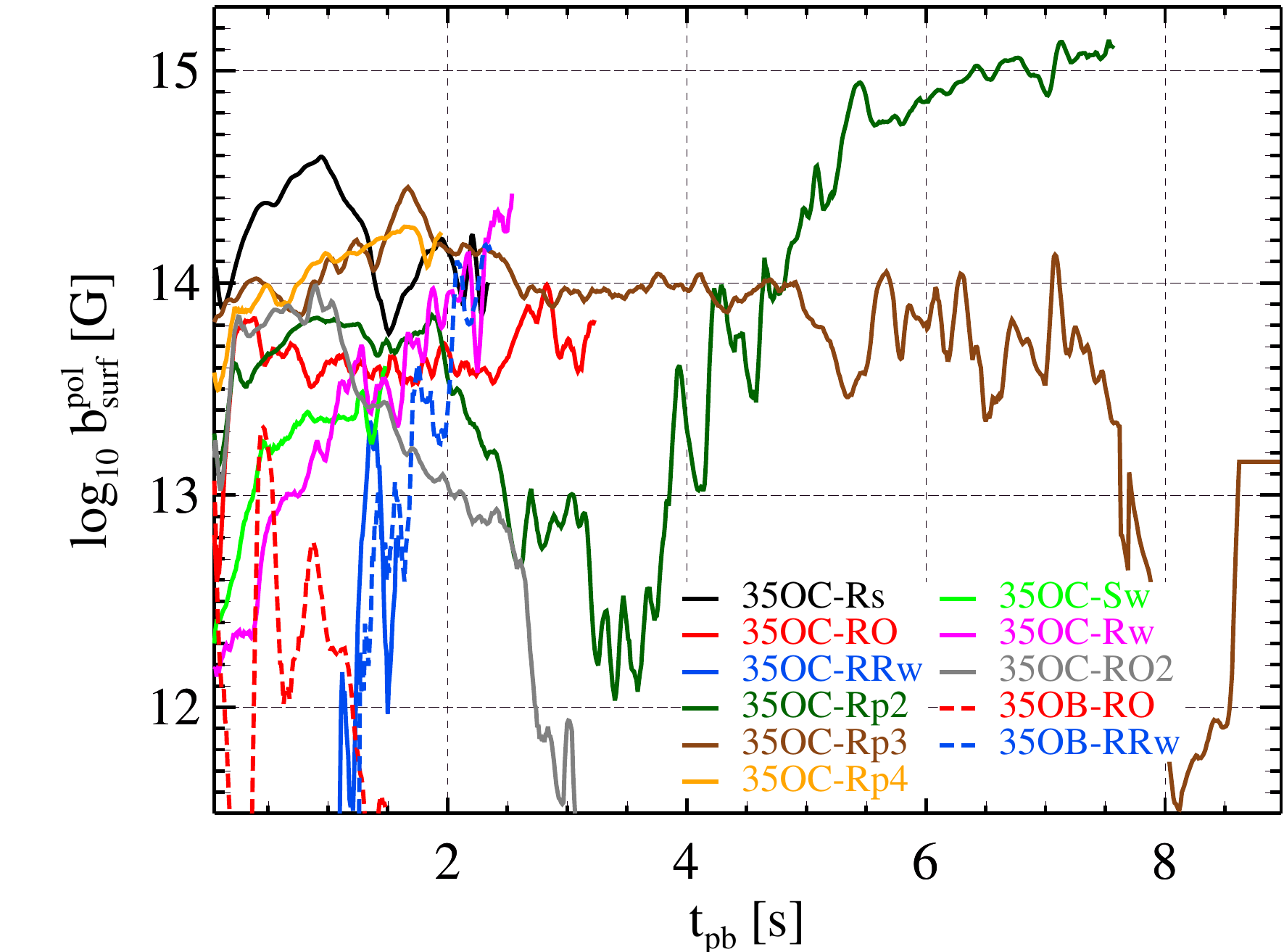}
  \includegraphics[width=0.32\linewidth]{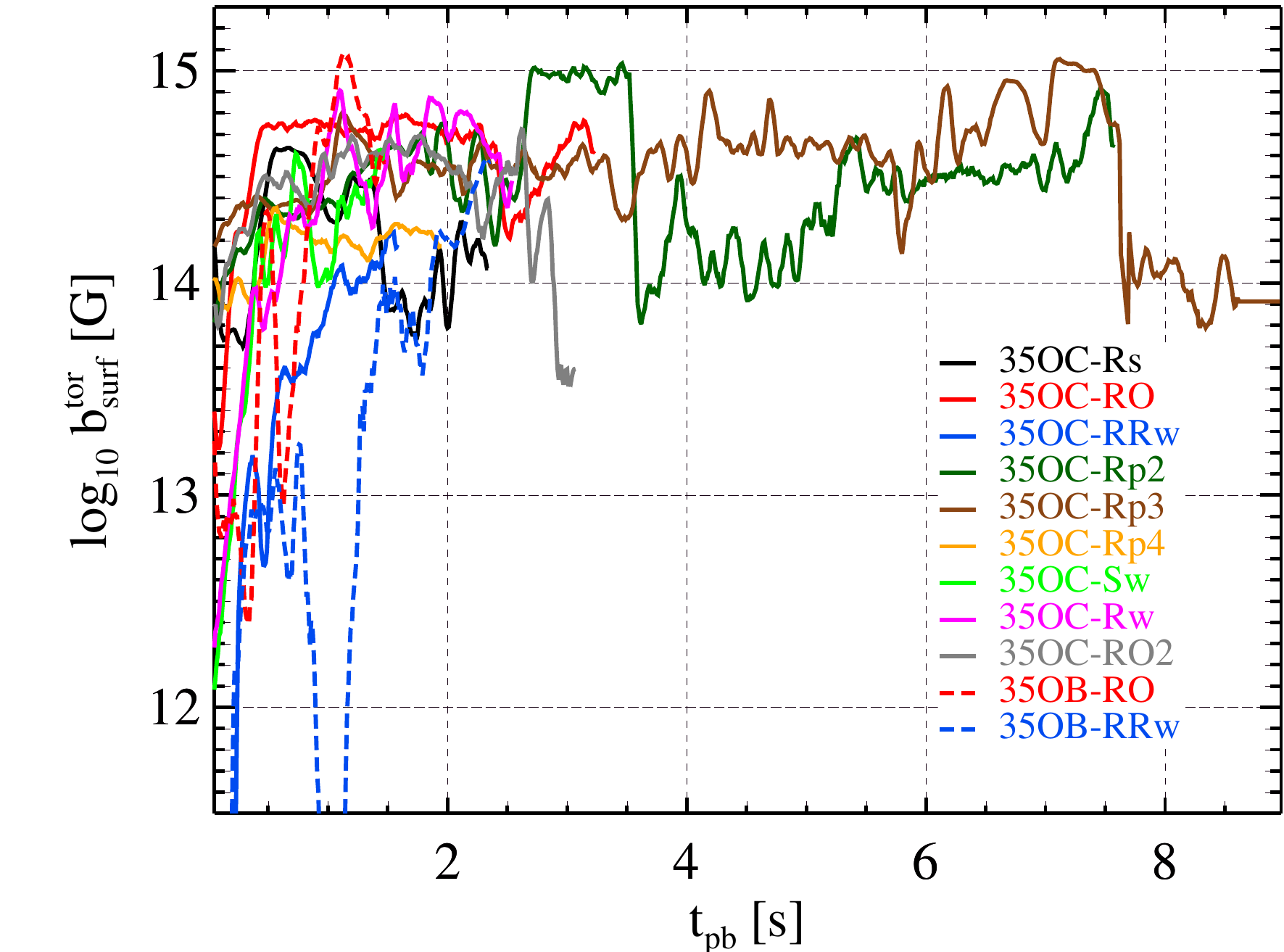}
  }
  }
     }%
   \node[fill=white, opacity=1, text opacity=1] at (-8.29,+3.99) {\large (a)};
   \node[fill=white, opacity=0, text opacity=1] at (-8.29,+0.49) {\large (d)};
   \node[fill=white, opacity=0, text opacity=1] at (-8.29,-3.1) {\large (g)};
   \node[fill=white, opacity=1, text opacity=1] at (-8.29,-6.7) {\large (j)};
   \node[fill=white, opacity=1, text opacity=1] at (-2.54,+3.99) {\large (b)};
   \node[fill=white, opacity=0, text opacity=1] at (-2.54,+0.49) {\large (e)};
   \node[fill=white, opacity=1, text opacity=1] at (-2.54,-3.1) {\large (h)};
   \node[fill=white, opacity=1, text opacity=1] at (-2.54,-6.7) {\large (k)};
   \node[fill=white, opacity=0, text opacity=1] at (+3.30,+3.99) {\large (c)};
   \node[fill=white, opacity=1, text opacity=1] at (+3.30,+0.49) {\large (f)};
   \node[fill=white, opacity=1, text opacity=1] at (+3.30,-3.1) {\large (i)};
   \node[fill=white, opacity=1, text opacity=1] at (+3.30,-6.7) {\large (l)};
   \node[fill=white, opacity=1, text opacity=1, rotate=90] at (+3.16,+4.99) {$\Erotpns/$};	
   \node[fill=white, opacity=1, text opacity=1, rotate=90] at (+3.13,-2.24) {$\Emag^\pnss/$};
  \end{tikzpicture}
  \caption{
    Selected global variables of our models. The variables shown are
    the time evolution of \banel{(a)} the mass of the PNS (marked
    with a grey horizontal line is the value of $\Mmax$), \banel{(b)}
    its mass accretion rate (with solid lines $\dot{M}_\pnss>0$,
    denoting mass gain by the PNS, and with symbols $\dot{M}_\pnss<0$,
    implying mass loss from the PNS), \banel{(c)} its rotational
    energy, \banel{(d)} the ratio rotational to gravitational energy
    of the PNS ($\betaG$; Eq.\,\eqref{eq:betaG}), \banel{(e)} the
    ratio magnetic to rotational energy ($\epsilonB$;
    Eq.\,\eqref{eq:epsilonB}), 
    \banel{(f)} the angular momentum of the PNS,
    \banel{(g)} the corresponding
    dimensionless spin parameter (Eq.\,\eqref{eq:aKerr}), \banel{(h)} the
    power-law index $q_\pnss$ (Eq.\,\eqref{eq:qpns}), \banel{(i)}
    the magnetic energy of the PNS , \banel{(j)} the PNS surface-averaged
    rotational frequency, \banel{(k)} the PNS surface-averaged
    poloidal magnetic field component, and \banel{(l)} the PNS surface-averaged
    toroidal magnetic field component.
  }
  \label{Fig:globalvars-1}
\end{figure*}

\subsection{BH forming models}
\label{sSek:BHforming}

Firstly, we show the most salient properties of the dynamical
evolution of models which either form a BH during the computed period
of time or may likely form one if they were evolved long enough. We
consider \emph{collapsar candidates} or proto-collapsars (PCs) all
models that collapse (or may shortly collapse) to a BH surrounded by
matter with sufficient angular momentum to form an accretion disc,
irrespective of the spin parameter of the BH itself.

\subsubsection{Model \modelname{35OC-RO} with the original magnetic field}
\label{sSek:35OC-RO}

\begin{figure}
  \centering
  \begin{tikzpicture}
  \pgftext{%
    \includegraphics[width=\linewidth]{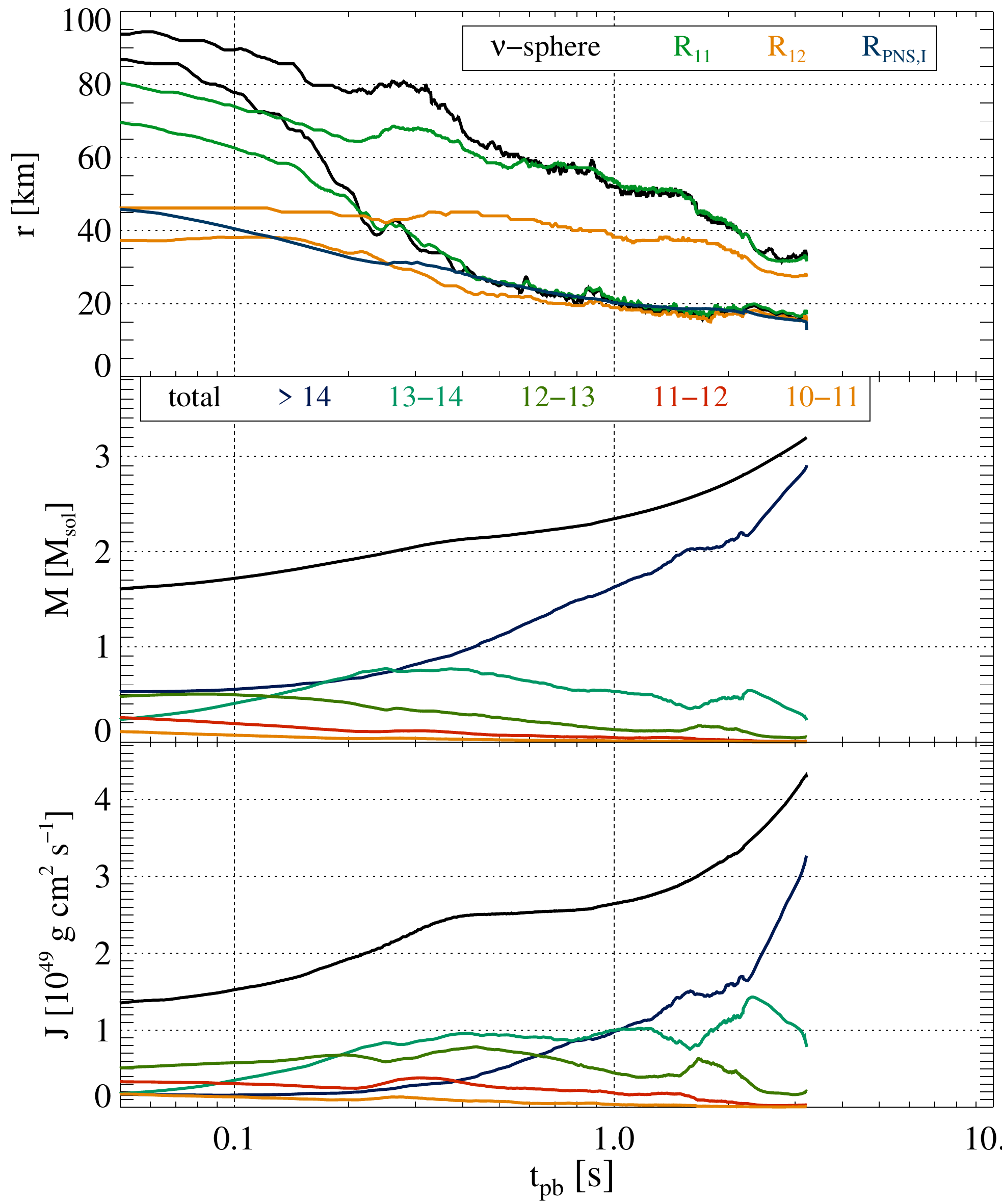}
       }%
   \node at (-4,+1.95) {\large (a)};
   \node at (-4,-1.1) {\large (b)};
   \node at (-4,-4.15) {\large (c)};
  \end{tikzpicture}
   \caption{
     Further global quantities of
     \modl{35OC-RO}.  
     \panel{a}: Evolution of three proxies for the PNS radii: the
     black, green, and orange pairs of lines display the maximum and
     minimum radii of the electron-\nusp, and the iso-density surfaces
     of $\rho = \zehn{11} \, \gccm$ and $\rho = \zehn{12} \, \gccm$,
     respectively.  We also include a (blue) line showing the
     evolution of $R_{\pnss\textsc{,i}}$ (Eq.\,\eqref{eq:Rpnssphequivalent}).
    \panel{b}: mass of all matter with densities in different
    ranges as indicated in the legend.
    \panel{c}: same as panel \panel{b}, but for angular
    momentum.
  }
  \label{Fig:35OC-RO-globals}
\end{figure}

\begin{figure}
  \centering
  \includegraphics[width=\linewidth]{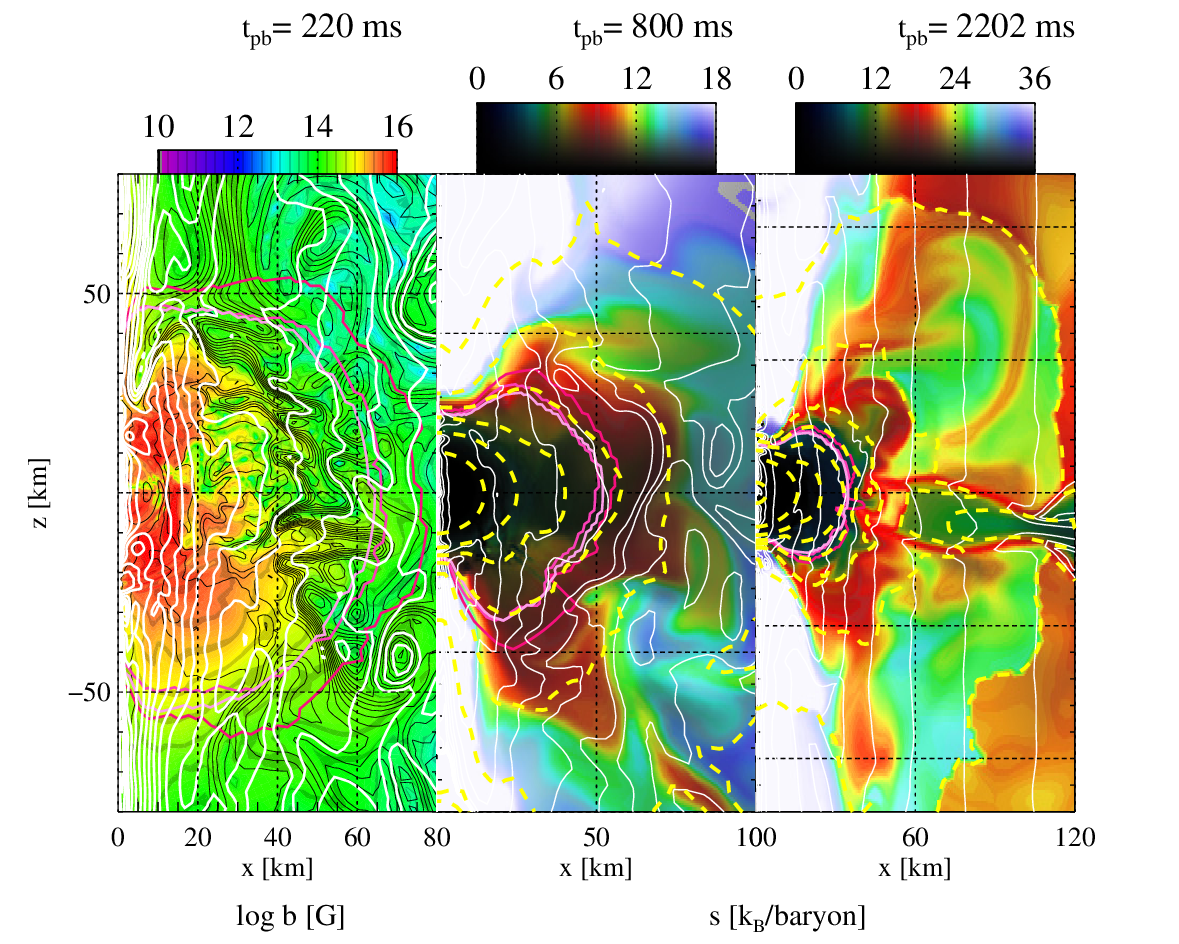}
  \caption{
    Structure of \modl{35OC-RO} at different times.  The left panel 
    shows maps of the logarithm of the magnetic field 
    strength, contours of the angular velocity (white lines), and
    magnetic field lines (black).  The colour in the central and right panels represents the specific entropy, while white
    and yellow lines show contours of the angular velocity and of the
    density (the innermost iso-density contour corresponds to
      $\rho=\zehn{14}\, \gccm$ and the rest of the contours are spaced by
      factors of 10 in decreasing order).
    The pink lines in all panels display the three neutrinospheres and are a proxy for the location of the PNS surface.
  }
  \label{Fig:35OC-RO-2d-3}
\end{figure}

We evolved \modl{35OC-RO}, using the rotational profile and the
magnetic field of the stellar evolution model, from the onset of
collapse to a final time of $\tpb = 3.22 \, \sek$ when the core forms
a BH.  The model launches an explosion relatively shortly after bounce
with a shock runaway that sets in at $\tpb \approx 150 \, \ms$.
Afterwards, the PNS continues to accrete mass until the end of the
simulation (see \figref{Fig:globalvars-1}\panel{a}).

\paragraph*{Overview of the post-bounce dynamics.}
The phase until $\tpb \approx 150 \, \ms$ is characterised by
accretion through the shock at all latitudes and, hence, a growth of
the PNS mass at an initially large, albeit decreasing rate of
$\dot{M}_\pnss > 1 \, \msolsec$ (see
\figref{Fig:globalvars-1}\panel{b}). After the onset of the explosion,
the mass accretion rate remains positive, though it decreases
considerably after the accretion of the surface of the inner core at a
mass coordinate of $M_{\mathrm{Fe}} \approx 2.1 \, \msol$
(\figref{Fig:globalvars-1}\panel{b}).  The PNS takes about
$1.5 \, \sek$ to reach a mass in excess of the maximum cold,
non-rotating PNS mass for our EOS $M_\pnss>\Mmax=2.45\Msol$
(\figref{Fig:globalvars-1}\panel{a}; see also below).  As the PNS
contracts and its density increases, most its mass resides in the
interior shells of higher density.  At $\tpb = 1.5 \, \sek$, for
instance, almost $80 \, \%$ of the PNS mass have a density higher than
$\zehn{14} \, \gccm$, and the mass contained in the surrounding shells
decreases with their density, but remains significant for all shells
throughout the evolution (\figref{Fig:35OC-RO-globals}\panel{b}).
The high rotational energy leads to a notable deformation after core
bounce as we displayed by the time evolution of the values of several
radii characterising the PNS, viz.~the electron-neutrino sphere and
the radii of the iso-density surfaces corresponding to
$\rho = 10^{11}$ and $10^{12} \, \gccm$
(\figref{Fig:35OC-RO-globals}\panel{a}). The evolution of the
electron-neutrino sphere radii parallels the evolution of the
iso-density surfaces corresponding to $10^{11} \, \gccm$, indicating
that both radii are excellent proxies of the actual PNS radius.
We also display in \figref{Fig:35OC-RO-globals}\panel{a} the time
evolution of the radius of a spherical and homogeneous configuration
with the same mass and moment of inertia than the PNS, defined as
\begin{align}
 R_{\pnss\textsc{,i}} := \left(\frac{5 I_\pnss}{2 M_\pnss}\right)^{1/2} .
  \label{eq:Rpnssphequivalent}
\end{align}
The time evolution of this radius shows a big similarity with the
polar radius tracking the isodensity surface of $10^{12} \, \gccm$,
specially after $t\sim 0.1\,$s. With relatively small variations, the
behaviour of $ R_{\pnss\textsc{,i}}$ is quite similar in nearly all
models considered in this paper.

Supported throughout the entire evolution partially by rotation, the
PNS possesses an oblate shape with maximum and minimum radii located
at the equator and at the poles, respectively,
(\figref{Fig:35OC-RO-2d-3}\panel{b} and \panel{c}).  The maximum
radius of the newly formed PNS exceeds its minimum radius by about
$10\,\%$. The asymmetry leaves an imprint in the neutrino burst, whose
total luminosity of $L_{\nu} \approx \zehnh{6.3}{53} \, \ergs$ shows
the same level of pole-to-equator asymmetry.
In the phase leading up to the onset of explosion, the deformation is
moderate, and so is the pole-to-equator difference of the neutrino
fluxes.  Nevertheless, this moderate degree of asymmetry (at
$\tpb = 150 \, \ms$, the neutrino flux along the poles exceeds that at
the equator by about $30\%$) is sufficient to focus enough of the
neutrino flux into cones around the poles and heat the gas efficiently
enough to revert the infall. From this moment on, the accretion onto
the PNS proceeds predominantly through the equatorial region.

The kinetic energy is dominated by the rotational energy in our
models.  As the PNS accretes matter, its rotational energy
($\Erotpns$) grows (\figref{Fig:globalvars-1}\panel{c}), leading to an
increase of the degree of asphericity.
The ratio of rotational ($\Erotpns$) to gravitational potential
binding energy ($\Egrav^\pnss$) in the PNS
\begin{align}
  \betaG:=\Erotpns/|\Egrav^\pnss| ,
  \label{eq:betaG}
\end{align}
tends to increase non-monotonically with time. The growth first levels
off approximately when $\betaG \sim 0.035 $, but later
($\tpb \gtrsim 2 \, \sek$) continues to a final value of
$\betaG \sim 0.04$ (see panel \panel{d} of \figref{Fig:globalvars-1}).
The magnetic energy ($\Emag$; \figref{Fig:globalvars-1}\panel{i})
increases at a rate smaller than that of $\Erotpns$, as can be
observed from the decreasing trend of the ratio $\epsilonB$
(\figref{Fig:globalvars-1}\panel{e}), defined as
\begin{gather}
  \epsilonB:=\Emag^\pnss /\Erotpns . \label{eq:epsilonB}
\end{gather}

\paragraph*{Stability of the hypermassive PNS.}
\label{par:stability}
The mass of the PNS at the time of BH formation by far exceeds,
$\Mmax$, value that our model \modelname{35OC-RO} exceeds after
$\tpb\sim 1.2\,$s. After that time, the PNS is stabilised by a fast
rotation.
We cannot disregard that the produced PNS may be unstable to
non-axisymmetric instabilities \citep{Andersson_2003CQGra..20R.105},
resulting in an earlier collapse to form a BH in 3D (the evolution
computed in 3D for this model is still insufficient to make any
forecast of its final fate; see Sec.\,\ref{Sek:3d}). Among the former
instabilities, the existence of dynamical bar modes, which happen when
the ratio $\betaG$ becomes large enough \citep[$\betaG>0.24-0.25$
according to][]{Shibata_2000ApJ...542..453} has been throughly studied
in the literature \citep[see also,
\eg][]{Watts_2005ApJ...618L..37,Saijo_2006MNRAS.368.1429}. However, in
magnetised, differentially rotating polytropes (with properties
resembling cold neutron stars), \cite{Franci_2013PhRvD..88j4028} find
that the dynamical bar mode instability is largely suppressed if the
magnetic field is able to grow to very large values
$\gtrsim \zehn{16}\,$G \citep[a result also found
by][]{Camarda_2009ApJ...707.1610}. Moreover,
\cite{Fujisawa_2015MNRAS.450.4016} points out that if a magnetised
polytrope possesses a high degree of differential rotation, the
toroidal magnetic field wound up from the poloidal one becomes highly
localised near the rotational axis. As a result, the ``low-$\betaG$''
instability, which happens for highly differentially rotating
non-magnetised stars for $\betaG\sim 0.14$
\citep[\eg][]{Centrella_2001ApJ...550L.193}, is more efficiently
suppressed than that of stars without differential rotation and
toroidal magnetic field. 
It must be noted that the toroidal magnetic field configurations considered in \cite{Fujisawa_2015MNRAS.450.4016} are strictly symmetric with respect to the equatorial plane, while we do not enforce equatorial symmetry in our models. This topological difference may impact the equilibrium states discussed by \cite{Fujisawa_2015MNRAS.450.4016}. Furthermore, 
the initial poloidal
magnetic field is $\sim 50$ times weaker than the toroidal one in \modl{35OC-RO}. 
Thus, the suppression of the ``low-$\betaG$'' instability is not guaranteed by
the mechanism suggested by
\cite{Fujisawa_2015MNRAS.450.4016}. 
Nevertheless, we observe a
significant rise of the toroidal field around the rotational axis
(where the poloidal and toroidal components reach typical values $\gtrsim \zehn{15}\,$G; see also \figref{Fig:35OC-RO-2d-3}).  This enhancement of the toroidal field
around the axis helps stabilising the 3d versions of models
\modelname{35OC-RO} and \modelname{35OC-Rs} (Sec.\,\ref{Sek:3d}). 
The general relativistic MHD simulations of
\cite{Muhlberger_2014PhRvD..90j4014} pointed out that when the total
magnetic to total kinetic energy ratio is as small as
$\epsilonB\simeq \zehnh{5.6}{-3}$ the ``low-$\betaG$'' instability is
significantly suppressed. In \modl{35OC-RO},
$\epsilonB \simeq \zehnh{9}{-4}$ at the end of the computed evolution,
with maximum values, $\epsilonB \lesssim 0.1$ reached soon after
collapse (\figref{Fig:globalvars-1}\panel{e}). Hence, in this
particular model it is likely that the ``low-$\betaG$'' instability is
significantly damped. 
As a matter of fact, the 3D version of \modl{35OC-RO} does not
  display signs of a strong instability during the first second of
  post-bounce evolution during which the values of $\betaG$ are even
  smaller than in 2D (see Sec.\,\ref{Sek:3d}). Moreover, the values of
  $\epsilonB$ in the 3D counterpart of \modl{35OC-RO} show that the
  PNS of the former is as magnetised as that of the latter. Hence, our
  results (both in 2D and in 3D) support to the possibility that
  sufficiently magnetised collapsed cores tend to hinder large changes of the PNS structure due to the "low-$\betaG$" instability.
\cite{Muhlberger_2014PhRvD..90j4014}  further found that the
\emph{numerical} growth rate of the instability is quite sensitive to
the formal order of the convergence of the numerical method:
high-order methods (\eg WENO5) render considerably slower growth rates
than low-order ones.  Since our models are computed also with a
fifth-order accurate intercell reconstruction, we also expect the
``low-$\betaG$'' instability to be marginally growing in our models.

\paragraph*{Properties of the remnant at the brink of BH collapse.}
The mass of the PNS at the time of BH formation,\footnote{Model
  \modelname{35OC-RO} has been run longer than in
  \citetalias{Obergaulinger_Aloy__2019}, until BH formation.}
$M_\pnss \approx 3.19 \, \msol$, can be taken as the initial BH mass,
which may latter grow due to fall-back accretion. We note that the
mass infall rate onto the PNS remains nearly constant during the last
$\approx 2\, \sek$ prior to BH collapse
(\figref{Fig:globalvars-1}\panel{b}). At the same time the
dimensionless spin parameter
\begin{gather}
  a_{\pnss} := c J_\pnss / (G M_\pnss^2),
  \label{eq:aKerr}
\end{gather}
moderately decreases up to a final value $a_{\pnss} \approx 0.48$
(\figref{Fig:globalvars-1}\panel{g}). This value is close to the
expectations of
\cite{Woosley_Heger__2006__apj__TheProgenitorStarsofGamma-RayBursts},
who predict $a\simeq 0.53$ for \modl{35OC-RO}.  Thus, the BH is not
extremely rapidly rotating.  The specific angular momentum of the
layers, still to be accreted by it, may nevertheless be sufficiently
large for an accretion torus to form after
$t_{\textsc{df}}\simeq 9.3\,$s (Tab.\,\ref{Tab:models}) and (likely)
increase the BH spin to significantly larger values. Hence, a
collapsar engine may eventually result from this model.

\begin{table*}
  \centering
  \begin{tabular}{l | cc | ccc | ll | lllllllll}
    \hline
    name & $t_{\rm max}$[s] & $t_{\bhs}$[s] &$\Omega_{\mathrm{surf}}$[Hz] & $\bar{\Omega}$[Hz] & $\bar{P}$[ms]&$b_{\mathrm{surf}\textsc{,14}}^{\mathrm{tor}}$
    & $b_{\mathrm{surf}\textsc{,14}}^{\mathrm{pol}}$& $\Tpns$ & $J^{\pnss}_{\textsc{49}}$& $a_{\pnss}$&$\beta_{\textsc{g,-2}}$ & $\epsilon_{\textsc{b,-2}}$&$I^{\pnss}_{\textsc{45}}$& $\frac{M_\pnss}{\msol}$
    \\
    \hline
    \modelname{35OC-RO} & $ 3.23 $ & $ 3.23 $ & $ 3003 $ & $ 7529 $ & $ 0.83 $ & $ 4.28 $ & $ 0.71 $ & $16.85 $ & $  4.31 $ & $ 0.48 $ & $ 3.99 $ & $ 1.16 $ & $ 5.72 $ & $ 3.19 $ \\
   \modelname{35OC-RO2} & $ 3.07 $ & $ 3.07 $ & $ 2828 $ & $ 6023 $ & $ 1.04 $ & $ 0.43 $ & $ 0.00 $ & $12.68 $ & $  3.04 $ & $ 0.41 $ & $ 3.88 $ & $ 0.04 $ & $ 5.04 $ & $ 2.90 $ \\
   \modelname{35OC-Rp2} & $ 7.57 $ & $ \times $ & $ 2087 $ & $ 2962 $ & $ 2.12 $ & $ 1.76 $ & $ 9.51 $ & $ 1.09 $ & $  0.70 $ & $ 0.15 $ & $ 0.42 $ & $ 0.60 $ & $ 2.35 $ & $ 2.33 $ \\
   \modelname{35OC-Rp3} & $ 8.96 $ & $ \times $ & $ 3712 $ & $ 5530 $ & $ 1.14 $ & $ 1.33 $ & $ 0.15 $ & $ 5.31 $ & $  1.91 $ & $ 0.42 $ & $ 3.12 $ & $ 0.04 $ & $ 3.46 $ & $ 2.28 $ \\
   \modelname{35OC-Rp4} & $ 1.95 $ & $ \times $ & $ 1539 $ & $ 2561 $ & $ 2.45 $ & $ 1.55 $ & $ 1.87 $ & $ 1.39 $ & $  0.98 $ & $ 0.33 $ & $ 1.46 $ & $ 0.32 $ & $ 3.84 $ & $ 1.85 $ \\
    \modelname{35OC-Rw} & $ 2.54 $ & $ >2.54$ & $ 1212 $ & $ 3185 $ & $ 1.97 $ & $ 3.03 $ & $ 4.09 $ & $ 5.74 $ & $  2.51 $ & $ 0.46 $ & $ 2.61 $ & $ 1.20 $ & $ 7.86 $ & $ 2.49 $ \\
    \modelname{35OC-Rs} & $ 2.34 $ & $ \times $ & $   -5 $ & $  180 $ & $34.8 $ & $ 1.38 $ & $ 1.30 $ & $ 0.22 $ & $  0.55 $ & $ 0.25 $ & $ 1.13 $ & $28.74 $ & $30.60 $ & $ 1.58 $ \\
    \modelname{35OC-Sw} & $ 1.48 $ & $ 1.48 $ & $ 1384 $ & $ 2516 $ & $ 2.50 $ & $ 3.66 $ & $ 0.43 $ & $ 0.93 $ & $  0.63 $ & $ 0.13 $ & $ 0.31 $ & $ 1.35 $ & $ 2.50 $ & $ 2.38 $ \\
   \modelname{35OC-RRw} & $ 1.59 $ & $\times$ & $ 2685 $ & $ 5051 $ & $ 1.24 $ & $ 0.92 $ & $ 0.05 $ & $11.78 $ & $  4.46 $ & $ 0.72 $ & $ 7.43 $ & $ 0.03 $ & $ 8.83 $ & $ 2.64 $ \\
\hline
    \modelname{35OB-RO} & $ 1.48 $ & $ 1.48$ & $ 2891 $ & $ 5344 $ & $ 1.18 $ & $ 3.11 $ & $ 0.00 $ & $ 3.96 $ & $  1.44 $ & $ 0.26 $ & $ 1.27 $ & $ 0.08 $ & $ 2.70 $ & $ 2.49 $ \\
   \modelname{35OB-RRw} & $ 2.36 $ & $ >2.36 $ & $  691 $ & $ 1649 $ & $ 3.81 $ & $ 3.06 $ & $ 1.31 $ & $16.83 $ & $ 10.22 $ & $ 1.01 $ & $ 8.36 $ & $ 1.92 $ & $61.94 $ & $ 3.39 $ \\
    \hline
  \end{tabular}
  \caption{Properties of the models we have built in this paper. From
    left to right the columns display the maximum computed post-bounce
    time ($t_{\rm max}$), the time of BH formation after core collapse
    (if no BH is expected, we use a $\times$ sign), the angularly
    averaged rotational speed of the PNS surface,
    $\Omega_{\mathrm{surf}}$, the rotational frequency of the PNS,
    $\bar{\Omega}=J_{\pnss}/I_{\pnss}$, the associated period,
    $\bar{P}=2\pi/\bar{\Omega}$, the toroidal and poloidal magnetic
    fields at the surface of the PNS in units of $\zehn{14}\,$G,
    $b_{\mathrm{surf}\textsc{,14}}^{\mathrm{tor}}$,and
    $b_{\mathrm{surf}\textsc{,14}}^{\mathrm{pol}}$, respectively, the
    dimensionless specific angular momentum of the PNS, $a_\pnss$
    (Eq.\,\eqref{eq:aKerr}), the rotational energy of the PNS in units
    of $\zehn{52}\,$erg, $\Tpns$, the ratios $\beta_{\textsc{g}}$
    (Eq.\,\eqref{eq:betaG}) and $\epsilonB$ (Eq.\,\eqref{eq:epsilonB})
    in units of $\zehn{-2}$, the moment of inertia of the PNS in units
    of $\zehn{45}\,$g\,cm$^{2}$ and the PNS mass.  All quantities have
    been measured at $\tpb=t_{\rm max}$. In case a model finishes with
    the formation of a BH, the listed values of $a_\pnss$ and
    $M_\pnss$ correspond to the initial values of these quantities for
    the just born BH.  }
  \label{Tab:model-properties}
\end{table*}

\subsubsection{Model \modelname{35OC-RO2} with twice stronger magnetic
field}
\label{sSek:35OC-RO2}

\begin{figure}
  \centering
  \includegraphics[width=\linewidth]{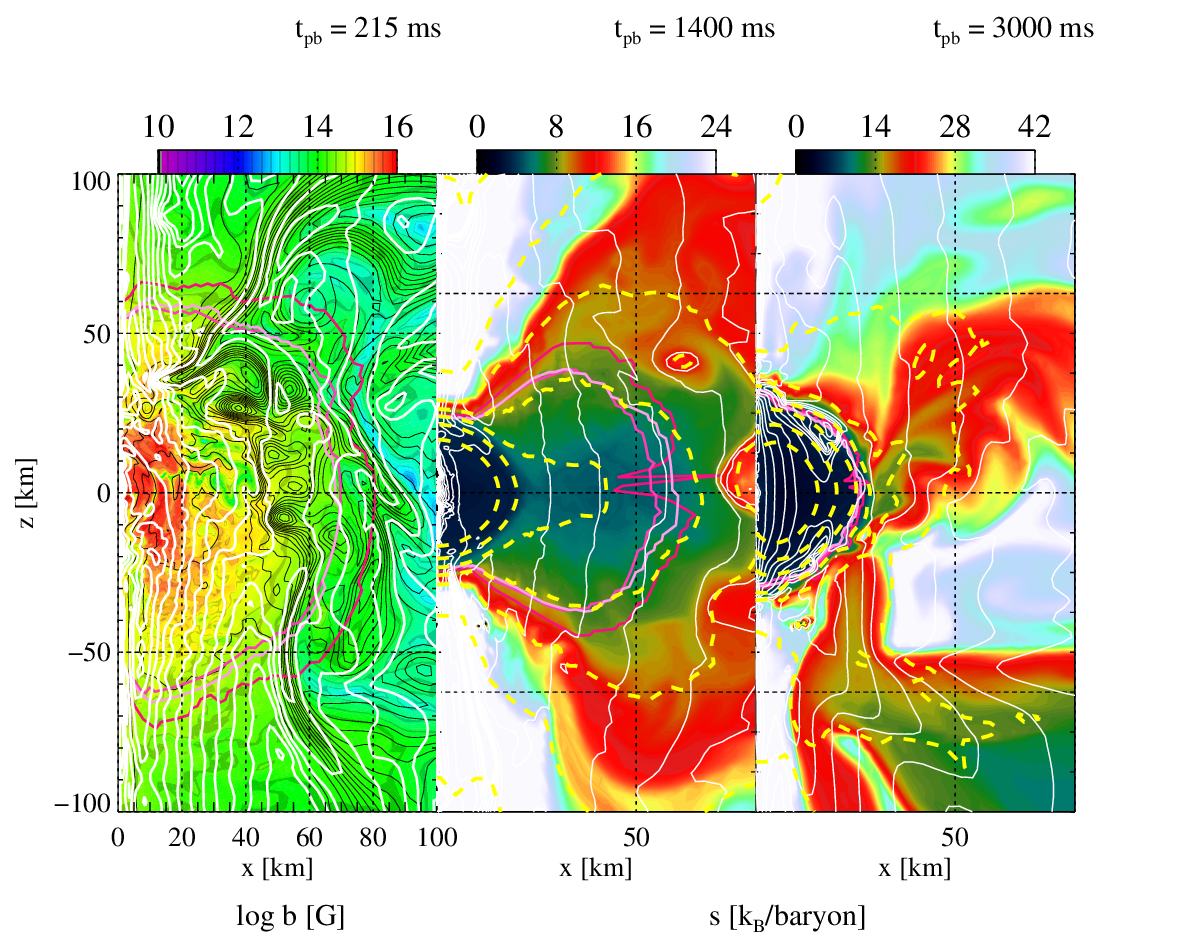}
  \caption{
    Same as in \figref{Fig:35OC-RO-2d-3}, but for
      \modl{35OC-RO2}.
    }
  \label{Fig:35OC-RO2-2dpns}
\end{figure}

Model \modelname{35OC-RO2} begins its time evolution with the same
rotational profile as the progenitor star 35OC, but with poloidal and
toroidal magnetic fields artificially increased by a factor 2 with
respect to the stellar progenitor (Tab.\,\ref{Tab:models}).  This
model forms a BH after $t_\bhs \simeq 3.07\,\sek$. By that time, the
PNS mass has grown well beyond $\Mmax$
(\figref{Fig:globalvars-1}\panel{a} and
Tab.\,\ref{Tab:model-properties}), and the mass accretion rate on the
PNS does not show signatures of saturation or decrease, maintaining a
level of $\gtrsim 0.3\msol \sek^{-1}$ during the whole evolution, and
even increasing above $\gtrsim 0.5\msol \sek^{-1}$ after
$\tpb \sim 2.5\,$s (\figref{Fig:globalvars-1}\panel{b}).

The time evolution of this model shares many similarities with model
\modelname{35OC-RO}, among them the fact that the explosion is
magneto-rotationally driven. However, the growth of the PNS mass after
the accretion of the Sillicon core happens at a smaller rate than in
the former model (\figref{Fig:globalvars-1}\panel{a}). Compared to
\modl{35OC-Rp2}, with the same poloidal initial field, the
post-collapse dynamics of both models is significantly different
(indeed, \modl{35OC-Rp2} does not form a BH during the computed
evolution; see Sec.\,\ref{sSek:35OC-Rp2-Rp3-Rp4}). Since the main
difference between models \modelname{35OC-RO} and \modelname{35OC-RO2}
is the twice larger toroidal magnetic field in the latter, we may
preliminary conclude that a moderate increase of the toroidal magnetic
field in the progenitor star does not alter the prospects for BH
formation.

A detailed look at the evolution of the structure of the core further
emphasises the parallelism to \modl{35OC-RO} (see
\figref{Fig:35OC-RO2-2dpns}): we find an early development of magnetic
channels ($\tpb = 215 \, \ms$, left) as well as a PNS that
intermittently has a very large equatorial extent
($\tpb = 1.4 \, \sek$, middle) and later contracts to a much smaller
size, while maintaining an oblate shape ($\tpb = 3 \, \sek$, right).

\subsubsection{\Modl{35OC-Sw}: slower rotation and weak field}
\label{sSek:35OC-Sw}

\begin{figure}
  \centering
  \includegraphics[width=\linewidth]{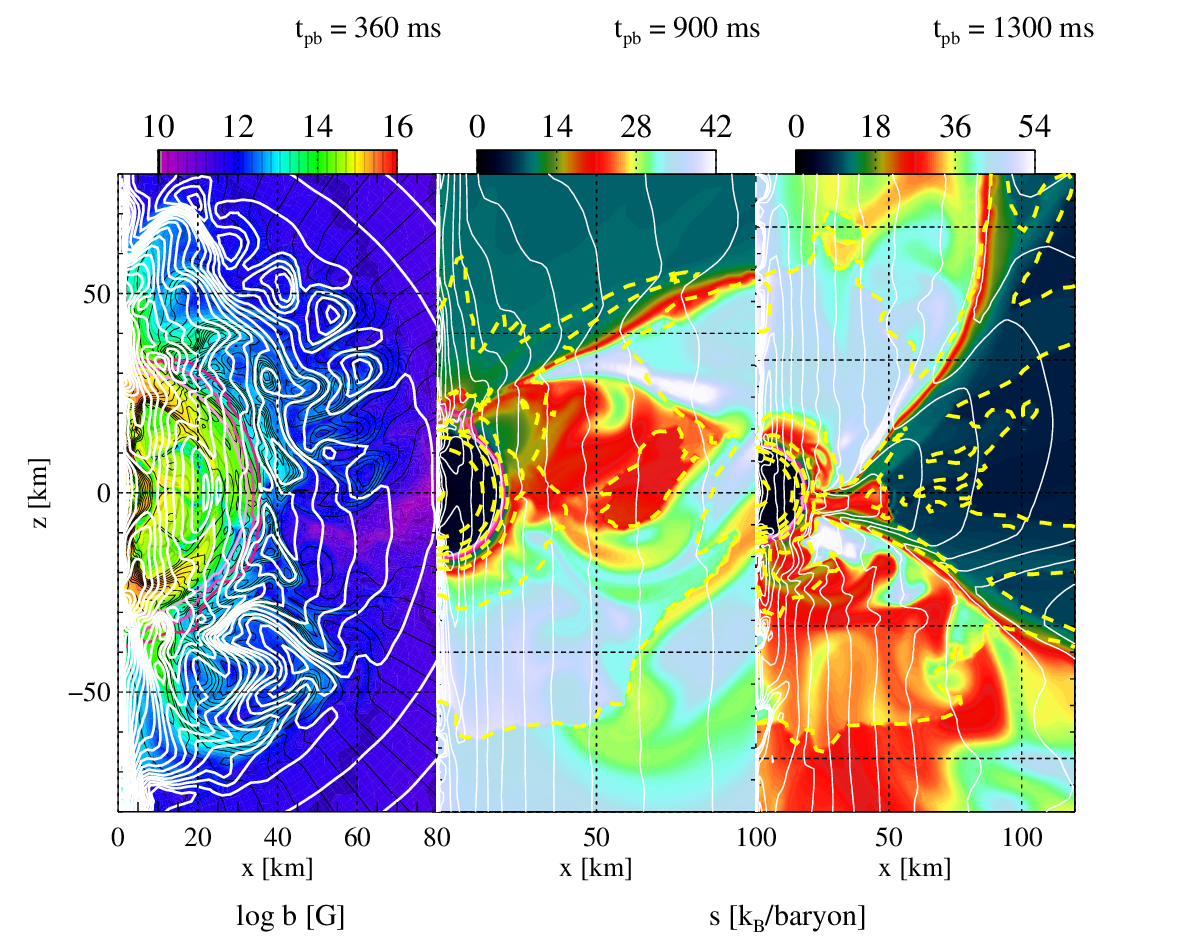}
  \caption{
    Same as \figref{Fig:35OC-RO-2d-3}, but for \modl{35OC-Sw}.
  }
  \label{Fig:35OC-Sw-2d}
\end{figure}

\Modl{35OC-Sw} combines a relatively weak field consisting of a global
dipole ($b^{\rm pol}_{\rm max}=\zehn{8}\,$G) and a toroidal component
($b^{\rm tor}_{\rm max}=\zehn{10}\,$G) with a comparably slow
rotational profile obtained by globally reducing the angular velocity
of the original stellar-evolution model by a factor 4. Admittedly,
this combination of slow rotation and weak magnetic field, departs
significantly from the original properties of the 35OC stellar
core. However, this model aims to \emph{complete} a region of the
parameter space uncovered with the rest of the models in this
paper. While this reduction does not make rotation irrelevant
($\Erotpns \approx \zehnh{9}{51} \, \erg$ at the end of the
simulation; see Tab.\,\ref{Tab:model-properties}), its influence is
far less notable, and the explosion mechanism differs considerably
from the models discussed above.  Furthermore, the magnetic field
remains mostly toroidal throughout the computed evolution, with only a
very weak addition of a poloidal field (panels \panel{k} and \panel{l}
of \figref{Fig:globalvars-1}). The reduced magnetic and rotational
energy of this model with respect to the original stellar progenitor
values induce a standard neutrino-driven supernova explosion aided by
hydrodynamic instabilities (mostly SASI), though with significantly
north-south asymmetric and collimated ejecta (see
\citetalias{Obergaulinger_Aloy__2019}).  The dynamics within the gain
region where the ejecta are accelerated are highly variable, much more
so than in the other models, translating into fluctuating fluxes of
mass and energy in each of the hemispheres and also fluctuating
locations of the accretion streams feeding the PNS.

The ongoing presence of strong accretion streams causes the PNS to
steadily grow in mass until it finally collapses to a BH at
$t_{\textsc{bh}}\simeq 1.48\,$s. At the time of BH formation the
compact remnant has a mass $M\simeq 2.4\Msol$
(Tab.\,\ref{Tab:model-properties}).  This value is slightly smaller
than $\Mmax$, and results as a consequence of the approximated
treatment of the general relativistic effects in our method.  The
minor influence of rotation prevents any of the effects that in other
models (\modls{35OC-RO} and \modelname{35OC-Rs}) lead to a rather
extended PNS envelope.  Instead, the PNS is comparably compact with a
radius at BH collapse of $R_{\mathrm{PNS}} \approx 18 \, \km$ and only
a small degree of asphericity.  The total angular momentum, strongly
concentrated at the very centre, corresponds to a spin parameter
$a_\pnss \approx 0.13$.  This is fairly small and we anticipate that
it is probably insufficient to yield a GRB progenitor. In this model,
the specific angular momentum of the star (see
\figref{Fig:progenitor-profile}\panel{b}) is insufficient to produce
an accretion disc within the first $\tpb \sim 40\, \sek$
($\tdf \approx 43\,\sek$; \tabref{Tab:models}). Hence, it is unlikely
that \modl{35OC-Sw} may produce a \emph{standard} collapsar engine.

\subsubsection{\Modl{35OB-RO}}
\label{sSek:35OB-RO}

\Modl{35OB}, based on a relatively high mass-loss rate, differs from
\modl{35OC} in the structure of the core.  The former one possesses a
fairly weak density and entropy jump located at a slightly higher mass
coordinate than the surface of the iron core of \modl{35OC-RO}.  In
contrast to core \modelname{35OC}, the rotational profile is
continuous there, leading to rapid rotation outside the density jump.

In \modl{35OB-RO} we also used the magnetic field from the pre-SN
progenitor.  This implies that the maximum poloidal and toroidal
magnetic field strengths in the initial model are
$b^{r}_{\rm max}=\zehnh{7}{10}\,$G and
$b^{\phi}_{\rm max}=\zehnh{9}{11}\,$G, respectively. The high mass
loss due to stellar winds translates into a comparably low rotational
energy. Hence, the influence of rotation on the evolution is
subdominant, leading to small degrees of anisotropy of the PNS and of
the neutrino emission; magnetic fields do not play any notable role.

The PNS accretes matter until it collapses to a BH at
$t_{\textsc{bh}} \approx 1.48 \, \sek$
(Tab.\,\ref{Tab:model-properties}).  With only a minor degree of
rotational support, its final mass before BH collapse is
$M_\pnss\simeq 2.49\Msol \gtrsim \Mmax$.  The angular momentum ceases
to grow in the inner core at densities $\rho > \zehn{14} \, \gccm$ at
$\tpb \approx 0.8 \, \sek$.  Afterwards, we observe a decrease of the
total angular momentum of the PNS
(\figref{Fig:globalvars-1}\panel{f}), leading to a spin parameter
$a_\pnss \approx 0.26$ at BH collapse
(Tab.\,\ref{Tab:model-properties}).

\subsubsection{Models \modelname{35OB-RRw}  and \modelname{35OC-RRw} with supra-stellar rotation}
\label{sSek:35OC-RRw}

\begin{figure}
  \centering
  \includegraphics[width=\linewidth]{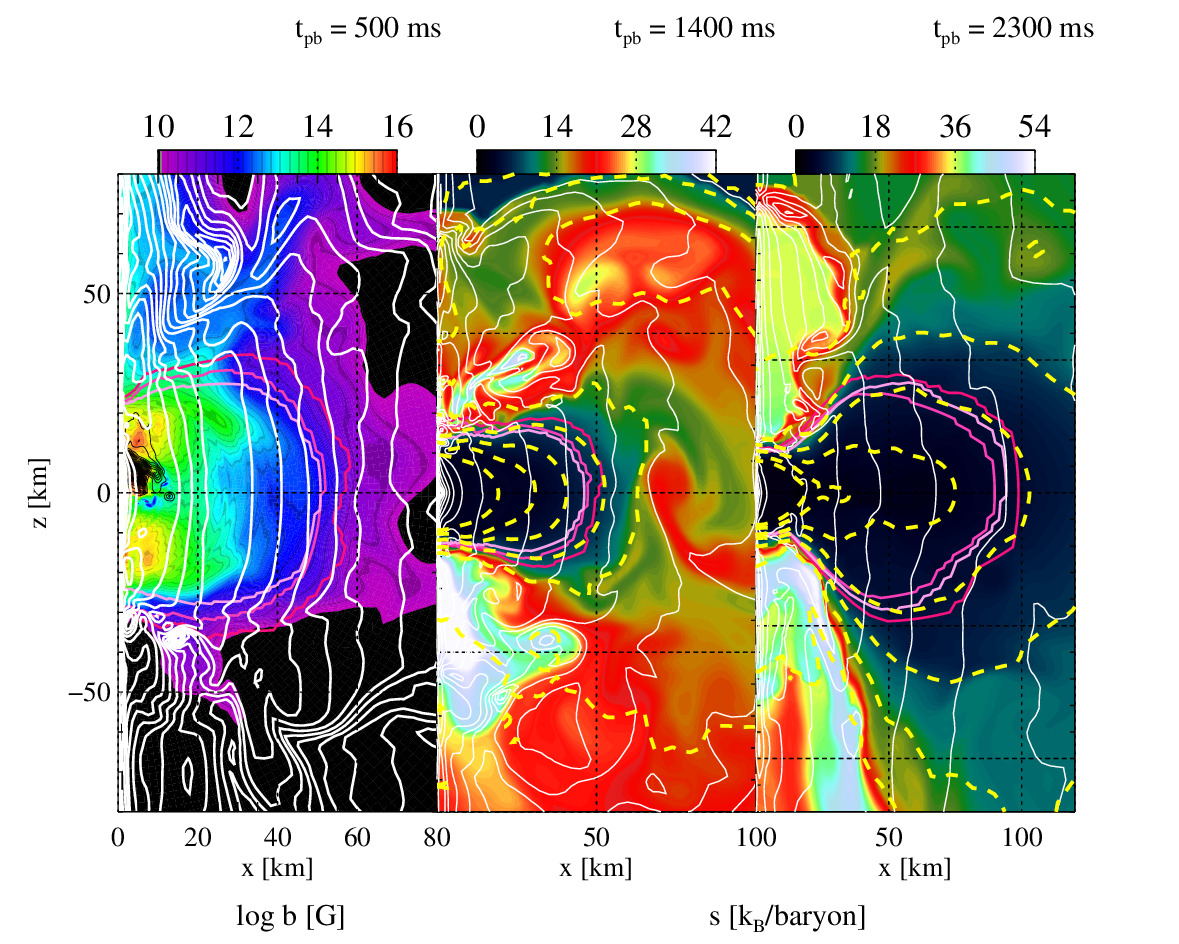}
  \caption{
    Same as in \figref{Fig:35OC-RO-2d-3}, but for
      \modl{35OB-RRw}.
  }
  \label{Fig:35OB-RRw-2dpns}
\end{figure}

We allow for another relatively small variation of the progenitor
stars 35OB and 35OC by considering a factor 2 and 1.5 increase of its
rotational speed in models \modelname{35OB-RRw} and
\modelname{35OC-RRw}, respectively. In order to allow for such an
increase in the rotational speed, we decrease drastically the magnetic
field strength, somehow mimicking the effect reduced magnetic torques
could operate in the stellar progenitor (Tab.\,\ref{Tab:models}). The
reduction of the magnetic field strength also serves to prevent a
prompt magnetorotational explosion (indeed, \modelname{35OB-RRw}
results in a failed SN explosion). Alternatively, one may explain the
increase of the rotational frequency of the progenitor core as the
result of, \eg a reduction of the mass loss rate during the stellar
evolution \citep[][built model 35OA with an iron core period that is
half the value of model 35OC by setting to zero the mass loss
rate]{Woosley_Heger__2006__apj__TheProgenitorStarsofGamma-RayBursts},
or the incorporation of the effects of wind anisotropies up to the
He-burning stage, since in this case
\cite{Meynet_Maeder_2007A&A...464L..11} predict faster rotating cores
than
\cite{Woosley_Heger__2006__apj__TheProgenitorStarsofGamma-RayBursts}. Interestingly,
there is an active debate on whether the evolution after the
He-burning phase may or may not significantly reduce the angular
momentum of the core \citep[see Sect.\,2 of][and references
therein]{Meynet_Maeder_2007A&A...464L..11}. Hence, the artificial
modification that we have included in the rotational speed of the iron
core, which may evolve somewhat \emph{detached} from the outer stellar
envelope is not totally inconceivable. This manipulation intends to
probe the impact of rotation on the dynamics.

The consequence of the faster rotation and hence enhanced rotational
support against gravity is that matter does not fall as deeply into
the gravitational well of the core as when using the original
rotational profile.  This effect, most pronounced in the equatorial
plane, leads to an oblate PNS with a particularly high axis ratio and
to a reduction of the neutrino luminosity induced by the accretion
onto the PNS, in comparison to \modl{35OB-RO}.  Rotation focuses the
neutrino emission strongly along the symmetry axis, creating
favourable explosion conditions there.  Nevertheless, even
taking into account the focusing effect, neutrino heating fails to
meet the conditions for shock revival.  The failure is in part owing
to the high mass accretion rate (larger for \modl{35OB-RRw} and for
\modl{35OC-RRw} throughout most of the evolution
\figref{Fig:globalvars-1}\panel{b}) and in part to the redistribution
of the energy deposited at the poles by lateral convective flows
acting on time scales below that of neutrino heating.  As a result, no
explosion takes place for more than $\sim2.3\,$s ($1\,$s) in the case
of \modl{35OB-RRw} (\modelname{35OC-RRw}) and the model exhibits a
steady increase of the PNS mass (\figref{Fig:globalvars-1}\panel{a}).
The very high mass reached by \modl{35OB-RRw},
$M(\rho \ge \zehn{12} \, \gccm) \approx 3 \, \msol$, by the end of the
simulation does not translate into an immediate collapse to a BH,
since the core is supported by centrifugal forces.  The development of
its structure can be followed in \figref{Fig:35OB-RRw-2dpns}.  The PNS
has a large axis ratio early on ($\tpb = 0.5 \, \sek$) that is
maintained as it contracts ($\tpb = 1.4 \, \sek$).  Later
($\tpb = 2.3 \, \sek$), the polar radius of the PNS is virtually
unchanged, but the equatorial region is characterised by an extended
dense torus-like configuration with an approximately cylindrical
rotational profile (thin white lines) surrounding the PNS.  In terms
of the thermodynamical state, the torus is a continuous extension of
the PNS, as we can see in the low entropies (colour scale).  This
structure corresponds to distributions in which the outer layers of
the PNS/torus system carry a relatively large and even increasing
amount of mass and angular momentum and to the largest spin parameters
($a_{\pnss} \ge 0.72$).

Based on the late stages of the simulation, we expect that the ongoing
accretion onto the PNS will eventually lead to the formation of a BH
in both models.  We cannot, however, provide the collapse time without
running the simulations (much) longer. It is, nevertheless, clear that
the moderate increase of the rotational rate of these two models has
not changed the type of compact remnant expected within seconds after
bounce.

\subsubsection{Model \modelname{35OC-Rw} with weak magnetic field}
\label{sSek:35OC-b0}

\begin{figure}
  \centering
  \begin{tikzpicture}
  \pgftext{%
  \includegraphics[width=\linewidth]{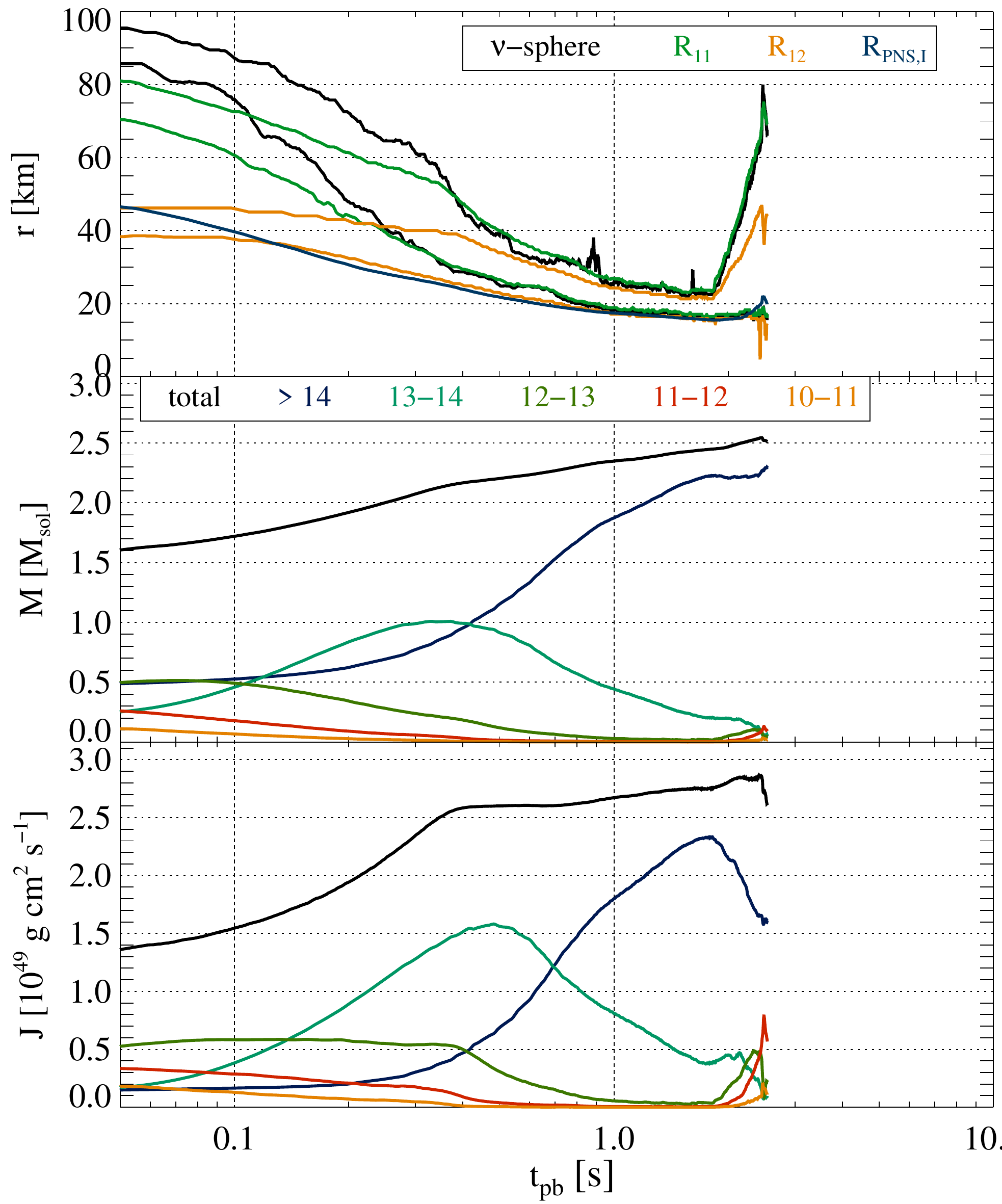}
     }%
   \node at (-4,+1.95) {\large (a)};
   \node at (-4,-1.1) {\large (b)};
   \node at (-4,-4.15) {\large (c)};
  \end{tikzpicture}
  \caption{Same as \figref{Fig:35OC-RO-globals}, but for \modl{35OC-Rw}.}
  \label{Fig:35OC-Rw-globals}
\end{figure}

\begin{figure}
  \centering
    \includegraphics[width=\linewidth]{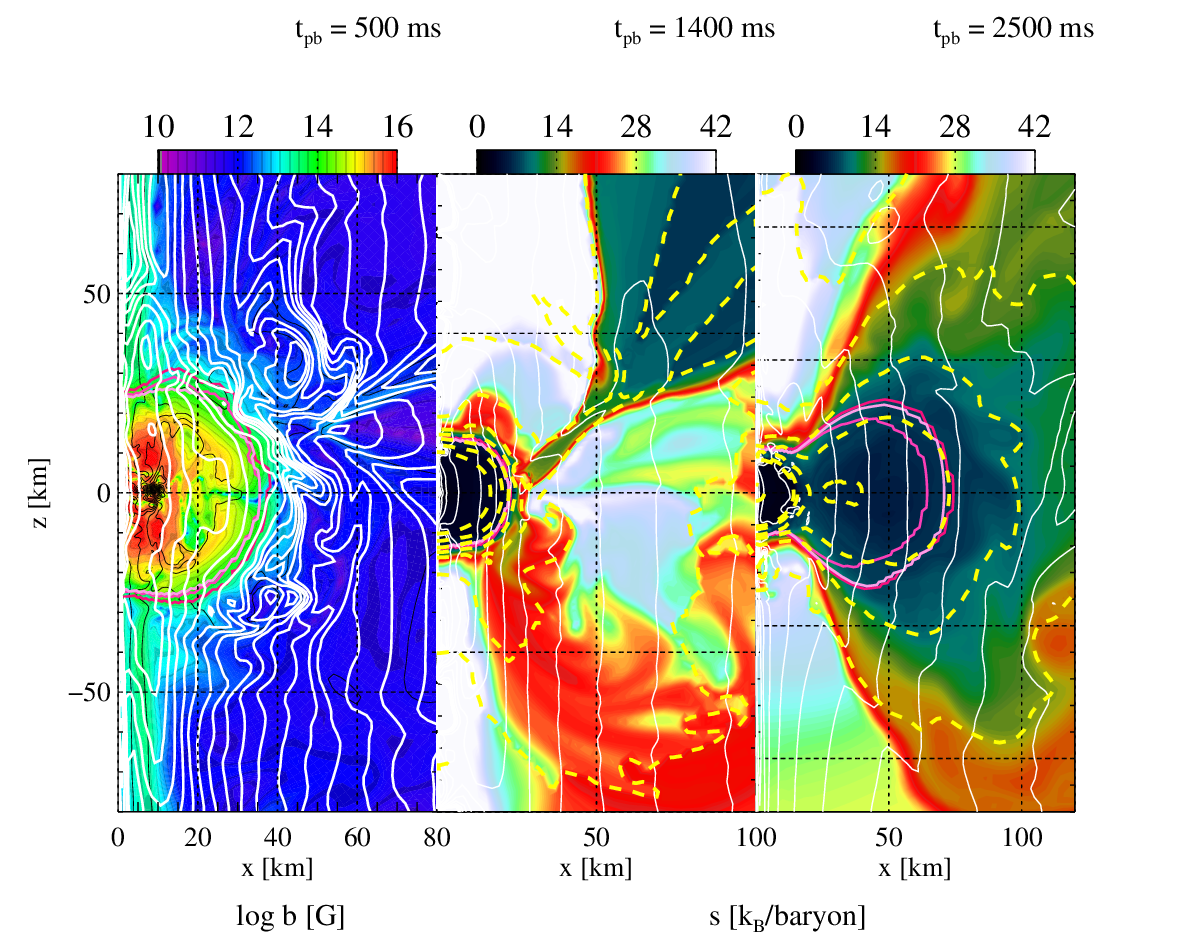}
  \caption{
    As to \figref{Fig:35OC-RO-2d-3}, but for \modl{35OC-Rw}. The gain layer
    is roughly enclosed by the outermost iso-density surface (dashed
    yellow line) corresponding to $\rho=\zehn{9}\gccm$.
  }
  \label{Fig:35OC-Rw-2d-3}
\end{figure}

\Modl{35OC-Rw} uses the same pre-collapse state as \modl{35OC-RO}, but
has a weaker magnetic field of dipolar geometry.  The core evolves
qualitatively similarly to \modl{35OC-RO}, reflecting the the magnetic
field is amplified to levels similar to the latter at the expense of
the rotational energy of the system
(\figref{Fig:globalvars-1}\panel{c}, \panel{i}). Although a BH has not
formed after $\tpb=2.54\,\sek$, by this time
$M_\pnss\simeq 2.5\,\Msol\gtrsim \Mmax$, and the core is loosing
rotational support (see below), enhancing the prospects for BH
collapse. However, this model enters into a phase of intermittent mass
shedding, which makes it difficult to predict the final
outcome. Hence, even if we have considered among the subset of our
models which may produce a BH long after the PNS formation,
\modl{35OC-Rw} is a borderline case between collapsar and PM forming
cases. Model \modelname{35OC-Rw} exhibits clear sloshing modes of the
SASI with large and small shock radii oscillating between the north
and south poles. Indeed, this model explodes by the combined action of
neutrino heating, the SASI, and rotation (see
\citetalias{Obergaulinger_Aloy__2019}), resulting in very asymmetric
(north/south) ejecta.

The profile of density and angular momentum of the two models are
equal, and thus the growth of the PNS is similar, though slightly
slower in \modl{35OC-Rw} after $\tpb\sim 1.3\,$s
(\figref{Fig:globalvars-1}\panel{f}).  Compared to \modl{35OC-RO},
there is virtually no redistribution of angular momentum from the
centre to the PNS envelope throughout most the evolution
($0.5\,\sek \lesssim \tpb \lesssim 1.5\,\sek$) as we can see in
\figref{Fig:35OC-Rw-globals}\panel{c}.  The layers with
$\zehn{12}\gccm <\rho < \zehn{13} \, \gccm$ transiently possess a
notable fraction of the total angular momentum of the PNS and quickly
lose most of it towards the inner regions.  As a consequence, they are
supported against gravity to a lower degree by centrifugal sources and
contract more than in \modl{35OC-RO}.  The resulting more compact
structure of the PNS is evident in the distribution of mass across
shells with the innermost layers
(\figref{Fig:35OC-Rw-globals}\panel{b}; blue line) containing almost
all of the mass.  Furthermore, the PNS has a less oblate shape than
that of \modl{35OC-RO} (\figref{Fig:35OC-Rw-2d-3}).  While the
polar/minimum radii of both models are similar, the equatorial/maximum
radii of \modl{35OC-Rw} decreases faster, leading to an axis ratio
approaching roughly $3:2$ after $\tpb\sim 1.5\,\sek$
(\figref{Fig:35OC-Rw-globals}\panel{a}).

The tendency of the outer layers to lose angular momentum is, however,
reversed at late times, $\tpb \gtrsim 1.5 \, \sek$ (red, orange and
olive green lines in the bottom panel of
\figref{Fig:35OC-Rw-globals}).  The magnetic field is then strong
enough to redistribute angular momentum from the interior to the PNS
surface.  This effect leads to a factor $\sim 3$ increase of the
equatorial radius and a very oblate shape of the PNS.  This
transition, extending the PNS surface into regions of low $Y_e$ allows
for very neutron-rich matter to be ejected and, hence, sets favourable
conditions for the generation of heavy elements
\citep[see][]{Reichert_etal_2020}. Furthermore, it may allow for the
formation of an extended, toroidally-shaped layer of nearly
centrifugally supported matter that may accrete onto the PNS on
relatively long timescales (\figref{Fig:35OC-Rw-2d-3} right panel).

\subsection{NS forming models}
\label{sSek:NSforming}

In the following subsections we describe the most salient properties
of the dynamical evolution of models which may not form a BH during a
significantly long time after collapse. These models are potential
hosts of PM central engines, and we may refer to them as
proto-magnetar candidates (PMCs). We take as criterion to include
models in this section that the final computed mass is below the
maximum mass allowed by the EoS (in the non-rotating and zero
temperature limit), $\Mmax\simeq 2.45\msol$. Certainly, the smaller
the value of $M_\pnss$, the better the prospects to produce a
\emph{long lived}, PM. We cannot dismiss the possibility that fall
back accretion from the stellar material not fully unbound by the SN
explosion may induce a subsequent BH collapse in a (much) longer
term. Aiming to understand transient activity during the first tens of
seconds post-bounce, it is relevant to understand whether the formed
PNS may survive and what are its properties.

\subsubsection{Models \modelname{35OC-Rp2}, \modelname{35OC-Rp3} and \modelname{35OC-Rp4} with supra-stellar magnetic field}
\label{sSek:35OC-Rp2-Rp3-Rp4}

\begin{figure}
  \centering
  \begin{tikzpicture}
  \pgftext{%
     \includegraphics[width=\linewidth]{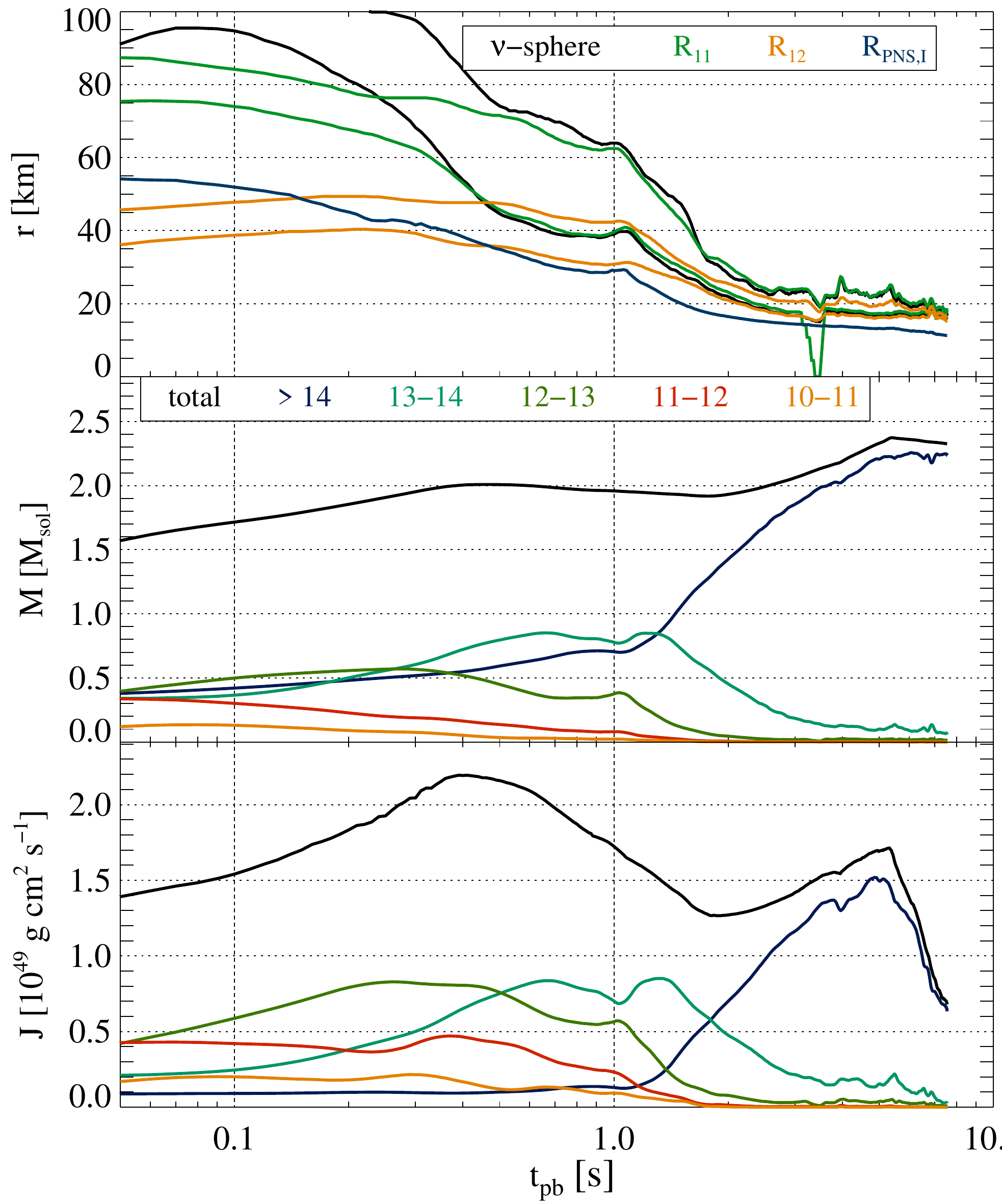}%
   }%
   \node at (-4,+1.95) {\large (a)};
   \node at (-4,-1.1) {\large (b)};
   \node at (-4,-4.15) {\large (c)};
  \end{tikzpicture}
  \caption{
    Same as \figref{Fig:35OC-RO-globals}, but for
      \modl{35OC-Rp2}.
  }
  \label{Fig:35OC-Rp2-globals}
\end{figure}

\begin{figure}
  \centering
  \begin{tikzpicture}
  \pgftext{%
    \includegraphics[width=\linewidth]{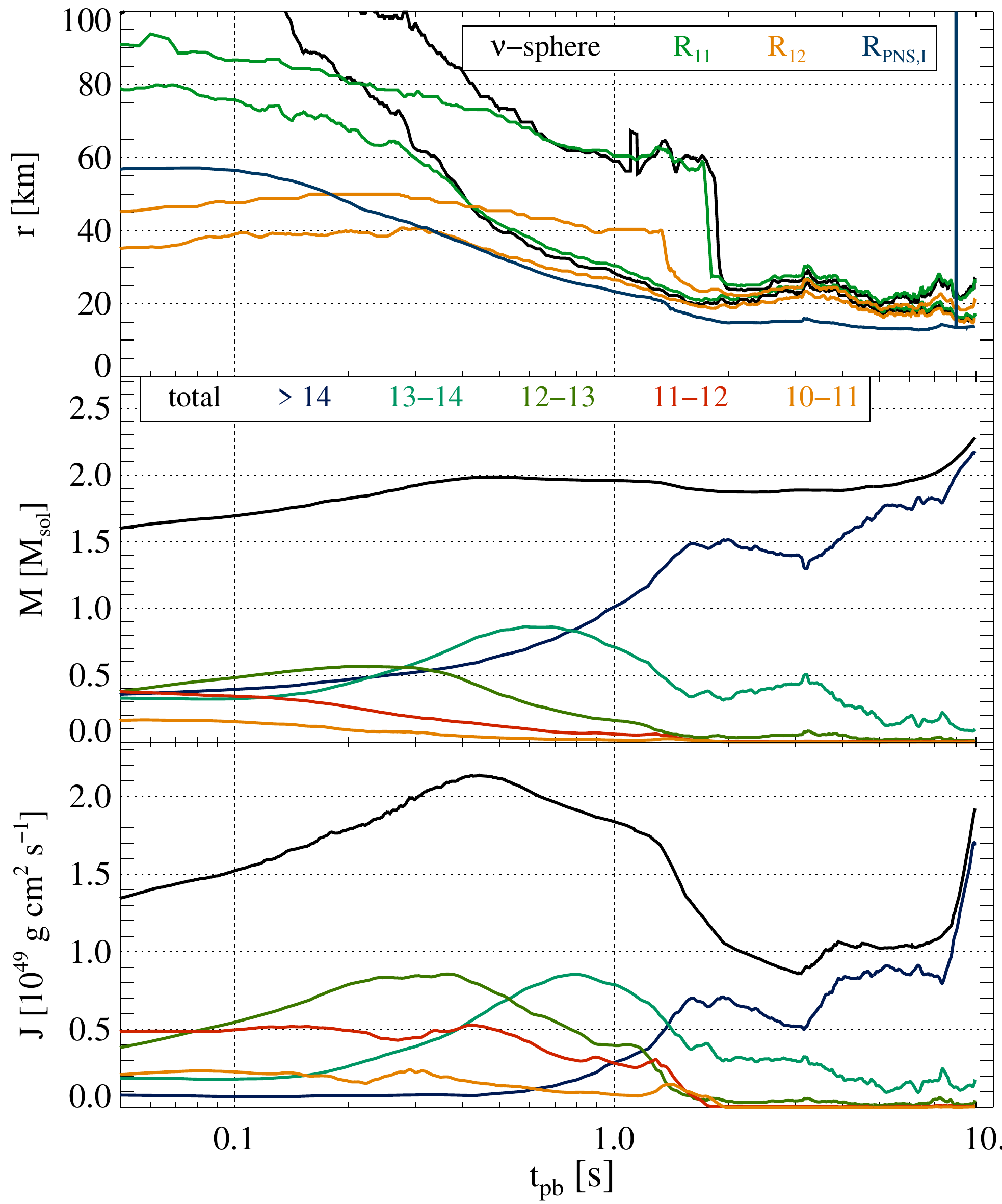}%
   }%
   \node at (-4,+1.95) {\large (a)};
   \node at (-4,-1.1) {\large (b)};
   \node at (-4,-4.15) {\large (c)};
  \end{tikzpicture}
  \caption{
    Same as \figref{Fig:35OC-RO-globals}, but for
    \modl{35OC-Rp3}.
  }
  \label{Fig:35OC-Rp3-globals}
\end{figure}

\begin{figure}
  \centering
  \includegraphics[width=0.98\linewidth]{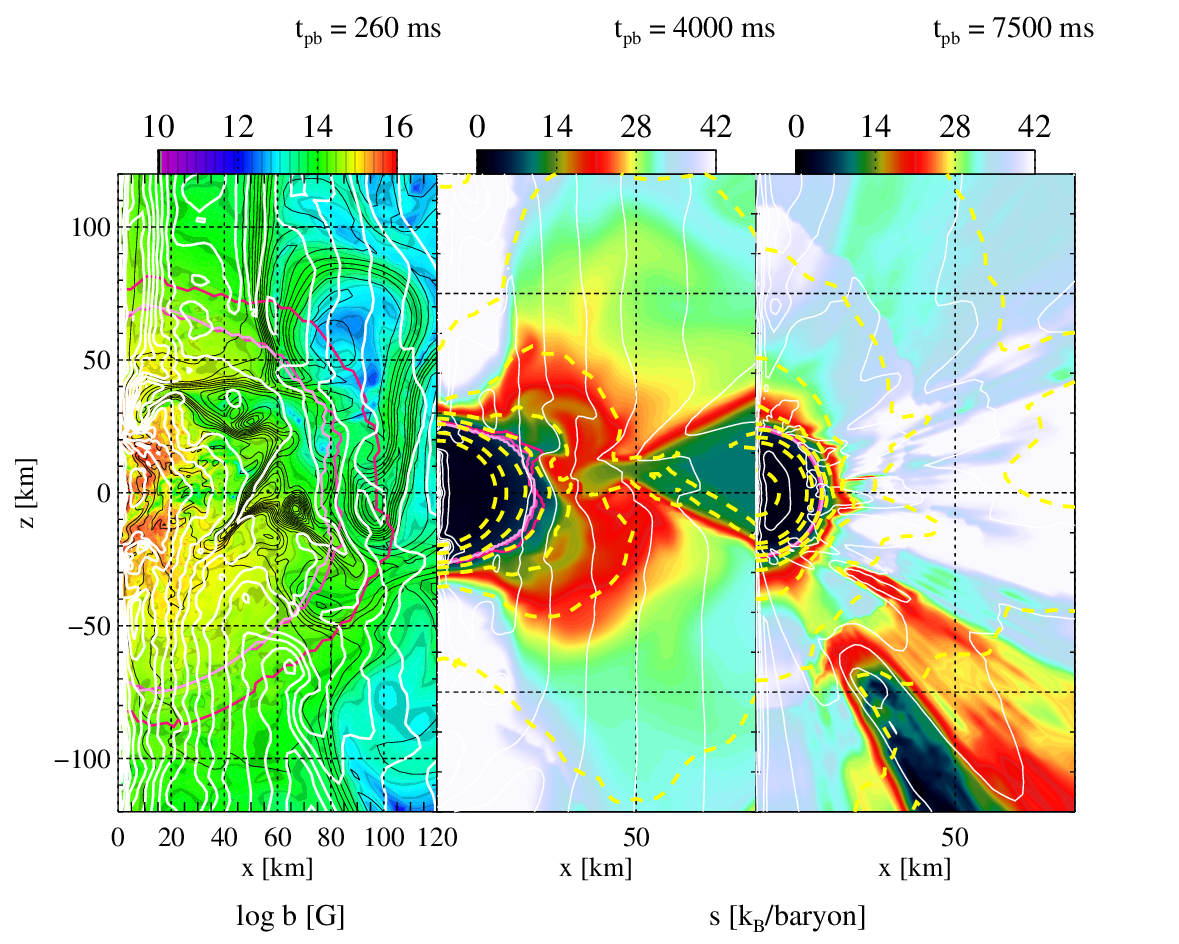}
  \includegraphics[width=0.98\linewidth]{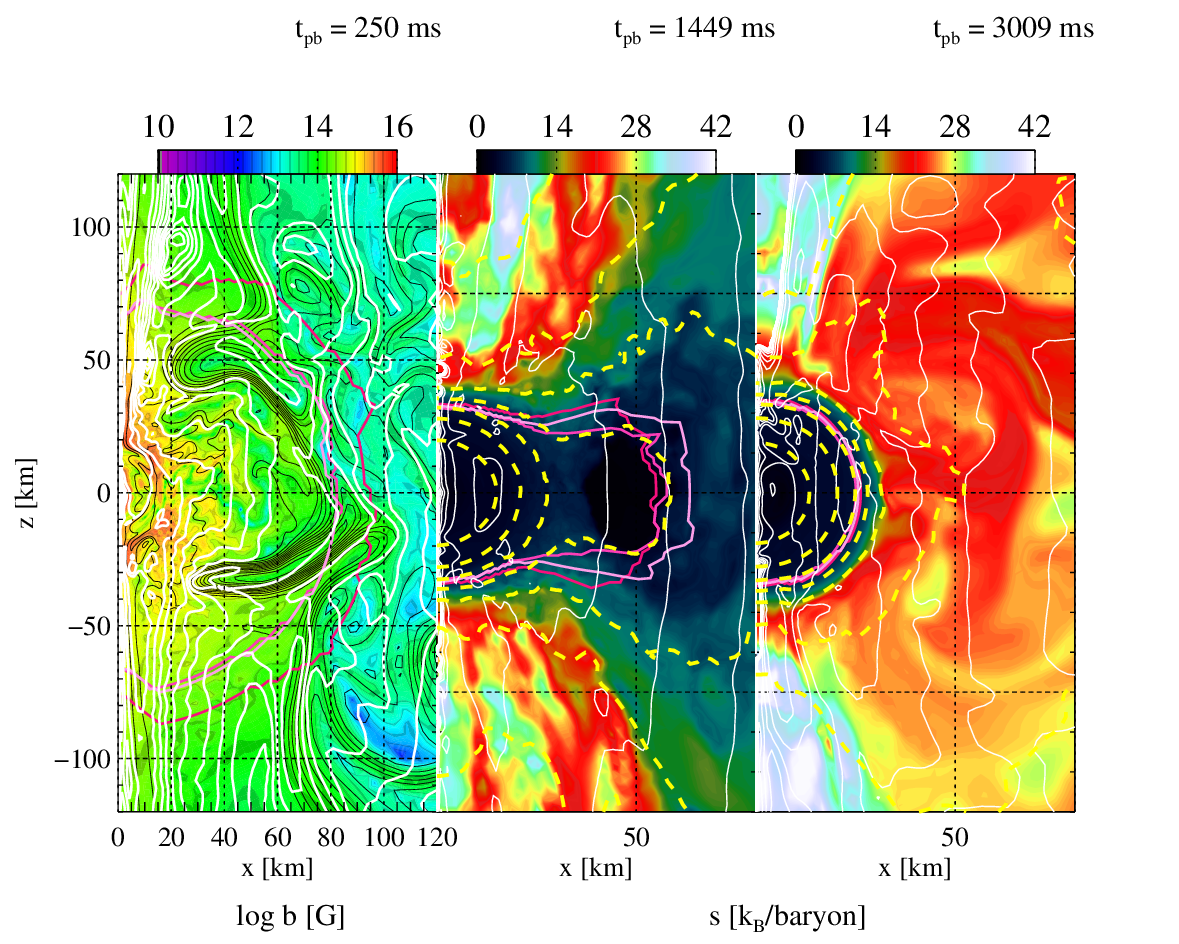}
  \includegraphics[width=0.98\linewidth]{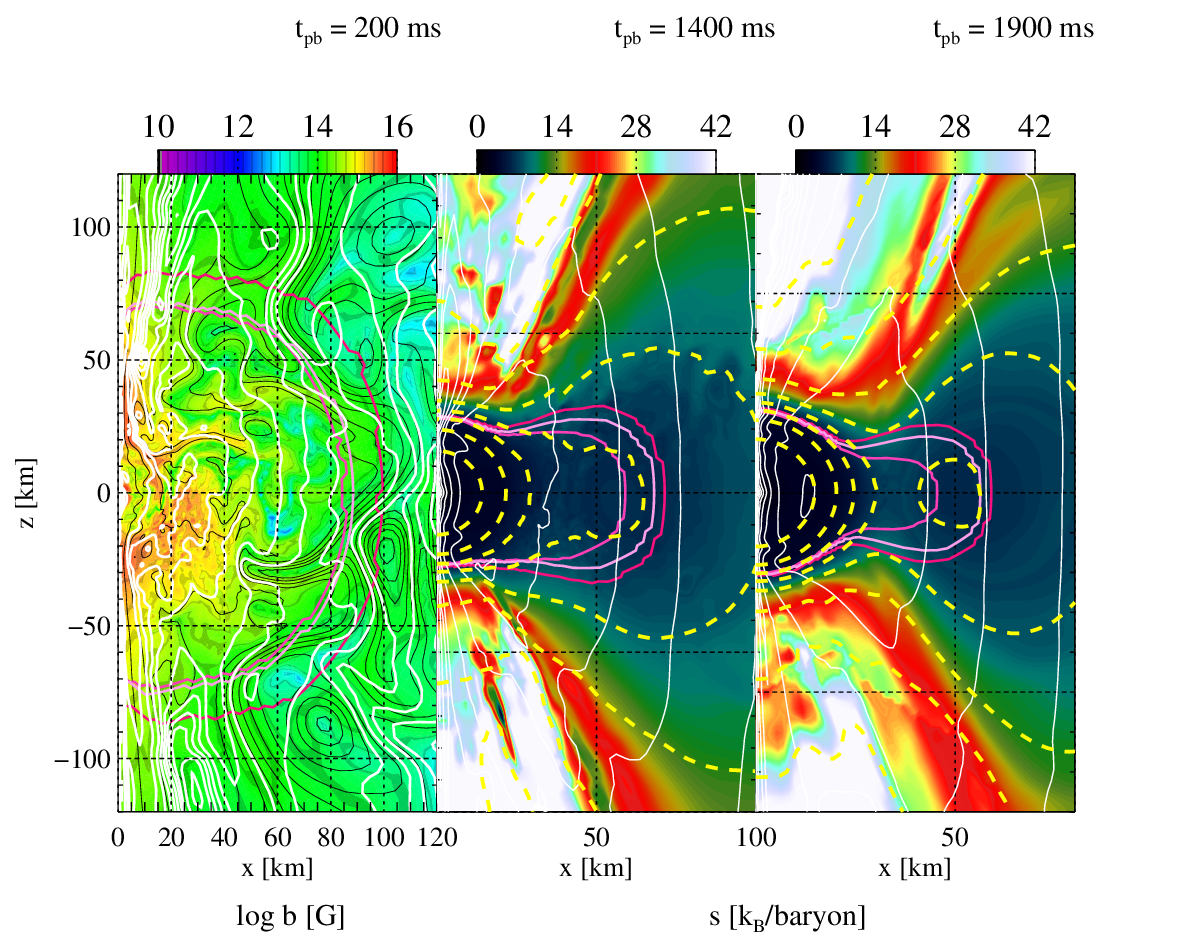}
  \caption{
   Same as \figref{Fig:35OC-RO-2d-3}, but for models
   \modelname{35OC-Rp2} (top panel), \modelname{35OC-Rp3} (mid panel)
   and \modelname{35OC-Rp4} (bottom panel).
  }
  \label{Fig:35OC-Rp2-2d}
\end{figure}

The series of models formed by \modelname{35OC-Rp2},
\modelname{35OC-Rp3} and \modelname{35OC-Rp4} differ only in the
strength of the poloidal magnetic field with respect to the original
pre-SN model \modelname{35OC-RO}. Considering these variations is
suggested by the diversity of post-bounce evolutions that differences
in the mapping of the poloidal magnetic field from the stellar
evolution model to our computational grid yield (see
App.\,\ref{sek:Variance}). In spite of the moderate increase of the
poloidal magnetic field strength in these model series, the fate of
their cores significantly departs from that of model
\modelname{35OC-RO}. They all form magnetised and significantly
massive PNSs and not BHs, as can be seen from the evolution of
$M_\pnss$ in \figref{Fig:globalvars-1}\panel{a}, as well as the
corresponding column of Tab.\,\ref{Tab:model-properties}. In all the
cases the PNS mass evolution is not monotonic and develops (one or
two) local maxima. The first of such local extrema happens at
$\tpb\sim 0.5\,\sek$ (\figref{Fig:globalvars-1}\panel{a}),
approximately when the whole iron core has been accreted.
\cite{Dessart_et_al__2008__apjl__TheProto-NeutronStarPhaseoftheCollapsarModelandtheRoutetoLong-SoftGamma-RayBurstsandHypernovae}
found a similar qualitative behaviour. The fact that the compact
remnant does not increase further its mass (but indeed, tends to
reduce it), has been interpreted as an indication that the PNS will
not collapse further to a BH in a foreseeable time
\citep{Dessart_et_al__2008__apjl__TheProto-NeutronStarPhaseoftheCollapsarModelandtheRoutetoLong-SoftGamma-RayBurstsandHypernovae,Obergaulinger_Aloy__2017__mnras__Protomagnetarandblackholeformationinhigh-massstars}\footnote{Fall-back
  accretion from the reverse shock of magnetised SN ejecta may bring
  less mass to the PNS than in an SN explosion. In magnetised ejecta,
  forming a strong reverse shock is more difficult, as it has been
  shown, \eg in the context of the dynamics of GRB ejecta
  \citep{Zhang_2005ApJ...628..315,Giannios_2008A&A...478..747,Mimica_2009A&A...494..879}. Furthermore,
  inasmuch as the magnetic structure of the PNS and surrounding is
  maintained, most of the mass falling at low latitudes will be
  shuffled towards the polar regions and contributing to the channeled
  outflow, not to the growth of the PNS mass.}. However, in previous
works the post bounce evolution was computed up to
$\tpb\lesssim 2\,$s. Here, we have gone (much) further, computing
until almost $\tpb\sim10\,\sek$ in some models. As a result, we
observe that after the first local maximum, the mass may grow and
reach a second global maximum as, \eg in the case of \modl{35OC-Rp2}
at $\tpb\simeq 5.4\,$s. \Modl{35OC-Rp3} displays a second phase of
mass increase starting at $\sim 2\,$s and continuing until the end of
the computed time
(\figref{Fig:globalvars-1}\panel{a}). \Modl{35OC-Rp4} has not been
evolved for so long as \modl{35OC-Rp3}, but one can guess that a
similar, long-lasting PNS mass evolution may happen. We note that the
mass accretion rate becomes negative (i.e. mass is extracted from the
PNS) episodically in these models, and when it is positive,
$\dot{M}_{\pnss}\lesssim 0.2\,\Msol\,\sek^{-1}$
(\figref{Fig:globalvars-1}\panel{b}). Since in all these models
$M_{\pnss}<\Mmax$ (at the end of the computed time) and a further
collapse to a BH does not seem imminent, we find it justified to refer
to these models as PMCs.

The root of the different fate of PMCs and other BH forming cases,
singularly with respect to the potential collapsar forming
\modl{35OC-RO}, is the (slightly) larger poloidal magnetic field of
the former models. This larger poloidal field drastically changes the
post-bounce accretion dynamics, significantly reducing the mass
accretion rate onto the PNS. Tightly linked to the reduced mass gain
of the PNS is the smaller rotational energy and angular momentum
attained in the long term evolution of PMCs. In
\figref{Fig:globalvars-1}\panel{c} we observe that PMCs may develop
$\Erotpns\lesssim \zehnh{5}{52}\,$erg, whereas values
$\Erotpns > \zehnh{6}{52}\,$erg are reached by BH forming models
developing either from the original 35OC core or variants thereof with
enhanced stellar rotational speed or reduced magnetic
fields. Associated to the smaller PNS rotational energy are the
smaller values of the rotational to gravitational binding potential
energy of PMCs, for which $\betaG\lesssim 0.03$
(\figref{Fig:globalvars-1}\panel{d}). Considering its larger
magnetisation and the reduced value of $\betaG$, PMCs may potentially
be less perturbed by the ``low-$\betaG$'' instability (see
Sec.\,\ref{par:stability}). PMCs also posses a PNS angular momentum
significantly smaller
($J_\pnss \lesssim \zehnh{2}{49}\,$g\,cm$^2$\,s$^{-1}$) than BH
forming models (\figref{Fig:globalvars-1}\panel{f}).

Another consequence of the increased poloidal field in PMCs is that
the shape of the PNS is less oblate than that of typical BH forming
cases. The PNS of PMCs posses both larger equatorial and polar radii
initially ($\tpb\lesssim 1.5\,$s), which tend to become similar
(i.e. the shape becomes less oblate) on longer time scales (compare
Figs.\,\ref{Fig:35OC-Rp2-globals} and \ref{Fig:35OC-Rp3-globals} with
\ref{Fig:35OC-RO-globals}).
The angular momentum in the PNS of PMCs concentrates in the denser
parts of the remnant ($\rho>\zehn{14}\, \gccm$) more effectively after
$\tpb\gtrsim 1.1\,$s (see lower panels of
Figs.\,\ref{Fig:35OC-Rp2-globals}-\ref{Fig:35OC-Rp3-globals}). Once
the inner core angular momentum dominates the overall PNS angular
momentum the fraction of the latter retained by layers with
$\zehn{11}\, \gccm< \rho<\zehn{14}\, \gccm$ is smaller in PMCs than in
BH forming models. However, we note that PMCs concentrate a larger
fraction of the PNS mass in the inner core ($\rho>\zehn{14}\, \gccm$)
than PCs (compare Figs.\,\ref{Fig:35OC-RO-globals}\panel{b} with
\ref{Fig:35OC-Rp2-globals}\panel{b}). Interestingly, the radius
$R_{\pnss,\textsc{i}}$ is $\sim 20\%-40\%$ smaller than the polar
radius of the PNS tracked with any other criteria (density isosurfaces
or neutrinospheric radius; compare, \eg
Fig.\,\ref{Fig:35OC-RO-globals}\panel{a} with
Fig.\,\ref{Fig:35OC-Rp2-globals}\panel{a}), implying that the moment
of inertia is more concentrated in PMCs than in PCs.

Model \modelname{35OC-Rp3} displays a fairly abrupt transition from a
very oblate shape (\figref{Fig:35OC-Rp2-2d}\panel{b},
$\tpb \approx 1.45 \, \sek$) to a more spherical one at
$\tpb \approx 2 \, \sek$, and remains so for a long time (see the
panel corresponding to $\tpb \approx 3 \, \sek$).  Slightly less than
$0.1 \, \msol$ of rapidly rotating matter is released from the PNS
surface and the axis ratio drops from about $3:1$ to almost unity
(\figref{Fig:35OC-Rp3-globals}\panel{a} and \panel{b}) and then
remains at a similar value for the following 7\,s of evolution.  The
transition of the shape and structure is accompanied by a decrease of
the angular momentum in the outer layers and $\Omega_{\rm surf}$.  The
loss of mass and rotational energy of the PNS partially contributes to
the outflow and enhances its energy flux with respect to, \eg
\modl{35OC-RO} which does not show the same intermediate, transitory
spin-down.  Apart from this effect, the stronger magnetic fields lead
to higher explosion energies and larger explosion masses (see
\citetalias{Obergaulinger_Aloy__2019} and Tab.\,\ref{Tab:models})
compared to the original field taken from the stellar evolution
progenitor by virtue of a greater Maxwell stress accelerating the gas.

\subsubsection{Model \modelname{35OC-Rs} with strong magnetic field}
\label{sSek:35OC-bmoltfort}

\begin{figure}
  \centering
\begin{tikzpicture}
\pgftext{%
\includegraphics[width=\linewidth]{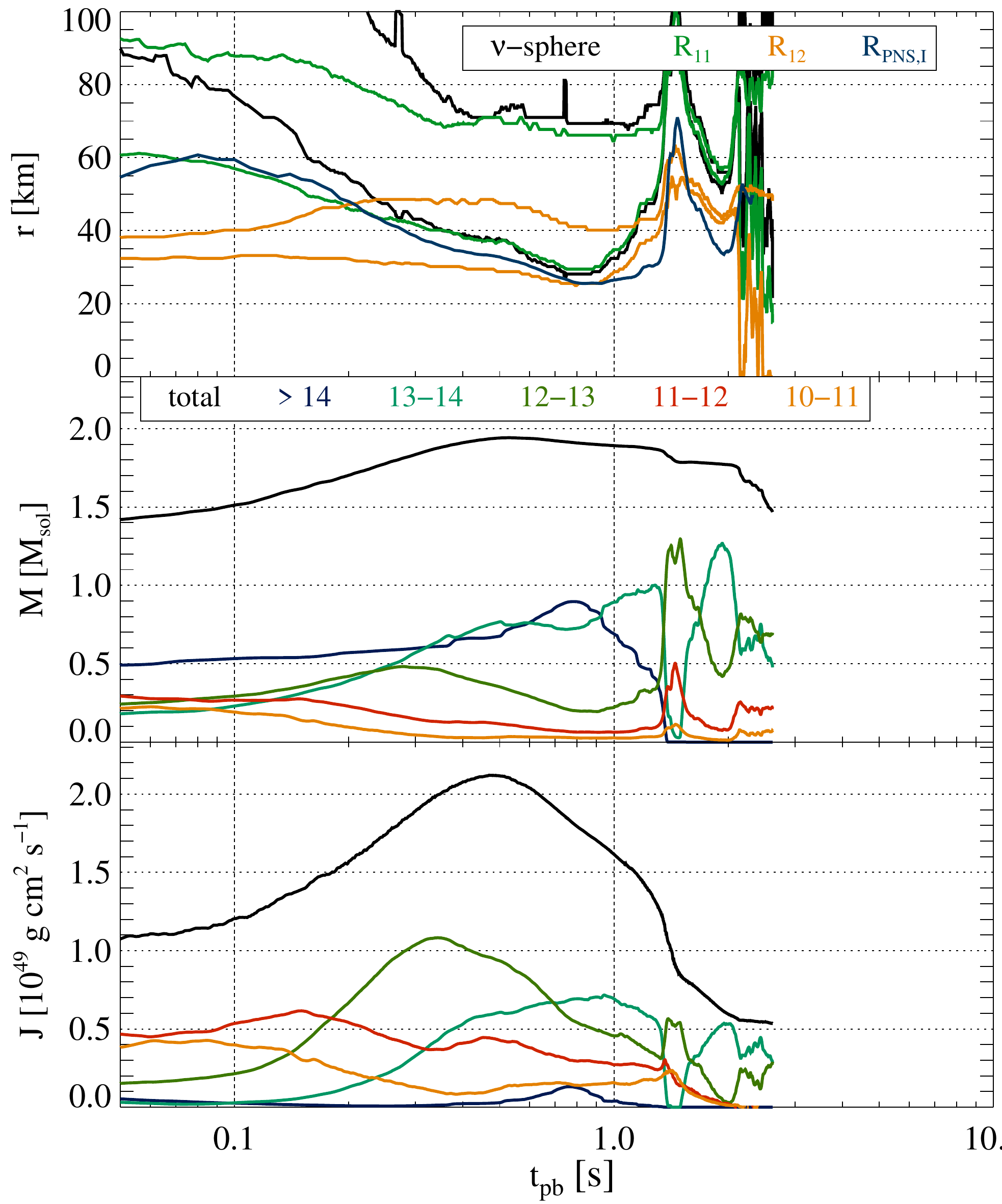}%
}%
\node at (-4,+1.95) {\large (a)};
\node at (-4,-1.1) {\large (b)};
\node at (-4,-4.15) {\large (c)};
\end{tikzpicture}
  \caption{
Same as \figref{Fig:35OC-RO-globals}, but for
      \modl{35OC-Rs}. 
    }
  \label{Fig:35OC-Rs-globals}
\end{figure}

\begin{figure}
  \centering
  \includegraphics[width=\linewidth]{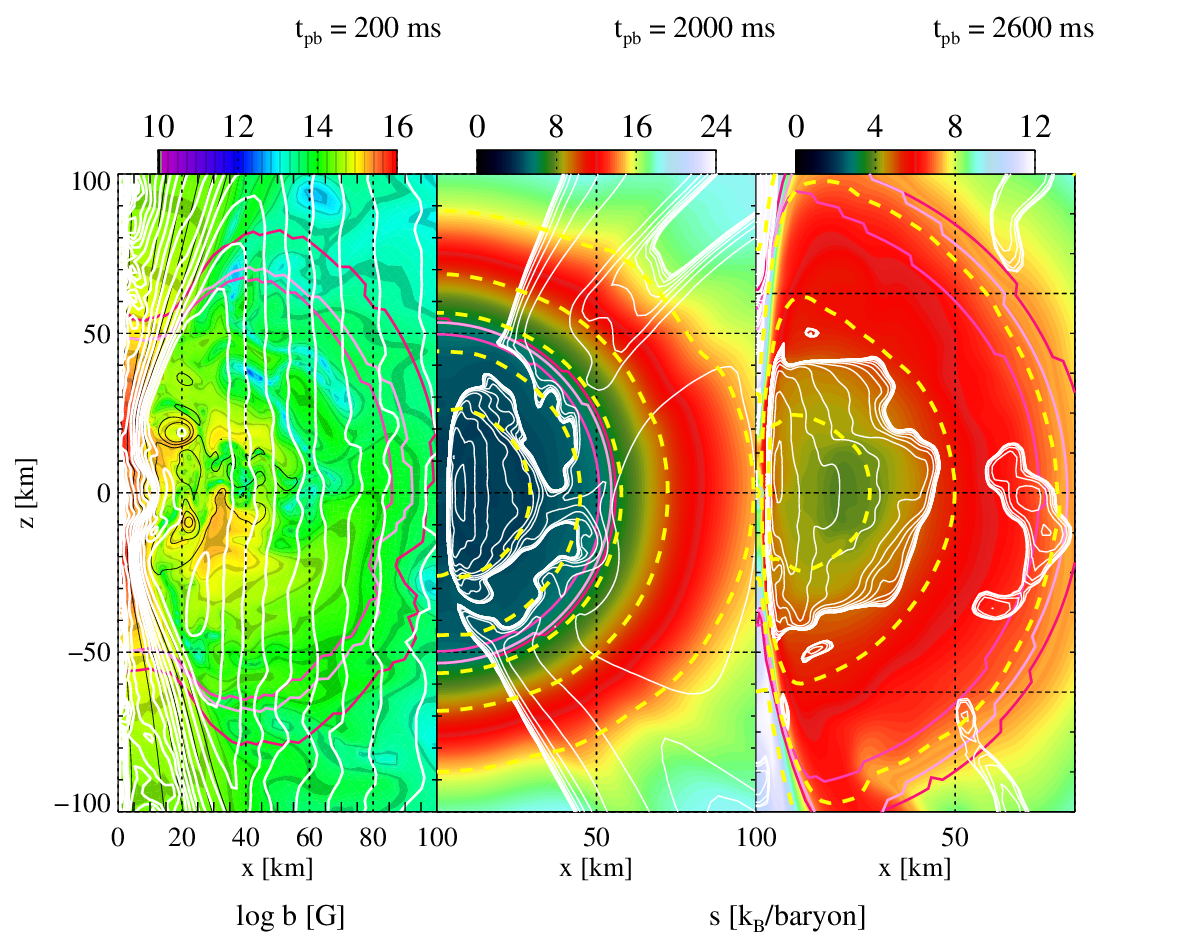}
  \caption{
    Same as \figref{Fig:35OC-RO-2d-3}, but for \modl{35OC-Rs}.
  }
  \label{Fig:35OC-Rs-2d}
\end{figure}

\Modl{35OC-Rs} contains a magnetic field consisting of a dipole and a
toroidal component in equipartition and with maximum values
$b^\phi=b^r=\zehn{12}\,G$. Compared to the maximum magnetic fields of
\modl{35OC-RO}, $b^r_{\rm max}=\zehnh{5}{10}\,$G and
$b^\phi_{\rm max}=\zehn{12}\,$G, the toroidal magnetic field is
roughly the same in both models, but the poloidal one is 50 times
larger in \modl{35OC-Rs}. This means that whereas it represents an
energetically almost negligible component in \modl{35OC-RO}, the
poloidal field constitutes about half of the total magnetic energy in
\modl{35OC-Rs} (Tab.\,\ref{Tab:models}).
We may justify the increase of the magnetic field strength with
respect to the stellar progenitor resorting to the limited ability of
numerical models to resolve the magnitude of MRI-amplified magnetic
fields. If the fastest growing MRI modes were resolved, a rough
equipartition between the toroidal and poloidal magnetic field
components may develop after core collapse
\citep[\eg][]{Obergaulinger_et_al__2006__AA__MR_collapse_TOV,Obergaulinger_2006A&A...457..209,Obergaulinger_etal__2009__AA__Semi-global_MRI_CCSN,Dessart_et_al__2008__apjl__TheProto-NeutronStarPhaseoftheCollapsarModelandtheRoutetoLong-SoftGamma-RayBurstsandHypernovae}. Under
the conservative assumption that our numerical resolution may limit
the poloidal magnetic field amplification (but see
Sec.\,\ref{sSek:amplification}, and note the growth by several orders
of magnitude of the surface-averaged poloidal magnetic field in
\figref{Fig:globalvars-1}\panel{k}), we adopt larger values of the
initial magnetic field component in this model.

The initial magnetic energy of \mRs is basically the same as that of
\mRpfour (Tab.\,\ref{Tab:models}), from which it mostly differs in the
magnetic topology. The poloidal component of the magnetic field in
\mRs is smoother on larger length scales than in \mRpfour (see
App.\,\ref{sek:Variance}). Indeed, there exists a significant
difference between the magnetic topology of \mRs and other models
inheriting directly the magnetic structure of the \modelname{35OC}
core. In \modl{35OC-RO} there are unmagnetised layers, corresponding
to convective regions of the progenitor star, where there are no
recipes to incorporate the action of magnetic torques in the stellar
evolution. Contrarily, magnetic field lines thread the whole stellar
progenitor in \modl{35OC-Rs}, some of which connect the stellar core
with the stellar surface and, hence, have a larger potential impact on
the post-bounce dynamics \citep{Bugli_2020MNRAS.492...58}. The
large-scale dipolar field explains why this model produces an almost
immediate magnetorotational explosion characterised by a pair of
polar, well collimated outflows, where neutrinos do not play a
significant role (see \citetalias{Obergaulinger_Aloy__2019}).

The post-shock gas is almost at rest and therefore the growth of the
PNS ceases at a maximum mass of
$M_{\mathrm{PNS}} \approx 1.9 \, \msol$ (a bit smaller than the iron
core mass) after a time of $\tpb \approx 450 \, \ms$.  Afterwards, the
mass of the PNS starts to decrease slowly as parts of its matter end
up in the polar jets. This behaviour reproduces what we have found for
models \modelname{35OC-Rp2}, \modelname{35OC-Rp3}, and
\modelname{35OC-Rp4}, but the PNS mass reaches a smaller maximum and
the reduction in the subsequent evolution is stronger (see
below). Given the differences in the magnetic field topology and
between \modl{35OC-Rs} (equipped with a large scale magnetic dipole)
and the former models, we directly attribute the significant change in
the PNS mass evolution to the enhancement of the poloidal magnetic
field in the iron core of PMCs compared to BH forming models.  Thus,
\modl{35OC-Rs} also shows optimal properties to produce a PM in the
mid term (namely, after a few seconds, once the PNS contraction
approximately ceases).

The PNS geometry transitions from toroidal initially to very oblate
around the time by which the maximum PNS mass is reached, since, as in
\modl{35OC-RO}, the equatorial surface ceases to contract at quite
large radii (\figref{Fig:35OC-Rs-globals}\panel{a}).  In that respect,
both models evolve very similarly during the early phases of the
evolution ($\tpb \lesssim 1\,\sek$), though \modl{35OC-Rs} has a
slightly larger radius at the equator.  The surface-averaged angular
velocity is similar as well until $\tpb\simeq 0.9\,\sek$, but later
on, the PNS of \modl{35OC-Rs} undergoes a rapid magnetic braking,
which effectively stops the rotation of the surface layers
(\figref{Fig:globalvars-1}\panel{j} and
\figref{Fig:35OC-Rs-globals}\panel{c}). This behaviour comes as a
consequence of a structural change that happens at $\tpb\gtrsim 2.1\,\sek$ 
in the PNS, whose shape changes
from an oblate ellipsoid (with maximum density at $r=0$) to a toroidal
structure (with maximum density off centre; see the dashed yellow
iso-density contours in \figref{Fig:35OC-Rs-2d} at $\tpb=2.6\,\sek$).
In this morphological transition matter close to the rotational axis
(whose density decreases significantly to values
$< \zehn{13}\,$gr\,cm$^{-3}$) is pulled away by the magnetic field,
which dominates (by far) the total pressure in this region.  As a
result, mass is ejected from the axial regions at rates that,
intermittently, can be $\gtrsim 1\,\Msol\,\sek^{-1}$
(\figref{Fig:globalvars-1}\panel{b}). The action of this sort of
\emph{interchange} instability, where the pressure support is provided
by the magnetic fields to a larger degree than by the baryons,
produces a flux tube along the rotational axis where matter
counter-rotates. Part of this counter rotating matter enters the polar
outflows, and the rest falls towards lower latitudes, slipping around
the PNS surface. Since in our procedure to compute surface-averaged
values we shall consider regions with a finite radial extension around
the neutrinospheres, positive and negative values of $\Omega$ add up
and yield as a net result that $\Omega_{\rm surf}\rightarrow 0$ (and
even $\Omega_{\rm surf}<0$; Tab.\,\ref{Tab:model-properties}). In coincidence with the morphological transition, the neutrino luminosity raises and displays a local maximum, where it reaches a value $\lesssim 40\%$ of the peak luminosity in the post-bounce neutrino burst. The morphology shift leaves, therefore, a signal in the neutrino luminosity (dominated by the combined contributions of $\mu$ and $\tau$ neutrinos and antineutrinos) and, presumably, in the gravitational wave emission.

\subsection{Magnetic field amplification}
\label{sSek:amplification}

\begin{figure*}
  \centering
  \begin{tikzpicture}
  \pgftext{
  \includegraphics[width=0.32\linewidth]{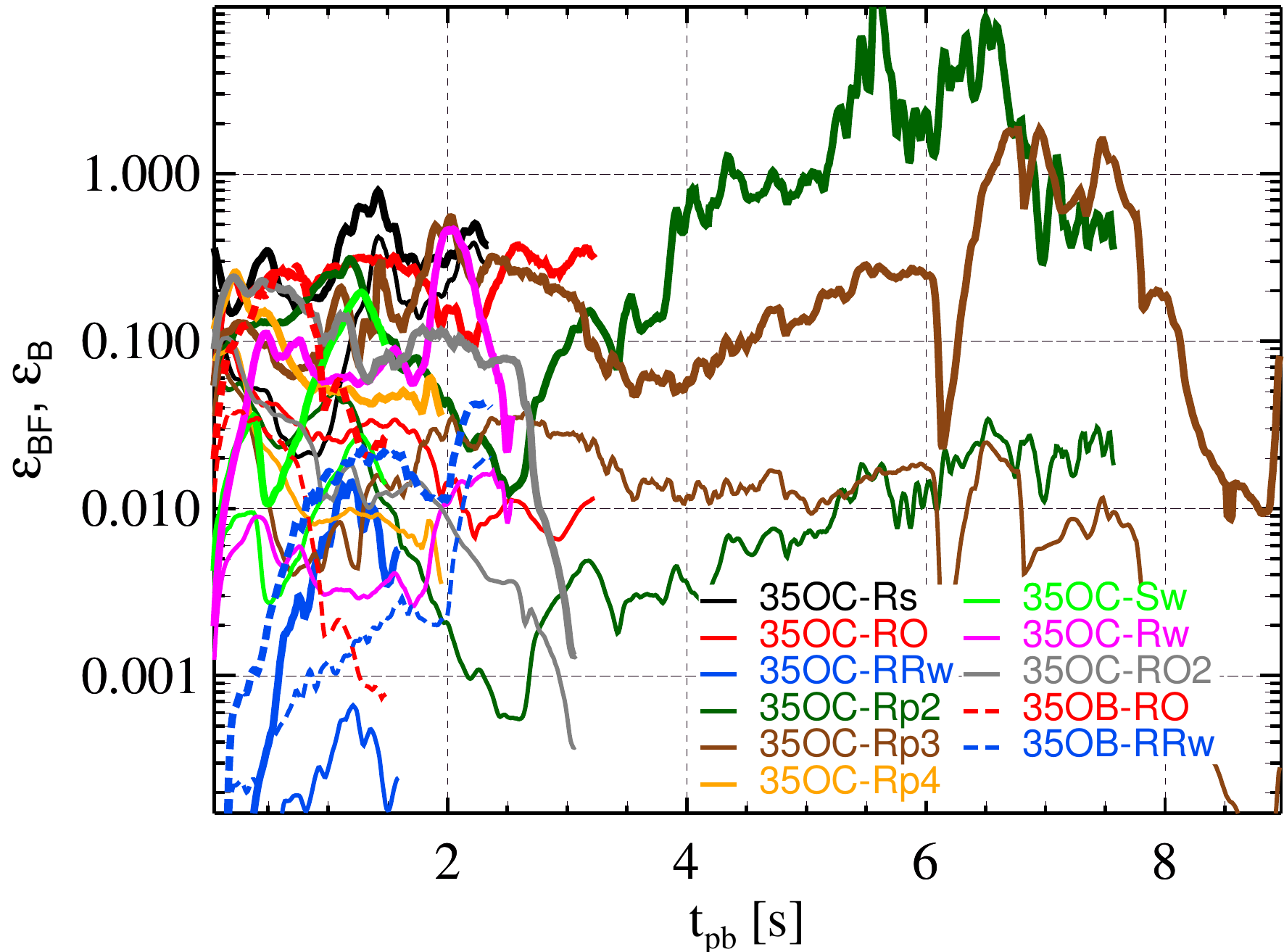}
  \includegraphics[width=0.32\linewidth]{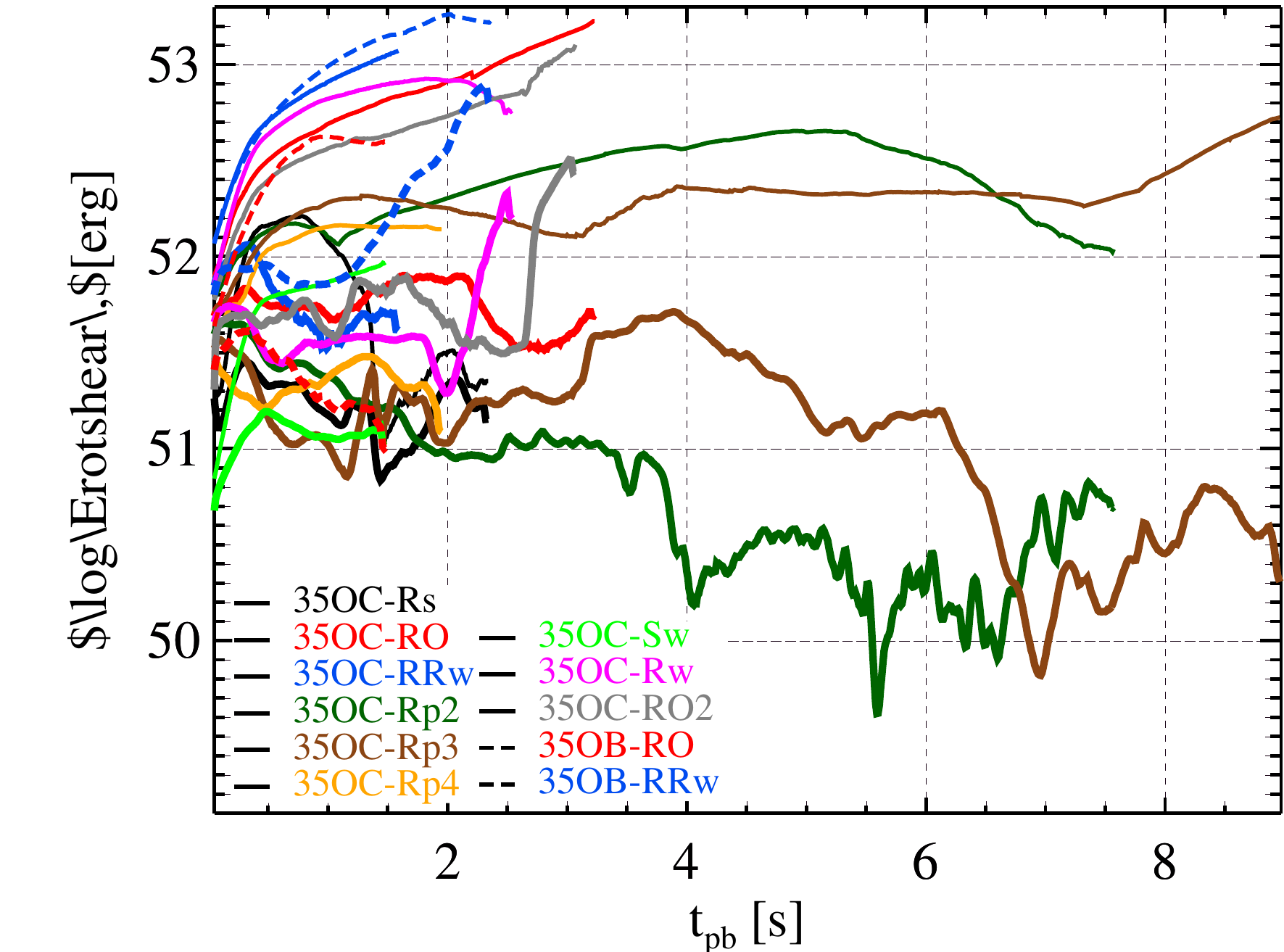}
  \includegraphics[width=0.32\linewidth]{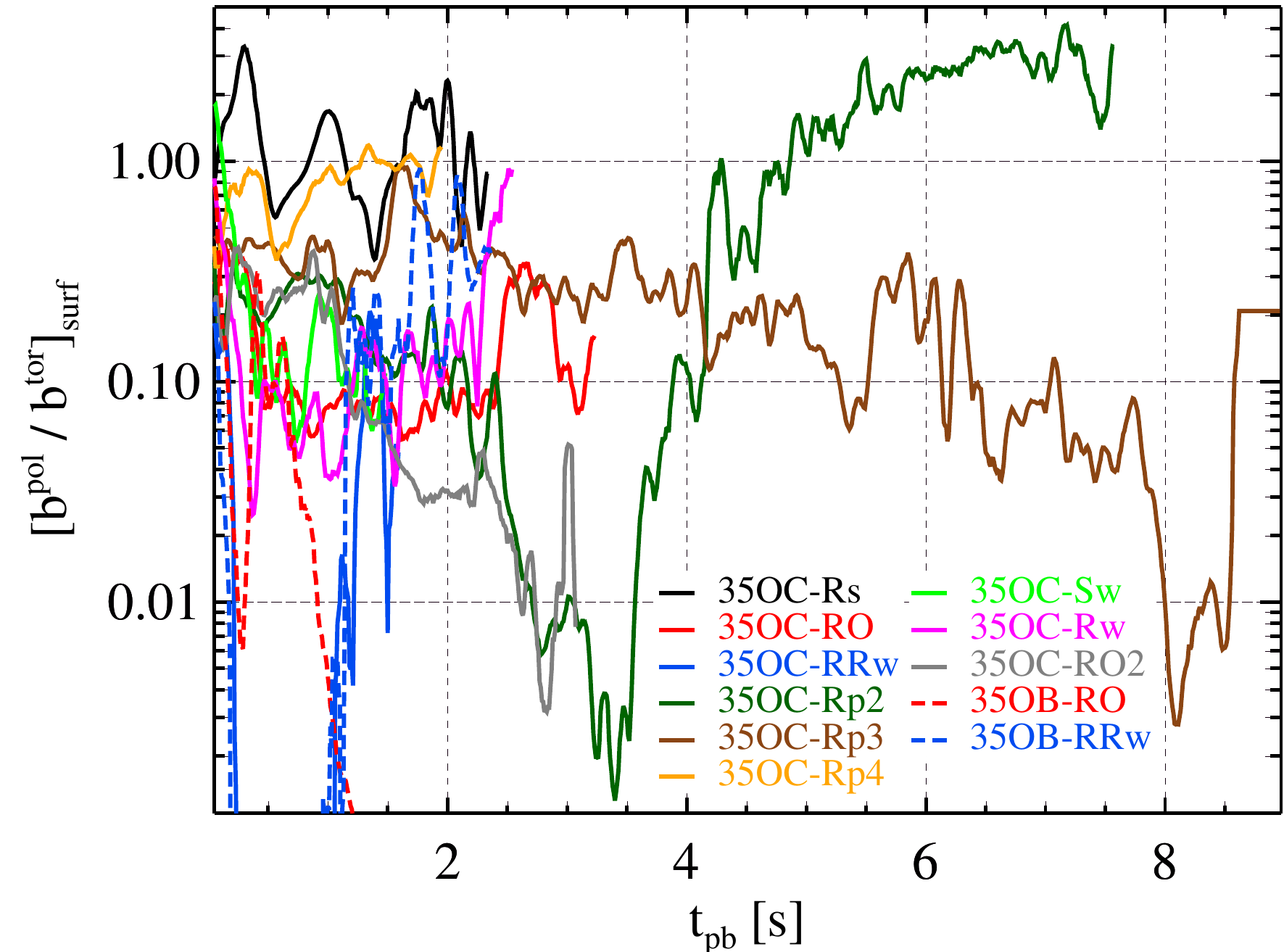}
}
   \node[fill=white, opacity=1, text opacity=1] at (-8.07,-1.8) {\large (a)};
   \node[fill=white, opacity=1, text opacity=1] at (-2.34,-1.8) {\large (b)};
   \node[fill=white, opacity=1, text opacity=1] at (+3.40,-1.8) {\large (c)};
   \node[fill=white, opacity=1, text opacity=1, rotate=90] at (-2.44,0.4) {$\log\Erotshearpns\,$[erg],\, $\log\Erotpns\,$[erg]};
  \end{tikzpicture}
  \caption{
    Time evolution of variables characterising the magnetic field,
    rotational velocity, and structure of the PNS of the models
    indicated in the legends.  The panels display: \panel{a} the
    ratios of magnetic to rotational energy ($\epsilonB$; thin solid
    lines) and magnetic to free energy in differential rotation
    ($\epsilonBF$; thick solid lines) contained in the PNS; \panel{b} the
    available free energy of differential rotation ($\Erotshearpns$;
    thick solid lines) and the rotational energy ($\Erotpns$; thin solid lines)
    of the PNS; \panel{c} the ratio of poloidal to toroidal field on
    its surface.}
  \label{Fig:globalvars-2}
\end{figure*}

The magnetic field in the PNS is amplified by the advection of
magnetic flux from the surrounding regions onto its surface, by
compression as the PNS contracts, and by differential rotation.
Compression yields a growth of the magnetic energy as a result of
magnetic flux conservation, but it also produces a comparatively
larger growth of the rotational energy resulting from the angular
momentum conservation, hence, reducing the ratio $\epsilonB$. The
growth of rotational energy resulting from compression may also induce
the growth of the magnetic energy, but in any case, the magnetic
energy growth is limited to a fraction of equipartition with the
available kinetic energy in the system.  The PNS dynamics in all our
models yields a (non-monotonic) growth of $\epsilonB$, whose maximum
values are $\epsilonB < 1$ in the case of \modl{35OC-Rs}
(Fig.\,\ref{Fig:globalvars-1}\panel{e}). In order to quantify more
precisely the origin of the magnetic field growth, we compute the
available free energy of differential rotation as the difference in
rotational energy of the PNS and the rotational energy of a uniformly
spinning PNS of the same angular momentum, $J$, and moment of inertia,
$I$,
  \begin{equation}
    \Erotshear := \Erot - \frac{J^2}{2I},
    \label{eq:Erotshear}
  \end{equation}
and we further define the fraction
\begin{gather}
  \epsilonBF :=\Emag^\pnss /\Erotshear^\pnss, \label{eq:epsilonBf}
\end{gather}
in analogy to $\epsilonB$ (Eq.\,\eqref{eq:epsilonB}).  The definition
of $\Erotshear$ is adapted from
\cite{Dessart_et_al__2012__apj__TheArduousJourneytoBlackHoleFormationinPotentialGamma-RayBurstProgenitors}
(c.f. their Eq.\,4), which is inspired, in its turn, by the fact that
solid-body rotation corresponds to the lowest energy state for fixed
total angular momentum, and this is the state any rotating fluid will
reach on a \emph{secular} time scale, if it may redistribute its
angular momentum.  In our models, the inner parts of the PNS (with
$\rho\gtrsim \zehn{14}\,$gr\,cm$^{-3}$) are rotating nearly rigidly
(though, differently from
\cite{Dessart_et_al__2012__apj__TheArduousJourneytoBlackHoleFormationinPotentialGamma-RayBurstProgenitors},
the innermost $\sim 5 - 10\,$km may develop a positive
$\Omega$-gradient), while the outer parts, with about one fifth of the
PNS mass, concentrate the differentially rotating shells and thus,
most of the free available rotational energy.

We have already pointed out (Sec.\,\ref{sSek:35OC-Rp2-Rp3-Rp4}) that
PMCs typically show smaller values of $\Erotpns$ than PCs. This is
also the case for $\Erotshearpns$ in models which begin from the same
rotational profile (\figref{Fig:globalvars-2}\banel{(b)}; solid
lines). The difference in $\Erotshearpns$ grows with time among PMCs
and PCs, and at, \eg $\tpb\simeq 2\,$s, PMCs display values of
$\Erotshearpns$ about 2 - 8 times smaller than the corresponding to
PCs.  It is important noticing that the free available rotational
energy is significantly smaller than the rotational energy (observe
that the ratio $\Erotpns/\Erotshearpns \sim 20$ for $\tpb\gtrsim 4\,$s
in \figref{Fig:globalvars-2}\banel{(b)}), which limits the prospects
for magnetic field amplification to values
$b^{\rm pol}_{\rm surf}\approx \zehn{15}\,$G
($b^{\rm pol}_{\rm surf}\lesssim \zehn{14}\,$G) in the case of
\modl{35OC-Rp2} (\modl{35OC-Rp3}); see
\figref{Fig:globalvars-1}\banel{(k)}.

We observe (\figref{Fig:globalvars-2}\panel{a}) that the PMCs possess
ratios of magnetic to rotational energies well in excess of the
estimates of
\cite{Metzger_et_al__2011__mnras__Theprotomagnetarmodelforgamma-raybursts}
(\citetalias{Metzger_et_al__2011__mnras__Theprotomagnetarmodelforgamma-raybursts}
in the following), who assume that the previous fraction is
$\sim \zehn{-3}$ (c.f.\, their Eq.(4)).  In our PMCs, the magnetic
field typically accounts for a fraction $\epsilonBF > 0.01$ of the
free rotational energy during the computed time evolution. The
previous fraction reduces to $0.001<\epsilonB \lesssim 0.01$ if we
compare the magnetic and rotational energies of the PNS.  We note that
also PCs, though weaker magnetised in relative terms, also exceed the
aforementioned estimate. The fact that $\epsilonBF$ reaches values
$ \sim 1$ for models \modelname{35OC-Rp2} and \modelname{35OC-Rp3}
suggests that the magnetic field growth happens at the expense of the
free rotational energy and not at the expense of the (significantly
larger) rotational energy of the PNS \citep[in agreement with,
\eg][]{Duncan_Thompson__1992__ApJL__Magnetars}.  As can be seen in
\figref{Fig:globalvars-2}\panel{a} (solid lines), $\epsilonBF$ is
significantly larger than 1 in the interval
$5.2\,\sek\lesssim \tpb \lesssim 7\,\sek$ for \modl{35OC-Rp2}, which
seems to contradict that $\mathcal{B}$ feeds off
$\Erotshear$. However, this is a consequence of various
factors. Firstly, the definition Eq.\,\ref{eq:Erotshear} is not
perfect in an accreting PNS whose surface is only approximately
defined by the location of the $\nu_e$-sphere. As illustrated by
\figref{Fig:globalvars-1}\panel{a}, after $\tpb\simeq 5.2\,\sek$ the
PNS mass of \modl{35OC-Rp2} decreases slightly. This is due to the
fact that the large specific angular momentum of the stellar layers
being accreted makes them only loosely bound to the PNS. Hence, a
fraction of them can be unbound and incorporated to the explosion
ejecta, explaining the reduction of the PNS mass and rotational
energy.  Second, the slow down of the PNS is not uniform, it is more
important in the outer layers, which decreases the degree of
differential rotation and, thereby, explains the increase of
$\epsilonBF$ also after $\tpb\sim 2.5\,\sek$
(\figref{Fig:globalvars-2}\panel{b}).

In the stellar model 35OC there is a gap of $\sim 3\Msol$ between the
inner magnetised core and the ensuing stellar shell containing
magnetised matter (Fig.\,\ref{Fig:progenitor-profile}). Qualitatively,
the same comment applies to models resulting from the stellar core
35OB. Hence, in the simulations that include either the original
magnetic field of the progenitor star or small variations there off,
the layers accreted by the PNS after $\tpb \sim 2 - 3\,\sek$ evolution
are not magnetised. As a result, the magnetic field amplification
observed in models like \modelname{35OC-Rp2} and \modelname{35OC-Rp3}
$\sim 2\,\sek$ after bounce does not result from the (practically
unmagnetised) accretion flow onto the PNS. Instead, the magnetic field
amplification results from local amplification processes (\eg
compression and MRI).  The amplification of the magnetic field in
\modl{35OC-Rs}, especially regarding the surface-averaged poloidal
component, is less intense than in other PMCs and also than in
\modl{35OC-RO} (\figref{Fig:globalvars-1}\panel{k} and
\figref{Fig:globalvars-1}\panel{l}). Indeed, \modl{35OC-Rs} shows a
poloidal field similarly strong (several $\zehn{14} \, \Gauss$) as the
toroidal one (Tab.\,\ref{Tab:model-properties}), due to the fact that
its initial values are large, reducing the prospects for further
amplification in the post bounce phase.
 
Considering first the case of \modl{35OC-RO}, we find typical
convective speeds in the PNS around
$v_{\mathrm{conv}} \le \zehn{8} \, \cms$, and the pattern of
convection is considerably modified by rotation.  We observe
convective cells aligned parallel to the rotational axis with a small
extent in $\varpi$-direction.  Differently from non-rotating
magnetised collapsing cores
\citep[\eg][]{Obergaulinger_et_al__2014__mnras__Magneticfieldamplificationandmagneticallysupportedexplosionsofcollapsingnon-rotatingstellarcores},
convection is not the main agent driving magnetic field
growth. Instead, the PNS is unstable against the MRI, which is the
main catalyst for magnetic field amplification.  Since the initial
poloidal field is sufficiently strong, we are able to numerically
resolve the growth of the MRI in the form of channel modes
\citep[\figref{Fig:35OC-RO-2d-3}, left panel; see
also][]{Rembiasz_et_al__2016__mnras__Terminationofthemagnetorotationalinstabilityviaparasiticinstabilitiesincore-collapsesupernovae,Rembiasz_et_al__2017__apjs__OntheMeasurementsofNumericalViscosityandResistivityinEulerianMHDCodes},
which develop between $\tpb \approx 150 \, \ms$ and
$\tpb \approx 240 \, \ms$ at cylindrical radii around
$\varpi \approx 50 \, \km$ and lead to an exponential growth of the
energy of the poloidal field component
(\figref{Fig:globalvars-1}\panel{k}).

The MRI activity is transient only.  After the end of the exponential
growth, the channels gradually fade away into fields dominated by
small-scale structures.  The energy of the poloidal field decays over
half a second, though the differential rotation is maintained
throughout the evolution and the angular velocity even increases due
to the accretion of rapidly rotating matter
(\figref{Fig:globalvars-1}\panel{j}).  Large regions of the interior
inside the \nusps show a cylindrical rotational profile (see the white
contours in the left panel of \figref{Fig:35OC-RO-2d-3}). Assuming
that the rotational frequency can be locally parametrised by a
power-law of the form $\Omega(r)\propto r^q$, a measurement of the
rotational profile is the power-law index, $q$, precisely defined as
\begin{align}
  q(r)=d\ln{\Omega(r)}/d\ln{r} .
  \label{eq:qrot}
\end{align}
We show in \figref{Fig:globalvars-1}\panel{h} the time evolution
of $q_\pnss$ computed in the outer layers of the PNS, precisely,
\begin{gather}
  q_\pnss := \frac{3}{R_\nu}\int_{2R_\nu / 3}^{R_\nu} {\rm d}r \, q(r),
  \label{eq:qpns}
\end{gather}
\ie $q_\pnss$ is a radial average value over the outermost one third
of the PNS radius on the equatorial plane. Since the inner part of the
PNS is rigidly rotating, the MRI will not develop there. However, the
outer (differentially rotating) layer of the PNS is much better suited
for the magnetic field growth due to the MRI
\citep[\eg][]{Guilet_et_al__2015__mnras__Neutrinoviscosityanddrag:impactonthemagnetorotationalinstabilityinprotoneutronstars,
  Rembiasz_et_al__2016__mnras__Terminationofthemagnetorotationalinstabilityviaparasiticinstabilitiesincore-collapsesupernovae}. Differentially
rotating profiles with $-2<q_\pnss<0$ permit the development of the
MRI. Model \modelname{35OC-RO} displays values $-2.5<q_\pnss<-1.6 $
during most of its evolution, with typical values $q_\pnss\simeq -2$,
which precisely may render the fastest growth rates for MRI,
$\gamma_{\textsc{mri}}\simeq\Omega$ \citep[in general, the growth rate
of the fastest growing MRI mode is
$\gamma_{\textsc{mri}}\simeq
|q_\pnss|\Omega/2$;][]{Pessah_Chan__2008__apj__Viscous_Resistive_Magnetorotational_Modes}.
Anyway, in \modl{35OC-RO} it is clear that the action of the magnetic
fields is insufficient to significantly flatten the rotational profile
of the outer regions of the PNS, since $q_\pnss$ is significantly
smaller than zero (indeed, $q_\pnss<q_{\textsc{k}}=-1.5$,
corresponding to a Keplerian profile). Besides, accretion keeps adding
angular momentum at high rates to the outer PNS layers, acting against
the development of a rigid rotational profile.

In other BH-forming models we also observe episodes of exponential
magnetic growth driven by the MRI. For instance, one of them starts at
$\tpb \approx 100...200 \, \ms$ in \modl{35OC-RO2}
(\figref{Fig:globalvars-1}\panel{k}) and, somewhat later
($\tpb \approx 400 \, \ms$) and showing a longer lasting episode of
growth for \modl{35OC-Rw}.  Both variants of the 35OC stellar core
share the same initial rotational profile, which should make both
models equally susceptible to the MRI.  The fact that we do not
observe it growing at the same magnitude can be attributed to the
weaker initial field of \modl{35OC-Rw}, which may shift the typical
MRI modes to wavelengths below the grid resolution.
We do not find indications of the MRI in the relatively weakly
magnetised models of progenitor \modelname{35OB}, and neither in the
\modl{35OC-Rw}.  In \modl{35OB-RO}, which develops fairly strong total
magnetic fields, the MRI is numerically damped due to the weak
poloidal field, which is far less intense than that of \modl{35OC-RO}.
The same holds for \modl{35OC-Rw}.  In this case, however, we find a
late phase of amplification of the poloidal component sustained for
half a second after $\tpb \approx 1.8 \, \sek$.  The amplification
occurs in the equatorial region near the PNS surface where a
convective layer develops, at the top of which the field grows most
rapidly.

The magnetic field growth in supra-stellar rotation models \modelname{35OC-RRw} and \modelname{35OB-RRw} represents an amazing  example of exponential amplification without the concourse of MRI (though favoured by the imposed axial symmetry). It is due to the development of axisymmetric convective cells around the rotational axis, where the field is essentially vertical and initially very weak.  \cite{Spruit__2013__ArXive-prints__EssentialMagnetohydrodynamicsforAstrophysics} shows that in planar symmetry a passive magnetic field can be exponentially amplified under the action of a converging-diverging, incompressible flow. This concept can be extended to axial symmetry. Near the axis and close to the PNS surface of models \modelname{35OC-RRw} and \modelname{35OB-RRw}, vertically stacked convective zones have an equivalent configuration to that of \cite{Spruit__2013__ArXive-prints__EssentialMagnetohydrodynamicsforAstrophysics}. There, the magnetic and velocity fields  can be roughly approximated by $\mathbf{b}\approx b_z(t)\hat{z}$, $\mathbf{v}\approx-a\varpi/2 \hat{\varpi}+v_\phi(t)\hat{\phi}+az\hat{z}$, where $\hat{\varpi}, \hat{\phi},$ and $\hat{z}$ are the unit vectors in the cylindrical coordinate directions $\varpi,\phi,$ and $z$, while $a$ is a constant. Under this configuration, one expects that the (initially passive) vertical magnetic field is exponentially amplified (for the approximate conditions stated above as $\mathbf{b}(t)\approx b_{z}(0)e^{at}\hat{z}$) until the vertical magnetic field becomes dynamically relevant and there is a back reaction on the fluid flow. The amplification shows signs of saturation in \modl{35OC-RRw} after $\tpb\approx 1\,\sek$ , but seems still ongoing for \modl{35OB-RRw} after $\tpb \approx 2.4\,\sek$ (Fig.\,\ref{Fig:globalvars-1}\banel{(i)}).

PMCs with a supra-stellar magnetic field (models \modelname{35OC-Rp2},
\modelname{35OC-Rp3}, and \modelname{35OC-Rp4};
Sect.\,\ref{sSek:35OC-Rp2-Rp3-Rp4}) also develop a rotational gradient
that could (potentially) allow for the growth of MRI (but see
below). Nevertheless, the values of $q_{\pnss}$ are larger (smaller in
absolute value) than in PCs (see next section), which allows for
different evolutions. While the poloidal magnetic field in
\modl{35OC-Rp2} initiates a sustained growth after
$\tpb\sim 3.5\,\sek$, which levels off after $\sim 6\,\sek$ at values
$b^{\rm pol}_{\rm surf}\approx 10^{15}\,$G
(Fig.\,\ref{Fig:globalvars-1}\panel{k}), in \modl{35OC-Rp3} it raises
very early to values $b^{\rm pol}_{\rm surf}\lesssim 10^{14}\,$G, and
maintains this level until $\tpb\sim 7\,\sek$, after which it sinks
steeply. In parallel to the magnetic field decline, the mass accretion
rate grows, highlighting the correlation between the poloidal magnetic
field strength and the ability to maintain the PNS mass below $\Mmax$.
We find during the $\sim 2\,\sek$ starting at $\tpb\sim 3.5\,\sek$
that the PNS of \modl{35OC-Rp2} develops vigorous convection, with
convective overturn times $\tau_{\rm conv} \sim 0.02\,\sek$. The
convective cells are forced into vertical cylinders by the rapid
rotation. They end abruptly at the PNS surface, where the magnetic
field accumulates first at the top, and then expands towards the
centre. Since the up- and down-flows within the PNS are aligned with
the rotational profile (instead of being perpendicular to it), we
attribute the large amplification of the poloidal field in the PNS of
\modl{35OC-Rp2} to the convection rather than to the MRI. An efficient
dynamo may result if the Rossby number $R_{\rm O}$, defined as the
ratio of the convective overturn time to the rotational period, $P$,
is of order unity or less
\citep{Duncan_Thompson__1992__ApJL__Magnetars}. The PNS is
differentially rotating and, therefore, the rotational period depends
(non-monotonically) on the distance to the rotational axis. In the
case of \modl{35OC-Rp2}, $P$ ranges from $\approx 40\,$ms (close to
the rotational axis) to $\approx 1\,$ms (at about 10\,km off
centre). This yields a broad range of Rossby numbers inside the PNS,
$R_{\rm O}\in[0.05, 2]$, such that $R_{\rm O}\lesssim 1$ between
$r\sim 5\,$km and the PNS surface. The large amplification factor
\citep[$\sim 1000$; even larger than the predictions
of][]{Duncan_Thompson__1992__ApJL__Magnetars}, by which the poloidal
field grows from $\sim 10^{12}\,$G to $\sim 10^{15}\,$G, contrasts
with the very moderate growth that convection produces in the hot
bubble surrounding the PNS in non-rotating, magnetised models
\citep{Obergaulinger_et_al__2014__mnras__Magneticfieldamplificationandmagneticallysupportedexplosionsofcollapsingnon-rotatingstellarcores}. An
approximate equipartition between $b^{\rm pol}_{\rm surf}$ and
$b^{\rm tor}_{\rm surf}$ in \modl{35OC-Rp2} is reached after
$\tpb\sim 4.2\,\sek$ (Fig.\,\ref{Fig:globalvars-2}\panel{c}).
The late fall-down of the poloidal magnetic field of \modl{35OC-Rp3}
is triggered by the accretion of unmagnetised stellar matter (see the
positive and increasing mass accretion rate of this model after
$\sim 5\,\sek$ in Fig.\,\ref{Fig:globalvars-1}\panel{b}), which partly
buries the magnetic field of the PNS surface. It is accompanied by the
(one order of magnitude) decrease of the toroidal magnetic field
component at $\tpb\sim 7.5\,\sek$
(Fig.\,\ref{Fig:globalvars-1}\panel{l}). The different dynamics of the
accretion flow onto the PNS of the previous models is connected to the
feedback between the SN ejecta and the stellar progenitor layers. In
model \modelname{35OC-Rp4} we do not observe the action of MRI,
except, perhaps very early after its core collapses. As we have
commented for \modl{35OC-Rs}, the initial (relatively large) strength
of the poloidal magnetic field of \modl{35OC-Rp4} may hamper the
development of the MRI due to the dynamical back-reaction of the
magnetic field onto the background flow.

\subsection{Angular momentum redistribution and stability of the hypermassive PNS}
\label{sSek:angmomredistribution}
We first discuss the redistribution of angular momentum in
\modl{35OC-RO}.  In parallel with the mass growth, the angular
momentum of the PNS increases due to accretion.  The angular momentum
is fairly evenly distributed among the different shells of the PNS
(see bottom panel of \figref{Fig:35OC-RO-globals}). The angular
momentum of the innermost layers of the PNS rises alongside their
increase in mass. The outer three shells between
$\rho = \zehn{10} \, \gccm$ and $\rho = \zehn{12} \, \gccm$ possess
rather high specific angular momentum around
$j \gtrsim \zehnh{1.5}{16} \, \cm^2 \, \sek^{-1}$.  This distribution
is caused by magnetic stresses removing angular momentum from the
interior of the PNS to its envelope, thereby countering the inward
transport by purely hydrodynamic flows.  The resulting rotational
support of the envelope of the PNS limits the degree to which its
concentration towards the centre can go on and contributes to maintain
the PNS stability against BH collapse once $M_{\pnss}>\Mmax$. This
effect explains why lower-density shells retain a comparably low, but
non-negligible fraction of the mass.  It also accounts for the
strongly prolate shape of the PNS and the large equatorial radii, with
a pole-to-equator axis ratio of about 10:18 by the end of the
simulation.

We found that magnetic redistribution of angular momentum from the
centre tends to increase the radius of the core.  At first, it might
be natural to expect the exact opposite outcome, viz.~a contraction
triggered by the loss of rotational support at the centre, analogously
to the case of an accretion disc where outward angular-momentum
transport enables accretion.  To understand our result, we have to
take into account that the angular momentum that is removed from the
inner regions of the core does not immediately leave the PNS.  Its
efficient transport is limited by two effects: firstly, transport is
restricted to a region where the magnetic fields are sufficiently
strong, and, secondly, it has to act against the infall of matter from
the post-shock region.  As a consequence of these effects, angular
momentum removed from the centre does not go beyond the outer layers
of the PNS, where it increases the centrifugal support and hence
causes an expansion. The effect that we have discussed for
\modl{35OC-RO} applies to nearly all the models in this paper. As a
general trend we observe that a stronger field reduces the total
rotational energy (\figref{Fig:globalvars-1}\panel{c}), as could be
expected. It does, however, deform the PNS to a more, rather than
less, oblate geometry, a shape typical for higher, rather than lower,
rotational energy.

We may compare the enhancement of the angular momentum of the shells
at $\rho < \zehn{13} \, \gccm$, first, in models with different
initial toroidal fields (models \modelname{35OC-Rw},
\modelname{35OC-RO}, \modelname{35OC-RO2)}. We find that the increase
of $J$, especially in the layer
$\zehn{12}\,\gccm <\rho < \zehn{13} \, \gccm$, is more pronounced for
models with initially smaller toroidal field, being \modl{35OC-Rw} the
one with the largest increase (\figref{Fig:35OC-Rw-globals}\panel{c})
and \modl{35OC-RO2} the one with the smaller increase (below the
values reached by \modl{35OC-RO}; \figref{Fig:35OC-RO-globals}).

We now turn to the effects of poloidal fields of increased strengths
on the angular momentum redistribution inside the PNS.  The angular
momentum redistribution in models \modelname{35OC-Rp2}, and
\modelname{35OC-Rp3} is imprinted in the rotational profile of the the
outer PNS layers. In Fig.\,\ref{Fig:globalvars-1}\panel{h}, we see
values of $q_{\pnss}$ above $\sim -1$ episodically for both of these
models. Indeed, during short time intervals $q_{\pnss}\sim 0$,
signalling epochs in which the outer third of the PNS rotates nearly
rigidly. These values contrast with $q_{\pnss}\lesssim -1.5$ for the
rest of the models, most of the time. The previous comment also
applies to \modelname{35OC-Rp4}. The time at which the raise of
$q_{\pnss}$ above $-1.5$ happens is inversely correlated with the
initial poloidal magnetic field strength. While for models
\modelname{35OC-Rp2} and \modelname{35OC-Rp3} these times are
$\sim 0.5\,\sek$ and $\sim 1.9\,\sek$, respectively, for
\modl{35OC-Rp4}, the increase of $q_{\pnss}$ above $-1.5$ does not
happen during the computed evolution time ($t_{\rm max}=1.95\,\sek$). Model \modelname{35OC-RO} only displays a short episode ($0.5\,\sek \lesssim \tpb\lesssim 0.7$\,s)  when $q_\pnss \gtrsim-1.5$ (Fig.\,\ref{Fig:globalvars-1}\panel{h}).

The PNS of \modl{35OC-Rs} is strongly affected by the redistribution
of angular momentum.  Its outer shells contain a significant fraction
of the total angular momentum on roughly the same level as the central
layers (\figref{Fig:35OC-Rs-globals}\panel{c}).  While the contraction
leads to an increase of the fraction of the total mass that resides at
the highest densities (until $\tpb\sim 0.8\,\sek$), the low-density
envelope continues to hold a constant mass (panel \panel{b}).  The
specific angular momentum of these layers is so large
($j \ge \zehnh{2.5}{16} \, \cm^2 \sek^{-1}$) that they can be
self-sustained against the gravitational pull of the PNS. The
morphological evolution of this model, whose PNS develops a toroidal
shape with a maximum density off-centre
(Sec.\,\ref{sSek:35OC-bmoltfort}) is instigated by the large amount of
angular momentum transported outwards. We note that nearly all the
angular momentum of the PNS in \modl{35OC-Rs} is concentrated at
densities below the nuclear saturation density
(\figref{Fig:35OC-Rs-globals}\panel{c}). Differently from models with
an initially weak field (\eg \modl{35OC-Rw}), in this case, the
initial field is so strong that no additional, MRI-driven
amplification is required to cause the effects described above.

\begin{figure*}
  \centering  
  \begin{tikzpicture}
  \pgftext{\hbox{
  \includegraphics[width=.33\linewidth]{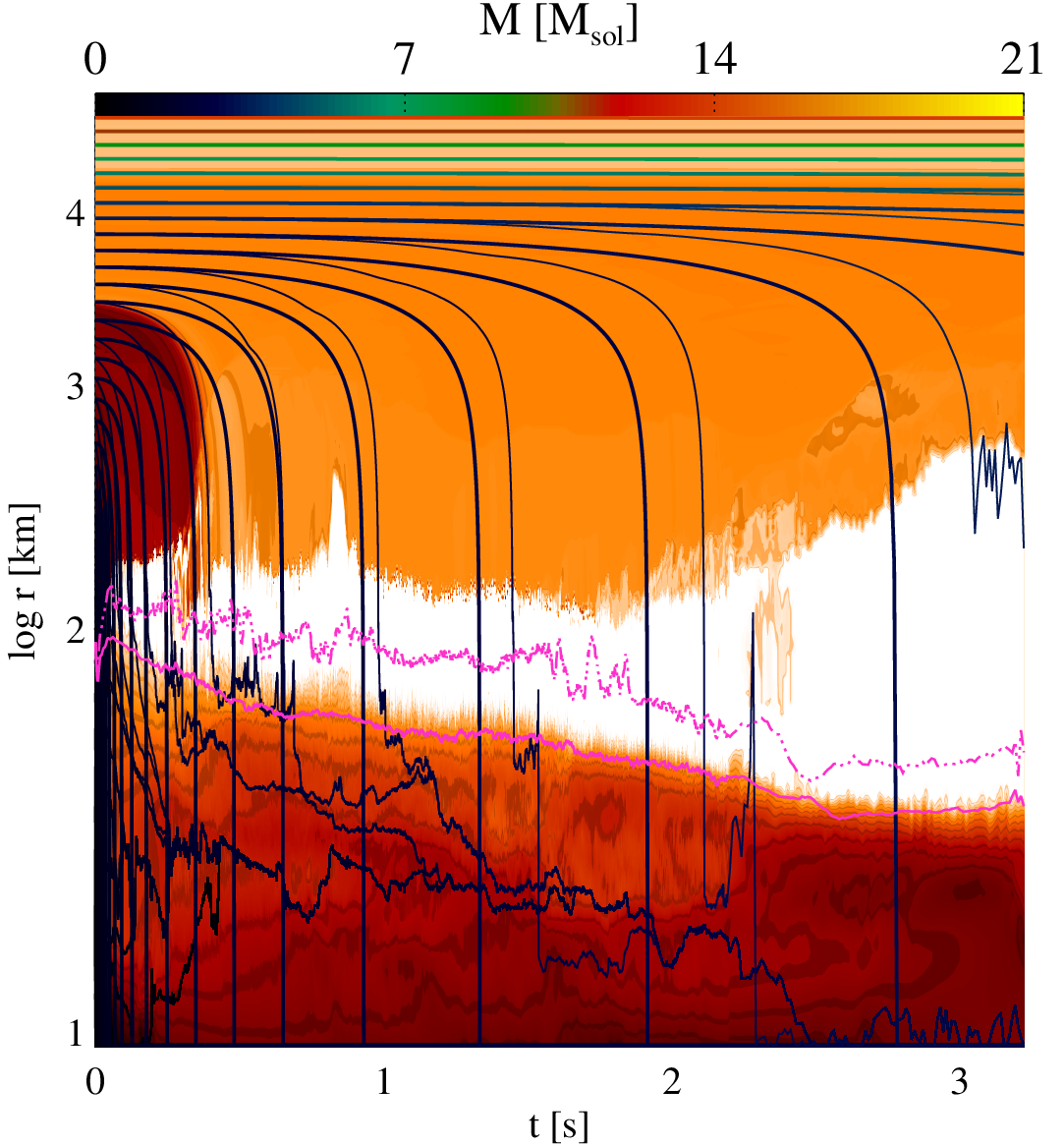}
  \includegraphics[width=.33\linewidth]{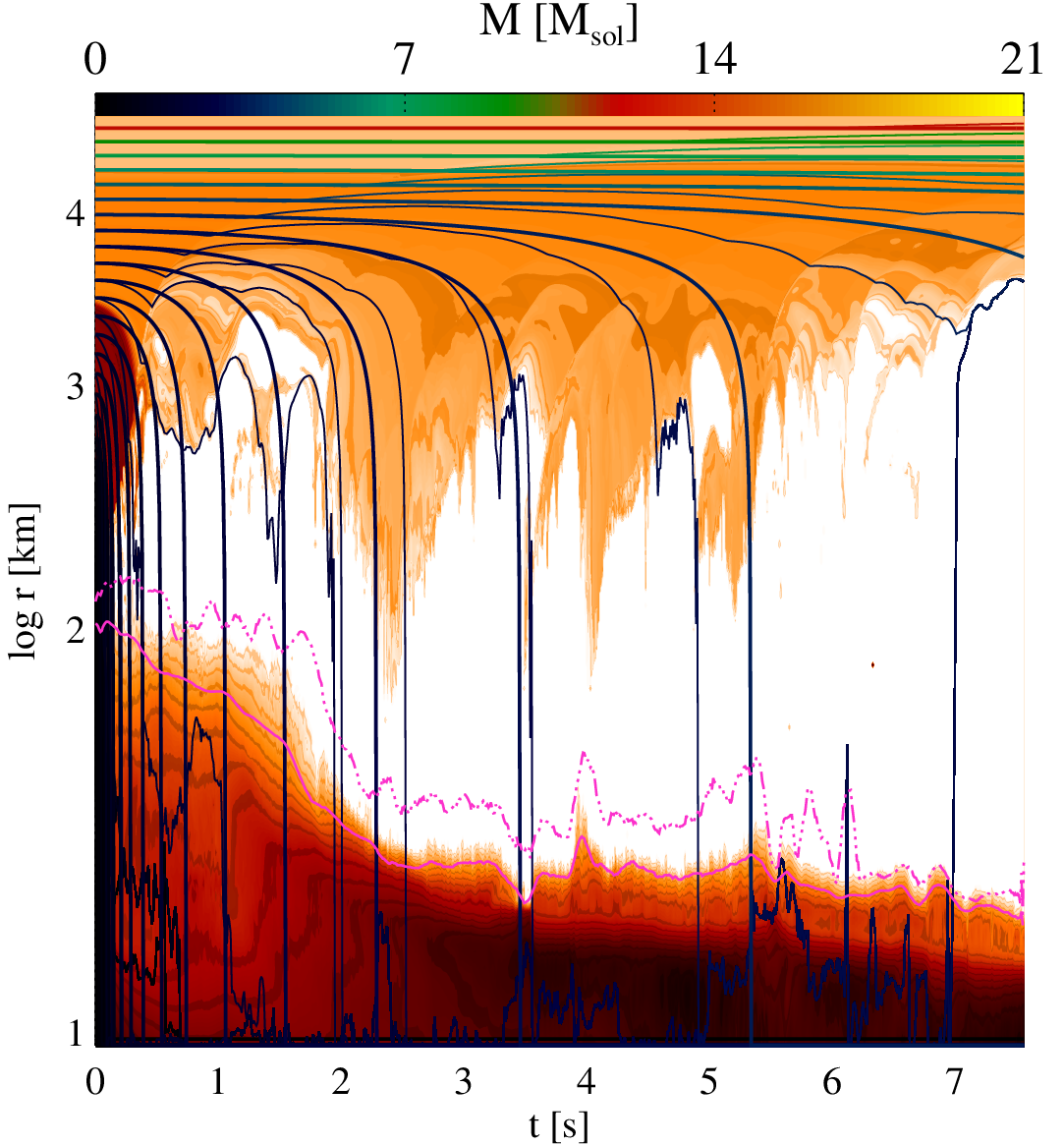}
  \includegraphics[width=.33\linewidth]{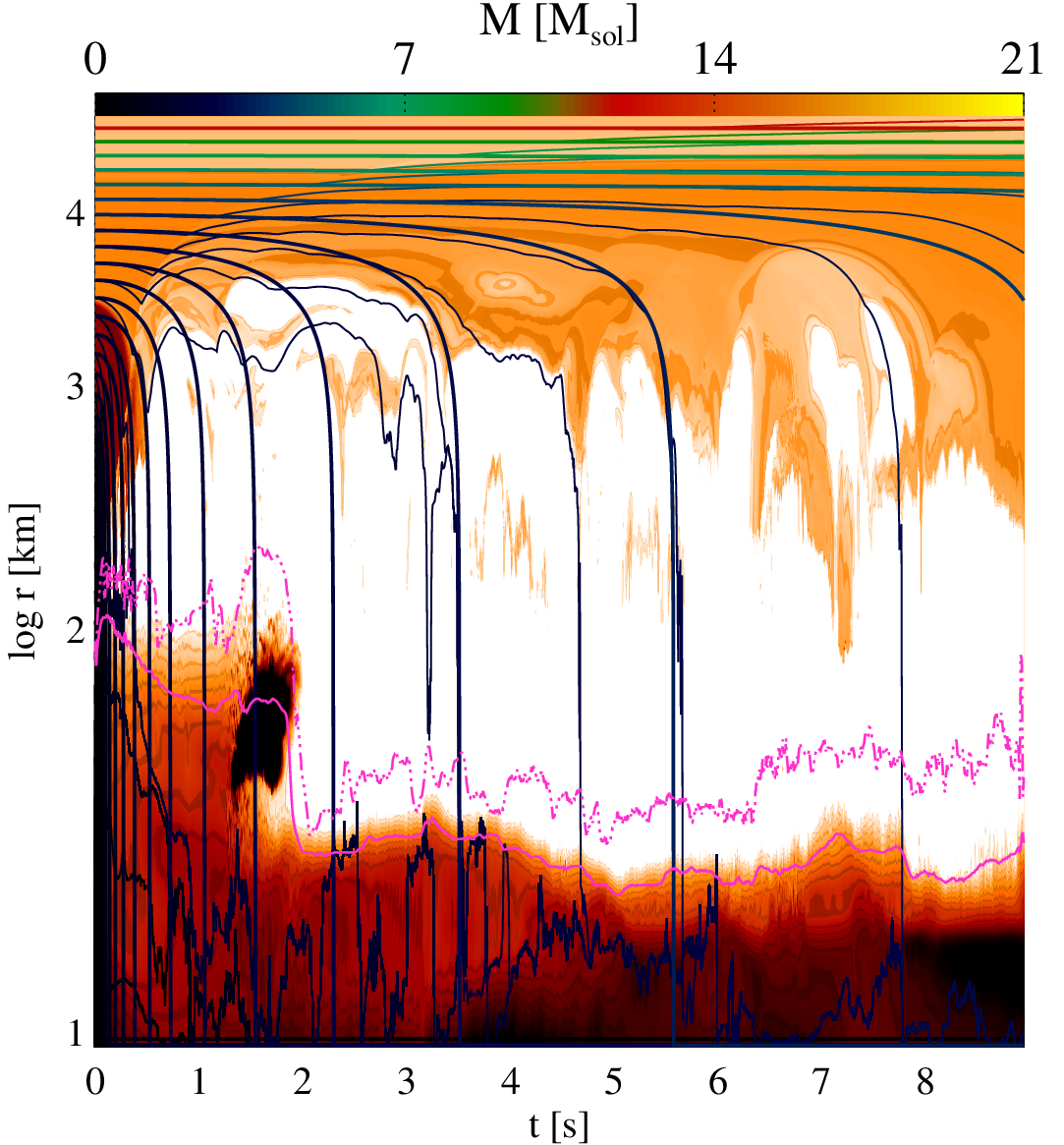}
                          }
               }%
   \node[fill=white, opacity=0, text opacity=1] at (-8.76,-2.7) {\large (a)};
   \node[fill=white, opacity=0, text opacity=1] at (-2.84,-2.7) {\large (b)};
   \node[fill=white, opacity=0, text opacity=1] at (+3.10,-2.7) {\large (c)};
  \end{tikzpicture}
  \caption{
    The background (orange shades) represents the entropy per baryon (lighter shades correspond to higher entropy)
    as a function of radius and time along the equator. Overlaid, we 
    represent with thick solid lines the space time
    trajectories corresponding to mass shells falling from a certain
    mass coordinate, $M$, according to an effective gravitational
    potential $-GM/(4r)$ (without accounting for any centrifugal
    effect), such that the fall time from $r(M)$ to $r=0$ corresponds
    to a time $t_\textsc{ff}(M)$ as given by Eq.\,\ref{eq:tdf}
    replacing $M_\textsc{df}$ by $M$. The colours of the lines
    correspond to the mass from which each mass-shell falls (see upper
    colour bar). The thin solid lines represent the actual space-time
    trajectories followed by the mass-shells along the equator. The solid and dashed-tripple- doted pink lines in each panel correspond to the equatorial PNS radius and to the equatorial gain radius, respectively. The
    \panel{a}, \panel{b}, and \panel{c} panels correspond to models
    \modelname{35OC-RO}, \modelname{35OC-Rp2}, and
    \modelname{35OC-Rp3} respectively. 
  }
  \label{Fig:free-fall}
\end{figure*}

\section{Evolution of the remnant}
\label{Sek:PNS-GRB}

\begin{figure*}
  \centering
    \begin{tikzpicture}
  \pgftext{\vbox{
  \hbox{
\includegraphics[width=0.32\linewidth]{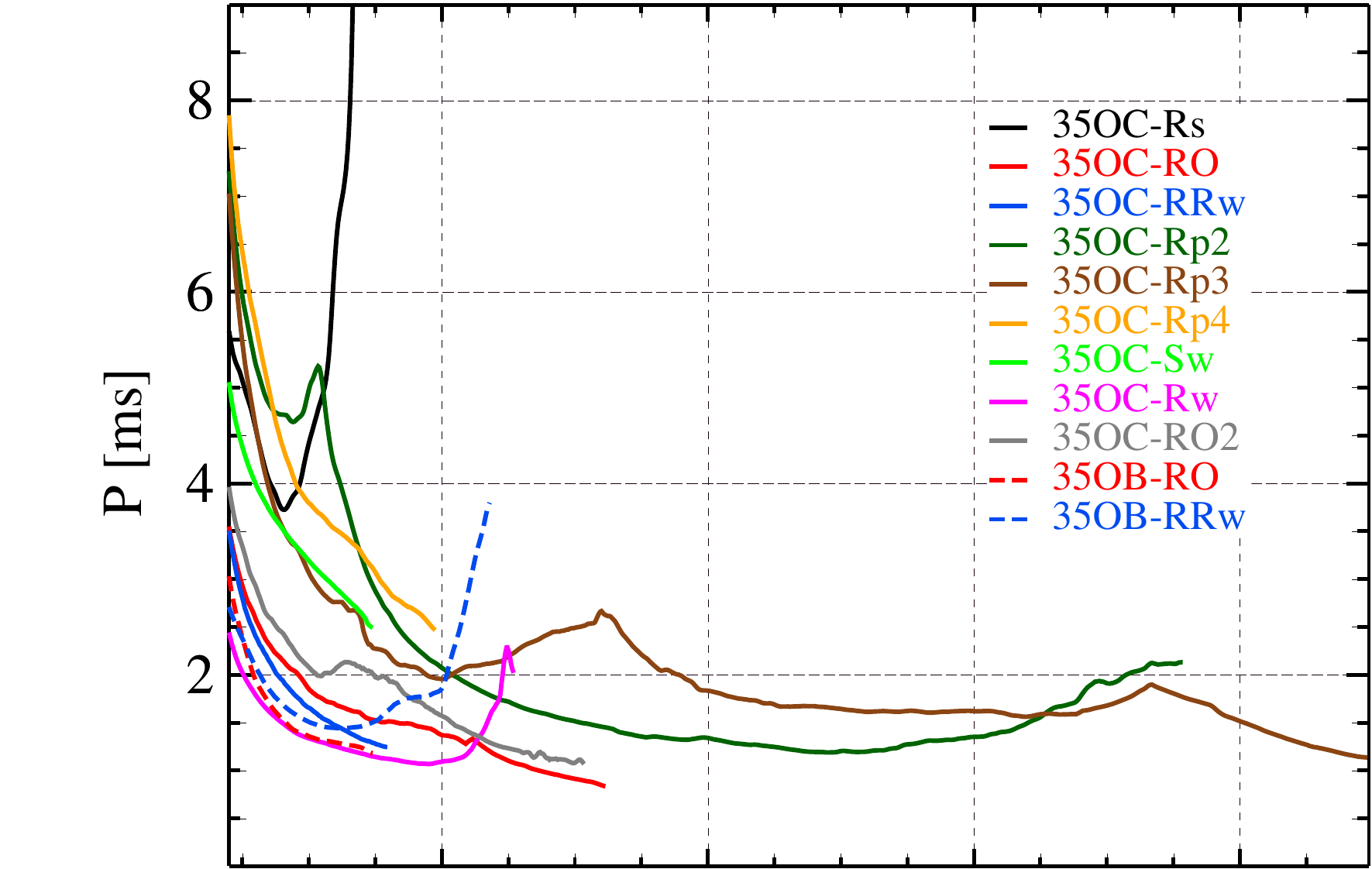}
\includegraphics[width=0.32\linewidth]{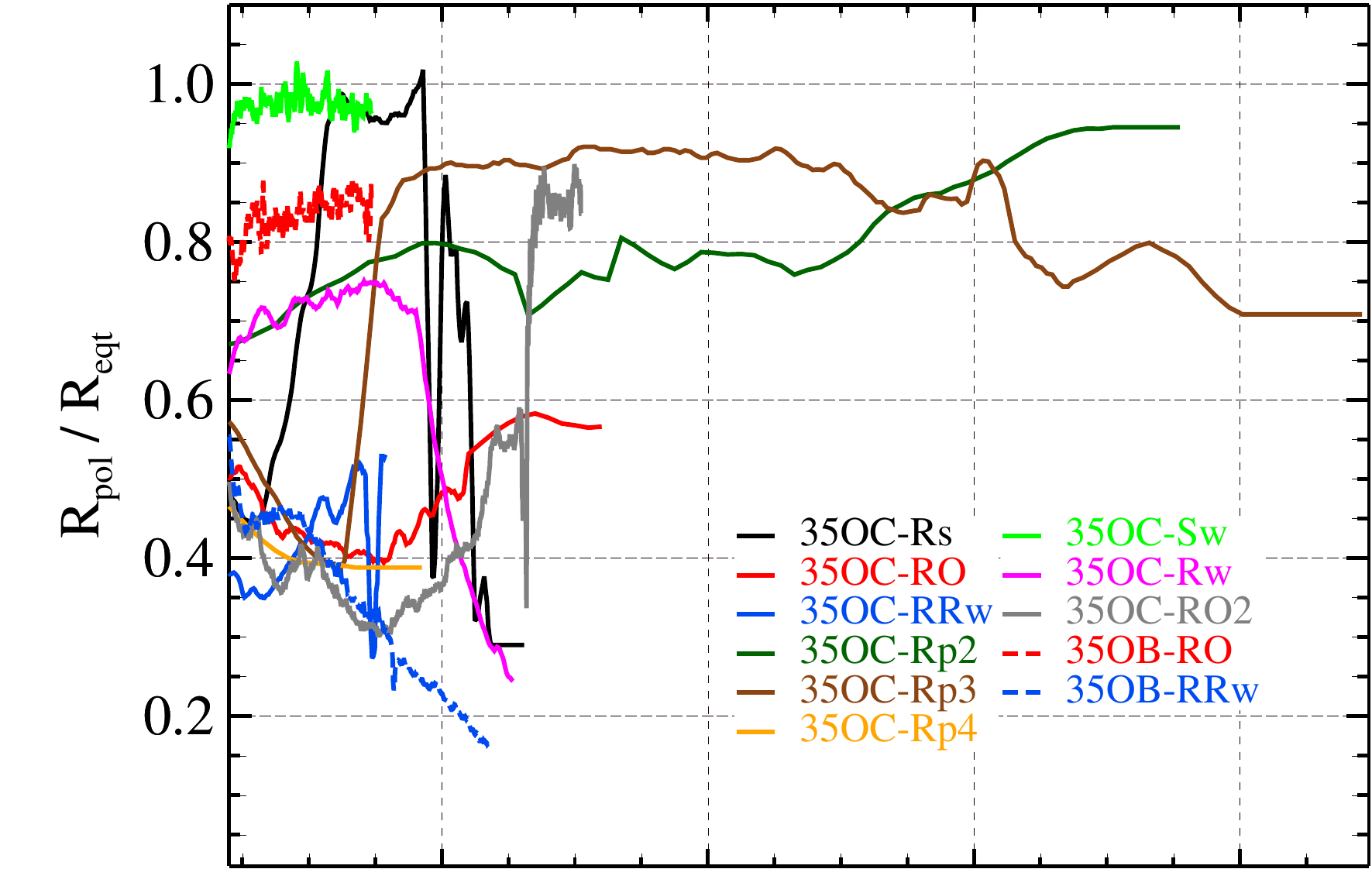}  
\includegraphics[width=0.32\linewidth]{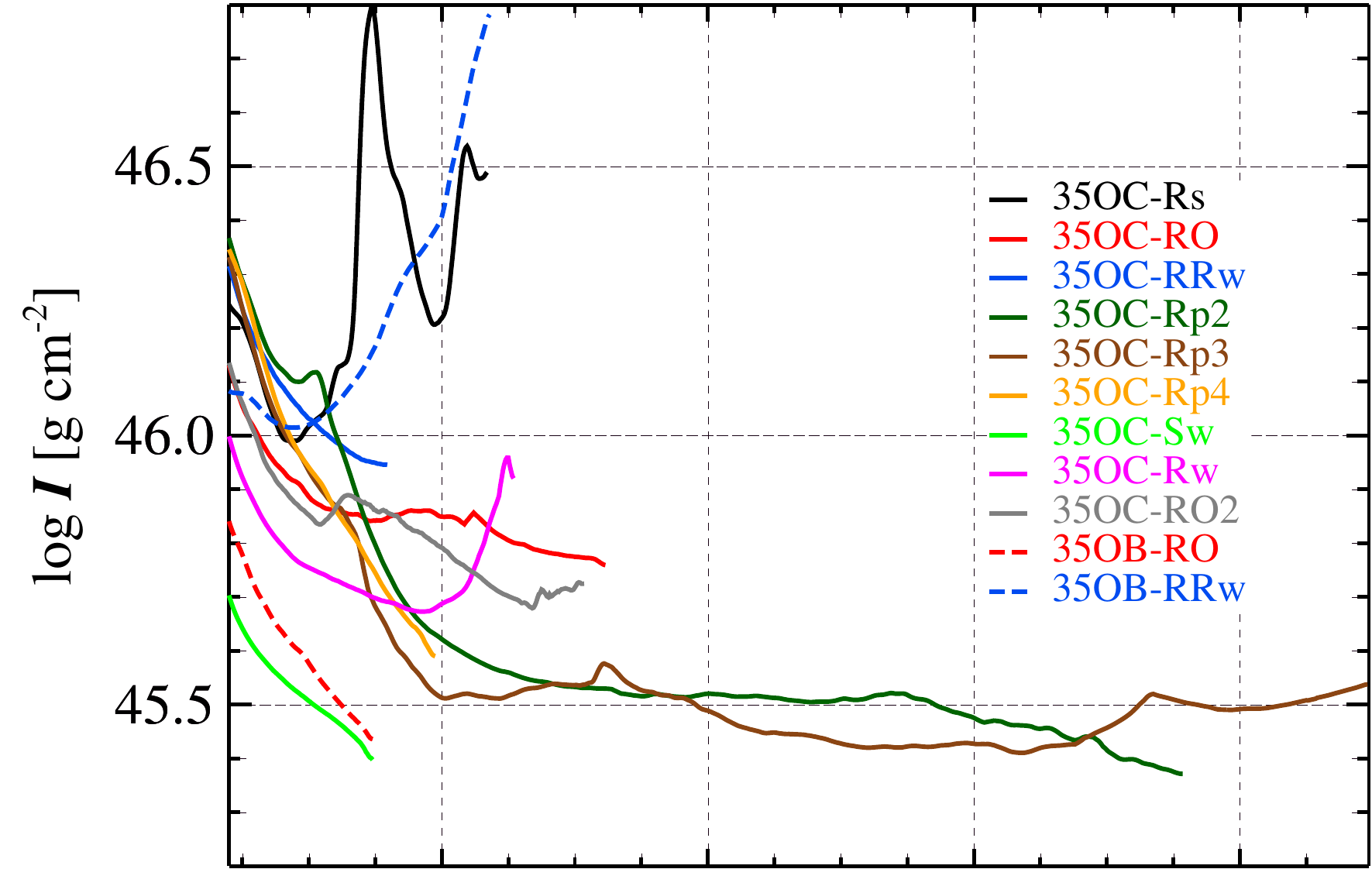}
         }
\vspace{-0.08cm}\hbox{
\includegraphics[width=0.32\linewidth]{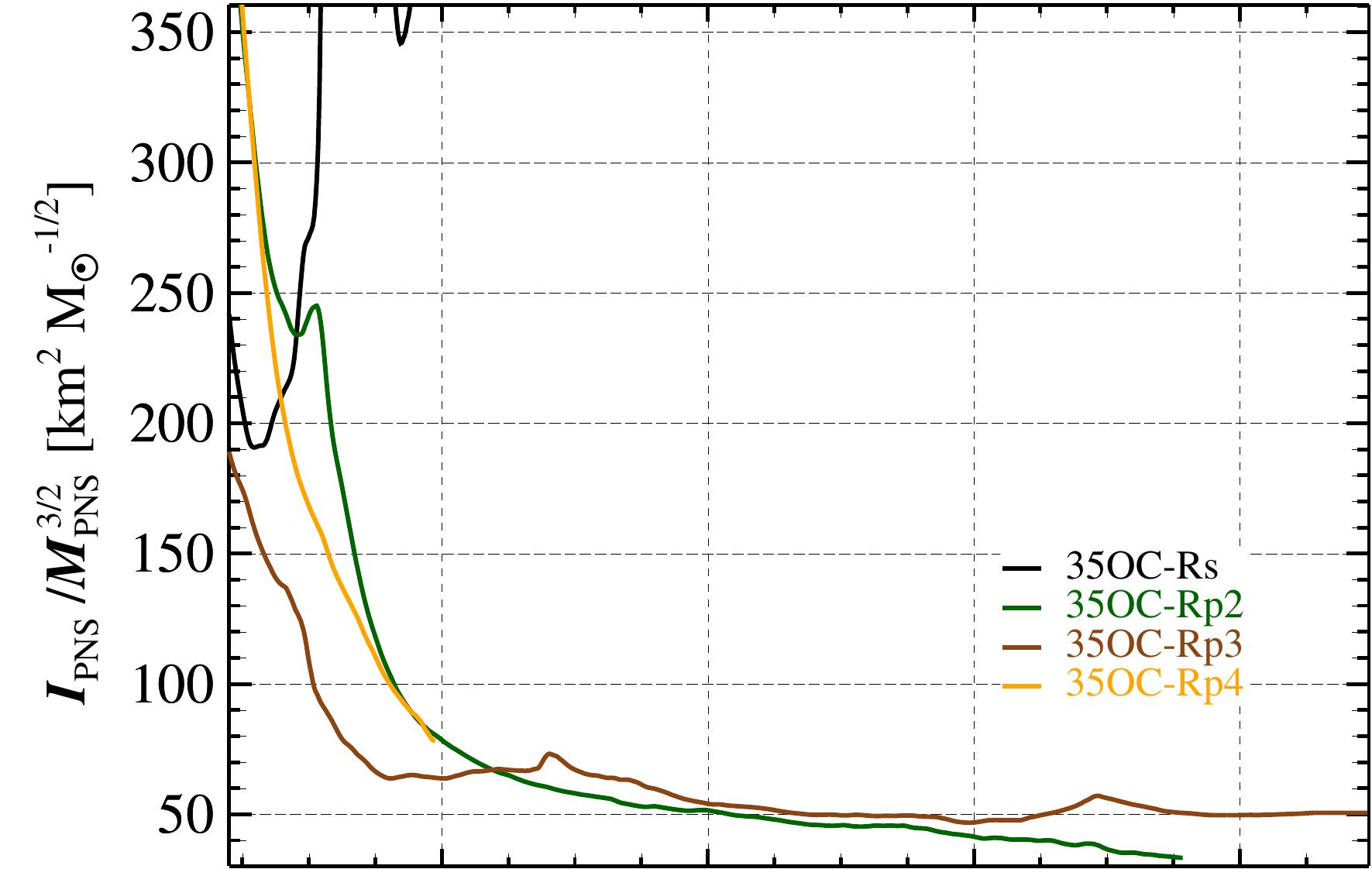}
\includegraphics[width=0.32\linewidth]{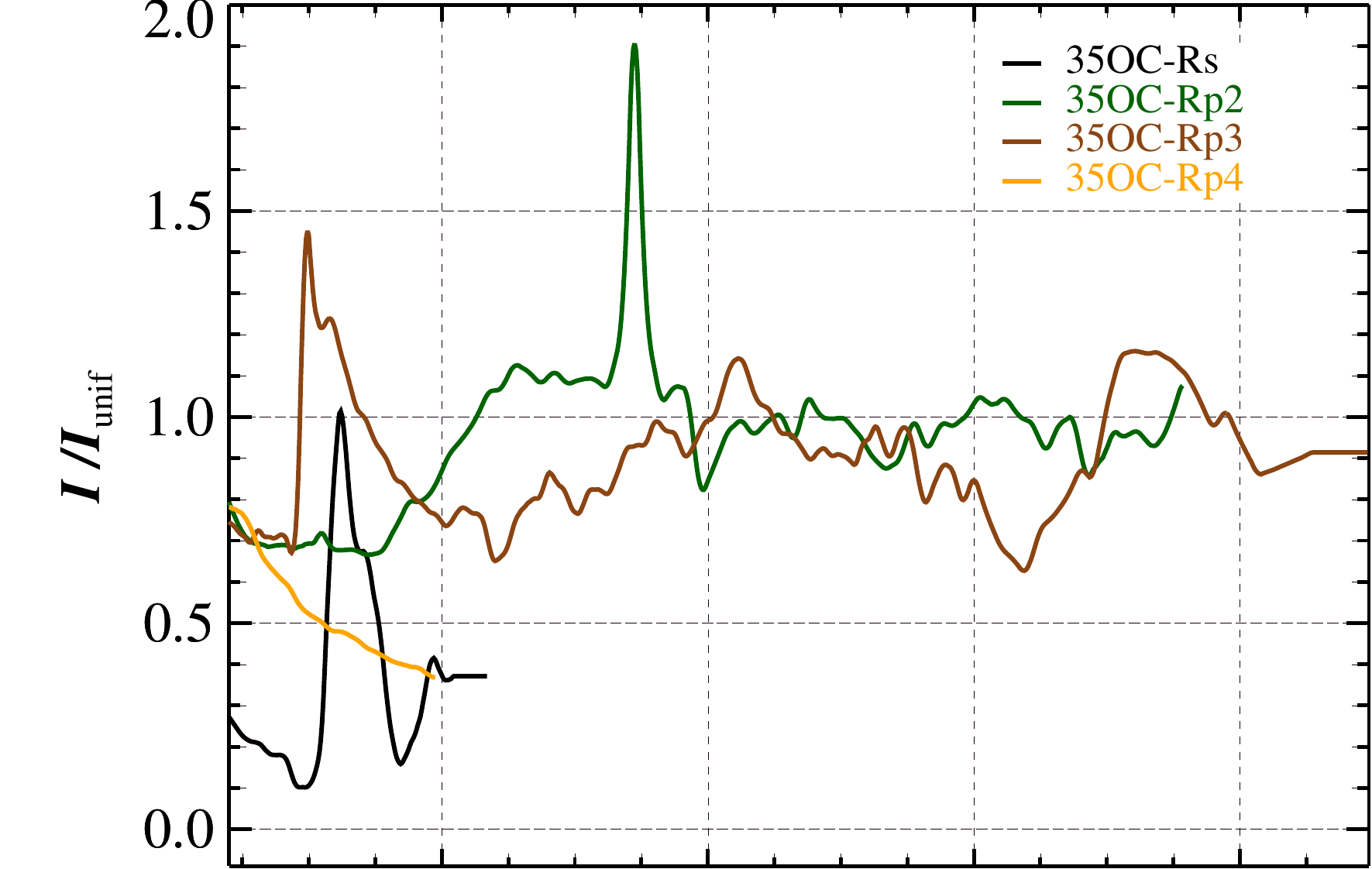}
\includegraphics[width=0.32\linewidth]{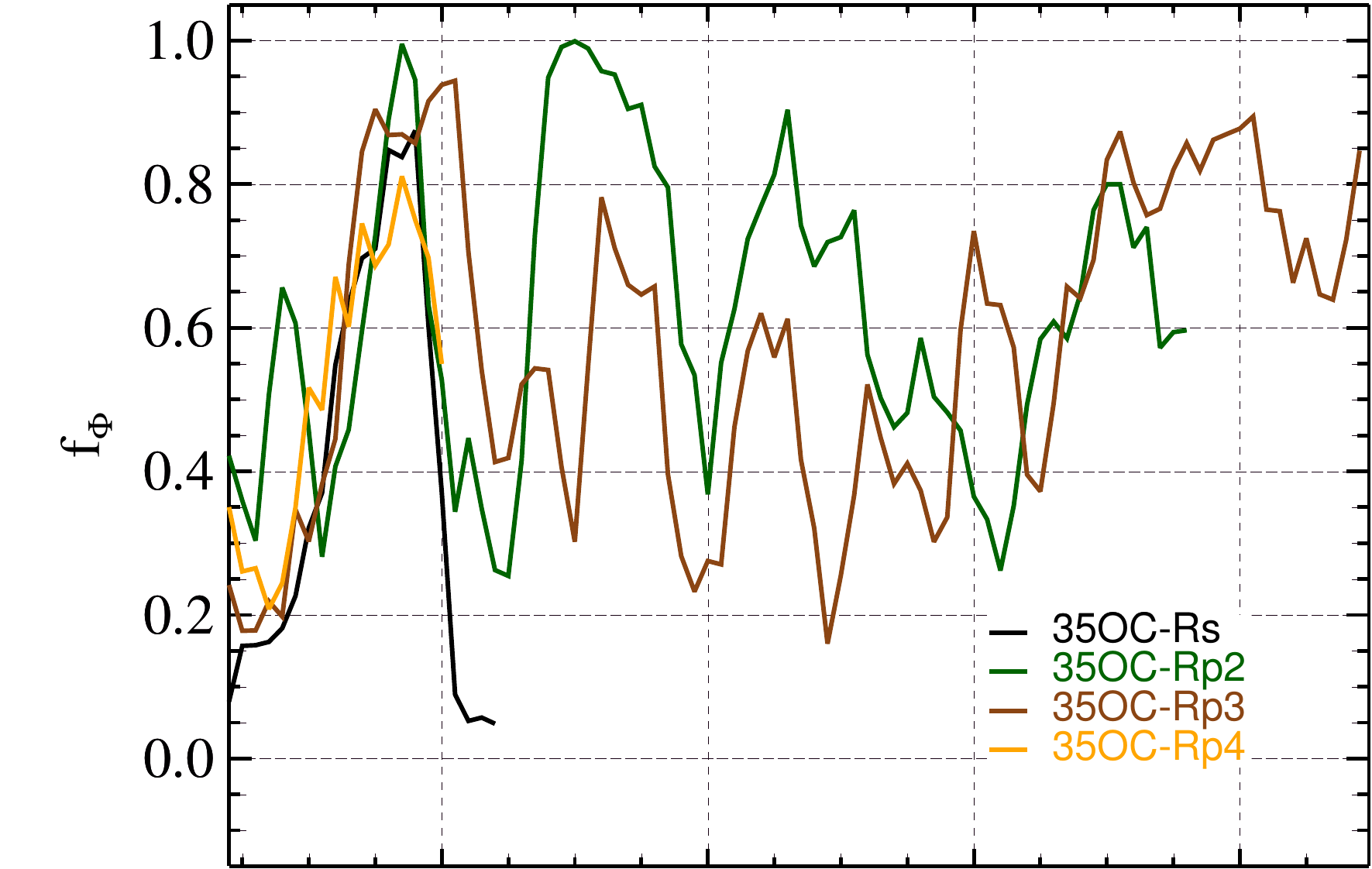}
         }
\vspace{-0.08cm}\hbox{
\includegraphics[width=0.32\linewidth]{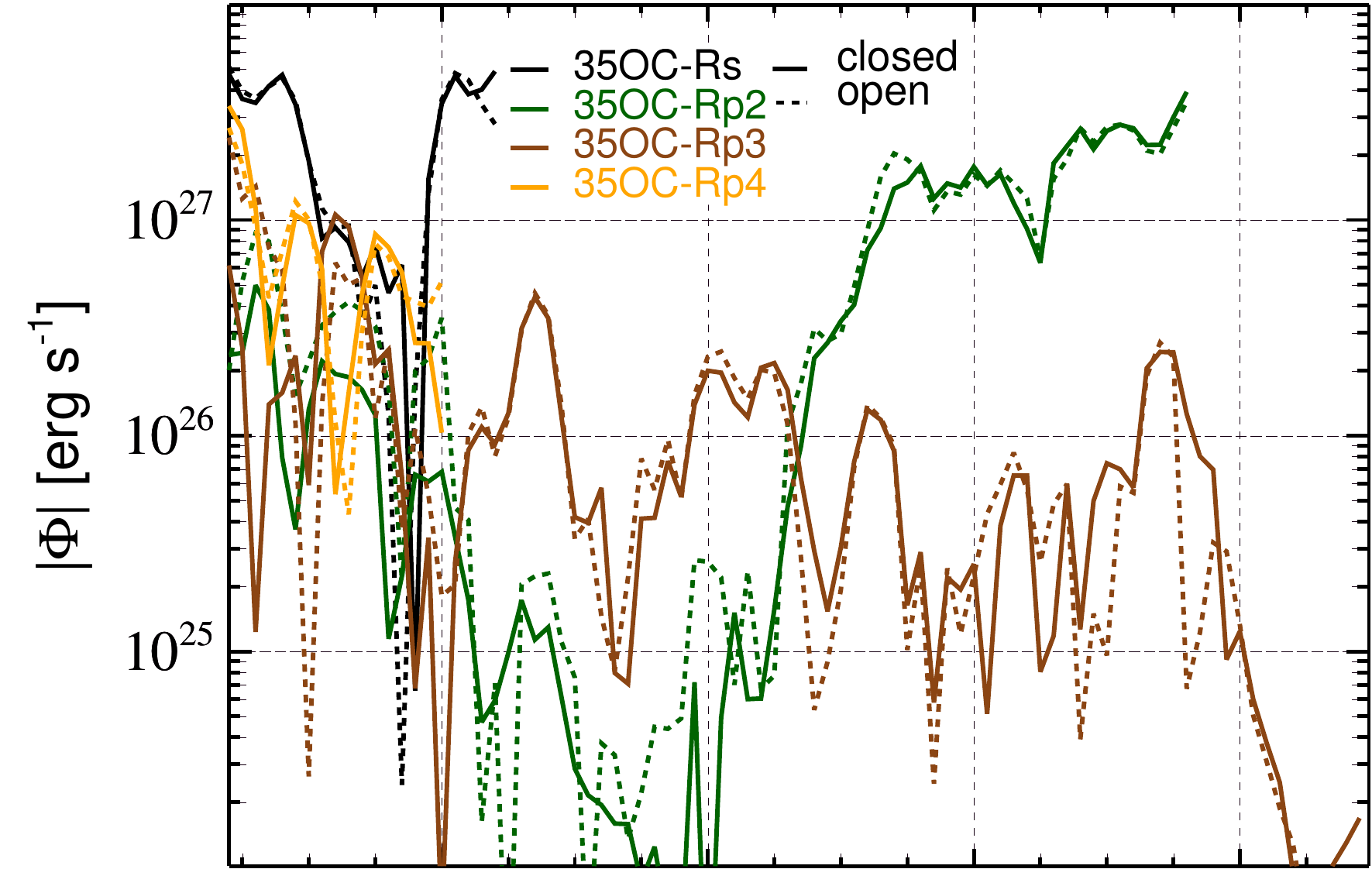}
\includegraphics[width=0.32\linewidth]{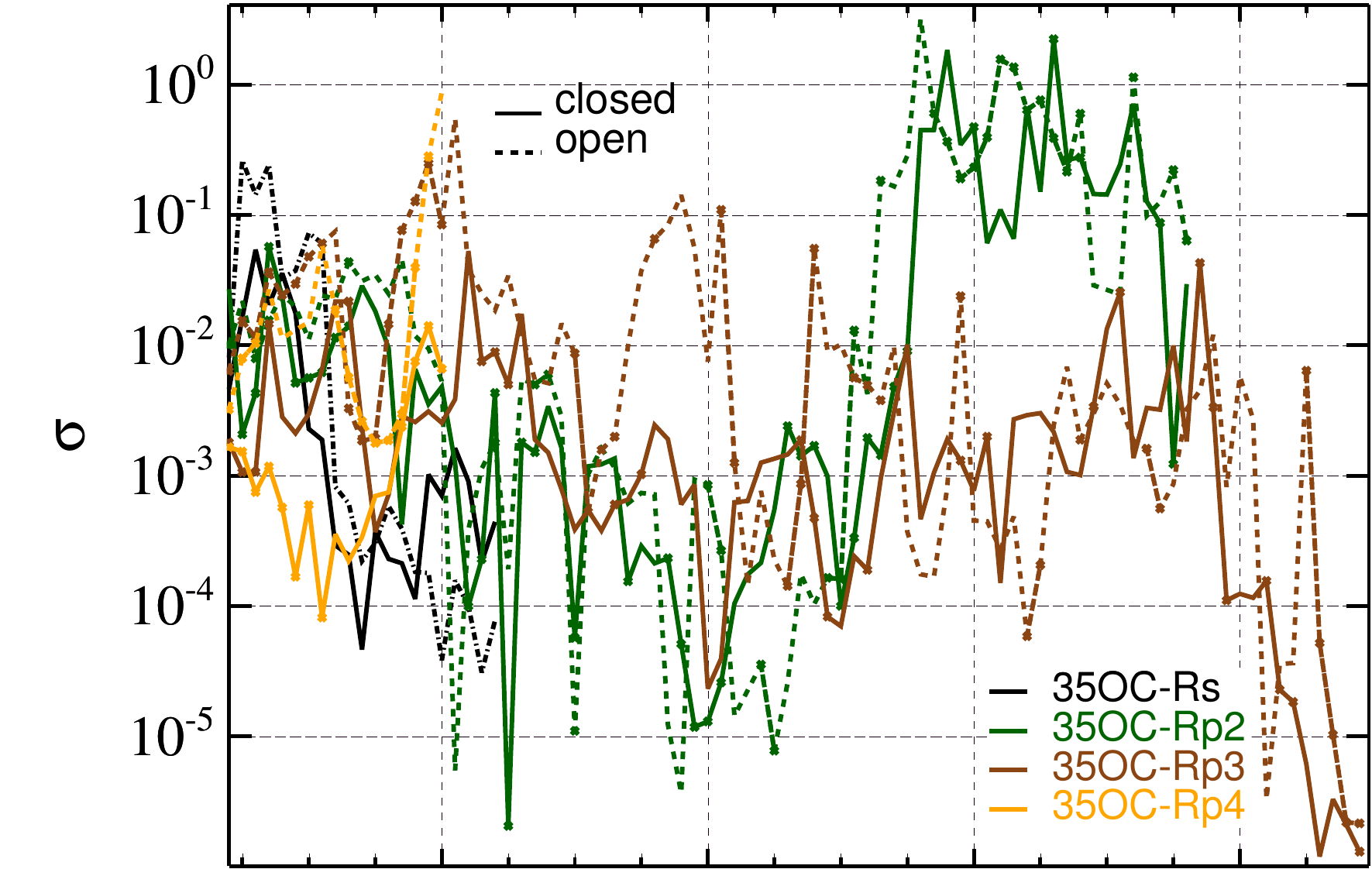}
\includegraphics[width=0.32\linewidth]{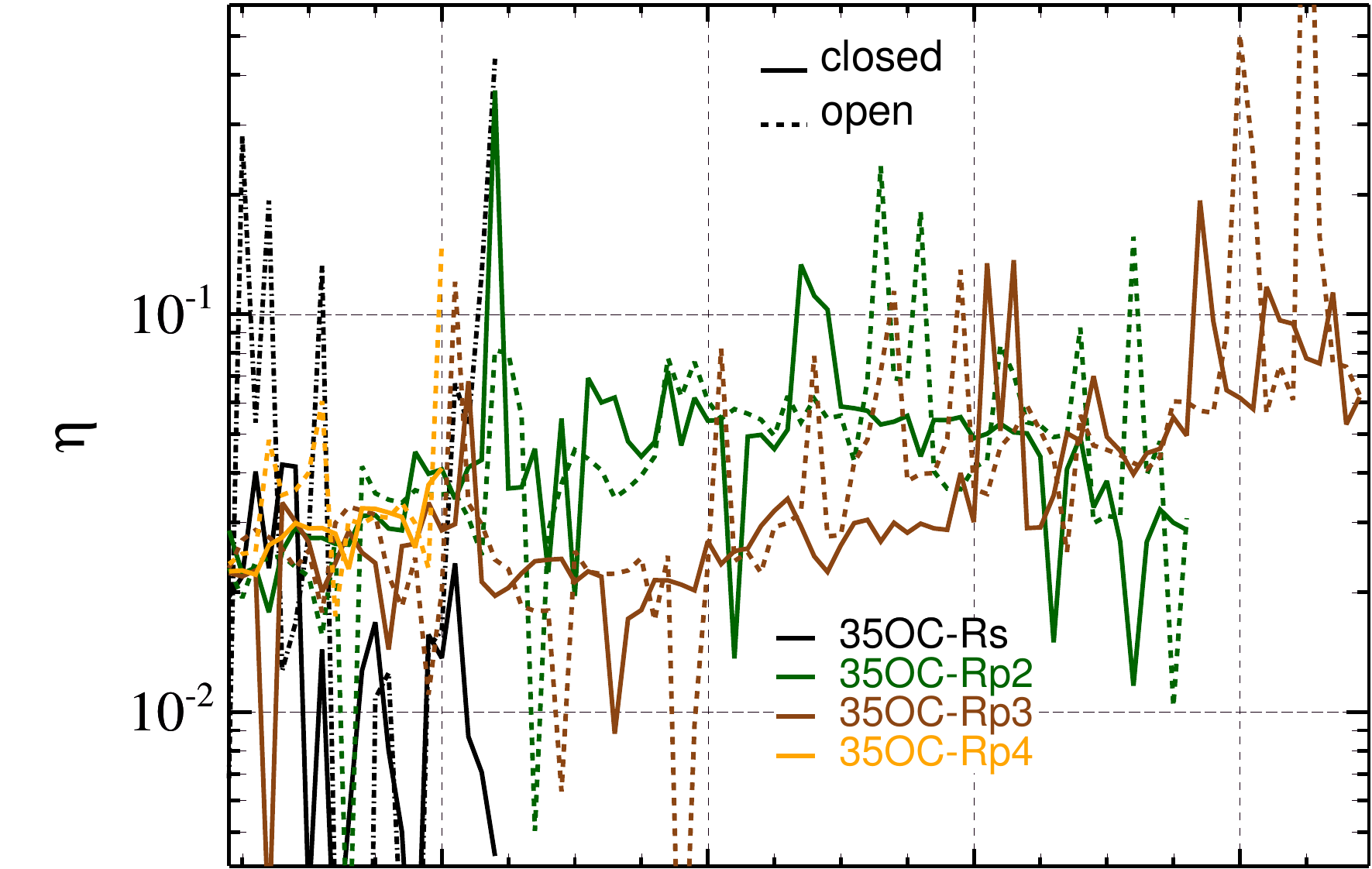}
         }
\vspace{-0.08cm}\hbox{
\includegraphics[width=0.32\linewidth]{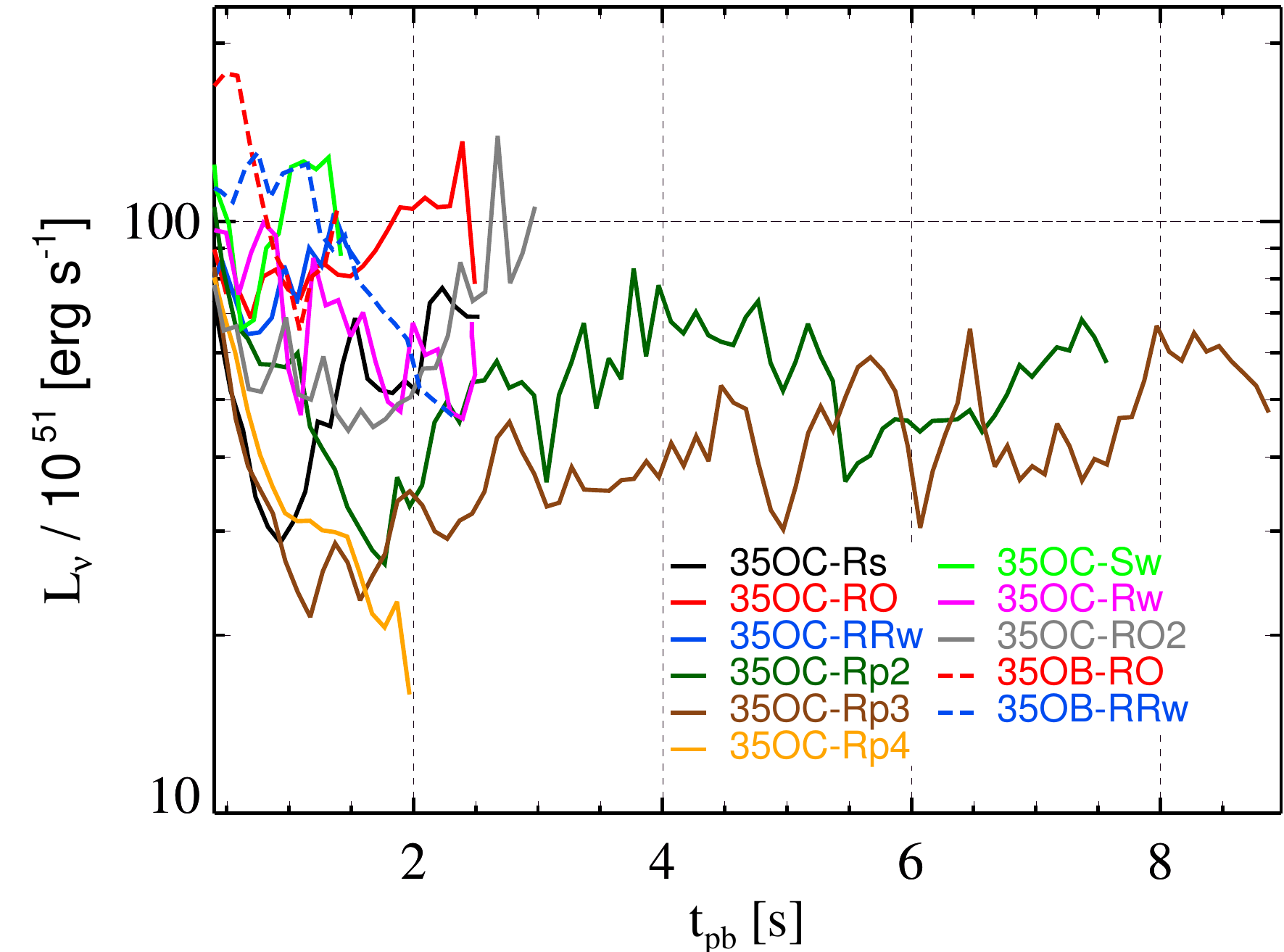}
\includegraphics[width=0.32\linewidth]{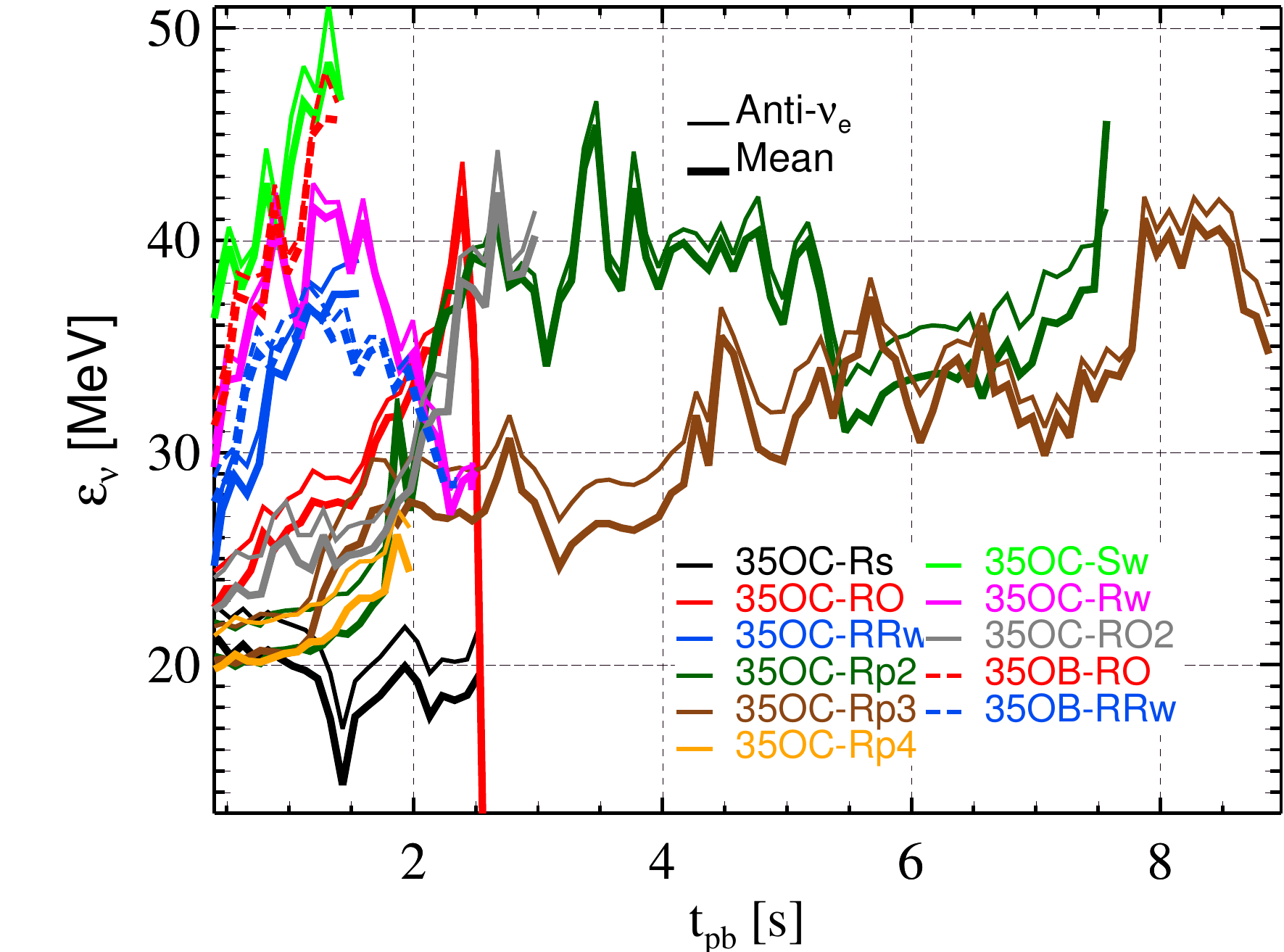}
\includegraphics[width=0.32\linewidth]{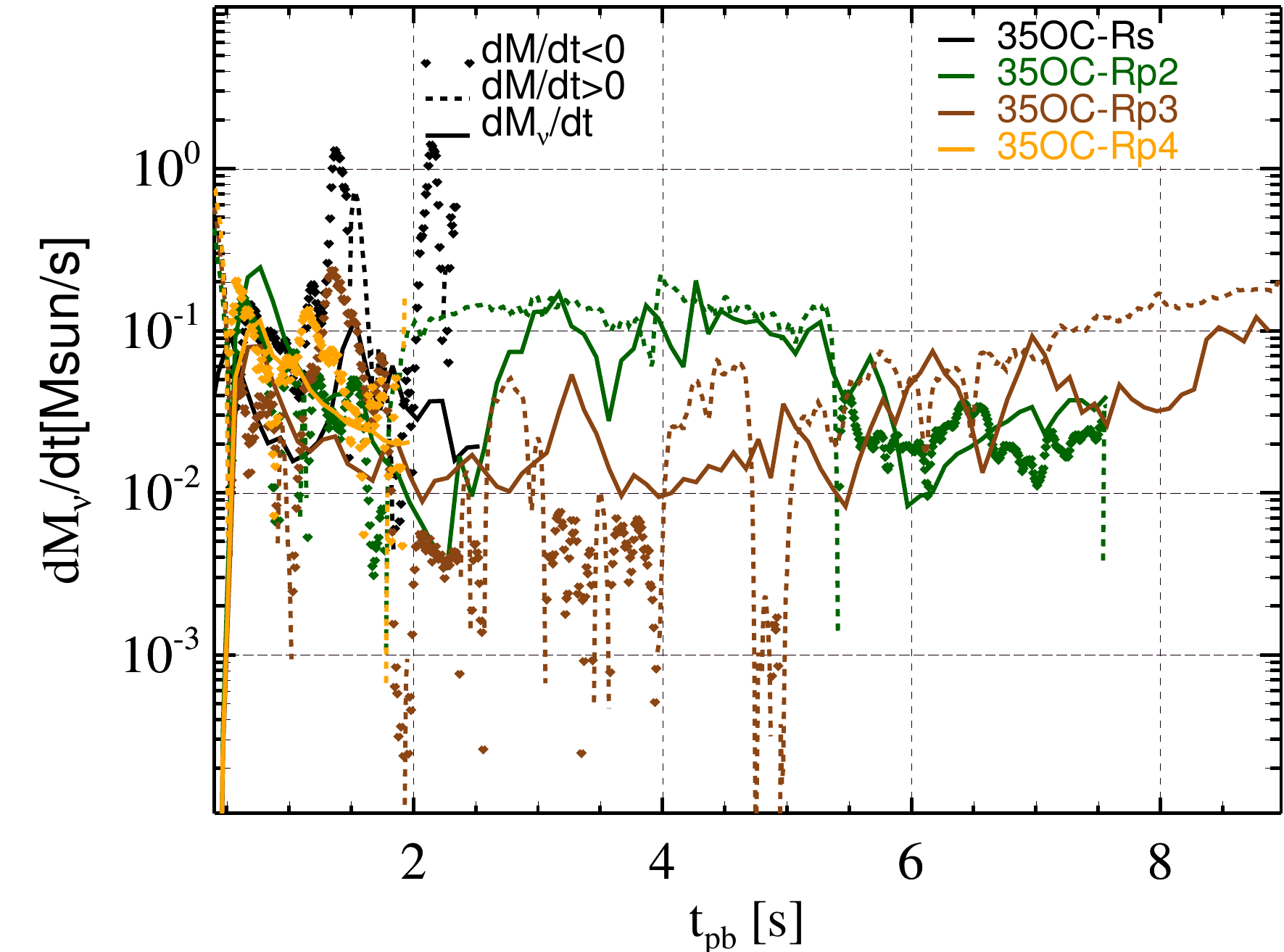}
         }
}
}
   \node[fill=white, opacity=0, text opacity=1] at (-8.29,+3.99) {\large (a)};
   \node[fill=white, opacity=0, text opacity=1] at (-8.29,+0.49) {\large (d)};
   \node[fill=white, opacity=0, text opacity=1] at (-8.29,-3.1) {\large (g)};
   \node[fill=white, opacity=1, text opacity=1] at (-8.29,-6.7) {\large (j)};
   \node[fill=white, opacity=0, text opacity=1] at (-2.54,+3.99) {\large (b)};
   \node[fill=white, opacity=0, text opacity=1] at (-2.54,+0.49) {\large (e)};
   \node[fill=white, opacity=1, text opacity=1] at (-2.54,-3.1) {\large (h)};
   \node[fill=white, opacity=1, text opacity=1] at (-2.54,-6.7) {\large (k)};
   \node[fill=white, opacity=0, text opacity=1] at (+3.30,+3.99) {\large (c)};
   \node[fill=white, opacity=0, text opacity=1] at (+3.30,+0.49) {\large (f)};
   \node[fill=white, opacity=1, text opacity=1] at (+3.30,-3.1) {\large (i)};
   \node[fill=white, opacity=1, text opacity=1] at (+3.30,-6.7) {\large (l)};
  \node[fill=white, opacity=1, text opacity=1, rotate=90] at (-8.20,+5.65) {\small $\bar{P}\,\text{[ms]}$};	
  \node[fill=white, opacity=1, text opacity=1, rotate=90] at (+3.15,-5.0) {\small $\text{dM}_{\rm o}\text{/dt}\,[\text{M}_\odot\,\sek^{-1}]$};	
  \end{tikzpicture}
  \caption{
    Time evolution of different variables of the models annotated in
    the legends.  The panels display: \panel{a} Period of the
    PNS. \panel{b} Ratio of polar to equatorial radii of the
    PNS. \panel{c} Moment of inertia of the PNS. \panel{d} Ratio of
    the moment of inertia to $M_\pnss^{3/2}$. \panel{e} Ratio of the
    moment of inertia of the PNS to the moment of inertia that a
    uniform sphere with and effective radius
    $R_\pnss=(R_{\rm pol}R_{\rm eqt}^2)^{1/3}$ and the same mass would
    have (Eq.\,\ref{eq:Iunif}). \panel{f} Fraction, $f_\Phi$, of the
    PNS surface threaded by open magnetic field lines. \panel{g}
    Absolute value of the magnetic flux in the open and closed
    magnetospheric regions. \panel{h} Magnetization, $\sigma$, in the
    open (Eq.\,\ref{eq:sigmaopen}) and closed
    (Eq.\,\ref{eq:sigmaclosed}) magnetospheric regions. \panel{i}
    Baryon loading, $\eta$, in the open (Eq.\,\ref{eq:etaaopen}) and
    closed (Eq.\,\ref{eq:etaaclosed}) magnetospheric
    regions. \panel{j} Neutrino mean luminosity
    (Eq.\,\ref{eq:Lnumean}). \panel{k} Neutrino mean energy
    (Eq.\,\ref{eq:Enumean}). \panel{l} Evolution of the mass loss
    ($\dot{M}_\pnss<0$) and mass gain ($\dot{M}_\pnss>0$) compared
    with the theoretical prediction of Eq.\,\eqref{eq:Mth}.}
  \label{Fig:GRB-figs}
\end{figure*}

In the following, we will discuss several aspects of our models
relevant to the formation of the central engines of GRBs within either
the collapsar or the PM model.  The two models rely on the formation
of either a BH or a PNS at the centre of the star, respectively, and,
in either case, the presence of high angular momentum and strong
magnetic fields.  Despite a number of important studies on the subject
such as the theoretical work by, \eg
\cite{Thompson_et_al__2004__apj__Magnetar_Spin-Down_Hyperenergetic_SNe_and_GRBs,Metzger_et_al__2011__mnras__Theprotomagnetarmodelforgamma-raybursts,Metzger_et_al__2015__mnras__Thediversityoftransientsfrommagnetarbirthincorecollapsesupernovae,Metzger_etal_2018ApJ...857...95},
and simulations
\citep[\eg][]{Bucciantini_et_al__2007__mnras__Magnetar-driven_bubbles_and_the_origin_of_collimated_outflows_in_GRBs,
  Bucciantini_et_al__2012__mnras__Shortgamma-rayburstswithextendedemissionfrommagnetarbirth:jetformationandcollimation,
  Burrows_etal__2007__ApJ__MHD-SN}, the specific requirements for both
models are not known to the level of detail that would allow for
reliable predictions about the evolutionary path of a given stellar
progenitor.

Formally, the subset of models that collapse to a BH might form a GRB
central engine, if they may also surround the central compact object
with a suitable accretion disc.  Our simulations make it abundantly
clear that at most moderately relativistic outflows are generated
during the fairly long phase of up to more than $2\,\sek$ in which an
PNS exists (before collapsing to a BH), in some cases even strongly
magnetised and rapidly rotating.  The same is true for the models
without a final BH collapse within the time scales of our runs (in one
model $\sim 9\,$s have been computed), which we consider potential PM
cases. Among these cases, the final fate of the compact remnant will
depend upon the amount of mass accreted onto the PNS on timescales
significantly longer than we have been able to compute so far. It is,
however, clear that models in which the PNS mass has stopped growing
(or the mass growth is small after the whole iron core has collapsed,
namely, $\langle \dot{M}\rangle< 0.05 \msol\,\sek^{-1}$; where
$\langle \dot{M}\rangle$ denotes a time averaged value) before
reaching the instability threshold set by the EoS are potential
candidates to host a PM central engine.  Hence, for all cases, we must
extrapolate our simulation results to later times in order to infer
the possibilities of a subsequent GRB engine.

\subsection{Collapsar candidates}
\label{sSek:BHevol}
For initial models with stellar or sub-stellar magnetic field,
centrifugal forces cause several of the cores to develop a strongly
oblate shape.  Although the processes accompanying the formation of a
BH at the centre of the core will certainly induce perturbations of
this structure, the long term survival of the outer PNS layers beyond
BH formation seems likely. This is because its stability against
gravity is provided by centrifugal forces to a much higher degree than
by the gas and neutrino pressure gradients.  Consequently, even a
sudden reduction of the thermal support would not lead to a prompt
accretion of these layers.  Consisting of matter with specific angular
momentum in excess of $j > \zehnh{1.5}{16} \cm^2\,\sek^{-1}$ and
undergoing infall of gas exceeding this value, they are very likely to
orbit the newly formed BH for many dynamical time scales, only to be
accreted gradually as a result of the slower processes governing the
redistribution of angular momentum. Besides the fate of the high
specific angular momentum of the aforementioned outer layers of the
PNS, there are stellar layers (located at mass coordinates
$> 7.5\msol$; Fig.\,\ref{Fig:progenitor-profile}) whose specific
angular momentum is large enough to be able to form an accretion
disc. We note that models with a successful SN explosion do not halt
completely the accretion process (see below). Therefore, we deem most
of our BH-forming models promising collapsar candidates. The only
likely exception to this estimation is \modl{35OC-Sw}, which due to
its low specific angular momentum may hardly form an accretion disk
around the formed BH (\figref{Fig:progenitor-profile}\panel{b}).

\paragraph*{Outflows.}
In \modelname{35OC-RO} polar outflows onto the PNS coexist with
equatorial downflows. As discussed in
\citetalias{Obergaulinger_Aloy__2019}, the success of polar, as
opposed to equatorial, shock revival in many of our models (and
singularly in \modl{35OC-RO}) is rooted in strong magnetic fields
concentrated along the rotational axis.  In addition, the pronounced
anisotropy of the neutrino emission caused by the rotational
flattening of the PNS and, in particular, the \nusps contribute to
launching the explosion.

The successful supernova explosion occurs in the form of collimated
jets of a fairly high energy. The outflow velocities ($<0.5c$) as well
as the propagation speed of the jet head are sub-relativistic
($\lesssim 0.15c$). For stellar progenitors as compact as 35OC (whose
radius is $R_{\ast}\approx \zehnh{5.3}{10}\,$cm), this means the
extremely well collimated outflow that we have identified with the SN
ejecta in \citetalias{Obergaulinger_Aloy__2019}, may break out of the
surface of the star within less than $t_{\textsc{bo}} \sim
12\,\sek$. Towards the end of the simulation, the mass density at the
polar region just outside the \nusp, where the outflows are generated,
remains roughly constant.  Hence, the mass loading of the jets does
not drop significantly for the velocities to increase drastically.  At
that point, this region contains magnetic fields close to
equipartition with the internal energy of the gas.  The associated
Lorentz forces could continue jet launching independently of neutrino
heating.  Hence, in case of a strong decrease of the mass density, the
energy injection could continue, potentially increasing the outflow
velocity to relativistic speeds.

\paragraph*{Accretion disc formation.} The (baryon-rich, moderately
magnetised, and sub-relativistic) SN ejecta must eventually be caught
up by the baryon-free, ultrarelativistic ejecta, which is responsible
for the GRB itself. The alluded ultrarelativistic outflow is the
sought for byproduct of the collapsar central engine. However, the
formation of the collapsar requires that the accretion disc
forms. According to the estimate of Eq.\,\eqref{eq:tdf},
$\tdf \sim 9.3\,\sek$ in models bearing the original stellar rotation
profile (Tab.\,\ref{Tab:models}). Hence, in our models,
$t_{\textsc{bo}}-\tdf \approx 2 - 3\,\sek$. It is, nevertheless, not
unlikely that the disc formation time be longer than twice the
free-fall time from a given mass shell in the star if strong 
magnetorotational explosions take place.  In
Fig.\,\ref{Fig:free-fall}\panel{a}, we show the space-time
trajectories of mass-shells along the equator that would fall to
$r\approx 0$ from its initial location $r=r(M)$ on a time,
$t_\textsc{ff}(M)$, equal to the expression of the disc formation time
Eq.\,\eqref{eq:tdf} but replacing $M_\textsc{df}$ by $M$, in the
progenitor \modelname{35OC-RO} (coloured, thick lines), as well as the
actual trajectories computed from the same selected subset of
mass-coordinates (black, thin lines). We note that for \modl{35OC-RO}, the estimated value of
$t_\textsc{ff}(M)$ (roughly equal to the point where the coloured,
thick lines intersect the horizontal axis) overestimates the actual
fall time. However, for \modls{35OC-Rp2} and \modelname{35OC-Rp3} the opposite is true: $t_\textsc{ff}(M)$ underestimates the actual falling time (\figref{Fig:free-fall}\panel{b},
  \panel{c}). This is because the ram-pressure of the explosion ejecta
partly counterbalances the free-fall of the outer stellar layers, very
specially, in a broad wedge around the rotational axis, but also along
the equator. The deviation between $t_\textsc{ff}(M)$ and, hence of
$\tdf \equiv t_\textsc{ff}(M_\textsc{df})$ and the true fall time
increases as the SN shock progresses along the equator (the location
of this shock roughly corresponds to the transition between white and
orange shades). As a result, we estimate that $t_{\textsc{bo}}<\tdf$,
i.e.  we find it very plausible that the SN ejecta breaks out of the
stellar surface before the accretion disc forms and, hence, before the
GRB jet is launched. If this happens the minimum luminosity that may
yield a GRB jet able to break through the star and the SN ejecta may
be (significantly) lowered, since the SN ejecta partly clears out the
way to the GRB jet \citep[\eg][]{Aloy_etal_2018MNRAS.478.3576}. The
observational consequences of the GRB jet breaking through the SN
outside of the original stellar progenitor are beyond the scope of
this paper. Nevertheless, we anticipate that they will strongly depend
on the optical thickness of the medium outside of the progenitor star.

\subsection{NS forming models}
\label{sSek:PMevol}

Next, we assess the viability of the PM mechanisms looking to some of
the properties that are expected to be fulfilled by the PMCs
considered in Sec.\,\ref{sSek:NSforming}.

\subsubsection{Rotational Period and shape evolution}
\label{ssSek:PMRotation}
The PM and collapsar candidates show a similarly parallel evolution in
terms of the their spin period, $\bar{P}:=2\pi/(J_\pnss/I_\pnss)$,
during the first second of evolution (\figref{Fig:GRB-figs}\panel{a}).
In most cases, the PNS contraction yields a decrease of the rotational
period. The exception to this behaviour is \modl{35OC-Rs}, whose
surface rotational period begins to increase after an initial phase of
decrease (note that this is also the case for the PCs
\modelname{35OC-Rw} and \modelname{35OB-RRw}). PMCs tend to develop
surface rotational periods $P:=2\pi/\Omega_{\rm surf}<5\,\ms$ and
smaller spin periods $\bar{P}\lesssim 2\,\ms$. Models evolved longer
(\modelname{35OC-Rp2} and \modelname{35OC-Rp3}) display a
non-monotonic evolution of the spin period, reaching long term values
$1\,\ms \lesssim P\lesssim 2\,\ms$. The PM model of
\citetalias{Metzger_et_al__2011__mnras__Theprotomagnetarmodelforgamma-raybursts}
demands that the PNS develops a millisecond period after its
contraction ceases.  Thus, taking $\bar{P}$ and not $P$ as an
estimator of the spin period "at birth", formally our models satisfy
the period requisites of the PM model. We note that previously in the
literature
\citep[\eg][]{Heger_et_al__2000__apj__Presupernova_Evolution_of_Rotating_Massive_Stars.I.Numerical_Method_and_Evolution_of_the_Internal_Stellar_Structure,
  Heger_et_al__2005__apj__Presupernova_Evolution_of_Differentially_Rotating_Massive_Stars_Including_Magnetic_Fields,
  Fryer_Heger__2000__apj__Core-CollapseSimulationsofRotatingStars,Fryer_Warren__2004__apj__The_Collapse_of_Rotating_Massive_Stars_in_Three_Dimensions},
$\bar{P}$ has been used to estimate the final spin period. However,
given the non-monotonic evolution of $\bar{P}$ in the most evolved
models (\eg $\bar{P}$ grows by a factor of 2 between $\sim 5\,\sek$
and $\sim 7\,\sek$ for \modl{35OC-Rp2}), $\bar{P}$ seems only a
predictor of the spin period "at birth" of the NS with a factor of
$2-5$ for the models at hand. Alternative forms of estimating the spin
period
\citep[\eg][]{Ott_et_al__2006__apjs__The_Spin_Periods_and_Rotational_Profiles_of_Neutron_Stars_at_Birth}
are similarly inaccurate since (i) equatorial accretion is ongoing and
(ii) magnetic stresses are exchanging angular momentum between the PNS
and its surrounding medium.  The volume of the PNS, traced by the
effective radius $R_{\rm vol}=(3V_{\pnss}/4\pi)^{1/3}$ is still
(slowly) decreasing after $\sim 7.5\,\sek$, a behaviour that is
modulated by some small amplitude variations.  Thus, formally, models
with supra-stellar magnetic field and with a poloidal magnetic field
below equipartition with the toroidal field strength satisfy the PM
model requisites on the rotational period. The model with initial
equipartition between the toroidal and poloidal magnetic field
components does not yield a millisecond surface period.

Two of the PMCs (\modls{35OC-Rs} and \modelname{35OC-Rp2}) show some
signs of period increase, but for different reasons. The surface
rotation of \modl{35OC-Rs} seems to nearly cease after
$\sim 0.9\,\sek$. As explained in Sec.\,\ref{sSek:35OC-bmoltfort},
this is, in part, due to the fact that counter-rotating matter ejected
from the PNS poles slides down the PNS surface towards the
equator. Besides, the average rotational rate over the whole PNS
(defined as $\bar{\Omega}:=J_\pnss/I_\pnss$) is much smaller in this
model than in any other (Tab.\,\ref{Tab:model-properties}), due to
including in the averaging counter-rotating regions close to the axis.
Another reason for the large period of this model are the
morphological changes that it experiences, leading to a toroidally
shaped PNS by the end of the computed time (note the large variations
in the aspect ratio $R_{\rm pol}/R_{\rm eqt}$ in
\figref{Fig:GRB-figs}\panel{b}). It is, anyway, the (very) strong
poloidal field of \modl{35OC-Rs} the responsible for these
effects. The analysis of the spin-down of \modl{35OC-Rp2} after
$\sim 6\,\sek$ is deferred to Sec.\,\ref{sSek:spindown}.

The evolution of the shape of a self-gravitating body under the action
of dynamically relevant magnetic fields can be interpreted resorting
to the virial theorem. According to it, sufficiently large average
magnetic field tends to flatten a self-gravitating body in
magneto-hydrostatic equilibrium
\citep[][Sec.\,IV]{Chandrasekhar_Fermi_1953ApJ...118..116}. In spite
of the fact that our models are not in equilibrium, the magnetic
fields that they develop tend to flatten the PNS (as mentioned in
Sec.\,\ref{sSek:angmomredistribution}), thus formally obeying the
results of the virial theorem. However, the oblateness of the PNS
cannot be directly mapped to the magnetic field of the progenitor,
since the amplification of the magnetic field is the result of the the
non-linear (and fairly complex) interplay of different dynamical
effects (rotation, convection, MRI, etc.;
Sect.\,\ref{sSek:amplification}). Taking aside \modl{35OC-Rs},
\figref{Fig:GRB-figs}\panel{b} suggest that the pole-to-equator radius
tends to grow in the PMCs that have been run longer, but the late
morphological changes in \modl{35OC-Rp3}, do not allow to give a clear
prediction of its longer term value. \Modl{35OC-Rp2} hints towards an
increase of the aspect ratio (developing
$R_{\rm pol}/R_{\rm eqt}\lesssim 1$), after the model begins its
spin-down ($\tpb\sim 5.5\,\sek$; see Sec.\,\ref{sSek:spindown}).

\paragraph*{Moment of inertia.} The moment of inertia is a relevant
quantity not only in the PM model, but also to constrain the equation
of state of nuclear matter. The time evolution of the moment of
inertia is not monotonic for our models, though it typically decreases
in time (\figref{Fig:GRB-figs}\panel{c}). Changes in time of
$I_{\pnss}$ reflect the variations in the shape and the mass
distribution of the PNS, being the PMC \modl{35OC-Rs} and the PC
\modl{35OB-RRw} the ones showing the most abrupt and larger amplitude
modulations. Restricting the analysis to the PMCs with longer computed
evolution, the moment of inertia remains roughly constant
($I_{\pnss}\sim \zehnh{3}{45}\,$cm$^{2}$\,gr$^{-1}$) after
$\sim 3\,\sek$ (\figref{Fig:GRB-figs}\panel{c}).
For a spherically symmetric mass distribution, a relation of the form
$I_{\pnss}\propto M_{\pnss}R_{\pnss}^{2}$ must hold.  For simplicity,
\citetalias{Metzger_et_al__2011__mnras__Theprotomagnetarmodelforgamma-raybursts} assume that the moment of inertia of the PNS corresponds to that
of a uniform sphere, namely $I_{\rm
  unif}=\frac{2}{5}M_{\pnss}R_{\pnss}^{2}$. However, the PNS in our
PMCs is more an heterogeneous, oblate ellipsoid or a toroid than a
uniform sphere. Hence, characterising the distribution of the mass
around the rotational axis with a single \emph{effective} radius
(``$R_{\pnss}$'') is neither too accurate, nor unanimously
defined. After some experimentation, we find that the PNS moment of
inertia can be approximated by
\begin{equation}
 I_{\rm unif}=\frac{1}{6} M_{\pnss}R_{\pnss}^{2} ,
 \label{eq:Iunif}
\end{equation}
where we use as effective PNS radius,
$R_{\pnss}=(R_{\rm pol}R_{\rm eqt}^{2})^{1/3}$, with $R_{\rm pol}$ and
$R_{\rm eqt}$ being the polar and the equatorial radii of the PNS,
respectively. After $\sim 3\,\sek$, $I_{\rm unif}$ approximates
$I_{\pnss}$ with deviations smaller than $\sim 30\%$
(\figref{Fig:GRB-figs}\panel{e}).\footnote{The large peak at
  $t\approx 3.4\,\sek$ in \modl{35OC-Rp2} is due to an inaccurate
  determination of the polar radius for a short period of time; see
  the sudden fall-down of $R_{11}$ in \figref{Fig:35OC-Rp2-globals}.}
Compared to the moment of inertia of a uniform sphere with the same
effective radius, Eq.\,\eqref{eq:Iunif} is $2.4$ times smaller.

\cite{Lattimer_Schutz_2005ApJ...629..979} show that the ratio
$I/M^{3/2}$ remains approximately constant for typical neutron star
masses. Different equations of estate yield different constant values
though. Using this property,
\cite{Metzger_et_al__2015__mnras__Thediversityoftransientsfrommagnetarbirthincorecollapsesupernovae}
infer that
$I_{\textsc{ns}}\approx \zehnh{1.3}{45}(M_{\textsc{ns}}/1.4
\msol)^{3/2}\,$g\,cm$^{2}$, for a typical value
$I/M^{3/2}\approx 50\,$km$^{2}$\,$\msol^{-1/2}$ and protomagnetar of
mass $M_{\textsc{ns}}\approx 1.4\msol$. Applied to the mass of the PNS
of models \modelname{35OC-Rp2} and \modelname{35OC-Rp3},
i.e. $M_{\pnss}\simeq 2.3\msol$, we obtain
$I_{\pnss}\approx \zehnh{2.3}{45}(M_{\pnss}/2.3
\msol)^{3/2}\,$gr\,cm$^{2}$, which approximates the actual value of
the moment of inertia within less than $40\%$ error at the end of the
computed time (see also Tab.\,\ref{Tab:models}). We note, however,
that the ratio $I_{\pnss}/M_{\pnss}^{3/2}$ evolves non-monotonically
in our PMCs (\figref{Fig:GRB-figs}\panel{d}). After sufficient time
($\sim 3\,\sek$), the relative change in $I_{\pnss}/M_{\pnss}^{3/2}$
significantly decreases. For \modl{35OC-Rp3},
$I_{\pnss}/M_{\pnss}^{3/2}$ settles to a value of
$\sim 50\,$km$^{2}$\,$\msol^{-1/2}$, while in \modl{35OC-Rp2}, it is
moderately decreasing by the end of the computed time, when it reaches
a value $\approx 30\,$km$^{2}$\,$\msol^{-1/2}$.

\subsubsection{Surface magnetic fields}
\label{ssSek:PMB-fields}

Based on stability arguments,
\citetalias{Metzger_et_al__2011__mnras__Theprotomagnetarmodelforgamma-raybursts}
argued that the field of the PNS is dominated by a toroidal component
about an order of magnitude stronger than the poloidal one.  As the
averages of the ratio between field components on the PNS surface
(identified here with the $\nu_e$-sphere;
\figref{Fig:globalvars-2}\banel{(c)}) show, not all our PMCs exactly
agree with this estimate.  \Modl{35OC-Rs} owes its extraordinarily
strong poloidal field to the initially very strong magnetic field with
equipartition in the poloidal and toroidal components (equipartition
that is preserved during the computed evolution). Thus, \modl{35OC-Rs}
displays $b^{\rm pol}/b^{\rm tor}\gtrsim 1$ throughout most of its
evolution. Also in \modl{35OC-Rp4}, the two components of the magnetic
field reach equipartition very early on and stay at that level during
the computed evolution. On the other extreme, for \modl{35OC-Rp2}
(with the smallest initial poloidal field of all PMCs) the poloidal
field decreases during the interval
$0.4\,\sek \lesssim \tpb\lesssim 3.4\,$s
(\figref{Fig:globalvars-2}\banel{(c)}; green line), while the toroidal
component attains a level
$\zehnh{3}{14} \lesssim b^{\rm tor}\lesssim \zehn{15}\,$G
(\figref{Fig:globalvars-1}\banel{(l)}). The dynamics radically changes
after $\tpb\gtrsim 3.4\,$s, highlighting the need of performing very
long term computations of the post-collapse remnant. After that time,
\modl{35OC-Rp2} exhibits a nearly exponential growth of the surface
poloidal field component (see Sect.\,\ref{sSek:amplification}). This
behaviour is connected to the much longer accretion time of the iron
core in PMCs than in PCs. For instance, in the case of \modl{35OC-Rp2}
the magnetised iron core (see \figref{Fig:progenitor-profile}) is
finally accreted after $\approx 2\,\sek$. This episode of accretion
can be traced by the mass-shell that at $\tpb=0$ is located at
$r\approx 2700\,$km. It falls down to $\sim 700\,$km and then is
lifted up and accreted again twice, until it begins falling down more
precipitously at $\tpb\sim 2\,\sek$ (\figref{Fig:free-fall}\panel{b}).
The change in the growth of the poloidal field of the PNS is, in part,
reflecting the magnetic structure of the pre-SN star, with a weak
poloidal component limited to the iron core and a couple of shells
located much further away from the centre
(\figref{Fig:progenitor-profile}).  Consequently, during an extended
interval of time after the accretion of the iron core, very little
additional field is accreted onto the PNS. During this period, the
vigorous convection in the PNS amplifies the poloidal field, while
rotational winding into toroidal field continues and increases the
latter component. This outcome might be modified in the presence of a
genuinely three-dimensional dynamo, as our preliminary 3D models show
(there convection can be found, though no similar growth in the
surface field).
Model \modelname{35OC-Rp3} reaches a value
$b^{\rm pol}/b^{\rm tor}\sim 0.1$ after about $4\,\sek$, though with
significant variations after $7.5\,\sek$. This model would roughly fit
within the parameterization of
\citetalias{Metzger_et_al__2011__mnras__Theprotomagnetarmodelforgamma-raybursts}. However,
since relatively small variations in the poloidal field strength of
the progenitor result in significantly different values of the ratio
$b^{\rm pol}/b^{\rm tor}$ (compare, \eg the evolution of models
\modelname{35OC-Rp2} and \modelname{35OC-Rp3} in
\figref{Fig:globalvars-2}\banel{(c)}), we cannot robustly confirm the
assumptions in
\citetalias{Metzger_et_al__2011__mnras__Theprotomagnetarmodelforgamma-raybursts}
on the poloidal to toroidal field strength ratio.

An important parameter regulating the rotational energy loss as well
as the mass loss rate in isolated, magnetised neutron stars is the
fraction of the neutron star surface threaded by open magnetic flux,
$f_\Phi$ \citep[\eg
\citetalias{Metzger_et_al__2011__mnras__Theprotomagnetarmodelforgamma-raybursts};][]{Margalit_etal_2018MNRAS,Metzger_etal_2018ApJ...857...95}.
The fraction of the PNS surface threaded by open field lines displays
large changes during the evolution
(Fig.\,\ref{Fig:GRB-figs}\panel{f}). For instance, in models
\modelname{35OC-Rp2} and \modelname{35OC-Rp3} $f_\Phi$ fluctuates
between $\sim 0.2$ and 1 for $\tpb\gtrsim 2\,\sek$ as a consequence of
the variable accretion down flows that hit the PNS surface. That
unsteady mass flow onto the PNS also limits the accuracy of our
prescription to distinguish between open and closed magnetic field
lines, namely, that the field line extends for more than 200\,\km~in
the radial direction or that it traverses unbound matter.  Hence, the
computed values of $f_\Phi$ should be taken with some care and using,
\eg $f_\Phi\sim 0.5$ \cite[as in][]{Metzger_etal_2018ApJ...857...95}
is accurate within a factor $\sim 1.5$.

\citetalias{Metzger_et_al__2011__mnras__Theprotomagnetarmodelforgamma-raybursts}
further assume that the contraction of the PNS happens at constant
magnetic flux, $\Phi_{\rm c}$, through the region of the PNS surface
threaded by closed magnetic field lines. They argue that this is a
good approximation if the field growth occurs rapidly, via MRI or the
action of convective dynamos. We find that the approximation of
constant magnetic flux in either the closed or the open magnetospheric
region is only roughly fulfilled, within an order of magnitude, by
\modl{35OC-Rp3} for $2\,\sek\lesssim \tpb \lesssim 8\,\sek$, but not
so much by the rest of the PMCs (\modl{35OC-Rp4} may have not evolved
enough to draw a strong conclusion; see
Fig.\,\ref{Fig:GRB-figs}\panel{g}).  The rough qualitative agreement
of the magnetic flux in the closed magnetosphere of \modl{35OC-Rp3}
with the
\citetalias{Metzger_et_al__2011__mnras__Theprotomagnetarmodelforgamma-raybursts}
assumption happens because the magnetic field in this model is
amplified very soon after core bounce. Later on, it shows neither
significant variations of the magnetic energy in the whole PNS
(Fig.\,\ref{Fig:globalvars-1}\panel{i}) nor in the surface magnetic
field (Fig.\,\ref{Fig:globalvars-1}\panel{k},\panel{l}). Hence, its
magnetic flux in the closed magnetospheric region is roughly constant
and relatively small,
$\Phi_{\rm c}\lesssim \text{few }\times\zehn{25}\,$G\,cm$^2$. In the
case of \modl{35OC-Rp2} we have identified a period of significant
(poloidal) magnetic field growth
($3.5\,\sek\lesssim \tpb \lesssim 6\,\sek$) driven by vigorous
convection (Sect.\,\ref{sSek:amplification}). During a significant
fraction of the convective growth of this model
($2.2\,\sek\lesssim \tpb \lesssim 4.5\,\sek$), the magnetic flux in
closed field lines is quite small,
$\Phi_{\rm c}\lesssim \zehn{25}\,$G\,cm$^2$ and roughly constant
(within one order of magnitude) too. However, later on, both
$\Phi_{\rm c}$ and the magnetic flux in the open magnetospheric
region, $\Phi_{\rm o}$, grow towards values
$\sim \zehnh{3}{27}\,$G\,cm$^2$. We find significant that the magnetic
flux for \modl{35OC-Rp2} is reasonably constant for
$5\,\sek\lesssim \tpb \lesssim 6.5\,\sek$, since during that time
interval this model shows an episode of electromagnetic spin-down
accompanied by wind ejection (see Sec.\,\ref{sSek:spindown}).

We define the magnetisation parameter, $\sigma$, as the ratio of
Poynting ($\mathcal{P}$) to mass ($\dot{\mathcal{M}}c^2$) flux at the
PNS surface. Its time evolution for PMCs is shown in
Fig.\,\ref{Fig:GRB-figs}\panel{h}. Due to the very complex interplay
between accretion and ejection of mass onto/from the PNS surface,
$\sigma$ has large variations with latitude and time. We display with
different line styles the value of $\sigma$ in parts of the surface
threaded by either open or closed field lines. Precisely, we define
\begin{align}
\sigma_{\rm o}:=& \left(\displaystyle\sum_{i_{\rm o}j_{\rm o}} \mathcal{P}_{i_{\rm o}j_{\rm o}} \right)\displaystyle / \left(\displaystyle\sum_{i_{\rm o}j_{\rm o}} \dot{\mathcal{M}}_{i_{\rm o}j_{\rm o}}c^2\right),
\label{eq:sigmaopen}\\
\sigma_{\rm c}:=& \left(\displaystyle\sum_{i_{\rm c},j_{\rm c}} \mathcal{P}_{i_{\rm c}j_{\rm c}} \right)/\left(\displaystyle\sum_{i_{\rm c}j_{\rm c}} \dot{\mathcal{M}}_{i_{\rm c}j_{\rm c}}c^2\right),
\label{eq:sigmaclosed}
\end{align}
where the subscripts $i_{\rm o}$ ($i_{\rm c}$) and $j_{\rm o}$
($j_{\rm c}$) annotate the radial, $r_{i_{\rm o}}$ ($r_{i_{\rm c}}$),
and polar, $\theta_{j_{\rm o}}$ ($\theta_{j_{\rm c}}$) discrete
locations on the $\rho=\zehn{10}\,\gccm$ isodensity surface (slightly
above the PNS surface) threaded by open (closed) magnetic field
lines. The apparent trend is that smaller values of the initial
poloidal magnetic field yield larger values of the magnetisation at
the PNS surface in the mid term. Typical values
$\zehn{-4}\lesssim \sigma \lesssim \zehn{-1}$ alternate with
relatively short episodes in which $\sigma$ rises very significantly
in most PMCs. For instance, \modl{35OC-Rp2} shows a prolonged episode
of relatively large values $\sigma\gtrsim \zehn{-3}$ after
$\tpb\sim 5\,\sek$. During the episodic rise of $\sigma$ we find
values $0.1\lesssim \sigma\lesssim 10$, in qualitative agreement with
\citetalias{Metzger_et_al__2011__mnras__Theprotomagnetarmodelforgamma-raybursts}
for a similar evolutionary time after bounce. Stated differently, we
observe a qualitative agreement with the conditions of the PM model,
namely that in coincidence with the episode of spin-down of
\modl{35OC-Rp2} (and also preceding it), $\sigma$ soars
quickly. Typically, the values of $\sigma_{\rm o}$ are, within the
same order of magnitude as $\sigma_{\rm c}$, although episodically
$\sigma_{\rm o}$ may be 100 times larger than $\sigma_{\rm c}$ (\eg in
\modl{35OC-Rp3} between $3\,\sek\lesssim \tpb \lesssim 4\,\sek$).

We have also monitored the baryon loading, $\eta$, in our our
models. The baryon loading is defined as the ratio of kinetic
($\mathcal{K}$) to mass flux \citep[\eg][]{Bugli_2020MNRAS.492...58}:
\begin{align}
\eta_{\rm o}:=& \left(\displaystyle\sum_{i_{\rm o}j_{\rm o}} \mathcal{K}_{i_{\rm o}j_{\rm o}} \right)\displaystyle / \left(\displaystyle\sum_{i_{\rm o}j_{\rm o}} \dot{\mathcal{M}}_{i_{\rm o}j_{\rm o}}c^2\right),
\label{eq:etaaopen}\\
\eta_{\rm c}:=& \left(\displaystyle\sum_{i_{\rm c},j_{\rm c}} \mathcal{K}_{i_{\rm c}j_{\rm c}} \right)/\left(\displaystyle\sum_{i_{\rm c}j_{\rm c}} \dot{\mathcal{M}}_{i_{\rm c}j_{\rm c}}c^2\right)
\label{eq:etaaclosed}
\end{align}
Values of $\eta\lesssim 1$ highlight the mildly relativistic character
of the outflow in our models. Even more than in the case of $\sigma$,
the values of the baryon loading are similar for regions of the
magnetosphere enclosing open and closed field lines
(Fig.\,\ref{Fig:GRB-figs}\panel{i}). This is remarkable in view of the
fact that the open field lines are associated to regions of outflow
above the poles of the PNS, while the closed field lines more closely
trace the regions of equatorial inflow.

\subsubsection{Neutrino cooling evolution}
\label{ssSek:Neutrinocooling}

An important quantity that determines the mass loss rate and the
temperature evolution of the PNS is the neutrino luminosity. Following
\citetalias{Metzger_et_al__2011__mnras__Theprotomagnetarmodelforgamma-raybursts},
we define the average neutrino luminosity, $L_\nu$, including the
contributions of neutrinos and antineutrinos weighted by their own
spectral energies averaged over the neutrino
($|\epsilon_{\nu_{\rm e}}|$) and antineutrino
($|\epsilon_{\bar{\nu}_{\rm e}}|$) absorption cross-section as
\begin{align}
L_\nu\epsilon_\nu^2 := L_{\nu_{\rm e}}|\epsilon_{\nu_{\rm e}}|^2 +  L_{\bar{\nu}_{\rm e}}|\epsilon_{\bar{\nu}_{\rm e}}|^2,
\label{eq:Lnumean}
\end{align}
where $\epsilon_\nu$ is the neutrino mean energy 
\begin{align}
\epsilon_\nu:=\frac{\sum_{j}\left(|\epsilon_{\nu_{\rm e}}|(\theta_j)L_{\nu_{\rm e}}(\theta_j)+|\epsilon_{\bar{\nu}_{\rm e}}|(\theta_j)L_{\bar\nu_{\rm e}}(\theta_j)\right)}{\sum_{j} \left(L_{\nu_{\rm e}}(\theta_j)+L_{\bar\nu_{\rm e}}(\theta_j)\right)}.
\label{eq:Enumean}
\end{align}
The sum extends over all polar angles $\theta_j$ in our models, and
the quantities in the last expression are computed for each angular
cell corresponding to the iso-density surface $\rho=\zehn{10}\,\gccm$.
Figures\,\ref{Fig:GRB-figs}\panel{j} and \ref{Fig:GRB-figs}\panel{k}
show the evolution of the average neutrino luminosity and of the
neutrino mean energy, respectively.  We observe that the values of
$L_\nu$ and of $\epsilon_\nu$ in all our models are correspondingly
larger than the ones shown in Fig.\,A1 of
\citetalias{Metzger_et_al__2011__mnras__Theprotomagnetarmodelforgamma-raybursts},
which includes the cooling curves from (non-rotating) models of
\cite{Pons_1999ApJ...513..780} and
\cite{Hudepohl_et_al__2010__PRL__NeutrinoSignalofElectron-CaptureSupernovaefromCoreCollapsetoCooling}. The
comparison to the previous models of PNS cooling employed in the PM
model is not straightforward, as usual calculations of the cooling
evolution begin in the so-called Kelvin-Helmhotz (KH) phase, when
accretion and convection have ceased.  The KH phase of non-rotating
PNS produced after the collapse of less massive progenitors typically
begins within 0.5\,s after bounce
\citep{Burrows_1988ApJ...334..891,Burrows_Goshy_1993__apjl__ATheoryofSupernovaExplosions}. In
our PMCs the KH phase has not strictly begun since accretion and also
convection are both still fully operational in our models (even after
more than 8\,s in the case of \modl{35OC-Rp2}). These large values of
$L_\nu$ result from the competition between two opposite effects. On
the one hand, the mass of the PNS of our models is larger than the
largest PNS mass included in
\citetalias{Metzger_et_al__2011__mnras__Theprotomagnetarmodelforgamma-raybursts}
(namely, $2\Msol$), and larger PNS mass yields larger $L_\nu$ and
$\epsilon_\nu$, but most importantly, accretion onto the PNS has not
ceased. Indeed, the equatorial accretion feeds the emission of
neutrinos significantly, compared with isolated neutron stars. On the
other hand, our models are rotating very fast, which should yield a
reduction in both the mean neutrino luminosity and its mean energy,
accompanied by a correspondingly longer cooling time scale
\citep[since rapid rotation lowers the inner temperature of the PNS;
\eg][]{Thompson_Quataert_Burrows__2004__ApJ__Vis_Rot_SN}.
We find that, in spite of the fast rotation of our models, the effect
of the mass accretion onto the PNSs explains the larger values of
$L_\nu$ of our models with respect to the most massive models of
\citetalias{Metzger_et_al__2011__mnras__Theprotomagnetarmodelforgamma-raybursts}. Also,
mean neutrino energies in excess of 30\,MeV are observed for the more
evolved PMC models (\modelname{35OC-Rp2} and
\modelname{35OC-Rp3}). Comparatively, PCs display a faster growth of
the mean neutrino energy and relatively larger values of $L_\nu$ than
PMCs within the first $\sim 1.5\,\sek$ of evolution. This behaviour is
connected to the typically smaller polar radii of PCs compared to PMCs
(Fig.\,\ref{Fig:GRB-figs}\panel{b}), which makes that the
neutrinospheres are located closer to the centre in the polar regions,
and thus, the temperature at last scattering neutrino-matter surface
is consistently larger.

The values of $L_\nu$ and $\epsilon_\nu$ are important to set the mass
loss rate from the PNS in general, and particularly, in the model of
\citetalias{Metzger_et_al__2011__mnras__Theprotomagnetarmodelforgamma-raybursts}. It
is well known that neutrinos may carry away a sizeable fraction of the
PNS rest-mass energy that, in our case, owed to the large PNS mass may
be $\sim 0.43\Msol c^2 (M_\pnss/2.3\Msol)^2$
\citep{Lattimer_Prakash_2001ApJ...550..426}. We note that our PMCs run
longer display typical mean luminosities
$\sim \zehnh{4}{52}\,\text{erg}\,\sek^{-1}$ after $\tpb\sim 2\,\sek$
and extending over more than $\sim 6\,\sek$, implying an emission of
equivalent rest-mass energy $\sim 0.13\Msol c^2$ during the late
post-bounce evolution. In order to obtain a theoretical estimate of
the mass loss rate induced by neutrinos, one may employ the results of
\cite{Qian_Woosley_1996ApJ...471..331}, valid for spherically
symmetric, non-rotating PNSs:
\begin{align}
\begin{split}
\dot{M}_{\nu}=\:& 2.5 \times 10^{-5} \mathrm{M}_{\odot} \mathrm{s}^{-1}\left(\displaystyle\frac{L_{\nu}}{10^{52}\, \mathrm{erg}\,\mathrm{s}^{-1}}\right)^{5 / 3} \left(\displaystyle\frac{\epsilon_{\nu}}{10\,\mathrm{MeV}}\right)^{10 / 3}\\
& \times\left(\displaystyle\frac{M_\pnss}{2\, \mathrm{M}_{\odot}}\right)^{-2}\left(\displaystyle\frac{R_\pnss}{10\, \mathrm{km}}\right)^{5 / 3}\left(1+\epsilon_{\mathrm{es}}\right)^{5 / 3}
\end{split},
\label{eq:dotMnu}
\end{align}
with $\epsilon_{\rm es}\simeq 1$ is a small correction that accounts
for the extra heating due to inelastic electron scattering
\begin{align}
\epsilon_{\rm es}=\:& 0.29 \left(\displaystyle\frac{M_\pnss}{2\, \mathrm{M}_{\odot}}\right)
\left(\displaystyle\frac{R_\pnss}{10\, \mathrm{km}}\right)^{-1}\left(\displaystyle\frac{\epsilon_{\nu}}{10\,\mathrm{MeV}}\right)^{-1} .
\end{align}
Next, $\dot{M}_\nu$ is modified to account for the effects of rotation
and the fact that mass loss resulting into unbound matter is only
possible along the fraction of the magnetosphere threaded by open
field lines, $f_\Phi$ (open field lines approximately span the two
polar caps in the range $0\le \theta \le \theta_{\rm open}/2$). The
centrifugal force enhances the mass loss rate approximately by a
factor (see
\citetalias{Metzger_et_al__2011__mnras__Theprotomagnetarmodelforgamma-raybursts})
\begin{align}
f_{\rm cent}=\exp[(P_{\rm c}/P)^{1.5}] \left(1-\exp{(-\zeta})\right) + \exp{(-\zeta)}
\end{align}
where
\begin{align}
P_{\mathrm{c}} &\simeq 1.8 \sin \theta_{\rm open} \left(\frac{R_{\pnss}}{10\, \mathrm{km}}\right)^{3 / 2}\left(\frac{M_{\pnss}}{2\, \mathrm{M}_{\odot}}\right)^{-1 / 2} \mathrm{ms},\\
\zeta & \simeq \left(\frac{\sigma_{\rm o} c^3}{G M_\pnss\bar{\Omega}}\right)^{1/3}.
\end{align}
Hence, the overall mass-loss rate in the
\citetalias{Metzger_et_al__2011__mnras__Theprotomagnetarmodelforgamma-raybursts}
model is
\begin{align}
\dot{M}_{\rm M11}=\left\{\begin{array}{lr}
\dot{M}_{\nu} f_\Phi, & \theta_{\text {open }} / 2 \ll \pi / 2 \\
\dot{M}_{\nu} f_\Phi f_{\text {cent }}, & \theta_{\text {open }} / 2 \gtrsim \pi / 2 
\end{array}\right. 
\label{eq:dMM11}
\end{align}
In Eq.\,\eqref{eq:dMM11} we have omitted the branch corresponding to
mass accretion rates below the Goldreich-Julian rate, since it may
only apply in the very long term evolution, not reached by our models.
In Fig.\,\ref{Fig:GRB-figs}\panel{l}, we show (solid lines) the
instantaneous values of
\begin{align}
\dot{M}_{\rm o}=\dot{M}_\nu f_\Phi
\label{eq:Mth}
\end{align}
compared with the actual time evolution of $\dot{M}_\pnss$,
distinguishing between episodes of mass ejection (displayed with
symbols) and accretion (using dashed lines). Including the factor
$f_{\rm cent}$ as in expression\,\eqref{eq:dMM11} yields a mass loss
rate orders of magnitude above the values measured in our PMCs and,
besides, it is not totally justified in our models, where most of the
mass loss happens along the polar caps of the PNS. Remarkably, the
theoretical prediction for the mass loss rate employing
Eq.\,\eqref{eq:dMM11} agrees, within factors $\lesssim 3$, with the
computed mass loss rate during the episodes of net mass ejection from
the PNS for \modls{35OC-Rp2}, \modelname{35OC-Rp3}, and
\modelname{35OC-Rp4}. The agreement is surprisingly good during the
epochs of net mass accretion onto the PNS. This is likely because the
mass loss rate analytically estimated for spherically symmetric
neutrino driven winds (Eq.\,\ref{eq:dotMnu}) is derived from
continuity arguments that apply identically to wind outflows and mass
accretion.

We furthermore check another of the requirements of the
\citetalias{Metzger_et_al__2011__mnras__Theprotomagnetarmodelforgamma-raybursts}
model.  These authors state that the neutrino-driven mass loss rate is
unaffected by the magnetic field for dipole (poloidal) fields
$B_{\mathrm{dip}} \lesssim \zehnh{3}{16} \, \Gauss$. None of our
models reaches values of the poloidal magnetic field larger than a
few\,$\times\zehn{15}\,\Gauss$. However, the mass loss rate does not
fit with the theoretical estimates for \modl{35OC-Rs} that is endowed
with the largest initial dipolar field (still significantly below the
aforementioned threshold value). This discrepancy can be seen
comparing the black (solid) line to the dashed line and the black
symbols in Fig.\,\ref{Fig:GRB-figs}\panel{l}. Other PMCs whose PNSs
reach poloidal magnetic fields smaller than that of \modl{35OC-Rs},
seem to roughly fit (again, within one order of magnitude) with the
assumption that the neutrino-driven mass loss rate is not notably
affected by the presence of the magnetic field.

\subsection{PMC spin-down}
\label{sSek:spindown}
The PNS may exchange angular momentum with the surroundings, either
incorporating it by mass accretion, releasing it by mass ejection or
by the action of Maxwell stresses. Likewise, the energy exchange
between the PNS and the medium surrounding it is mediated by magnetic
fields, mass exchange and neutrinos. Among the PMCs we have identified
one case in which an incipient spin-down phase has begun, and we
discuss it here in detail for its interest as central engine of a
long-duration GRB.

\begin{figure}
  \centering  
\begin{tikzpicture}
  \pgftext{\vbox{
      \includegraphics[width=0.99\linewidth]{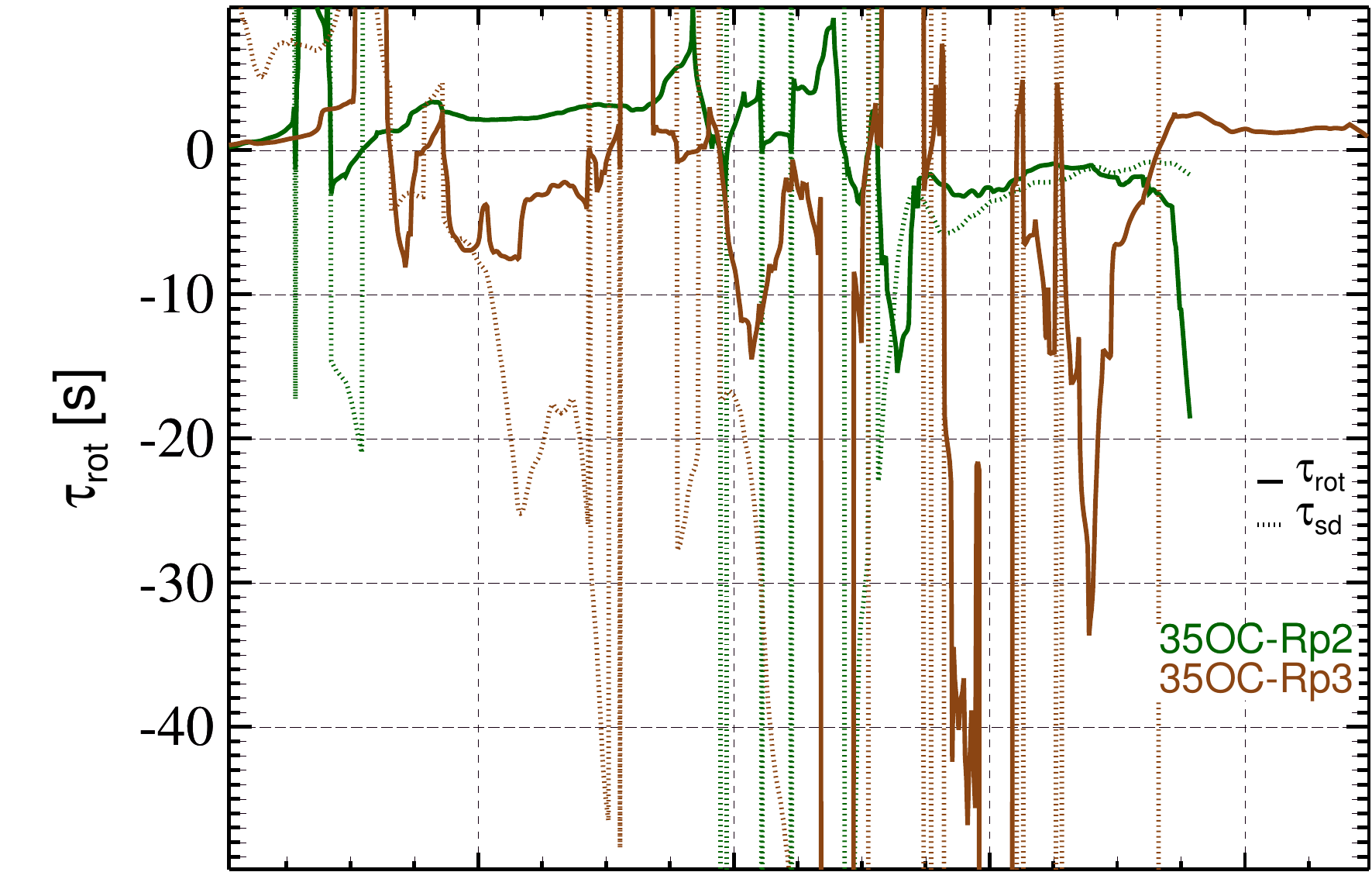}\\ 
      \includegraphics[width=0.99\linewidth]{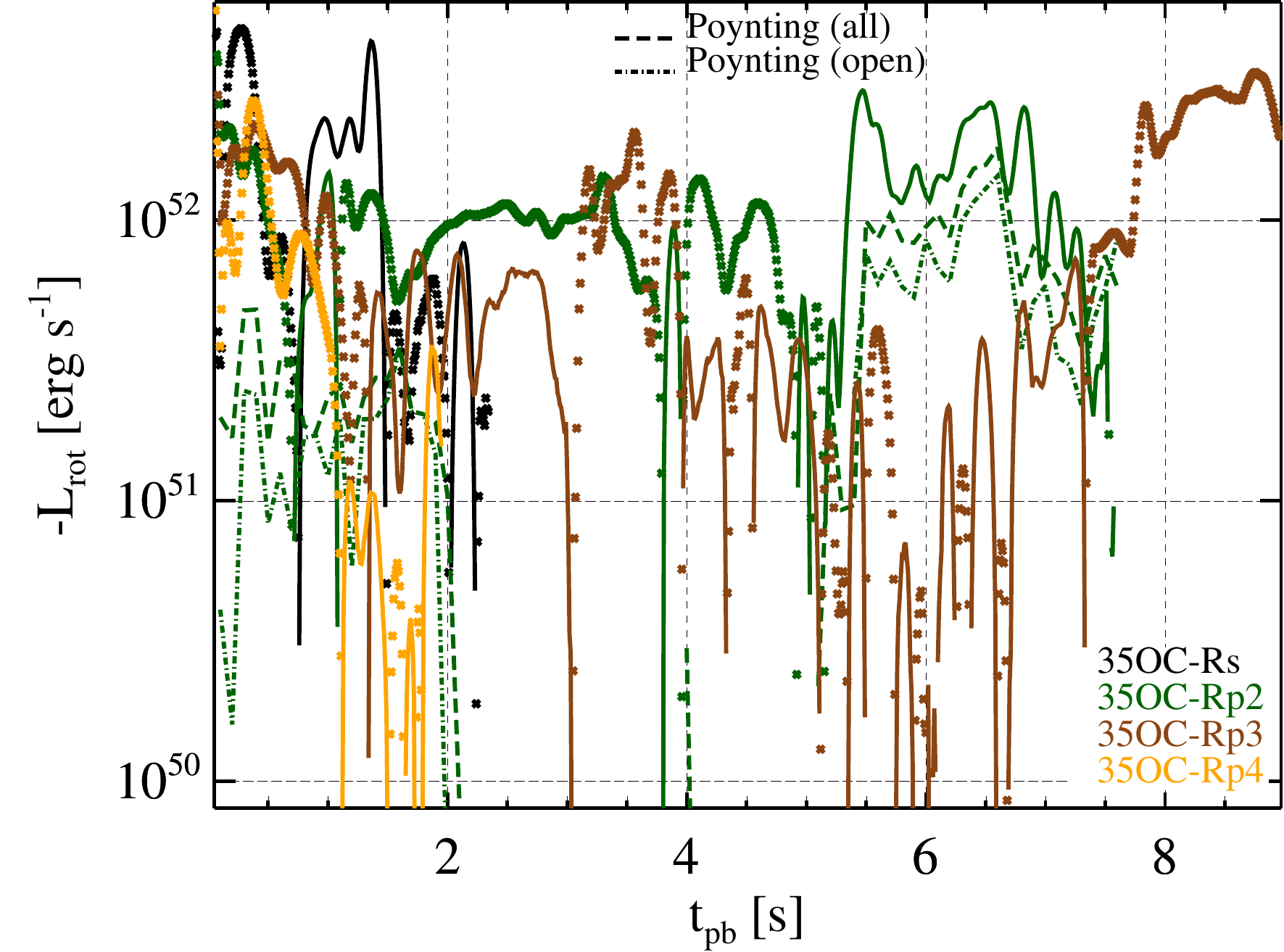}}}
   \node[fill=white, opacity=1, text opacity=1] at (-3.70,+0.6) {\Large (a)};
   \node[fill=white, opacity=0, text opacity=1] at (-3.70,-5.0) {\Large (b)};
   \node[fill=white, opacity=1, text opacity=1, rotate=90] at (-3.80,-2.4) {\large $-\dot{\Erotpns}\,[\text{erg}\,\sek^{-1}]$};	
  \end{tikzpicture}
  \caption{
    \panel{a} Evolution of the rotational time scale $\tau_{\rm rot}$
    computed taking averages of the local value of
    $\Erotpns/\dot{\Erot}^{\pnss}$ over intervals of $25\,$ms. Note
    that $\tau_{\rm rot}<0$ ($\tau_{\rm rot}>0$) correspond to times
    in which $\dot{\Erot}^{\pnss}<0$ ($\dot{\Erot}^{\pnss}>0$),
    corresponding to a spin-down (spin up) of the PNS. \panel{b}
    Rotational luminosity, $-\dot{\Erot}^{\pnss}$. The instants of
    time in which $-\dot{\Erot}^{\pnss}<0$ (the PNS spins up) are
    represented with symbols, while solid lines correspond to times in
    which $-\dot{\Erot}^{\pnss}>0$ (the PNS spins down). With
    dashed-dotted (dashed) green lines the Poynting flux enclosed by open
    (or all, open and closed) field lines is displayed for
    \modl{35OC-Rp2}. }
  \label{Fig:Spindown}
\end{figure}

\begin{figure}
  \centering
  \includegraphics[width=\linewidth]{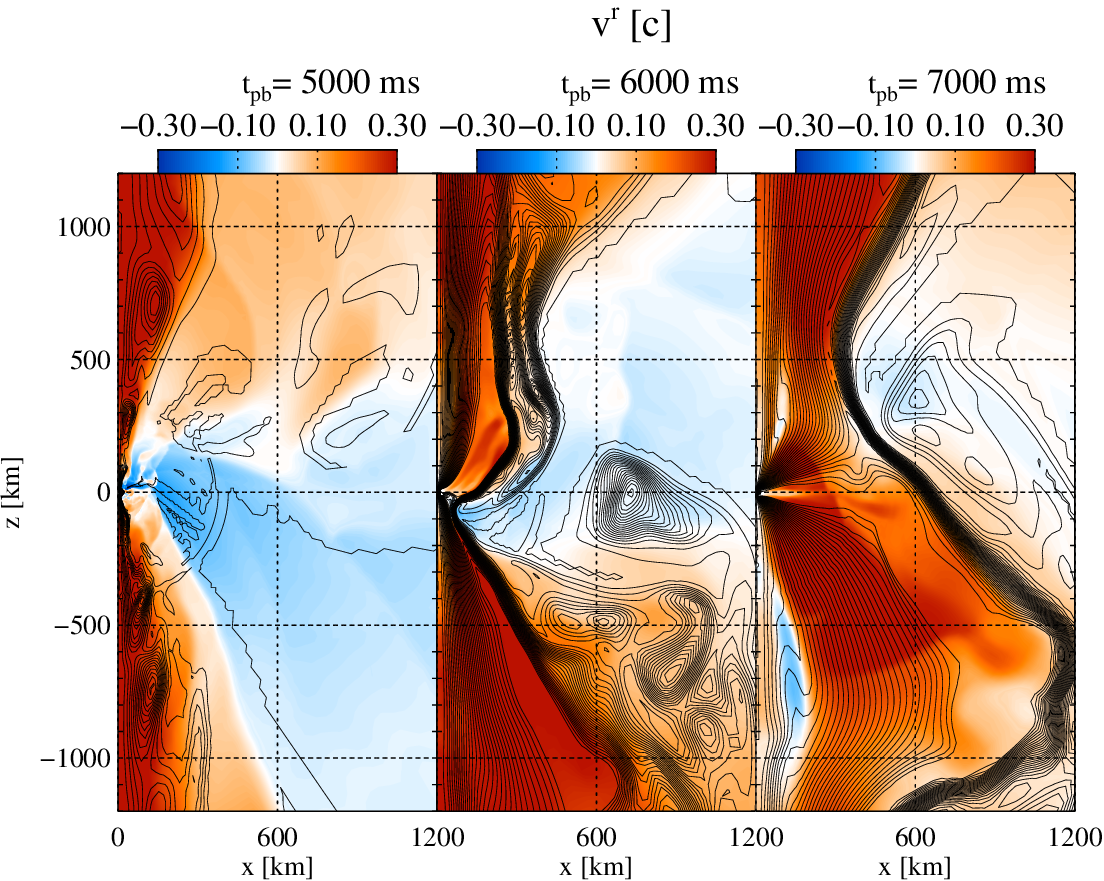}
  \includegraphics[width=\linewidth]{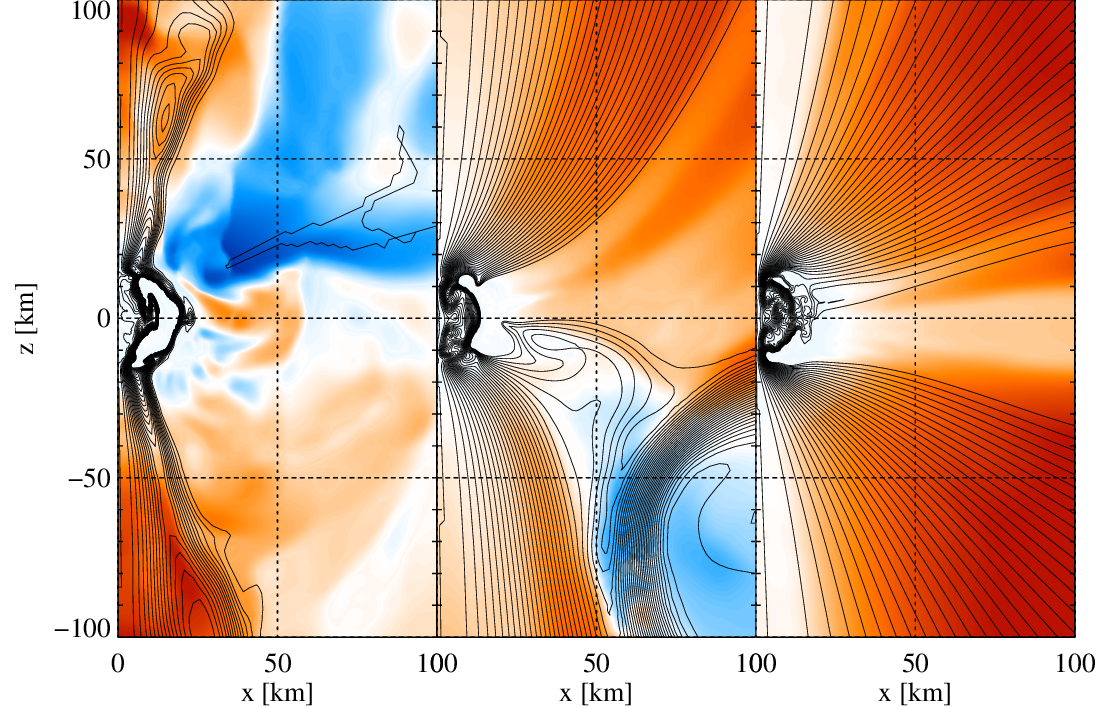}
  \caption{
    Generation of the late-stage rotationally-driven outflow of
    \modl{35OC-Rp2}.  We show the radial velocity in units of the
    speed of light and magnetic field lines.  Top and bottom panels
    show a region of 1200 km and a zoom onto the innermost 100 km, respectively.
  }
  \label{Fig:35OC-Rp2-2}
\end{figure}

Starting around $\tpb \sim 5.5 \, \sek$, \modl{35OC-Rp2} launches an
outflow powered by the spin-down of the PNS.  During this period,
angular momentum is extracted at high rates by the strong magnetic
field.  Consequently, the rotational energy decreases to about $25\% $
of its maximum value and the surface average of the angular velocity
drops by half.
We approximately quantify a rotational spin-down timescale by
computing the quantity $\tau_{\rm rot}:=
\Erotpns/\dot{\Erot}^{\pnss}$. In order to avoid the noise associated
to the numerical evaluation of $\dot{\Erot}^{\pnss}$
(\figref{Fig:35OC-Rp2-2}\panel{b}), we take averages of the the ratio
$\Erotpns/\dot{\Erot}^{\pnss}$ over intervals of $25\,$ms. In
\figref{Fig:Spindown}\panel{a} we show the time evolution of
$\tau_{\rm rot}$ (solid lines) for models \modelname{35OC-Rp2} and
\modelname{35OC-Rp3}. Values of $|\tau_{\rm rot}|\sim 2 - 3\,\sek$ are
a typical during the the spin-down phase $5.3\,\sek \lesssim \tpb
\lesssim 7.6\,\sek$ of \modl{35OC-Rp2}.  More variable, but typically
longer spin-down timescales ($|\tau_{\rm rot}|\sim 10 - 40\,\sek$) are
observed in \modl{35OC-Rp3} in the period $4\,\sek\lesssim \tpb \lesssim
7.2\,\sek$. Thus, a $50\%$ increase in the initial
poloidal field of \modl{35OC-Rp3} with respect to \modl{35OC-Rp2}, yields an
order of magnitude larger spin-down timescales.

Also in \figref{Fig:Spindown}\panel{a} we show the theoretical
prediction of \cite{Metzger_etal_2018ApJ...857...95} for the spin-down
time scale, $\tau_{\rm sd}$, for accreting PNSs (dotted lines). More
precisely, we define
\begin{equation}
\tau_{\rm sd} := \text{sign}(\tau_{\rm rot})\frac{\Erotpns}{L_{\rm sd}},
\label{eq:tausd}
\end{equation}
with
\begin{equation}
L_{\rm sd}=\left\{
\begin{array}{ll}
L_{\rm em} \left( R_{\rm lc} / R_\pnss \right)^2 , & \dot{M} \gtrsim \dot{M}_{\mathrm{ns}} \\
L_{\rm em} \left( R_{\rm lc} / R_{\rm m}\right)^2 , & \dot{M}_{\mathrm{lc}} \lesssim \dot{M} \lesssim \dot{M}_{\mathrm{ns}} \\
L_{\rm em}, & \dot{M} \lesssim \dot{M}_{\mathrm{lc}}
\end{array}
\right.
\label{eq:Lsd}
\end{equation}
where $L_{\rm em}:=b_{\rm pol}^2 R_\pnss^6 \Omega^4/c^3$, $R_{\rm
  lc}=c/\Omega$ is the light-cylinder radius, and $R_{\rm
  m}:=\left(1.5 b_{\rm pol}^2 R_\pnss^6 /
  (\dot{M}\sqrt{GM_\pnss})\right)^{2/7}$ is the Alfv\'en radius
\citep[see][Eq.\,(13)]{Metzger_etal_2018ApJ...857...95}. The three
branches in Eq.\,\eqref{eq:Lsd} result from different accretion
thresholds, namely,
\begin{align}
\dot{M}_{\mathrm{ns}} \simeq 0.086 b_{{\rm pol},15}^{2} M_{1.4}^{-1 / 2} M_{\odot} \mathrm{s}^{-1}, 
\label{eq:dotMns}
\end{align}
that limits the accretion rate when $R_{\mathrm{m}}=R_\pnss$, and
\begin{equation}
\dot{M}_{\mathrm{lc}} \simeq 6.9 \times 10^{-4} b_{{\rm pol},15}^{2} P_{\mathrm{ms}}^{-7 / 2} M_{1.4}^{-1 / 2} M_{\odot} \mathrm{s}^{-1},
\label{eq:dotMlc}
\end{equation}
that operates enhancing the mass accretion rate when the Alfv\'en
radius is smaller than the light cylinder radius (hence,
Eq.\,\eqref{eq:dotMlc} results when
$R_{\mathrm{m}}=R_{\mathrm{lc}}$). We point out that the absolute
value of the mass accretion rate onto the PNS never falls below the
threshold set by Eq.\,\eqref{eq:dotMlc} in our models. Hence, the
lower branch of Eq.\,\eqref{eq:Lsd} is never met. This is within our
expectations, since the spin-down luminosity is mostly as that of a
magnetic dipole spin-down for \emph{effectively} non-accreting neutron
stars (which is not the case in our models).  In the previous
equations, $M_{1.4}$ is the mass of the PNS is units of $1.4\Msol$ and
$P_{\rm ms}$ is the rotational period measured in milliseconds. We
have included the $\text{sign}(\tau_{\rm rot})$ in Eq.\eqref{eq:tausd}
in order to easy the comparison with $\tau_{\rm rot}$.

The agreement between between $\tau_{\rm rot}$ and $\tau_{\rm sd}$
during the epoch of spin-down for \modl{35OC-Rp2} is remarkable,
reinforcing our interpretation that we are observing a transient PM
spin-down episode in that model. The agreement between the theoretical
spin-down time scale and $\tau_{\rm rot}$ is only within one order of
magnitude in the late (very mild) spin-down episode of \modl{35OC-Rp3}
($4\,\sek\lesssim \tpb\lesssim 7.3\,\sek$) and within a factor 3
during the first extended spin-down episode of this model
($1.3\,\sek\lesssim \tpb\lesssim 3\,\sek$).

The spin-down phase is characterised by a layer of very strong field
surrounding the rotational axis at a distance of several hundreds of
kilometres and separating the outflow from falling matter outside
(compare the late stages, $\tpb = 6, 7 \, \sek$ in
\figref{Fig:35OC-Rp2-2} to the earlier one at $\tpb = 5 \, \sek$).  In
this layer, the magnetic field dominates over the internal and kinetic
energies, wherefore it is able to prevent the accretion onto the PNS.
Approximately in spatial coincidence with the aforementioned layer we
may find the Alfv\'en surface. This Alfv\'en surface is located at a
distance from the rotational axis which exhibits large variations in
time, specially close to the equator. There it is not uncommon that it
episodically shrinks until it hits the PNS surface. The epoch of
spin-down is strongly correlated with periods in which the Alfv\'en
surface is located further away along the equator
(\figref{Fig:35OC-Rp2-2}), hence enabling an efficient
magneto-rotational braking \citep[as the ejected matter from the PNS
is forced to nearly corotate -specially along closed magnetic field
lines- with the angular frequency of the PNS surface;
cf.][]{Thompson_et_al__2004__apj__Magnetar_Spin-Down_Hyperenergetic_SNe_and_GRBs}. Indeed,
the spin-down is so efficient during the late-time episode in
\modl{35OC-Rp2} that it seems to be caused by a propeller mechanism
\citep{Romanova_2004ApJ...616L.151}.  However, our models do not enter
the propeller regime, though they are not too far off.  The magnetic
field does not fully enforce corotation in the magnetically dominated
layer around the PNS, but is not too far from that. Thus, rotation is
not fast enough to expel matter by centrifugal forces.

Towards the end of the simulation, the PNS is at the centre of a
magnetic field whose poloidal component is predominantly radial and
has a monopolar geometry that transfers rotational energy to the
surrounding gas.  The field launches a wind that is, in contrast to
earlier epochs and to all other models, highly isotropic.  Only after
several hundred km, the gas is collimated by the surrounding stellar
core into a pair of jets streaming along the axis, in qualitative
agreement with the findings of
\cite{Uzdensky_MacFadyen__2007__apj__Magnetar-Driven_Magnetic_Tower_as_a_Model_for_Gamma-Ray_Bursts_and_Asymmetric_Supernovae}
and
\cite{Bucciantini_et_al__2007__mnras__Magnetar-driven_bubbles_and_the_origin_of_collimated_outflows_in_GRBs,
  Bucciantini_et_al__2008__mnras__Relativistic_jets_and_lGRBs_from_the_birth_of_magnetars}.

\citetalias{Metzger_et_al__2011__mnras__Theprotomagnetarmodelforgamma-raybursts}
assume that the contraction of the NS happens at constant angular
momentum. As we can see from Fig.\,\ref{Fig:globalvars-1}\panel{f},
after $\sim 2\,\sek$, the angular momentum of the PNS displays a
moderate time evolution, that drives changes in $J_{\pnss}$ of less
than a factor 2 within $\sim 5 - 7\,\sek$. Thus, even if the
assumption of of
\citetalias{Metzger_et_al__2011__mnras__Theprotomagnetarmodelforgamma-raybursts}
does not strictly hold, it is broadly compatible with our
results. Although neutrinos may contribute to reduce the angular
momentum of the PNS by $\sim 43\%$ in favourable cases
\citep[\eg][]{Janka__2004__YoungNeutronStarsandTheirEnvironments__Neutron_Star_Formation_and_Birth_Properties},
here the main driver of angular momentum loss is the magnetic braking
combined with the ejection of mass from the PNS surface.

\section{Outlook: three-dimensional models}
\label{Sek:3d}

\begin{figure}
  \centering
    \begin{tikzpicture}
  \pgftext{
  \includegraphics[width=0.95\linewidth]{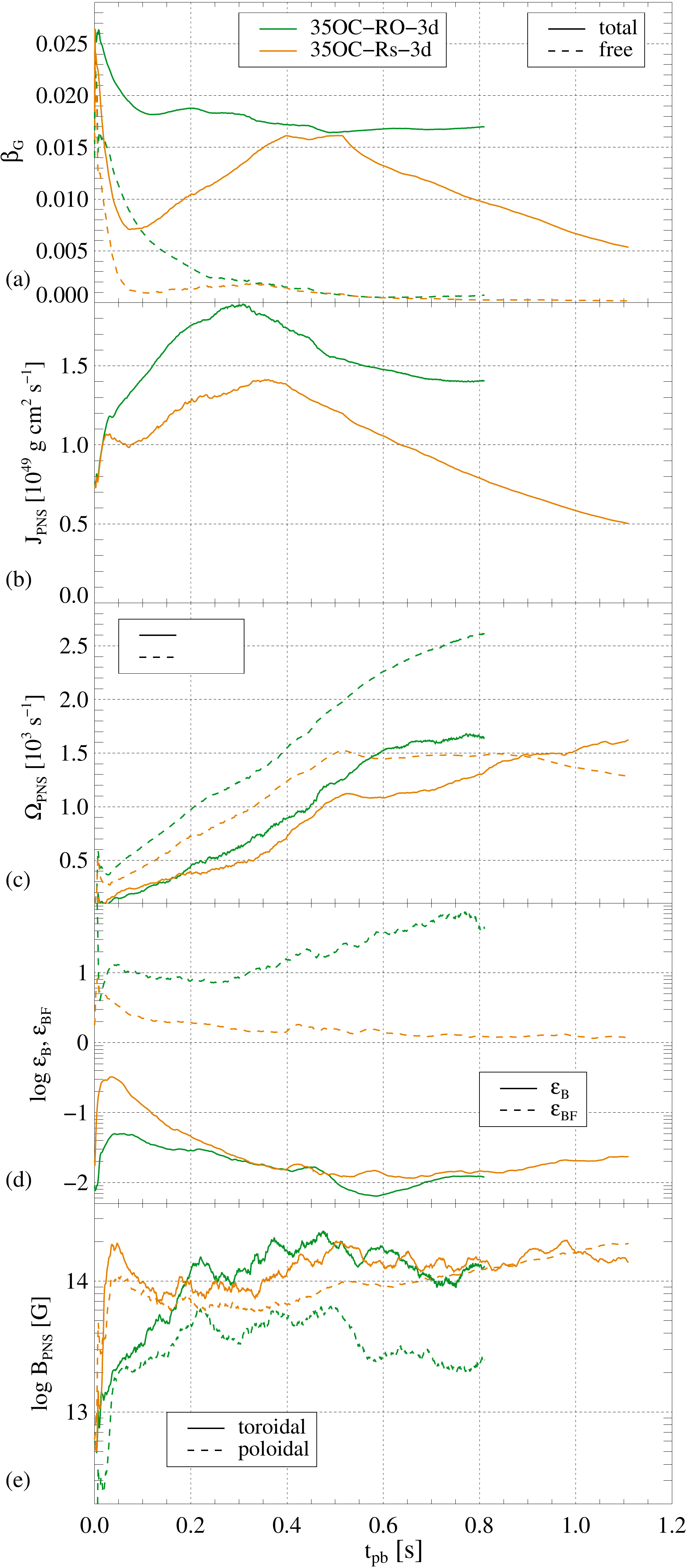}
     }%
   \node[fill=white, opacity=1, text opacity=1, rotate=90] at (-3.54,-6.65) {\small $\log\Emag^\pnss\,\text{[G]}$};
   \node[fill=white, opacity=0, text opacity=1] at (-1.75,1.47) {\small $\bar{\Omega}$};
   \node[fill=white, opacity=0, text opacity=1] at (-1.55,1.72) {\small $\Omega_{\rm surf}$};	
  \end{tikzpicture}
  \caption{
    Global quantities of the PNSs of the 3D versions of \modls{35OC-RO} and
    \modelname{35OC-Rs}: from top to bottom, the panels show the ratio
    of the total and free rotational energy to the gravitational
    energy, the angular momentum, the angular frequency on the surface and averaged over the
    PNS, the ratio of the magnetic energy to the total and free
    rotational energy,  and the average surface magnetic field strength.
  }
  \label{Fig:3dglobals}
\end{figure}

Deferring a thorough investigation of our three-dimensional models to
a subsequent article, we conclude this section by placing them in the
context of the present study.  Key quantities discussed above for the
axisymmetric models are presented in \figref{Fig:3dglobals}.  As
described in \citetalias{Obergaulinger_Aloy__2019}, \modl{35OC-RO-3d}
undergoes a delayed shock revival driven by magnetic fields and
neutrino heating, while \modl{35OC-Rs-3d} develops a prompt explosion
powered by the strong magnetic field.  Hence, despite quantitative
differences, the models behave similarly to the axisymmetric versions.
The similarities extend to the jet-like morphology of the ejecta
propagating at moderately relativistic speeds along the rotational
axis.  Accretion onto the PNSs, on the other hand, is weaker in 3D
than in 2D, causing their masses to stop growing already within the
first half second after bounce. Though the simulations could only be
run for a shorter time, a collapse to a BH seems unlikely on
timescales similar to those observed in 2D.  Hence, the parameter
space for collapsars formation is more restricted in 3D.

The evolution of the PNSs puts both models into the regime of PMCs.
Their rotational energy corresponds to up to $\betaG \lesssim 2 \, \%$
with the gravitational energy, \ie, of the same order as in
axisymmetry. Hence the development of dynamical instabilities seems
limited in our 3D models, backing up our long-term 2D calculations.
While \modl{35OC-RO-3d} approaches a roughly constant value of
$\betaG \approx 1.8 \, \%$, the rotational energy of \modl{35OC-Rs-3d}
enters a rapid decline after peaking at $\betaG \approx 1.6 \, \%$.
Both tendencies parallel the evolution of the axisymmetric versions,
though at a more quantitative level the early end of the accretion
onto the PNS limits the 3D models to values that are below the 2D
versions.  Like in 3D, only a small fraction of the total rotational
energy is in the form of free rotational energy.  Both models reach
similar levels $\Erotshearpns \approx 10^{51} \, \erg$ corresponding
to a fraction $\Erotshearpns/|\Egrav|\sim 10^{-3}$ of the
gravitational energy.  Similarly to the rotational energy, the angular
momentum grows, but not quite as much as in 2D.  Both models reach a
maximum of $J_\pnss$ around the time the PNS mass ceases to increase.
In the case of \modl{35OC-RO-3d}, the subsequent decline eventually
slows down and $J_\pnss$ levels off, whereas \modl{35OC-Rs-3d} does
not reach a stationary value by the end of the simulation.  The PNSs
develop rotational frequencies exceeding $10^{3} \, \isek$.  The
angular velocity on the PNS surface of \modl{35OC-RO-3d} exceeds the
volume average, $\bar{\Omega}$, by at least $30 \, \%$ throughout the
evolution, indicating a high degree of differential rotation.  In
\modl{35OC-Rs-3d}, the two measures of the rotational velocity become
similar after $\tpb \approx 0.7 \, \sek$, pointing toward a more rigid
internal rotational profile.  In both models, the growth of
$\bar{\Omega}$ is only very gradual at late times.

Both PNSs possess strong magnetic fields, which, despite the large
differences in pre-collapse magnetisation, converge to similar
magnitudes. Towards the end of the simulations, their energies account
for around $1 \, \%$ of the total rotational energy. They are in
equipartition with the free rotational energy (\modelname{35OC-Rs-3d})
or surpass it by one order of magnitude.  Hence, the ratio
$\mathcal{B} / \mathcal{F}$ is larger than in 2D.  The surface field
strength slightly exceeds
$b_{\mathrm{surf}} \gtrsim 10^{14} \, \Gauss$ during most of the
evolution with only the poloidal component of \modl{35OC-RO-3d}
falling short of this value by a factor $\approx 3$.  In
\modl{35OC-Rs-3d}, on the other hand, both components develop the same
strength.

\begin{figure}
  \centering
  \includegraphics[width=\linewidth]{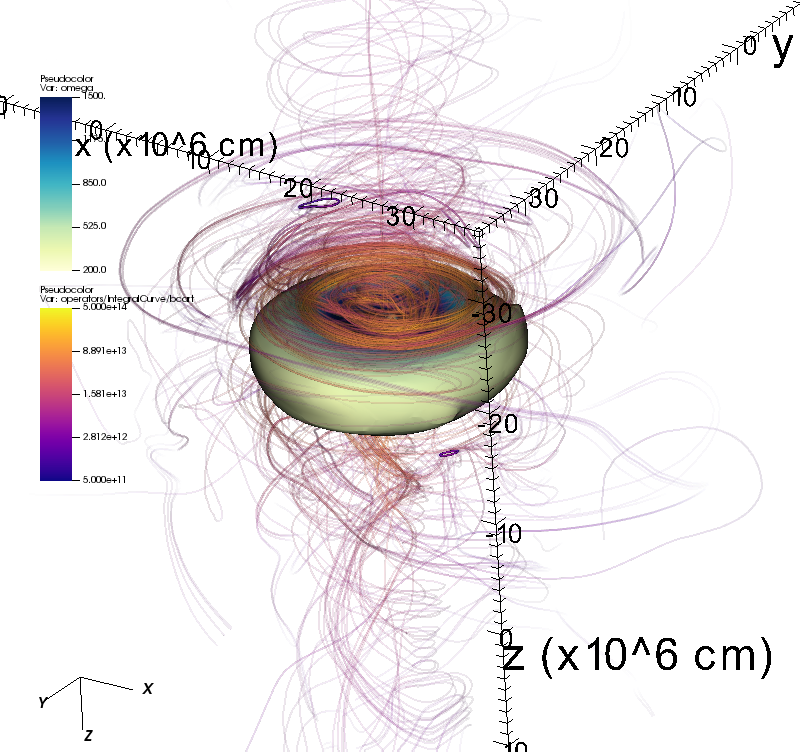}
  \includegraphics[width=\linewidth]{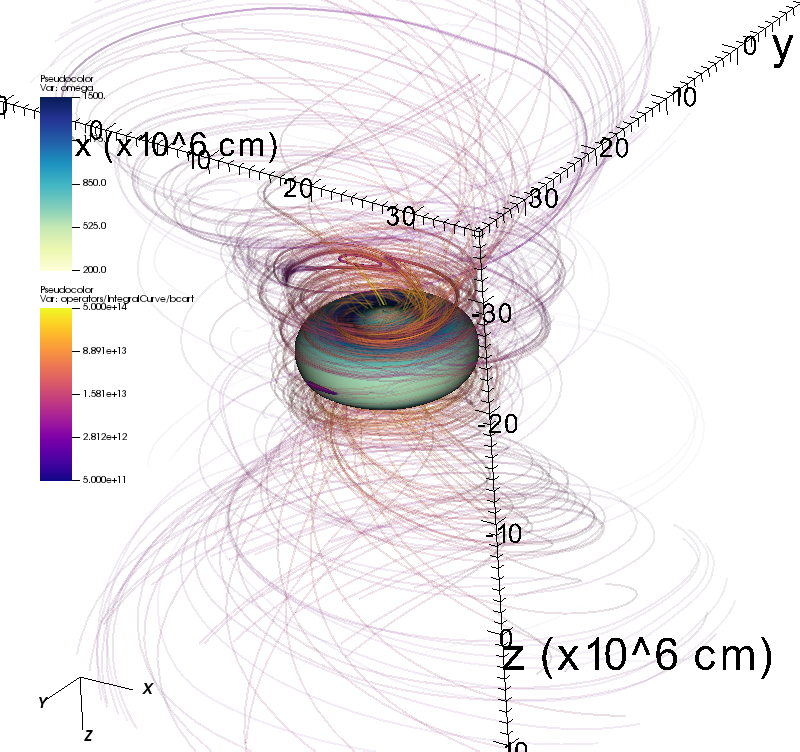}
  \caption{
    Structure of the PNS and its immediate surroundings in the
    three-dimensional versions of models \modelname{35OC-RO} ($\tpb
    \approx 0.81 \, \sek$, top) and \modelname{35OC-Rs} ($\tpb \approx
    1.15 \, \sek$, bottom).  Both panels present the angular velocity
    of the gas on isodensity surfaces of $\rho = 10^{10} \, \gccm$ and
    magnetic field lines with the colour scale showing the field
    strength.
  }
  \label{Fig:3dmodels}
\end{figure}

The structure of the PNSs, visualised for late times in
\figref{Fig:3dmodels}, is characterised by a high degree of rotational
flattening (cf.~the iso-density surface in the two panels).
Deviations from an axisymmetric shape are minor. The PNS surface shows
a strong differential rotation between the equatorial bulge rotating
at sub-kHz frequencies and much faster rotation at higher latitudes.
The magnetic field, strongest near the polar axis, forms a helix
around the $z-$axis.  Close to the PNS, the field widens more in
\modl{35OC-Rs-3d} than in \modelname{35OC-RO-3d}.

The results indicate that the structure of the PNS as well as its
magnetic field are sufficiently similar to the axisymmetric versions
of the two models for the 3D models to be possible GRB progenitors.
The dimensionality may, however, have an influence on the threshold
separating collapsars from PMCs, as suggested by the cessation of the
growth of the PNS mass in \modl{35OC-RO-3d} in contrast to the
monotonic increase of $M_{\mathrm{PNS}}$ in axisymmetry.

\section{Discussion and conclusions}
\label{Sek:SumCon}

We followed the post-collapse, long-term evolution of the cores of
several rotating and magnetiSed stars with zero-age-main-sequence mass
$M_\ZAMS = 35 \, \msol$ and subsolar metallicities. Our main goal is
to asses the robustness of the predictions stating that these models
may form a collapsar or any other type of possible central engine of
long GRBs. In particular, we aim at assessing the variance of the
possible outcomes resulting from relatively small changes in the
magnetic field strength and topology of the pre-collapse star, as well
as on modifications of the rotational profile.
Given the uncertainties still existing in (one-dimensional) stellar
evolution, predicting whether a given massive, compact core may yield
a BH after its gravitational collapse is still adventurous \citep[but
see the notable advances in cases where rotation and magnetic fields
are less important, \eg][employing a one dimensional
approach]{Ugliano_et_al__2012__apj__Progenitor-explosionConnectionandRemnantBirthMassesforNeutrino-drivenSupernovaeofIron-coreProgenitors,Sukhbold_et_al__2016__apj__Core-collapseSupernovaefrom9to120SolarMassesBasedonNeutrino-poweredExplosions,Ertl_2020ApJ...890...51,Woosley_2020ApJ...896...56}. We
point out that, specifically, the topology of the magnetic field in
the mapping from 1D to multiple dimensions is not unanimously
defined. Hence, variations in the ratio of poloidal to toroidal
magnetic field components may be a source of variance in
multidimensional initial models, since the 1D stellar progenitors only
provide the toroidal and radial magnetic field components (see
App.\,\ref{sek:Variance}). Due to the much larger computational costs
of multidimensional models (compared to 1D models), we cannot explore
systematically the influence of the stellar progenitor properties on
the final outcomes. Instead, we have restricted to a couple of
high-mass, low-metallicity cores and considered what variations of the
magnetic field and rotational profile produce BHs or PNSs as compact
remnants.  We have performed multidimensional (mostly 2D, but also
some preliminary 3D) simulations coupling special-relativistic MHD
with a neutrino-transport scheme based on the two-moment formulation
of the spectral transport equation (see
\citetalias{Obergaulinger_Aloy__2019}, Sect.\,2).
Our initial cores (i.e. the progenitor stars) map into 2D or 3D the 1D
stellar evolution models \modelname{35OC} and \modelname{35OB} of
\cite{Woosley_Heger__2006__apj__TheProgenitorStarsofGamma-RayBursts},
which explicitly include rotation and magnetic fields, although in a
spherical approximation.  They are, therefore, a subset of the stellar
progenitors considered in \citetalias{Obergaulinger_Aloy__2019}. These
models have been considered as a potential progenitors of long GRBs,
due to the possibility of forming a collapsar engine, since their Kerr
parameter is $a>0.3$ at $3\,\Msol$ and the angular momentum increases
outwards
\citep{Woosley_Heger__2006__apj__TheProgenitorStarsofGamma-RayBursts}.

The high mass and fairly large compactness of our initial models
\citep{OConnor_Ott__2011__apj__BlackHoleFormationinFailingCore-CollapseSupernovae}
causes their evolution after core bounce to transit along a borderline
between producing either a BH or a PNS, with a broad range of
intermediate possibilities in which the PNS dodges its collapse to a
BH for a (very) long time. This final fate does naturally depend on
the ability to minimise the post-bounce mass accretion rate and,
hence, on the complex interplay of the explosion dynamics and the
compact remnant. Ultimately, the possible evolutionary paths depend on
variations in the pre-collapse cores.

We performed eleven simulations: nine versions of core
\modelname{35OC}, and two of core \modelname{35OB}.  Most simulations
used the original rotational profile of the stellar-evolution
calculations, but a few control models were run with decreased and
increased angular velocities.  Some of the simulations of each core
were run with the original magnetic field, others with an artificial
magnetic field of mixed poloidal-toroidal topology and different
normalisation.  All simulations were run until the cores collapsed to
a BH or, if failing to do so, for various seconds post-bounce; in a
couple of models for more than $\sim 8\,\sek$. Our main results can be
summarised as follows:

\paragraph*{The key role of the pre-SN poloidal magnetic field.}
The strength and spatial smoothness of the poloidal field is decisive
to determine the lifetime of the PNS post-bounce. Two-dimensional
models run with the original progenitor magnetic field, \eg
\modelname{35OC-RO} and \modelname{35OB-RO}, produce BHs within less
that $\sim 3.3\,\sek$ after core collapse. The 3D version of the
former has not been evolved for a sufficiently long time to confirm
this possibility. However, its smaller mass growth rate after
$\sim 0.4\,\sek$ compared to the 2D version of the core
\modelname{35OC-RO} suggests that BH collapse may take even longer for
this model in 3D (see \citetalias{Obergaulinger_Aloy__2019}). Compared
to the toroidal component, the original stellar progenitor includes a
relatively weak poloidal field. A small change of a factor 2 of the
poloidal field strength in the pre-SN iron core while maintaining the
same toroidal field suffices to halt the growth of the PNS mass,
significantly delaying, if not completely preventing BH formation in
the series of models with supra-stellar magnetic fields (models
\modelname{35OC-Rp2}, \modelname{35OC-Rp3}, and
\modelname{35OC-Rp4}). We stress that this small change in the
poloidal field in axial symmetry brings a negligible increase of the
magnetic energy of the initial model.
Our goal is not to find in this paper an extremely accurate magnetic
field strength and topology which changes the fate of the original
stellar progenitor, eventually producing a PM instead of a BH and
(likely) a collapsar. There are (at least) two reasons for
that. Firstly, the very same mapping from the 1D stellar evolution
model to our multidimensional grids introduces variations in the
post-bounce evolution, which may significantly change the life time of
the PNS. Second, in 3D the aforementioned value may be changed
quantitatively and, hence, it does not pay off to explore with great
accuracy the threshold dividing the formation of a PM from a
collapsar. In a future work we will explore additional models for
longer post-bounce times in 3D to check our findings in axial
symmetry. Besides, we have not found a monotonic trend stating that
larger initial poloidal magnetic field guarantees the avoidance of BH
formation. While the possibility of dodging BH formation seems evident
in \modls{35OC-Rp2} and in \modelname{35OC-Rs}, the fate fo
\modl{35OC-Rp3} (with an initial poloidal field in between of the
former cases) is not clear. Its late time ($\gtrsim 7\,\sek$) increase
of the accretion rate above $\dot{M}_\pnss\sim 0.1\,\Msol\,\sek^{-1}$
may let it collapse to a BH within the next couple of seconds (after
the $\tpb\simeq 9\,\sek$ of computed evolution). However, attending to
the non-monotonic evolution of the PNS in our models (especially
regarding the PNS mass in PMCs), a precise forecast of the fate of
\modl{35OC-Rp3} is not possible.
  
\paragraph*{Variations in the magnetic topology.} The topology of the
magnetic field is more important than the strength of the field in
determining the evolutionary path of our models. Comparing the case in
which we double the strength of the magnetic field (multiplying by 2
both the -dominant- toroidal and poloidal components; \modl{35OC-RO2})
with the case in which we double the strength of the poloidal field
(\modelname{35OC-Rp2}), the former model forms a BH, while the latter
staves off it. Since the main difference between the model with the
original magnetic field and rotation (\modelname{35OC-RO}) and
\modl{35OC-RO2} is the twice larger toroidal magnetic field in the
latter, we conclude that a moderate increase of the toroidal magnetic
field in the progenitor star does not alter the prospects for BH
formation.  In line with the results of
\cite{Bugli_2020MNRAS.492...58}, dipolar configurations tend to
produce more collimated explosion ejecta and more oblate
PNSs. Episodes of PNS spin-down (several seconds after bounce) tend to
reduce the ellipticity of the models, increasing the
polar-to-equatorial radius ratio. Indeed, the model with the largest,
purely dipolar magnetic field (\modl{35OC-Rs}) eventually undergoes a
morphological transition from a revolution ellipsoid to a toroid, with
a maximum density off-centre for $\tpb \gtrsim 2.5\,\sek$.  It remains
to be confirmed that a similar morphology may be attained by the 3D
version of this model (which has, so far, been run up to
$\tpb \simeq 1.2\,\sek$).

\paragraph*{Variations in the rotational profile.} For the pre-SN core
\modelname{35OC} (\modelname{35OB}), increasing the rotational rate by
a factor $1.50$ ($2$) and, at the same time, reducing significantly
the magnetic field strength does not prevent BH formation. Models with
supra-stellar rotation tend to form centrifugally supported,
toroidally-shaped structures around the central PNS. These structures
may survive to BH collapse for a few seconds by virtue of their larger
specific angular momentum ($j > \zehnh{1.5}{16} \cm^2\,\sek^{-1}$; see
below).
This finding contrast with the expectations of
\cite{Dessart_et_al__2008__apjl__TheProto-NeutronStarPhaseoftheCollapsarModelandtheRoutetoLong-SoftGamma-RayBurstsandHypernovae},
who argued that stars with large angular momentum in the core may not
transition to a BH. These authors suggest that fast rotating cores
lead to magnetically driven, baryon-loaded, non-relativistic jets
without any GRB signature. Although we have only tested two
progenitors of the same ZAMS mass, our results hint towards a more
intricate interplay between the rotational structure and the magnetic
field dynamics, which hinders an unequivocal prediction of the
high-energy signatures our our models.  In line with the expectations,
a reduction of the rotational rate of the pre-SN core \modelname{35OC}
(\modl{35OC-Sw}) facilitates an early BH formation
($\tpb\lesssim 1.5\,\sek$).

\citetalias{Metzger_et_al__2011__mnras__Theprotomagnetarmodelforgamma-raybursts}
suggest that there is a division between magneto\-rotationally- and
neutrino-powered SNe on the basis of the ratio between the envelope
binding energy, $\Ebinding$, and the rotational energy of the PNS
$\Erotpns$. While $|\Ebinding|/\Erotpns>1$ would lead to
magnetorotational explosions, the complementary case,
$|\Ebinding|/\Erotpns<1$, may produce neutrino-driven SNe (see their
Fig.\,1). This basic division does not account for the role of the
magnetic fields except in a parametrised way (assuming that the
magnetic energy is some fraction of the core rotational energy). The
growth of magnetic fields partly happens at the expense of the
rotational energy (mostly) of the pre-SN core, but it is also driven
by vigorous convection in a bunch of our models, specially in PMCs
(see below). Besides, $\Erotpns$ is a time evolving quantity in the
collapsed core (Fig.\,\ref{Fig:globalvars-1}\panel{c}), while
$\Ebinding$ changes very moderately (within a factor 2) throughout the
post-bounce evolution. Interestingly, the ratio $|\Ebinding|/\Erotpns$
tends to decrease with time after bounce for BH forming models (since
$\Erotpns$ grows for most of them, while $\Ebinding$ decreases
slightly). The reduction of $|\Ebinding|/\Erotpns$ tends to be less
pronounced (or much shallower) for PMCs. We also observe exceptions to
the basic scheme of
\citetalias{Metzger_et_al__2011__mnras__Theprotomagnetarmodelforgamma-raybursts}. For
instance, \modls{35OC-RRw} and \modelname{35OB-RRw} have both four
times more rotational energy than the corresponding stellar evolution
\modls{35OC-RO} and \modelname{35OB-RO}, respectively. All these
models (\modelname{35OC-RO}, \modelname{35OB-RO},
\modelname{35OC-RRw}, and \modelname{35OB-RRw}) develop values
$|\Ebinding|/\Erotpns\ll 1$ in the post-bounce evolution. In contrast
to the latter models, the former ones produce either a neutrino driven
explosion aided by rotational effects or a mixed type explosion, which
is not clearly magnetorotationaly driven (see Tab.\,\ref{Tab:models}).

\paragraph*{Angular momentum transport.} Many of our models develop
strong enough magnetic fields, which enable angular momentum transport
from the inner regions of the PNS towards its surface, where it
accumulates. There, high specific angular momentum matter forms
extended toroidal structures with a low electron fraction. These
neutron-rich regions are only loosely bound and may be dragged along
the bipolar outflows that most models develop. Hence, the formation of
r-process nuclei is an interesting possibility that is currently under
investigation \citep{Reichert_etal_2020}. We cautiously suggest that
some GRB precursor activity might be observed in connection to the
accretion of these high-$j$ layers in BH forming cases. They may be
accreted a few seconds before the definitive accretion disc forms (and
hence fuels the outflow).  The aforementioned transport does not slow
down the PNS globally, because it does not reach the surrounding
region, and the angular momentum remains in the envelope of the PNS.
As a consequence, these layers expand, raising the axis ratio of the
PNS.

\paragraph*{Hypermassive NSs.} Temporarily stable PNSs of high mass
are formed by all our models. We find it significant that the $\betaG$
ratio (Eq.\,\eqref{eq:betaG}) maintains values smaller than $\sim 3\%$
for the longest run models. These low values of $\betaG$ may not allow
for the development of dynamical instabilities inducing collapse to
BH. The 3D models we have run appear to confirm these small values of
$\betaG$, albeit, so far, on a relatively short period of evolution
after stellar core collapse.  We find that PMCs systematically posses
smaller values of $\betaG$ than PCs, which enhances their prospects to
not collapse to BHs relatively soon after bounce, since PMCs may be
less perturbed by the ``low-$\betaG$'' instability.
As established in \citetalias{Obergaulinger_Aloy__2019}, and confirmed
here with much longer evolutions post-bounce, the mass of the PNSs
formed by our models are of the order of or larger than the masses of
the iron core of the respective pre-SN model. This means that in case
these PNSs do not collapse further to a BH (due to, \eg prolonged
episodes of late fall-back accretion on time scales of hours), very
heavy neutron stars (with masses
$1.85\,\Msol \lesssim M \lesssim 2.5\,\Msol$) may result. These
masses, especially the ones closer to the maximum mass allowed by the
EoS in the absence of rotation are only marginally consistent with the
current observational limits \citep{Ozel_Freire_2016ARAA..54..401},
and, in any case, they would belong to the $\sim 20\%$ fraction of the
population of \emph{massive} NSs \citep[with
$M>1.8\Msol$;][]{Antoniadis_2016arXiv160501665}. We would need to
compute a much longer time evolution (of minutes to hours) in order to
ascertain the kind of NS that will finally develop from our models.
The reason is that we cannot reliably know what fraction of the outer
layers will be blown away by the very aspherical SN explosions that
our models trigger and, consistently, which fraction may be accreted
onto the PNS. Besides, not all the matter hitting the PNS surface will
finally end up adding to its mass and angular momentum. During the
computed time of evolution, a large fraction of it drifts from low to
high latitudes to be incorporated to the outflow ejecta. Furthermore,
the non-monotonic increase in the mass and other properties of the PNS
(\eg their magnetic energy and angular momentum as well as the surface
magnetic fields) makes it difficult to extrapolate the properties of
our massive PNSs on time scales of hours. We note that this situation
departs significantly from the evolution of non-rotating, unmagnetised
cores, where relatively simple prescriptions for the mass evolution
can be given and, hence, a solid extrapolation of the properties of NS
at birth can be done \citep[\eg][for a recent
example]{Woosley_2020ApJ...896...56}. It seems, however, difficult
that the accretion of a few $0.01\Msol$ may bury the magnetic fields
already built at the end of their computed evolution, with values
$b^{\rm pol}_{\rm surf}$ larger than $\zehn{14}\,$G in our models with
long lasting PNSs \citep[cautiously extrapolating the results
of][]{Torres-Forne_et_al__2016__mnras}. Thus, magnetar field strengths
are expected in our high-mass PMCs. The rotational period is difficult
to predict as a result of the alternation of spin-down and spin up
periods. By the end of our most evolved models (more than $7.5\,\sek$
after core bounce), surface periods of $\sim 1.5 - 4\,$ms and polar
(equatorial) radii of $\sim 14-17\,$km ($\sim 20-30\,$km) are
observed. Interestingly, these radii, which trace the location of the
PNS neutrinosphere, tend to be larger than the radius of an equivalent
spherical and homogeneous configuration with the same mass and moment
of inertia than the PNS, with typical values
$R_{\pnss,1}\simeq 12 - 14\,$km at the end of the computed
evolution. The difference between these two radii is accounted for by
the layer of high specific angular momentum and relatively low density
surrounding the PNS that forms as a result of angular momentum
transport (see above). Thus, we find it difficult to include the
outcome of our PMCs within existing NS categories, but we tentatively
classify them as \emph{super-magnetars}
\citep[\eg][]{Rea_2015ApJ...813...92}.
Because of the uncertainties in the mass estimations obtained in
compact binary millisecond pulsars (black-widows and redbacks), they
have not been included in most global studies of the NS mass
distribution \citep{Linares_2019arXiv191009572}. However,
\emph{supermassive} NSs, with more than $2\Msol$ have been found in
these systems (\eg $2.3\,\Msol$ in PSR J2215+5135,
\citealt{Linares_etal_2018ApJ...859...54}; or $2.4\,\Msol$ in PSR
B1957+20, \citealt{vanKerkwijk_etal_2011ApJ...728...95}). Although our
progenitors are single stars (not binaries), our results suggest a
channel to produce supermassive NS also from isolated progenitors.

\paragraph*{Magnetic field amplification.} The strongly differentially
rotating cores fulfil the criterion for the MRI in wide regions, both
inside and outside the \nusp. We are able to identify episodic growth
of the MRI in various BH forming models.  For instance, we were able
to resolve the growth of the instability in the simulation of core
\modelname{35OC} with the original rotational profile and magnetic
field.  MRI channel modes grow in a single episode at about 150\,ms
post-bounce, most prominently just inside the PNS surface, yielding an
increase of the energy of the poloidal field by a factor of a few. In
contrast, NS forming models obtain their large magnetic fields chiefly
as a result of the vigorous convection in their fast rotating PNSs,
not because of MRI. This is because the Rossby number inside the PNS
is smaller than 1 in large regions.  Contrasting with the very
moderate growth that convection produces in non-rotating, magnetised
models, in our fast rotating models it yields large amplification
factors $\sim 1000$, even larger than the theoretical expectations of
\cite{Duncan_Thompson__1992__ApJL__Magnetars}.
  
\paragraph*{Spin of the formed BHs.} In general, all PNS possess
fairly rapid rotation.  The dimensionless spin parameter is
$\sim 0.3-0.5$ for models with the rotational profile from stellar
evolution.  These values turn out to be very similar to the formal
values computed from the mass and angular momentum of the inner
$3\Msol$ of the progenitor
\citep{Woosley_Heger__2006__apj__TheProgenitorStarsofGamma-RayBursts}. The
similarity between our measured $a_\pnss$ and the formal values
happens in spite of the complex accretion/ejection dynamics, which
imprints a non-monotonic evolution of $a_\pnss$. Furthermore, models
with relatively mild increases in the magnetic field strength (in
particular \modl{35OC-Rp2}) yield significantly smaller values of
$a_\pnss$ and no BH results in these cases. With the aforementioned
values of the spin parameter, and noting that the strength of the
magnetic field at the PNS surface in PCs is $\sim \zehnh{7}{14}\,$G
the \emph{initial} Blandford-Znajek luminosity of a potential long GRB
collapsar will be rather mild
$\sim
\zehnh{8}{49}(a_\pnss/0.4)^2(M_\textsc{bh}/3\Msol)^2(B/\zehnh{7}{14}\,\text{G})^2
\,\text{erg}\,\sek^{-1}$ \citep[using the estimates of][and assuming
that the spin of the BH coincides with the spin of the PNS at the
brink of collapse]{Mahlmann_2018MNRAS.477.3927}. This relatively small
value may be increased as the BH mass and spin increase due to the
ongoing accretion.

\paragraph*{Formation of collapsars.} Among the models that undergo BH
collapse, highly anisotropic explosions allow for continuing accretion
increasing the PNS mass beyond the instability threshold.  Collapse
occurs after more than a second post-bounce.  However, the exact time
when this happens is sensitive to the choice of equation of state
\citep[e.g.][and references
therein]{OConnor_Ott__2011__apj__BlackHoleFormationinFailingCore-CollapseSupernovae,Fischer_2011ApJS..194...39,Aloy_etal_2019MNRAS.484.4980}
and, in our case, to the approximate treatment of the general
relativistic gravitational field. The ram-pressure of the explosion
ejecta makes the fall of the stellar layers outside the central core
happen on time scales longer than the simple estimate for the disc
formation time as twice the free-fall time (Eq.\,\ref{eq:tdf};
$t_\textsc{df}\sim 9.3\,\sek$ for the \modelname{35OC} core). The
practical consequence of this fact is that the mildly relativistic,
collimated SN ejecta may break out of the stellar surface sooner than
or at about the same time as the accretion disc forms around the
central BH. Extrapolating their results,
\cite{Dessart_et_al__2008__apjl__TheProto-NeutronStarPhaseoftheCollapsarModelandtheRoutetoLong-SoftGamma-RayBurstsandHypernovae}
suggest that these ejecta may yield a weak precursor polar jet, which
may soon be overtaken by a baryon-free, collimated relativistic jet
\citep[see also][]{Aloy_etal_2018MNRAS.478.3576}. We basically agree
with that forecast in case of BH forming models, but in case of PMCs
the scenario may be different (see below).
Since our models that may potentially form a collapsar are computed to
the brink of BH formation, we cannot give a precise time after core
bounce when collapsar formation may take place and, strictly speaking,
whether a collapsar (understood as a BH girded by a suitable accretion
disc) may form. The ongoing, quite energetic explosion (see
\citetalias{Obergaulinger_Aloy__2019}) makes it difficult to estimate
the amount of mass that may be available for accretion in the mid term
and, indeed, whether an accretion disc with the properties required by
typical collapsar models
\citep[\eg][]{MacFadyen__2001__apj__Supernovae_Jets_and_Collapsars}
may form at all. Our results suggest that it may not be strictly
necessary to form an accretion disc in order to produce an
ultrarelativistic jet. Once the BH is formed, a fraction of its
rotational energy may be extracted by means of the Blandford-Znajek
mechanism with luminosities broadly compatible with those of long GRBs
(see above). There is no need of forming an accretion disc if
down-flows keep going on along the equatorial regions (extrapolating
the conditions at the brink of BH collapse in our models), even if the
magnetic field is relatively disordered and clumps into relatively
small scale structures \citep[see,
e.g.][]{Mahlmann_2020MNRAS.494.4203}. If this possibility could
materialise, it would open the prospects for a number of other stellar
evolution models to be considered as potential progenitors of long
GRBs.

The kind of collapsars that our models may form cannot be classified
in any of the types defined by
\cite{MacFadyen__2001__apj__Supernovae_Jets_and_Collapsars}. The
models that produce BHs relatively promptly yield also successful SN
explosions, which disqualifies them as Type I collapsars. Besides, if
any of the models with supra-stellar magnetic fields would finally
yield a BH by late-time, fall-back accretion, still the
magneto-rotational explosions produced are very energetic (with
energies in the hypernova range; see
\citetalias{Obergaulinger_Aloy__2019}). This could be the case of
\modl{35OC-Rp3} (see above) or models with supra-stellar poloidal
field in between of \modl{35OC-RO} and \modelname{35OC-Rp2}. Assuming
that these models were able to form a collapsar (namely, by assembling
an accretion disc around the new born BH), the expected explosion
energy does not allow to classify them as Type II collapsars. The
argument of
\cite{MacFadyen__2001__apj__Supernovae_Jets_and_Collapsars}, according
to which more massive helium cores may fail to eject all matter
outside the neutron star does not apply here. Even if it is true that
the gravitational binding energy of the helium core increases with
mass roughly quadratically, our magnetorotational explosions produce
very collimated ejecta, which prevent their failure (as it would
likely be the case under more isotropic explosion types). Given the
different explosion properties of our models which do not collapse
promptly to BH, we suggest a possible third type of collapsars
(\emph{Type III}) produced in the remnant of magnetorotational
explosions, tens of seconds after core collapse. In these models the
progenitor envelope may be exploded by a combination of a disk wind
and magnetorotational stresses, resulting in a hypernova-like SN with
potentially large luminosity if the amount of $^{56}$Ni mass produced
in the disk wind is large enough \citep[as suggested by
\eg][]{Nagataki_et_al__2007__apj__GRB_Jet_Formation_in_Collapsars,
  Dessart_et_al__2008__apjl__TheProto-NeutronStarPhaseoftheCollapsarModelandtheRoutetoLong-SoftGamma-RayBurstsandHypernovae}.

\paragraph*{Formation of protomagnetars.} Some of the models which do
not form a BH promptly seem promising for PM-driven SNe and GRBs.  The
pre-SN models that may originate PMCs combine the high rotational
energy available in the iron core and magnetic fields (a bit) stronger
than in the original stellar evolution model. PMCs posses a PNS
angular momentum significantly smaller
($J_\pnss \lesssim \zehnh{2}{49}\,$g\,cm$^2$\,s$^{-1}$) than BH
forming models.
In PMCs, BH collapse is prevented by very strong outflows that manage
to suppress mass accretion or even turn it into mass ejection, with a
significant loss of mass of the PNS.  A decreasing PNS mass has been
found in previous papers
\citep[\eg][]{Dessart_et_al__2008__apjl__TheProto-NeutronStarPhaseoftheCollapsarModelandtheRoutetoLong-SoftGamma-RayBurstsandHypernovae,
  Obergaulinger_Aloy__2017__mnras__Protomagnetarandblackholeformationinhigh-massstars},
but here, owing to the very long evolutionary times computed, we find
various episodic phases of mass decrease interleaved with a moderate
PNS mass growth. The angular momentum of the PNS parallels the
evolution of its mass, showing episodes of spin-down alternating with
spin up phases. This non-monotonic evolution shows that not only the
journey to BH formation is arduous in potential GRB progenitors
\citep{Dessart_et_al__2012__apj__TheArduousJourneytoBlackHoleFormationinPotentialGamma-RayBurstProgenitors},
but also the path to PM formation is tortuous.
In \modl{35OC-Rs} with a modified magnetic field of dipolar topology,
we also observe a decrease of the mass, of the rotational energy, and
of the angular momentum of the PNS during the first second
post-bounce. As in the case of models with supra-stellar poloidal
magnetic field, the energy-momentum and mass lost end up in the
jet-like outflows.  While those are not highly relativistic yet, we
consider this model also a potential PM central engine in its earliest
stage. As BH forming models, PMCs may launch magnetorotationally
powered jets. Initially, these jets are quite baryon loaded and only
mildly relativistic, but as time goes by the PM wind becomes
progressively more relativistic and baryon-free. Furthermore, the PM
wind has a predominantly radial geometry close to the PNS, which turns
into a paraboloidally shaped one after several hundred kilometres,
where it enters the highly collimated cavity blown by the ongoing SN
shock. Hence, it is not unlikely that there is a relatively smooth
transition from a mildly relativistic precursor ejecta to a
relativistic jet (in the spirit of the model of
\citetalias{Metzger_et_al__2011__mnras__Theprotomagnetarmodelforgamma-raybursts}). As
in the case of BH forming models, a disk wind may produce the required
amounts of $^{56}$Ni to accompany the GRB jet with a luminous SN
event.  While a detailed analysis of the nucleosynthesis of our models
will be the subject of a subsequent publication, we note that the
explosion energies, ejecta masses, and the thermodynamic conditions of
the ejecta are broadly compatible with the production of considerable
Ni masses.

In spite of the long time evolution span by our axisymmetric models
(nearly 10\,s), they have not fully entered the KH phase, since
accretion and convection are still on-going. This result is in strong
contrast with the standard assumptions that place the beginning of the
\emph{quasy-stationary} phase a few hundred milliseconds after core
collapse
\citep[\eg][]{Pons_1999ApJ...513..780,Hudepohl_et_al__2010__PRL__NeutrinoSignalofElectron-CaptureSupernovaefromCoreCollapsetoCooling},
and for models aiming to bridge from the post-bounce phase to the KH
phase \citep[\eg][]{Martinon_2014PhRvD..90f4026}. We cautiously note
that our results need further validation with full-fledged 3D models.

\paragraph*{PM spin-down.} Our models with supra-stellar poloidal
magnetic fields as well as \modl{35OC-Rs}, if run long enough, display
episodes of PNS spin-down. Confirming that we have not fully entered
the KH phase and that the evolution remains highly dynamic for nearly
10\,s after collapse (remarkably in the case of \modl{35OC-Rp3}), we
have not found a steady PM spin-down. However, it is reassuring that
many quantitative and qualitative facets of our results are similar to
the predictions of the PM model of
\citetalias{Metzger_et_al__2011__mnras__Theprotomagnetarmodelforgamma-raybursts}
during the episodic spin-down phases. It is, however, necessary to
consider even longer evolutions to sort out whether our most promising
PMCs may generate a PM after a longer evolution.  Whether or not the
evolution does indeed confirm this possibility depends on additional
factors eluding inclusion into an approximate model like the one of
\citetalias{Metzger_et_al__2011__mnras__Theprotomagnetarmodelforgamma-raybursts},
such as the geometry and efficiency of ejection of matter necessary
for evacuating the surroundings of the PNS and thereby reducing the
baryon-loading of the potential GRB jet.

The magnetic breaking of the compact remnant is linked to the
increased coupling between the magnetorotational evolution of the core
and the surrounding envelope. As could be expected, the core-envelope
coupling is more effective if large scale poloidal magnetic fields are
present in the pre-collapsing core. However, even without
\emph{artificial} large scale poloidal fields, the PNS spins down in
the long term by the ejection of magnetised winds from its surface. We
observe that during the periods of spin-down the Alfv\'en surface
moves a few hundred kilometres away from the rotational axis,
facilitating the transport of angular momentum towards the
surroundings of the PNS. Additional effects carrying away angular
momentum such as neutrino emission or magneto-rotational dynamical
instabilities (though in our axisymmetric simulations they may not be
relevant) carrying away angular momentum seem subdominant in our
models.  Spin down time scales as short as
$|\tau_{\rm rot}|= \Erotpns/|\dot{\Erot}^{\pnss}|\sim 2 - 3\,\sek$ or,
equivalently, $P/|\dot{P}|\sim 1-1.5\,\sek$ are typical during the the
late, more than $\sim 2\,\sek$ long spin-down phase of
\modl{35OC-Rp2}. These $P/|\dot{P}|$ time scales may not be maintained
for too long as they quickly rise to $\tau_{\rm rot}\lesssim 20\,\sek$
towards the end of the computed time for \modl{35OC-Rp2}. As with
other quantities, it is difficult to make a forecast for the value of
$P/|\dot{P}|$ at NS birth, but a (simplistic) extrapolation of our
results hints towards $P/|\dot{P}| \sim \text{few}\times 100\,\sek$ at
about $t\sim 10\,\sek$ post-bounce. According to the model of
\cite{Metzger_etal_2018ApJ...857...95}, with these spin-down time
scales and surface poloidal magnetic fields $\sim \zehn{15}\,$G, a
long GRB, but unlikely an ultra-long GRB \citep[with durations
$\gtrsim \zehn{3}\,\sek$;
\eg][]{Gendre_2013ApJ...766...30,Levan_2014ApJ...781...13}, can be
produced unless the evolution changes significantly. An order of
magnitude longer spin-down time scales,
$|\tau_{\rm rot}|\sim 10 - 40\,\sek$, and smaller surface magnetic
field strengths, $b^{\rm pol}_{\rm surf}\lesssim \zehn{14}\,$G, are
found in other potential PMCs differing by less than $50\%$ in the
initial poloidal field (\eg \modl{35OC-Rp3}).  This remarkable
variation of the spin-down time scale with the initial poloidal field
strength opens up the possibility that long GRBs with very different
durations may result from relatively small variations in the
properties of the pre-SN star. The episodic nature of the spin-down
periods computed in our models suggests that the subrelativistic
ejecta can be quite heterogeneous both in the radial and polar
directions. The propagation of a relativistic jet on these
heterogeneous environment is a source of variability that will be
blended with that imprinted by, \eg the development of instabilities
during the crossing of the stellar envelope
\citep{Aloy_et_al__2002__aap__Stability_analysis_of_relativistic_jets_from_collapsars_and_its_implications_on_the_short-term_variability_of_gamma-ray_bursts,
  Morsony_et_al__2010__apj__TheOriginandPropagationofVariabilityintheOutflowsofLong-durationGamma-rayBursts,
  Bromberg_Tchekhovskoy_2016MNRAS.456.1739,
  Aloy_etal_2018MNRAS.478.3576}.

As we have commented in \citetalias{Obergaulinger_Aloy__2019} of this
series, the main limitation of the present study is that most of our
models are axisymmetric.  The amplification of magnetic fields, the
dynamics of the explosion, and the development of several
instabilities can be quite different in three-dimensional geometry
(see \citetalias{Obergaulinger_Aloy__2019}).  In order to partly cross
check some of the conclusions drawn on the basis of axisymmetric
models, we have also presented preliminary results of
three-dimensional simulations with reduced grid resolution. These 3D
models show outflows that develop similarly to the axisymmetric
versions of the same models and thus seem to alleviate the concerns,
but some caution remains appropriate before drawing overarching
conclusions from the so far limited number of models.  Hence, our
efforts for improving upon this work should concentrate on simulating
models in full three-dimensional geometry, for which we are planning
to address selected issues in different stages of the evolution.

From our results, we can draw the conclusion that high-mass stars
offer a very wide range of potential post-collapse dynamics.
Furthermore, strong rotation favours the development of possible GRB
engines.  As a consequence, we consider most models promising
candidates for GRB engines in the long run after the end of our
simulations.  While our models do not cover the available parameter
space comprehensively, a collapsar scenario seems as viable as a PM
engine because the path to a PM is mostly dependant on relatively
small variations of the poloidal magnetic field in the pre-SN
core. The large variability observed in many of the variables after
collapse (singularly, the spin period, the mass and the rotational
energy) of the accreting PNS justifies our method for pushing as long
as possible the computed post-bounce evolution and encourages us
continuing pushing further in time our study in the future.

\section{Acknowledgements}
\label{Sek:Ackno}

This work has been supported by the Spanish Ministry of Science,
Education and Universities (PGC2018-095984-B-I00) and the Valencian
Community (PROMETEU/2019/071).  MO acknowledges support from the
European Research Council under grant EUROPIUM-667912, and from the
the Deutsche Forschungsgemeinschaft (DFG, German Research Foundation)
-- Projektnummer 279384907 -- SFB 1245 as well as from the Spanish
Ministry of Science via the Ram{\'o}n y Cajal programme (\miRyC).  We
furthermore thank for support from the COST Actions PHAROS CA16214 and
GWverse CA16104.  The computations were performed under grants
AECT-2016-1-0008, AECT-2016-2-0012, AECT-2016-3-0005,
AECT-2017-1-0013, AECT-2017-2-0006, and AECT-2017-3-0007,
AECT-2018-1-0010, AECT-2018-2-0003, AECT-2018-3-0010, and
AECT-2019-1-0009 of the Spanish Supercomputing Network on clusters
\textit{Pirineus} of the Consorci de Serveis Universitaris de
Catalunya (CSUC), \textit{Picasso} of the Universidad de M{\'a}laga,
and \textit{MareNostrum} of the Barcelona Supercomputing Centre,
respectively, on the clusters \textit{Tirant} and \textit{Lluisvives}
of the Servei d'Inform\`atica of the University of Valencia (financed
by the FEDER funds for Scientific Infrastructures; IDIFEDER-2018-063),
and under grant number 906 on cluster \textit{Lichtenberg} of the
Technical University of Darmstadt.

\vspace{-.5cm}
\section*{Data Availability}
The data underlying this article will be shared on reasonable request to the corresponding authors.
\vspace{-.5cm}

\bibliographystyle{mn2e}
\bibliography{bib2.bib}


\appendix

\section{Variance resulting from the magnetic field initial mapping}
\label{sek:Variance}
\begin{figure}
  \centering
    \begin{tikzpicture}
  \pgftext{
  \includegraphics[width=0.99\linewidth]{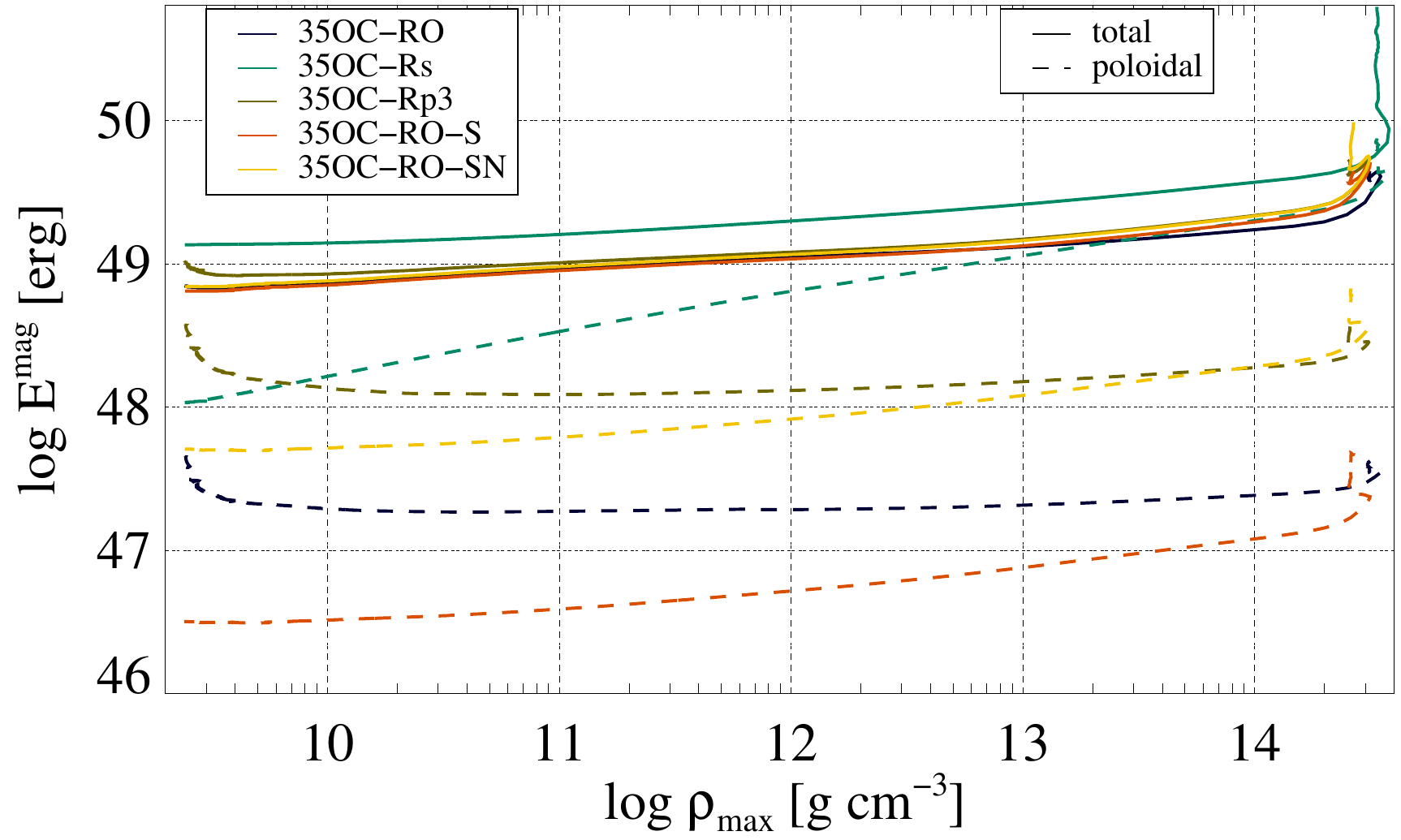}
         }%
    \node[fill=white, opacity=1, text opacity=1, rotate=90] at (-3.9,0.39) {\large$\text{ }\,\log\Emag\,\,\text{[erg]}$};	
  \end{tikzpicture}
  \caption{
    Evolution of  $\Emag$ and of  $\Emag^\mathrm{pol}$ (solid and dashed 
    lines, respectively) as a function of the maximum
    rest-mass density until core bounce in the variants of the
    \modl{35OC-RO}: \modelname{35OC-RO} with the standard mapping from
    the stellar evolution model to our 2D grid, \modelname{35OC-RO-S}
    with the smoothing of the poloidal magnetic field set by
    Eq.\,\eqref{Gl:Init-Heger-Br2} before the mapping to our grid, and
    \modelname{35OC-RO-SN}, which in addition to the smoothing
    includes a renormalisation of $\Emag^\mathrm{pol}$ to compensate the energy loss in
    the smoothing procedure. For comparison, \modls{35OC-Rp3} and
    \modelname{35OC-Rs} are also shown.
  }
  \label{Fig:35OC-variance}
\end{figure}

The initial grid of the stellar core \modelname{35OC}
(\modelname{35OB}) consists of 1053 (956) zones covering the range
$[0,R_\star]$, where $R_\star$ is the radius of the progenitor
star. We only map the inner $\sim \zehnh{1.2}{10}\,$cm into our grid,
i.e. only the inner 931 (867) zones of the stellar evolution model are
mapped into a grid of $n_r\times n_\theta=400\times 128$ zones (the
radial zones are not uniform; see Sect.\,\ref{Sek:Init}). As our
radial grid is coarser than that of the stellar evolution model,
direct interpolation dissipates part of the energy,
$\Emag^\mathrm{pol}$, stored in the smallest scales of the poloidal
magnetic field ($b^{\mathrm{pol}}_{\mathrm{pre}\textsc{sn}}$). This is
neither the case for the toroidal magnetic field
($b^{\mathrm{tor}}_{\rm preSN}$) nor for the rest of the physical
hydrodynamical variables. In the pre-SN core,
$b^{\mathrm{tor}}_{\mathrm{pre}\textsc{sn}}$ is much smoother than
$b^{\mathrm{pol}}_{\mathrm{pre}\textsc{sn}}$. The poloidal magnetic
field displays a large variability on small scales, close to the grid
resolution employed in the stellar evolution code. As a result a
sizeable fraction of $\Emag^\mathrm{pol}$ resides on scales below the
typical size of our numerical grid. This is a consequence of the fact
that the magnetic field in the stellar evolution of the models here
considered is not based on a consistent MHD modelling, which would
result in a smoother distribution of
$b^{\mathrm{pol}}_{\mathrm{pre}\textsc{sn}}$ due to the magnetic
solenoidal constraint.  In the following, we show how different
strategies to bridge from $b^{\mathrm{pol}}_{\mathrm{pre}\textsc{sn}}$
to the 2D axisymmetric grid introduce variegated paths in the
post-collapse evolution.

Our default procedure to obtain the initial magnetic field consists of
setting the $\phi$-component of the initial field as,
\begin{equation}
  \label{Gl:Init-Heger-Bphi}
  b^{\phi} = \beta_0^{\phi} b^{\mathrm{tor}}_{\mathrm{pre}\textsc{sn}},
\end{equation}
and computing the $r$-component from its poloidal component
\begin{equation}
  \label{Gl:Init-Heger-Br}
  b^{r} = \beta_0^{r} b^{\mathrm{pol}}_{\mathrm{pre}\textsc{sn}} \cos ( n^r
  \theta),
\end{equation}
where $\beta_0^{\mathrm{r/\phi}}$ and $n^r$ are dimensionless
parameters (normally, we set $\beta_0^{\mathrm{r/\phi}}=n^r=1)$.  The
$\theta$-component follows directly from the solenoidal condition.

The procedure to map the magnetic field from the 1D stellar evolution
models to our 2D (or 3D) computational grids is not unanimously
defined. In view of the fact that the small scale variability of
$b^{\mathrm{pol}}_{\rm preSN}$ is likely an artefact of the model to
include magnetic torques in stellar evolution, we have also considered
the possibility of smoothing the stellar evolution profile by taking
the running average over several neighbouring zones, \ie for the
radial magnetic field at each radial position, $r_i$, we consider
\begin{equation}
  \label{Gl:Init-Heger-Br2}
  b^\mathrm{pol}_{\mathrm{pre}\textsc{sn}, \mathrm{smth}}(r_i) = \frac{1}{2s+1}\sum_{j=i-s}^{i+s}b^{\mathrm{pol}}_{\mathrm{pre}\textsc{sn}}(r_j) \, ,
\end{equation}
with $s=10$. We compute $b^r$ using Eq.\,\eqref{Gl:Init-Heger-Br},
replacing $b^{\mathrm{pol}}_{\mathrm{pre}\textsc{sn}}$ by the smoothed
poloidal magnetic field profile,
$b^\mathrm{pol}_{\mathrm{pre}\textsc{sn}, \mathrm{smth}}$. With this
procedure we evolve a new setup dubbed \modelname{35OC-RO-S}.
\begin{figure}
  \centering
   \begin{tikzpicture}
  	\pgftext{
  	\includegraphics[width=0.99\linewidth]{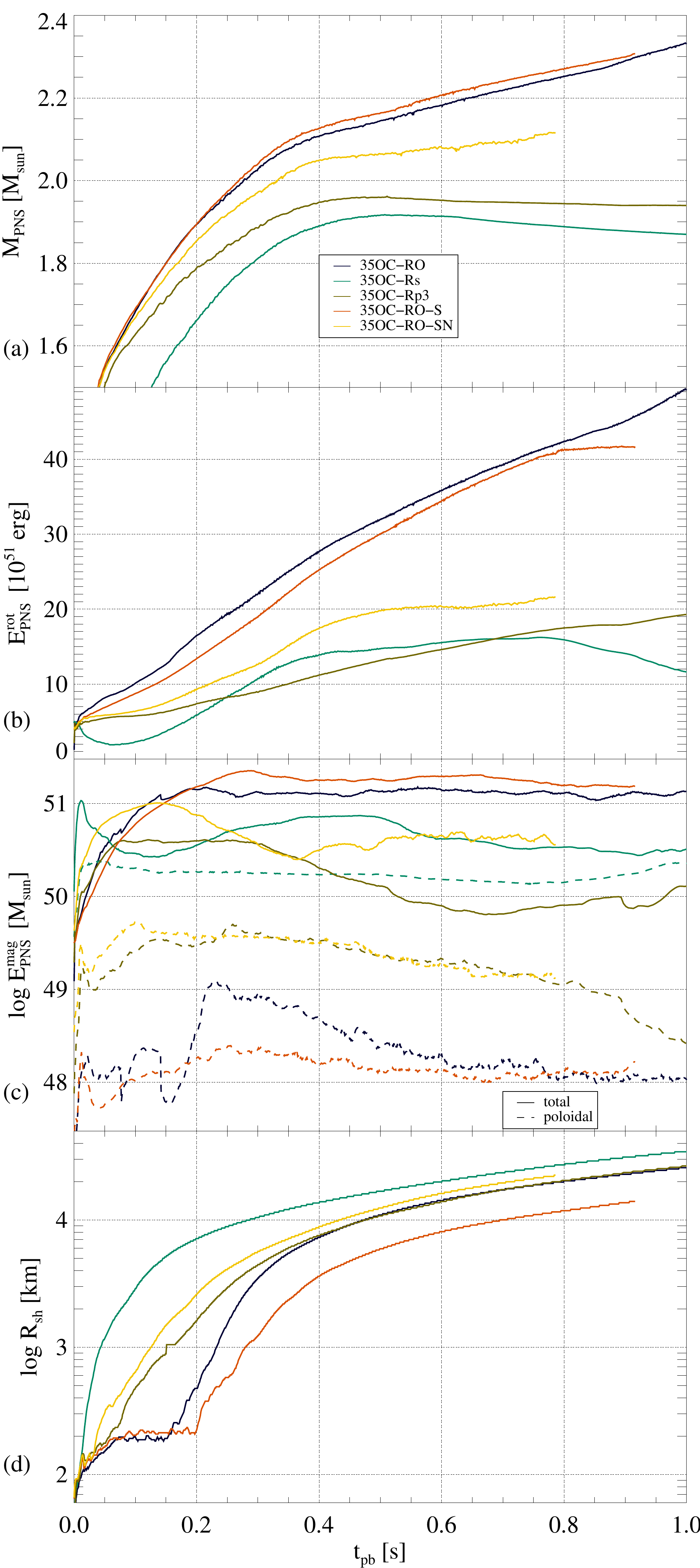}
	     }%
   \node[fill=white, opacity=1, text opacity=1, rotate=90] at (-3.95,-1.98) {\small $\text{ }\log\Emag^\pnss\,\text{[erg]}\text{ }$};
   \node[fill=white, opacity=1, text opacity=1, rotate=90] at (-3.95,2.50) {\small $\text{ }\Erot^\pnss\,\text{[}10^{51}\,\text{erg]}\text{ }$};
   \node[fill=white, opacity=1, text opacity=1, rotate=90] at (-4.03,6.94) {\small $\text{ }M_\pnss\,[\Msol]\text{ }$};
  \end{tikzpicture}
  \caption{
    Post-bounce evolution of different quantities of the PNS (from top to
    bottom: mass, rotational energy, and logarithm of the magnetic energy)
    as well as the shock radius for the same models as in
    \figref{Fig:35OC-variance}.
  }
  \label{Fig:35OC-variance2}
\end{figure}
The process of smoothing yields a loss of $\Emag^\mathrm{pol}$. In
order to avoid it, we renormalise the obtained results selecting
appropriately the factor $\beta_0^{\mathrm{r}}$, so that the initial
energy in the $b^\mathrm{pol}$ component equals the same quantity in
the stellar evolution model. In this way, we setup \modl{35OC-RO-SN}.

Figure\,\ref{Fig:35OC-variance} compares the evolution of the magnetic
energy as a function of the maximum density in the pre-bounce
phase. \Modl{35OC-RO} (blue dashed line) begins its evolution with the
same $\Emag^\mathrm{pol}$ as \modl{35OC-RO-SN} (yellow dashed line),
but after a quick initial readjustment phase, $\Emag^\mathrm{pol}$
levels off until the bounce takes place. At that time,
$\Emag^\mathrm{pol}$ is the same as in the model with the smoothed
poloidal magnetic profile (\modelname{35OC-RO-S}; red dashed line).
The total magnetic energy (dominated by the contribution of the
toroidal magnetic field) runs in parallel for all the variants of the
\modelname{35OC-RO} (yellow, blue and red solid lines). Thus, we
conclude that, during the collapse a significant fraction of
$\Emag^\mathrm{pol}$ is \emph{dissipated} in \modl{35OC-RO}, (note the
difference of a factor $\sim 16$ between the initial values of
$\Emag^\mathrm{pol}$ in \modls{35OC-RO} and \modelname{35OC-RO-S},
which disappears at the time of bounce). The dissipated energy
corresponds to the smallest scales mapped from the initial stellar
evolution model.

For comparison, we also display in \figref{Fig:35OC-variance}
\modl{35OC-Rp3}, which reaches nearly the same value of
$\Emag^\mathrm{pol}$ than \modl{35OC-RO-SN}, even having started with
an energy in the poloidal magnetic field component $\sim 10$ times
larger than the latter. The pre-bounce evolution of
$\Emag^\mathrm{pol}$ in \modl{35OC-Rp3} parallels (at a higher level
though) that of \modl{35OC-RO} and, hence, we also conclude that the
part of $\Emag^\mathrm{pol}$ of the former model stored in the
smallest scales has been dissipated as in the latter case.

Finally, \figref{Fig:35OC-variance} also illustrates the fact that the
much smoother poloidal magnetic structure of \modl{35OC-Rs} is more
efficiently amplified during collapse as that corresponding to models
with similar initial values of $\Emag^\mathrm{pol}$, but with (much)
more energy stored in smaller scales (\eg \modls{35OC-RO-Rp3},
\modelname{35OC-RO}). The growth of $\Emag^\mathrm{pol}$ in
\modl{35OC-Rs} (green dashed line) is even faster than in the smoothed
versions of \modl{35OC-RO}, whose growth until collapse is nearly
parallel (though starting from different initial values; compare
yellow and red dashed lines). Hence, we conclude that the dissipation
of $\Emag^\mathrm{pol}$ during collapse is closely connected to the
smoothness of the topology of
$b^\mathrm{pol}_{\mathrm{pre}\textsc{sn}}$.

The variegated evolutions of $\Emag^\mathrm{pol}$ resulting from
different initial mapping procedures of
$b^\mathrm{pol}_{\mathrm{pre}\textsc{sn}}$ onto our computational grid
yields also a significant variance in the post-bounce evolution, which
is illustrated in \figref{Fig:35OC-variance2}. There, we only consider
the first second post bounce and we can see that the PNS properties
are sensitively impacted by the initial mapping. While the model with
the smoothed initial profile follows an evolutionary path very close
to our default \modl{35OC-RO}, \modl{35OC-RO-SN} displays an smaller
PNS mass growth \panel{a}, and significantly smaller rotational
\panel{b} and magnetic \panel{c} energy. Not only the PNS properties
are modified, also the explosion properties, \eg the shock radius
evolution (\figref{Fig:35OC-variance2}\panel{d}). While the model with
initially smoothed poloidal magnetic field (red line) develops a
successful explosion later than \modl{35OC-RO}, the model with an
smoothed profile and renormalised initial $\Emag^\mathrm{pol}$ yields
and early magneto-rotational explosion akin to that of \mRpthree.

As a final note, the shown pre- and post-bounce evolution of the
variants of \modl{35OC-RO} with smoothing and with/without
renormalisation, fully justifies our choice of enhancing the poloidal
magnetic field component of the original \modelname{35OC} pre-SN core
and consider the evolution of \modls{35OC-Rp2}, \Rpthree and \Rpfour.


\bsp	
\label{lastpage}
\end{document}